\documentclass[a4paper,11pt]{report}

\usepackage[T1]{fontenc}

\usepackage{fullpage}

\usepackage{booktabs} 

\usepackage{amsmath}
\usepackage{amssymb}
\usepackage{amsfonts}

\usepackage{latexsym}
\usepackage{xurl}
\usepackage{listings}
\usepackage{xspace}
\usepackage[binary-units]{siunitx}

\usepackage{multirow}

\usepackage{cite}

\usepackage{imakeidx}

\usepackage{pgfplots}
\pgfplotsset{compat=1.17}

\usepackage{hyperref}

\makeindex

\newcommand{\Sec}{Section\@\xspace}

\newcommand{\Tab}{Table}

\newcommand{\marksec}{\protect{$^{\textstyle\star}$}}

\newcommand{\eg}{e.g.\@\xspace}
\newcommand{\ie}{i.e.\@\xspace}
\newcommand{\vs}{vs.\@\xspace}
\newcommand{\etc}{etc.\@\xspace}
\newcommand{\etal}{et al.\@\xspace}
\newcommand{\btw}{btw.\@\xspace}

\newcommand{\aseq}{\textsf{Seq}\xspace}
\newcommand{\apar}{\textsf{Par}\xspace}

\newcommand{\tseq}{T_{\mathsf{seq}}}
\newcommand{\tpar}[1]{T^{#1}_{\mathsf{par}}}
\newcommand{\tinf}{T\infty}
\newcommand{\wpar}[1]{W^{#1}_{\mathsf{par}}}

\newcommand{\SU}[2]{\mathrm{SU}_{#1}(#2)}
\newcommand{\SUR}[2]{\mathrm{SUR}_{#1}(#2)}
\newcommand{\EFF}[2]{\mathrm{E}_{#1}(#2)}
\newcommand{\parco}{Parallel Computing\index{Parallel Computing}\xspace}
\newcommand{\disco}{Distributed Computing\index{Distributed Computing}\xspace}
\newcommand{\conco}{Concurrent Computing\index{Concurrent Computing}\xspace}

\newcommand{\pthreads}{\texttt{pthreads}\xspace}
\newcommand{\openmp}{\textrm{OpenMP}\xspace}
\newcommand{\mpi}{\textrm{MPI}\xspace}
\newcommand{\cilk}{\textrm{Cilk}\xspace}

\newcommand{\gcc}{\texttt{gcc}\xspace}

\newcommand{\ompschedule}{\texttt{OMP\_\-SCHEDULE}\xspace}
\newcommand{\ompnumthreads}{\texttt{OMP\_\-NUM\_\-THREADS}\xspace}

\newcommand{\ompwtime}{\texttt{omp\_\-get\_\-wtime()}\xspace}

\newcommand{\cilkspawn}{\texttt{cilk\_\-spawn}\xspace}
\newcommand{\cilksync}{\texttt{cilk\_\-sync}\xspace}
\newcommand{\cilkfor}{\texttt{cilk\_\-for}\xspace}

\newcommand{\rank}{\mathtt{rank}}

\newcommand{\ceiling}[1]{\lceil #1\rceil}
\newcommand{\floor}[1]{\lfloor #1\rfloor}
\newcommand{\diam}{\mathrm{diam}}
\newcommand{\degree}{\mathrm{degree}}
\newcommand{\dist}{\mathrm{dist}}
\newcommand{\bisec}{\mathrm{bisec}}

\newtheorem{definition}{Definition}
\newtheorem{theorem}{Theorem}
\newtheorem{proposition}{Proposition}
\newtheorem{lemma}{Lemma}

\newcommand{\mpicc}{\texttt{mpicc}\xspace}
\newcommand{\mpich}{\texttt{mpich}\xspace}
\newcommand{\mvapich}{\texttt{mvapich}\xspace}
\newcommand{\openmpi}{\textrm{OpenMPI}\xspace}

\newcommand{\mpisuccess}{\texttt{MPI\_\-SUCCESS}\xspace}
\newcommand{\mpiundefined}{\texttt{MPI\_\-UNDEFINED}\xspace}

\newcommand{\mpiident}{\texttt{MPI\_\-IDENT}\xspace}
\newcommand{\mpicongruent}{\texttt{MPI\_\-CONGRUENT}\xspace}
\newcommand{\mpisimilar}{\texttt{MPI\_\-SIMILAR}\xspace}
\newcommand{\mpiunequal}{\texttt{MPI\_\-UNEQUAL}\xspace}
\newcommand{\mpidistgraph}{\texttt{MPI\_\-DIST\_GRAPH}\xspace}
\newcommand{\mpicart}{\texttt{MPI\_\-CART}\xspace}

\newcommand{\mpicommworld}{\texttt{MPI\_\-COMM\_\-WORLD}\xspace}
\newcommand{\mpicommself}{\texttt{MPI\_\-COMM\_\-SELF}\xspace}
\newcommand{\mpicommnull}{\texttt{MPI\_\-COMM\_\-NULL}\xspace}
\newcommand{\mpistatusignore}{\texttt{MPI\_\-STATUS\_\-IGNORE}\xspace}
\newcommand{\mpistatusesignore}{\texttt{MPI\_\-STATUSES\_\-IGNORE}\xspace}
\newcommand{\mpiinfonull}{\texttt{MPI\_\-INFO\_\-NULL}\xspace}
\newcommand{\mpiprocnull}{\texttt{MPI\_\-PROC\_\-NULL}\xspace}

\newcommand{\mpimaxversion}{\texttt{MPI\_\-MAX\_\-LIBRARY\_\-VERSION\_\-STRING}\xspace}
\newcommand{\mpimaxprocessor}{\texttt{MPI\_\-MAX\_\-PROCESSOR\_\-NAME}\xspace}

\newcommand{\mpicomm}{\texttt{MPI\_\-Comm}\xspace}
\newcommand{\mpidatatype}{\texttt{MPI\_\-Datatype}\xspace}
\newcommand{\mpiwin}{\texttt{MPI\_\-Win}\xspace}
\newcommand{\mpigroup}{\texttt{MPI\_\-Group}\xspace}
\newcommand{\mpirequest}{\texttt{MPI\_\-Request}\xspace}
\newcommand{\mpistatus}{\texttt{MPI\_\-Status}\xspace}
\newcommand{\mpiop}{\texttt{MPI\_\-Op}\xspace}
\newcommand{\mpierrhandler}{\texttt{MPI\_\-Errhandler}\xspace}
\newcommand{\mpiinfo}{\texttt{MPI\_\-Info}\xspace}

\newcommand{\mpisource}{\texttt{MPI\_\-SOURCE}\xspace}
\newcommand{\mpitag}{\texttt{MPI\_\-TAG}\xspace}
\newcommand{\mpierror}{\texttt{MPI\_\-ERROR}\xspace}

\newcommand{\mpitagub}{\texttt{MPI\_\-TAG\_\-UB}\xspace}

\newcommand{\mpiaint}{\texttt{MPI\_\-Aint}\xspace}

\newcommand{\mpibyte}{\texttt{MPI\_\-BYTE}\xspace}
\newcommand{\mpichar}{\texttt{MPI\_\-CHAR}\xspace}
\newcommand{\mpishort}{\texttt{MPI\_\-SHORT}\xspace}
\newcommand{\mpiint}{\texttt{MPI\_\-INT}\xspace}
\newcommand{\mpilong}{\texttt{MPI\_\-LONG}\xspace}
\newcommand{\mpifloat}{\texttt{MPI\_\-FLOAT}\xspace}
\newcommand{\mpidouble}{\texttt{MPI\_\-DOUBLE}\xspace}

\newcommand{\mpipacked}{\texttt{MPI\_\-PACKED}\xspace}

\newcommand{\mpiinit}{\texttt{MPI\_\-Init}\xspace}
\newcommand{\mpiinitialized}{\texttt{MPI\_\-Initialized}\xspace}
\newcommand{\mpifinalize}{\texttt{MPI\_\-Finalize}\xspace}
\newcommand{\mpifinalized}{\texttt{MPI\_\-Finalized}\xspace}
\newcommand{\mpiabort}{\texttt{MPI\_\-Abort}\xspace}

\newcommand{\mpirank}{\texttt{MPI\_\-Comm\_\-rank}\xspace}
\newcommand{\mpisize}{\texttt{MPI\_\-Comm\_\-size}\xspace}

\newcommand{\mpiwtime}{\texttt{MPI\_\-Wtime}\xspace}

\newcommand{\mpicommdup}{\texttt{MPI\_\-Comm\_\-dup}\xspace}
\newcommand{\mpicommsplit}{\texttt{MPI\_\-Comm\_\-split}\xspace}
\newcommand{\mpicommsplittype}{\texttt{MPI\_\-Comm\_\-split\_\-type}\xspace}
\newcommand{\mpicommcreate}{\texttt{MPI\_\-Comm\_\-create}\xspace}
\newcommand{\mpicommcreategroup}{\texttt{MPI\_\-Comm\_\-create\_\-group}\xspace}
\newcommand{\mpicommgroup}{\texttt{MPI\_\-Comm\_\-group}\xspace}
\newcommand{\mpicommfree}{\texttt{MPI\_\-Comm\_\-free}\xspace}

\newcommand{\mpicommcompare}{\texttt{MPI\_\-Comm\_\-compare}\xspace}
\newcommand{\mpigroupcompare}{\texttt{MPI\_\-Group\_\-compare}\xspace}

\newcommand{\mpigetcount}{\texttt{MPI\_\-Get\_\-count}\xspace}
\newcommand{\mpigetelements}{\texttt{MPI\_\-Get\_\-elements}\xspace}

\newcommand{\mpicartcreate}{\texttt{MPI\_\-Cart\_\-create}\xspace}
\newcommand{\mpicartrank}{\texttt{MPI\_\-Cart\_\-rank}\xspace}
\newcommand{\mpicartcoords}{\texttt{MPI\_\-Cart\_\-coords}\xspace}
\newcommand{\mpicartshift}{\texttt{MPI\_\-Cart\_\-shift}\xspace}
\newcommand{\mpicartget}{\texttt{MPI\_\-Cart\_\-get}\xspace}
\newcommand{\mpitopotest}{\texttt{MPI\_\-Topo\_\-test}\xspace}
\newcommand{\mpidimscreate}{\texttt{MPI\_\-Dims\_\-create}\xspace}

\newcommand{\mpidistgraphcreate}{\texttt{MPI\_\-Dist\_\-graph\_\-create}\xspace}

\newcommand{\mpisend}{\texttt{MPI\_\-Send}\xspace}
\newcommand{\mpirecv}{\texttt{MPI\_\-Recv}\xspace}
\newcommand{\mpisendrecv}{\texttt{MPI\_\-Sendrecv}\xspace}
\newcommand{\mpisendrecvreplace}{\texttt{MPI\_\-Sendrecv\_\-replace}\xspace}
\newcommand{\mpiisend}{\texttt{MPI\_\-Isend}\xspace}
\newcommand{\mpiirecv}{\texttt{MPI\_\-Irecv}\xspace}
\newcommand{\mpiisendrecv}{\texttt{MPI\_\-Isendrecv}\xspace}
\newcommand{\mpisendinit}{\texttt{MPI\_\-Send\_\-init}\xspace}
\newcommand{\mpirecvinit}{\texttt{MPI\_\-Recv\_\-init}\xspace}

\newcommand{\mpiprobe}{\texttt{MPI\_\-Probe}\xspace}
\newcommand{\mpiiprobe}{\texttt{MPI\_\-Iprobe}\xspace}

\newcommand{\mpianysource}{\texttt{MPI\_\-ANY\_\-SOURCE}\xspace}
\newcommand{\mpianytag}{\texttt{MPI\_\-ANY\_\-TAG}\xspace}

\newcommand{\mpiwincreate}{\texttt{MPI\_\-Win\_\-create}\xspace}
\newcommand{\mpiwinfree}{\texttt{MPI\_\-Win\_\-free}\xspace}
\newcommand{\mpiget}{\texttt{MPI\_\-Get}\xspace}
\newcommand{\mpiput}{\texttt{MPI\_\-Put}\xspace}
\newcommand{\mpiaccumulate}{\texttt{MPI\_\-Accumulate}\xspace}
\newcommand{\mpigetaccumulate}{\texttt{MPI\_\-Get\_accumulate}\xspace}
\newcommand{\mpiwinfence}{\texttt{MPI\_\-Win\_\-fence}\xspace}
\newcommand{\mpiwinpost}{\texttt{MPI\_\-Win\_\-post}\xspace}
\newcommand{\mpiwinstart}{\texttt{MPI\_\-Win\_\-start}\xspace}
\newcommand{\mpiwinwait}{\texttt{MPI\_\-Win\_\-wait}\xspace}
\newcommand{\mpiwintest}{\texttt{MPI\_\-Win\_\-test}\xspace}
\newcommand{\mpiwincomplete}{\texttt{MPI\_\-Win\_\-complete}\xspace}
\newcommand{\mpiwinlock}{\texttt{MPI\_\-Win\_\-lock}\xspace}
\newcommand{\mpiwinunlock}{\texttt{MPI\_\-Win\_\-unlock}\xspace}

\newcommand{\mpilockexclusive}{\texttt{MPI\_\-LOCK\_\-EXCLUSIVE}\xspace}
\newcommand{\mpilockshared}{\texttt{MPI\_\-LOCK\_\-SHARED}\xspace}

\newcommand{\mpibarrier}{\texttt{MPI\_\-Barrier}\xspace}
\newcommand{\mpibcast}{\texttt{MPI\_\-Bcast}\xspace}
\newcommand{\mpigather}{\texttt{MPI\_\-Gather}\xspace}
\newcommand{\mpigatherv}{\texttt{MPI\_\-Gatherv}\xspace}
\newcommand{\mpiscatter}{\texttt{MPI\_\-Scatter}\xspace}
\newcommand{\mpiscatterv}{\texttt{MPI\_\-Scatterv}\xspace}
\newcommand{\mpiallgather}{\texttt{MPI\_\-Allgather}\xspace}
\newcommand{\mpiallgatherv}{\texttt{MPI\_\-Allgatherv}\xspace}
\newcommand{\mpialltoall}{\texttt{MPI\_\-Alltoall}\xspace}
\newcommand{\mpialltoallv}{\texttt{MPI\_\-Alltoallv}\xspace}
\newcommand{\mpialltoallw}{\texttt{MPI\_\-Alltoallw}\xspace}
\newcommand{\mpireduce}{\texttt{MPI\_\-Reduce}\xspace}
\newcommand{\mpiallreduce}{\texttt{MPI\_\-Allreduce}\xspace}
\newcommand{\mpireducescatterblock}{\texttt{MPI\_\-Reduce\_\-scatter\_\-block}\xspace}
\newcommand{\mpireducescatter}{\texttt{MPI\_\-Reduce\_\-scatter}\xspace}
\newcommand{\mpiscan}{\texttt{MPI\_\-Scan}\xspace}
\newcommand{\mpiexscan}{\texttt{MPI\_\-Exscan}\xspace}

\newcommand{\mpireducelocal}{\texttt{MPI\_\-Reduce\_\-local}\xspace}
\newcommand{\mpiopcommutative}{\texttt{MPI\_\-Op\_\-commutative}\xspace}

\newcommand{\mpiibcast}{\texttt{MPI\_\-Ibcast}\xspace}

\newcommand{\mpiinplace}{\texttt{MPI\_\-IN\_\-PLACE}\xspace}

\newcommand{\mpitypeextent}{\texttt{MPI\_\-Type\_\-get\_\-extent}\xspace}
\newcommand{\mpitypetrueextent}{\texttt{MPI\_\-Type\_\-get\_\-true\_\-extent}\xspace}

\newcommand{\mpitypevector}{\texttt{MPI\_\-Type\_\-vector}\xspace}

\newcommand{\mpityperesized}{\texttt{MPI\_\-Type\_\-create\_\-resized}\xspace}

\newcommand{\mpitypecommit}{\texttt{MPI\_\-Type\_\-commit}\xspace}

\newcommand{\mpisum}{\texttt{MPI\_\-SUM}\xspace}
\newcommand{\mpiprod}{\texttt{MPI\_\-PROD}\xspace}
\newcommand{\mpimax}{\texttt{MPI\_\-MAX}\xspace}
\newcommand{\mpimin}{\texttt{MPI\_\-MIN}\xspace}
\newcommand{\mpiland}{\texttt{MPI\_\-LAND}\xspace}
\newcommand{\mpilor}{\texttt{MPI\_\-LOR}\xspace}
\newcommand{\mpilxor}{\texttt{MPI\_\-LXOR}\xspace}
\newcommand{\mpiband}{\texttt{MPI\_\-BAND}\xspace}
\newcommand{\mpibor}{\texttt{MPI\_\-BOR}\xspace}
\newcommand{\mpibxor}{\texttt{MPI\_\-BXOR}\xspace}

\newcommand{\mpimaxloc}{\texttt{MPI\_\-MAXLOC}\xspace}
\newcommand{\mpiminloc}{\texttt{MPI\_\-MINLOC}\xspace}

\newcommand{\mpiinitthread}{\texttt{MPI\_\-Init\_\-thread}\xspace}

\newcommand{\mpithreadsingle}{\texttt{MPI\_\-THREAD\_\-SINGLE}\xspace}
\newcommand{\mpithreadfunneled}{\texttt{MPI\_\-THREAD\_\-FUNNELED}\xspace}
\newcommand{\mpithreadserialized}{\texttt{MPI\_\-THREAD\_\-SERIALIZED}\xspace}
\newcommand{\mpithreadmultiple}{\texttt{MPI\_\-THREAD\_\-MULTIPLE}\xspace}

\newcommand{\impi}[1]{\emph{#1}\index{#1}}
\newcommand{\impisee}[2]{\emph{#1}\index{#1|see{#2}}}

\lstdefinestyle{SnippetStyle}{
  language=C,
  frame=lines,
  morekeywords={par,true,false,dynamic,guided},
  basicstyle=\small\ttfamily
}

\begin{document}

\title{Lectures on Parallel Computing\thanks{This script is scheduled to eventually appear as Springer LNCS 14600.}}
\author{Jesper Larsson Tr\"aff\\
  TU Wien\\
  Faculty of Informatics\\
  Institute of Computer Engineering\\
  Research Group Parallel Computing 191-4\\
  Treitlstrasse 3, 5th Floor, 1040 Vienna, Austria}
\date{June 30th, 2024}
\maketitle


\chapter*{Foreword}

These lecture notes are designed to accompany an imaginary, virtual,
undergraduate, one or two semester course on fundamentals of \parco as
well as to serve as background and reference for graduate courses on
High-Performance Computing, parallel algorithms and shared-memory
multiprocessor programming. They introduce theoretical concepts and
tools for expressing, analyzing and judging parallel algorithms and,
in detail, cover the two most widely used concrete frameworks \openmp
and \mpi as well as the threading interface \pthreads for writing
parallel programs for either shared or distributed memory parallel
computers with emphasis on general concepts and principles. Code
examples are given in a C-like style and many are actual, correct C
code. The lecture notes deliberately do not cover GPU architectures
and GPU programming, but the general concerns, guidelines and
principles (time, work, cost, efficiency, scalability, memory
structure and bandwidth) will be just as relevant for efficiently
utilizing various GPU architectures. Likewise, the lecture notes focus
on deterministic algorithms only and do not use randomization.  Slides
or blackboard drawings are imagined to be worked out for the actual
lectures by the lecturer, so the lecture notes deliberately do not
provide such important visual aid: some is available from the author
on request. Also the student of this material will find it instructive
to take the time to understand concepts and algorithms
visually. The exercises can be used for self-study and as
inspiration for small implementation projects in \openmp and \mpi that
can and should accompany any serious course on \parco. The student
will benefit from actually implementing and carefully benchmarking the
suggested algorithms on the parallel computing system that may or
should be made available as part of such a \parco course. In class,
the exercises can be used as basis for hand-ins and small programming
projects for which sufficient, additional detail and precision should
be provided by the instructor.

\section*{Acknowledgments}

These lecture notes have grown out of a bachelor course given at TU
Wien, Austria, since 2011, and have benefitted much from comments,
often severe criticism, weariness and occasionally very good questions
and suggestions by the students who have taken (and had to take) this
course over the years. The starred material is usually not covered in
the actual lecture. The lecture notes themselves were written starting
from March 2020. The author sincerely thanks everyone who has
contributed in spirit and materially, perhaps unbeknownst to
themselves. In particular, Sascha Hunold has over the years
significantly influenced the shape of the \parco course and the
thinking and presentation in these lecture notes.  Leonhard Patoschka
has done an extremely careful proof-reading which much improved the
presentation in the last phase. Enjoyable discussions with Christian
Siebert and Thom Fr\"uhwirth have likewise been of value over the
years. The
responsibility for the selection and presentation of the material as
well as any mistakes, errors or omissions in these lecture notes is
solely the author's.

\begin{flushright}
  Jesper Larsson Tr\"aff\\
  TU Wien, March 2020 --- April 2024
\end{flushright}

\chapter*{Deutsches Vorwort}

Dieses Skriptum ist als Lesehilfe für die Folien und den Vortrag
der Bachelorvorlesung ``Parallel Computing'' an der TU Wien gedacht.
Wir versuchen, auf die besonders wichtige Punkte aufmerksam zu machen
und die jeweiligen Vorlesungseinheiten zusammenzufassen. Erg\"anzende
Textb\"ucher, die Material enthalten, das nicht in der Vorlesung
besprochen wird, sind das Buch von Rauber und R\"unger~\cite{RauberRunger13},
das Buch von Grama \etal~\cite{GramaKarypisKumarGupta03} sowie das Buch von
Schmidt \etal~\cite{SchmidtGonzalezHundtSchlarb18}.  Umgekehrt enth\"alt die
Vorlesung auch viel Material, das nicht in diesen B\"uchern zu
finden ist. Das Skriptum ist auf Englisch verfasst.

Die mit $\star$ markierten Abschnitte sind nicht Teil des Stoffes f\"ur
die Bachelorvorlesung.

\tableofcontents


\chapter{Introduction to Parallel Computing: Architectures and Models}

\index{Amdahl's Law|see{Law}}
\index{Moore's Law|see{Law}}
\index{graphics processing unit|see{GPU}}
\index{Message-Passing Interface|see{MPI}}
\index{Posix threads|see{pthreads}}
\index{Parallel Random Access Machine|see{PRAM}}
\index{Random Access Machine|see{RAM}}
\index{Partitioned Global Address Space|see{PGAS}}
\index{High-Performance Computing|see{HPC}}
\index{exclusive prefix sums|see{prefix sums}}
\index{inclusive prefix sums|see{prefix sums}}
\index{Bulk Synchronous Parallel|see{BSP}}
\index{Breadth-First Search|see{BFS}}
\index{Depth-First Search|see{DFS}}
\index{Communicating Sequential Processes|see{CSP}}
\index{Symmetric MultiProcessing|see{SMP}}

The first third of the lectures on \parco deals with fundamental facts
of \parco which is distinguished from \disco and \conco.
Computational problems can be creatively and constructively explored
and studied with the PRAM model and judged by comparing against the best
(known) sequential baselines.  This naturally leads to the fundamental
concepts of (parallel) work, time, work- and cost-optimality,
speed-up, efficiency, and various notions of scalability. These concepts
are help- and meaningful, theoretically as well as practically and
empirically. Parallelization patterns that can help both in design and
analysis of parallel algorithms and programs are described. As concrete
examples, parallel algorithms for important problems with easy
linear-time, sequential algorithms are discussed at some length.

\section{First block (1--2 lectures)}

Parallel computers, meaning computers and computer systems with more
than one processing element, each capable of executing a program and
collaborating with other processing elements, are everywhere. The
number of processing elements, in modern terminology often called a
\impi{processor-core} or just \emph{core}, range from a few (embedded
systems, mobile devices) to tens and hundreds (desktops, servers), to
thousands, ten-thousands, and even millions in the largest
High-Performance Computing (HPC)\index{HPC} systems (see
\url{http://www.top500.org} for some such systems).  Every computer
scientist has to be aware of this fact and know something about
\parco.

Despite being an active area of research and also of commercial
developments of actual parallel computer systems in the mid-80s to
mid-90s of the last century, parallel computing was largely absent
from main stream computer science during the 90s to early in the $2000$
years.  This has had and still has dire consequences. The area was
largely missing from university curricula (\eg, parallel algorithms,
programming and software development), leading to a lack of
knowledgeable experts and professionals and to quite frequent
rediscovery of already known results and techniques: It still makes
much sense to read books and technical papers from the 1980ties and
90ties.

\subsection{``Free Lunch'' and Moore's Law}

One reason for all this was the ``free lunch''
phenomenon~\cite{SutterLarus05}, also sometimes called
\emph{Moore's Law}\index{Law!Moore's Law}: The performance
of sequential
computers was observed (and projected) to increase exponentially, with
a doubling rate of 18 to 24 months. To many, this made the more modest
performance improvements by the use of more processing elements seem
uninteresting and (commercially) irrelevant. This popular version of
this ``Law'' held from the 70s until the early- to mid-$2000$ years, but is not
exactly what Gordon Moore actually speculated~\cite{Moore65}. Nevertheless,
the exponential increase in sequential computer performance made building
and selling parallel computers commercially tough. Many ambitious and
well-founded companies
folded in the early $1990$ies, and other companies changed their
strategies: HPC\index{HPC} was one niche
where some companies could survive. Conversely, ``Moore's
Law''\index{Law!Moore's Law} exerted enormous pressure on
processor manufacturers; also
this had consequences leading, for instance, to many fantastic and
fantastically useless HPC systems being built.

In the early $2000$ years (say, $2005$) the ``free lunch'' was largely over.
The performance of sequential processors has not increased as dramatically
since then, as has been documented by many popular studies (that may
deserve a closer look)\footnote{see, for instance~\url{https://www.karlrupp.net/2018/02/42-years-of-microprocessor-trend-data/}}.
A way out to continue increasing nominal and possibly achieved
performance is to employ
\emph{parallelism}\index{nominal processor performance}.

\subsection{Performance of Processors}

For now, we define \emph{nominal processor performance} strictly
processor-centrically as the maximum (best-case) number of operations
(of some type, often: \emph{FLoating point OPerations per Second},
\impi{FLOPS}) that can be carried out per unit of time (second) by the
processor\index{nominal processor performance}.  The performance of a
single processor-core\index{processor-core}
is calculated as the product of the clock
frequency, number of ``ticks'' (cycles) per second, usually measured in
\si{\giga\hertz}, and the number of instructions that the processor
can complete per clock cycle (FLOPs/cycle). The number of instructions
per clock cycle is determined by the processor architecture: Number of
pipelines, depth of pipelines, number of functional units, types of
instructions (fused
multiply-add, for instance, other complex instructions), super-scalar
capabilities, vectorization (\impi{SIMD}) capabilities,
\etc~\cite{BryantOHallaron15,HennessyPatterson17}. The nominal
processor performance\index{nominal processor performance} provides an
optimistic upper bound on the performance that can actually be
achieved by real-world applications by assuming that all capabilities
of the processor can be utilized during the execution of the
application. We note that the FLOPS abbreviation is ambiguous and
quite unfortunate: sometimes the number of FLOPs are meant, sometimes
the FLOPs/second.

Whether the nominal performance of a processor can be reached depends
on at least two factors.  First, whether the program being
executed contains operations in the right mix and with the right
dependencies to allow full utilization of the components and features
of the\index{processor-core} processor-core.
For instance, a program solving a graph problem
may use integers and therefore executes $0$ FLOPs.
It does not exploit any of the
floating point capabilities of the processor (likely a major part). A
fused multiply-add instruction (and the related parts of the
processor) may be good for matrix--vector multiplication, but not for
many other tasks.  Second, the memory system must be able to supply
the data needed to keep all parts of the processor busy fast enough.
This is often a or even the major reason for observed, ``poor'' performance.

The ratio between processor performance and memory access time has not
improved at the pace processor performance has improved (Moore's
Law). The main idea to narrow the gap has been the introduction of
(larger and larger, hierarchically organized)
caches~\cite{BryantOHallaron15,PattersonHennessy20}. Caches and the
memory system play an important role in \parco and later in these
lectures (see \Sec~\ref{sec:cachelocality} and onwards).

With current terminology, a \impi{processor} (CPU) consists of
multiple \emph{(processor-)cores}, also called
\impi{processing elements} (PE) or processing units (PU): These are
the entities that are capable of executing a program. What is now
called cores used to be called processors. A processor with a smaller
number of cores (a handful, \eg, \numlist{4;8;10;16;24;32;48;64} which
is typical of current server processors) is termed a
\impi{multi-core processor}, and a processor with a large number of cores a
\impi{many-core processor}. The distinction is blurry and mostly
connotative.  The prototypical example of the latter is the
\emph{graphics processing unit} (GPU\index{GPU}),
which will play almost no role in
these lectures. We will use only the term \emph{multi-core} (where
needed). The \emph{nominal performance} of a multi-core processor is
calculated by multiplying the nominal per-core performance with the
number of cores\index{nominal processor performance}. For
several reasons, also the nominal multi-core processor performance is a very
optimistic upper bound on the performance that applications can
actually reach.

To make matters more complicated, many modern
processor-cores\index{processor-core} are capable
of executing a small number of independent instruction streams
(programs, processes, threads) \emph{simultaneously}, typically two to four,
with the purpose of exploiting the core's various functional units associated
with the core more efficiently and possibly also to be able to hide memory
access latencies by switching between streams. Such techniques implemented
in hardware are called \impi{hardware multi-threading}, \impi{hyperthreading}
or \impi{simultaneous multi-threading} (SMT). To the application programmer,
they make the multi-core processor\index{multi-core processor}
look as if it had two (or four) times
the number of hardware cores. Hardware multi-threading effectively improves the
number of instructions per clock and is thus accounted for in the nominal
processor performance as calculated above.
Hardware multi-threading can sometimes
improve the measured performance on the order of 10\%, but certainly not
by the number of supported hardware threads. Therefore, hardware threads
are not counted as cores.

Some recommended text books to check up on computer systems and
computer architecture
are~\cite{BryantOHallaron15,HennessyPatterson17,PattersonHennessy20,Tanenbaum12:sco}
(regularly updated).

\subsection{Parallel \vs Distributed \vs Concurrent Computing}

The focus of \parco is on using parallel resources (processors,
pro\-ces\-sor-co\-res) \emph{efficiently} for solving given
\emph{computational (algorithmic) problems}\index{computational problem}.
Towards this end, \parco is concerned with \emph{algorithms},
their \emph{implementation} in suitable \emph{programming languages}
that realize more or less explicitly formulated \emph{programming
models}\index{programming model} capturing the essentials
for analyzing and reasoning about
programs, and the structure and capabilities of the underlying actual
or imagined \emph{computer architecture}. We judge efficiency in all
these respects, both theoretically and
practically/experimentally. \parco is thus theoretical, practical, and
experimental Computer Science and much broader
in scope than just \emph{parallel programming}, which will also be treated
in these lectures
with C and \pthreads\index{pthreads}
and \openmp\index{OpenMP} and \mpi\index{MPI} as concrete
examples.

\parco is intimately related to the disciplines of distributed and
concurrent computing, and
distinguishing is a matter of what we are interested in (our focus). In
these lectures we propose and use the following definitions.

\begin{definition}[Parallel Computing]\index{Parallel Computing}
  \label{def:parco}
  The discipline of efficiently utilizing dedicated parallel resources
  for solving given computational problems\index{computational problem}.
\end{definition}

The focus of \parco is on \emph{problem solving efficiency}, and a
fundamental assumption is that the full computer system is at our
disposal (dedicated). Interesting parallel computing problems are
those that require significant interaction (communication, be it via
memory reads/writes, or explicit communication over some
interconnection network) between the parallel resources (cores), on
systems that actually provide significant inter-communication and
processing capabilities. Therefore, real parallel computers are
commonly not thought of as spatially (widely) distributed (the
internet)~\cite{BilardiPreparata95}.

\parco is related to and can benefit from results in
\emph{distributed} and \emph{concurrent} computing, by which 
the following is meant (our definitions, others may disagree).

\begin{definition}[Distributed Computing]\index{Distributed Computing}
  \label{def:disco}
  The discipline of making independent, non-dedicated resources
  available to cooperate toward solving specified problem complexes.
\end{definition}

The focus of \disco is on availability of resources
that are not readily at hand, may be spatially widely distributed, may
change dynamically, and may \emph{fail}. In \parco, \index{processor-core}
processor-cores
\emph{do not fail} (at least not in this lecture!). Specific,
individual problems or larger problem complexes may be studied.  A
central tenet in \disco is that there is no centralized
control. Example: Acquiring resources from the cloud, subject to
certain constraints and requirements, may, on the one hand, be a \disco
problem. Using the resources as a (virtual) parallel machine for
solving the problem we are interested in efficiently (for instance,
within given time constraints) is, on the other hand, a \parco problem.
Example: Routing data through a (dynamically changing) network while
sustaining a high (guaranteed) throughput and low latency with no possibility
of deadlock or lost data
can be viewed as a \disco problem, solutions to which are
obviously relevant for \parco.

\begin{definition}[Concurrent Computing]\index{Concurrent Computing}
  \label{def:conco}
  The discipline of managing and reasoning about interacting processes
  that may or may not progress simultaneously.
\end{definition}

The focus of \conco is on \impi{concurrency}, activities
that may or may not happen at the same time, are usually not
centrally coordinated, and therefore, on reasoning about and
establishing correctness (in a broad sense) in such situations (\eg,
by process calculi~\cite{Hoare85,Milner88}). In contrast, \parco is
specifically concerned with bounds on the performance that can be also
practically achieved, and typically make much more and stronger
assumptions about progress and actual concurrency in the system.

\subsection{Sample Computational Problems}

Some computational problems\index{computational problem} that will be
considered and used as examples throughout these lectures are:
\begin{itemize}
\item
  Computing sums and maxima over objects stored in arrays,
\item
  matrix--vector multiplication,
  matrix--matrix multiplication\index{matrix--matrix multiplication},
\item
  merging\index{merging} of ordered sequences of numbers and objects,
\item
  sorting numbers or objects from ordered sets by merging, by
  counting, by Quicksort\index{Quicksort}, \ldots and other methods,
\item
  performing reductions over sets of numbers and objects with
  given associative operators,
\item
  computing prefix sums\index{prefix sums} over arrays, compacting arrays,
\item
  listing prime numbers,
\item
  performing stencil computations\index{stencil computation} on matrices,
  and
\item
  graph search problems (\eg, Breadth-First Search\index{BFS} or
  Depth-First Search\index{DFS}).
\end{itemize}

Such computational problems that can be precisely and quantitatively
defined are routinely considered and solved in algorithms
courses~\cite{CormenLeisersonRivestStein22}. Most of them, \eg, the
matrix-computations from
\emph{basic linear algebra}\index{linear algebra} and the stencil
computations, are clearly important enough
by themselves. Almost all of them are crucial as building blocks in
more complex algorithms, \eg, sorting, prefix sums and the graph
search\index{BFS}\index{DFS} problems.
More importantly, the solutions illustrate general
patterns, approaches, and techniques for analyzing and solving similar
problems. We define the problems more precisely as we deal with them.
Some of the problems exhibit \impi{regular} computational patterns, \eg, the
matrix problems, that may even be \impi{oblivious} to (independent of)
the actual input. Some of the problems have more \impi{irregular}
computational patterns
that depend on the specific input, \eg, some sorting algorithms and
many graph search\index{BFS}\index{DFS} algorithms.
For the matrix-problems we will here
consider only so-called \impi{dense} variants that are solved by
regular, oblivious\index{oblivious} algorithms.
In other words, we will not in any way take the
(algebraic) structure of the matrices and vectors into account
(triangular matrices, diagonal matrices, block matrices, matrices with
many zero and one elements, \ldots). Doing so and dealing with
\impi{sparse} matrix-problems is considerably more challenging, for
sequential as well as for parallel algorithmics; but can sometimes
lead to faster solution.

\subsection{Models for Sequential and Parallel Computing}

For designing and analyzing parallel algorithms, a suitable model of
computation is needed.  A good model is one which makes it possible to
derive interesting algorithms and results, makes analysis tractable,
and bears enough resemblance to actual machines and systems that the
algorithms can be implemented and results predictive of, say,
performance.

A model with the last property is sometimes called a
\impi{bridging model} (we use the term in this fashion), a term originally
introduced by Les Valiant~\cite{Valiant90:bridge,Valiant11} who
proposed a specific model as bridge for \parco, the so-called
\emph{Bulk Synchronous Parallel} (\impi{BSP}) model. A minimum
requirement for a good bridging model is that if some algorithm $A$ is
shown to perform better than algorithm $B$ in the model, then a
(faithful) implementation of $A$ should perform better than an
(equally faithful) implementation of $B$ on the real machine
(``bridging'').  The (vague) notion of \impi{performance portability}
is related to the bridging idea. It says that the good performance of a
program can be preserved when moving from one system to another. This
is clearly a desirable property.

While there are various ``bridging models'' in sequential computing
with the \impi{RAM}, \emph{Random Access Machine}\index{RAM}
being the most important one, although it is not unproblematic and has
many restrictions, the situation is completely different for \parco. There
are many different parallel computer architectures (multi-core CPU \vs
GPU; distributed memory system \vs shared memory system, \etc), at
vastly different scales, and no model (so far) bridges them all to any
useful extent. The BSP\index{BSP} model has so far not been successful
(in finding universal or even widespread use). Also,
model assumptions that are desirable for the design of algorithms do,
to an even lesser extent than for sequential models, hold for parallel
computer systems. Many such assumptions are related to the memory
behavior. For instance, the assumption of unit-time, uniform memory
access of the RAM\index{RAM} is already problematic for sequential
computers, and even more so for large parallel systems with widely
distributed memory.

\subsection{The PRAM Model}

One extremely useful, but unrealistic model of parallel computing is
the \impi{Parallel Random Access Machine} (\impi{PRAM})~\cite{JaJa92},
a natural generalization of the equally useful and pervasive,
sequential \emph{Random Access Machine} (\impi{RAM}). Like the RAM,
the PRAM assumes a large (as large as needed) memory where processors
(as many as needed) can read and write words (addresses) in unit
time. A more concrete PRAM, closer to physical reality, would have a
certain, given number of processors.  These processors all execute
their own program, but do so in \emph{lock-step}: strictly
synchronized, all following the same, global clock and performing an
instruction in each \emph{time step}.  This means that the machine is
always in a well-defined \emph{state} comprised of the program counter
of the processors, contents of the memory and the processor
registers. State transitions happen instantaneously by the synchronous
clock ticks, and reasoning with state invariants, \index{invariant} as
done in sequential RAM\index{RAM} algorithms, is a way to prove properties. A
PRAM algorithm specifies what the processors are to do in each step.

With many processors operating in lock-step, it can potentially happen
that more than one processor is accessing some memory word in the same
time step. The PRAM model needs to define what happens in such
cases. First, a memory word can, in a step, be either read or written;
but not read by some processor(s) and written by another. For
potentially \emph{concurrent accesses} to a memory word in a step by
two or more processors, there are three main variations of the PRAM
that have been used in the literature:
\begin{itemize}
\item
  An \emph{EREW} (Exclusive Read Exclusive Write)
  \emph{PRAM}\index{PRAM!EREW} disallows accesses to the same
  memory word in the same step by more than one processor. It is the
  algorithm designer's responsibility to make sure that simultaneous
  accesses do not happen.
\item
  A \emph{CREW} (Concurrent Read Exclusive Write)
  \emph{PRAM}\index{PRAM!CREW} allows simultaneous (parallel, concurrent)
  reads to any memory word by more than one processor in a time step, but not
  simultaneous writes to a memory word in a time step.
\item
  A \emph{CRCW} (Concurrent Read Concurrent Write)\index{PRAM!CRCW}
  \emph{PRAM} allows both simultaneous reads and simultaneous writes to
  the same memory word in the same step; but not reads and writes to one
  and the same word in the same time step (many or all processors may read
  the same word; many or all processors may write to the same word).
  What happens when two or more processors write to a
  word in a step? In a \emph{Common CRCW PRAM}\index{PRAM!Common CRCW},
  it must be ensured that the writing processors all write the
  same value. In an \emph{Arbitrary CRCW PRAM}\index{PRAM!Arbitrary CRCW},
  either of the written values will survive in the memory
  word. A \emph{Priority CRCW PRAM}\index{PRAM!Priority CRCW} has
  some priority associated with the processors, and the writing
  processor with the highest priority will successfully write its value
  to the memory word.
\end{itemize}

What happens in case the EREW/CREW/CRCW constraints are violated by
our algorithm is just a matter of model design: perhaps the machine
breaks down, explodes, halts, delivers incorrect results, or some
other outcome. The important requirement is that the algorithm
designer has to make sure (prove!) that the constraints of the PRAM
variant at hand are never violated when the algorithm is executed.
Per definition, any algorithm that can be executed correctly on an
EREW PRAM, can execute on any of the, in that sense, stronger variants.

The PRAM is largely a purely theoretical construct; there have been
several attempts to realize emulated PRAMs in real hardware, but so
far none have been entirely or commercially successful~\cite{AbolhassanDrefenstedtKellerPaulScheerer93,AbolhassanKellerPaul99,Traff00:prambook,Forsell02,Vishkin03}.
We use it as an analytical tool to precisely describe and analyze
(fast) parallel algorithms with high parallelism: many processors
relative to the size or computational demands of the problem to be
solved. We can therefore freely invent convenient pseudo-code to
liberally express algorithms, as long as it is clear that the PRAM
model assumptions are satisfied. The goal is to be able to
characterize time (number of parallel steps) and effort (number of
processors used in the parallel steps) of parallel computations. For
this, we allow to freely choose, for each parallel step, the number of
PRAM processors to be used in that step. This can be a fixed number
(sometimes just one), a function of the input size or a free
parameter. On a physical PRAM with some fixed number of processors,
the allocated (virtual) processors would be emulated by the available,
possibly fewer physical processors.

In order to be able to describe interesting algorithms more concretely,
we introduce a pseudo-code construct for starting a set of processors,
each being assigned an identity (some integer) to which it can
refer. This is the \texttt{par}-construct that looks similar to a
C pseudo-code \texttt{for}-loop. This construct allows us to declare
a range or set of processors to start working. We will assume that starting a
reasonably specified set of processors can be done even on an EREW
PRAM in a constant number of operations, $O(1)$ per processor. This is
reasonable for simple ranges where each processor identity can be
computed by simple arithmetic.  On a physical PRAM realized in
hardware, it would be the task of the run-time system and compiler to
provide constructs for starting or allocating well-defined sets of
(virtual) processors with some well-defined (small) overhead. In order
to fulfill the lock-step assumption, correct pseudo-code will make
sure that the allocated processors in a \texttt{par}-construct all
perform the exact same number of instructions. This means that open
\texttt{while}-loops where the number of iterations may be different
for different processor identities are not allowed. Also,
\texttt{if}-statements have to be written in such a way that both
branches will have the same number of instructions to execute; but we
will here just leave it to the (virtual) compiler to pad branches with
the needed no-op instructions to ensure this. If it is not obvious how
this can be done, the algorithm-code should be rewritten.

Using the analytic PRAM, we can now give interesting
algorithms for many of our computational problems\index{computational problem},
for instance, for finding the maximum among $n$ numbers, and for doing
matrix--matrix multiplication\index{matrix--matrix multiplication}
of $m\times l$ and $l\times n$ matrices
into an $m\times n$ result matrix.

Our first, non-obvious PRAM algorithm expressed in PRAM-pseudo code
for finding the maximum in a set of numbers (stored in an array) is
given below, and the results are summarized in the theorems that
follow.

\begin{lstlisting}[style=SnippetStyle]
par (0<=i<n) b[i] = true; // a[i] could be maximum
par (0<=i<n, 0<=j<n) {
  if (a[i]<a[j]) b[i] = false; // this a[i] is not maximum
}
par (0<=i<n) if (b[i]) x = a[i];
\end{lstlisting}

\begin{theorem}
  \label{alg:fastmax}
  The maximum of $n$ numbers stored in an array can be found in $O(1)$
  parallel time steps, using $n^2$ processors and performing $O(n^2)$
  total operations on a Common CRCW PRAM\index{PRAM!Common CRCW}.
\end{theorem}

In the program, the input is stored in the $n$-element array
\texttt{a}, indexed C-style from $0$ to $n-1$.  The idea of this
fastest possible algorithm is to do all the $n^2$ pairwise element 
comparisons in one parallel step (actually, the $n(n-1)$ comparisons
with different element indices would suffice), and use the outcome to
knock out the elements that cannot possibly be the maximum. This is
done with the Boolean array \texttt{b}, which is used to mark each of the
$n$ elements as a candidate for being a maximum. As outcome of the
pairwise comparisons, elements that cannot be maximum by virtue of being smaller
than some other element are unmarked by one or more of the $n^2$ assigned
processors.  The three \texttt{par}-constructs start $n$, $n^2$, and
$n$ processors, respectively, first for initializing the
\texttt{b}-array, second for performing all the $n^2$ comparisons in
parallel, and finally for writing out the maximum to the result variable
\texttt{x}.  Since in one step, several (up to $n$) processors can
discover that some element \texttt{a[i]} cannot be a maximum
since \texttt{a[i]<a[j]},
concurrent writing to the same \texttt{b[i]} can happen. Whether and
at which indices this happens is dependent on the input. When several
processors write to a location \texttt{b[i]} or \texttt{x} in a step,
they, however, write the same value (\texttt{false}, or the maximum
value, respectively), and therefore a Common CRCW
PRAM\index{PRAM!Common CRCW} suffices for this algorithm.  This
is an interesting, maximally fast (there is nothing faster than
constant time, and the constants here seem to be small) algorithm: The
PRAM model is good for exposing the maximum amount of parallelism in a
problem. The time taken by the algorithm is the number of parallel time
steps (here three), and the number of processors used is the maximum
number of processors assigned in a parallel step (here $n^2$).

Simultaneous, concurrent writing to the same memory location or memory
module is a (too?) powerful capability of a parallel computer, which
should presumably be avoided if possible.  In order to avoid
concurrent writing in the maximum finding problem, a different
algorithmic idea is needed: Instead of doing all pairwise comparisons
in a step, do only up to $n/2$ comparisons in parallel between disjoint
pairs of elements. The pseudo-code below implements this idea.

\begin{lstlisting}[style=SnippetStyle]
nn = n; // number of elements per while-loop iteration
while (nn>1) {
  k = (nn+1)>>1; // ceil(nn/2) by shift
  par (0<=i<k) {
    if (i+k<nn) a[i] = max(a[i],a[i+k]);
  }
  nn = k;
}
\end{lstlisting}

\begin{theorem}
  \label{alg:logmax}
  The maximum of $n$ numbers stored in an array can be found in $O(\log
  n)$ parallel time steps, using a maximum of $n/2$ processors (but
  performing only $O(n)$ operations) on a CREW PRAM\index{PRAM!CREW}.
\end{theorem}

This algorithm goes through $\ceiling{\log_2 n}$ iterations, in each
one roughly halving the number of element pairs to compare. Elements
are stored in the \texttt{a}-array which as can be seen is
destructively updated. The resulting maximum ends up in location
\texttt{a[0]}. In each iteration, the algorithm performs comparisons
between $\floor{n/2}$ pairs only, in each of which the larger element
is stored.  This reduces the number of possible maximum elements for
the next iteration to $\ceiling{n/2}$.  The comparison step needs to
be iterated $\ceiling{\log_2 n}$ times, after which a maximum element
is left in \texttt{a[0]}.  As written, the algorithm requires
concurrent reading, namely of \texttt{k} and \texttt{nn}, but it can
be modified to run also on an EREW PRAM\index{PRAM!EREW}, and it is
a good exercise to do so.

The last example turns the definition of
matrix--matrix multiplication\index{matrix--matrix multiplication},
into parallel PRAM code. The $m\times n$ matrix product $C$ of
$m\times l$ and $l\times n$ input matrices $A$ and $B$ is defined as
\begin{eqnarray*}
  C[i,j] = \sum_{k=0}^{l-1}A[i,k]B[k,j]
\end{eqnarray*}
for $0\leq i<m, 0\leq j<n$. Since we do not (yet) know how to compute
the sum of $l$ elements (the $l$ element products in the sum), this
part of the definition is implemented as a sequential loop, but all
$mn$ sums are computed in parallel as specified by the outer
\texttt{par}-construct.

\begin{lstlisting}[style=SnippetStyle]
par (0<=i<m, 0<=j<n) {
  C[i,j] = 0;
  for (k=0; k<l; k++) {
    C[i,j] += A[i,k]*B[k,j];
  }
}
\end{lstlisting}

\begin{theorem}
  \label{thm:mmslow}
  Two $m\times l$ and $l\times n$ matrices can be multiplied into an
  $m\times n$ matrix in $O(l)$ time steps and $O(mnl)$ operations on a
  CREW PRAM.
\end{theorem}

The algorithm shown can also be improved to run on an EREW PRAM by
using extra space for intermediate results. It can be made faster by
employing a variant of the maximum finding algorithm to do the
summations in parallel.

The complexity properties of the PRAM algorithms so far were stated in
terms of the total number of parallel steps required for the given input,
the maximum number of processors needed in some parallel
step, the total number of operations carried out by all allocated
processors during the course of execution, and the PRAM model required
by the algorithm. The natural goal when studying the parallel
complexity of specific problems is to minimize these
requirements on all counts: as few parallel steps, as few total
operations, and as weak a PRAM model as possible. As the observations
and theorems above show, some of these goals seem contradictory and
not achievable simultaneously.
A strong Common CRCW PRAM model\index{PRAM!Common CRCW} made it
possible to find the maximum of $n$ numbers optimally fast (constant
time), but at the additional cost of a large number of (redundant)
operations (Theorem~\ref{alg:fastmax}). An algorithm for the weaker,
possibly less expensive CREW PRAM using less operations and processors
was given; but it uses more time (parallel steps)
(Theorem~\ref{alg:logmax}).  We elaborate on these measures and
trade-offs which will be a main theme in the following parts of these lectures.

The PRAM model has been productive in finding highly parallel, fast
algorithms for many interesting problems and also in establishing
lower bounds on how fast and with how many resources (processors)
they can be solved~\cite{JaJa92}. Whether the algorithms studied
so far are good or useful, will be elaborated on in the following.

Other theoretical models for \parco that we may encounter but will not use
here include comparator networks, systolic arrays, cellular automata,
\ldots. The theoretician (and computer architect, but with the
constraints of the world we live in) is free to invent
models that serve the purpose: such models have been productive in
establishing important results on how to do and not to do things.

\subsection{Shared \vs Distributed Memory Models and Systems}

The PRAM model is an example of a \parco model with a shared memory
from and into which processors can freely read and write data and thus exchange
information with each other, subject only to the EREW, CREW or CRCW
constraints of the particular PRAM\index{PRAM}.  The PRAM model allows
us to formulate algorithms using as many processors as needed. All
processors can access all words of a common shared memory which is
also as large as needed. Access to memory always takes unit time,
namely a single clock cycle, and this is independent of which location
is being accessed by which processor. The PRAM is the most extreme
case of a \impisee{Uniform Memory Access}{UMA} (UMA\index{UMA})
model. Access times are uniformly the same. Furthermore, memory
operations are fine-grained and done in units of single words (Bytes,
integers, doubles, \ldots). In addition, the PRAM makes the strong
assumption that processors operate synchronously in lock-step.

Real, shared memory systems are quite far from all these PRAM model
assumptions. Memory access times are not on the order of a single or a
few clock cycles, but take much longer than instructions carried out
by the processor-cores. More importantly, access times are not
uniform.  Access to registers in the small register bank memory of the
processor can indeed be fast, accesses to data stored in cache
memories already slower (see \Sec~\ref{sec:cachelocality}),
accesses to data in ``main memory'' again slower and so on.  Memories,
especially ``main memory'' is often divided into ``banks'' with some
banks being ``closer'' to some processor-cores\index{processor-core}
than to other
processor-cores, and accesses to data in different banks can take 
different times for different processors. Memory can even be local to
processors in the sense that some form of explicit communication is
required for one processor to access memory that is controlled by
another processor.  These characteristics are loosely called
\impisee{Non-Uniform Memory Access}{NUMA} (NUMA)\index{NUMA}.  More
realistic computational models that capture aspects of these realities
are much harder to formalize and use.

A third of these lectures is devoted to models, aspects, and concrete
programming of so-called shared memory (multi-core) systems. The
processor-cores in such systems exchange informations and solve
computational problems by reading and writing from and to a quite
large, but finite shared memory, somewhat like the PRAM.
But the processors are not really synchronized
and memory access times are both NUMA and higher than
operations done by the processor. Memory is managed at different
granularities. More about this will follow in Chapter~\ref{chp:sharedmemory}.

Another third of these lectures is devoted to models, aspects, and
concrete programming of so-called distributed memory (multi-node,
multi-core) systems. Each multi-core processor-node has memory that is
local to that node, and explicit communication between processors on
different multi-core nodes is needed for exchanging information and
solving computational problems. Communication is facilitated by a dedicated
communication network. All this in Chapter~\ref{chp:distributedmemory}.

\subsection{Flynn's Taxonomy}

A different, frequently used, less architecture-oriented and rather
crude characterization of parallel machines, systems and even
programs is the so-called \impi{Flynn's taxonomy}~\cite{Flynn72}.
This taxonomy looks at the instruction and data stream(s) of the
computing system.  A \emph{Single-Instruction, Single-Data}
(\impi{SISD}) system is a sequential computer: one program is executed
and the instructions operate on a single stream of data. This is, of
course, a naive and simplified notion of the workings of a modern
processor. A \emph{Single-Instruction, Multiple-Data} (\impi{SIMD})
system is one in which a single instruction can operate on a larger
batch of data, like, for instance, a whole vector (array) of some
size. Thus, classical \emph{vector computers}\index{vector computer}
that operate on long vectors, or modern processors with capabilities to
operate on short vectors of a few words (with AVX or SSE instruction sets)
are typical SIMD systems. A PRAM machine would be classified as
\emph{Multiple-Instruction, Multiple-Data} (\impi{MIMD}), since each
processor can execute its own instruction stream, each operating on its
own stream of data. Finally, but not obviously, a
\emph{Multiple-Instruction, Single-Data (\impi{MISD})} system could be
a deeply pipelined system where a single stream of data passes through
several processing stages. Many say that such systems do not exist, \ie,
that this taxon in the taxonomy does not make sense.

Flynn's taxonomy is sometimes also used to characterize
\emph{programming models}\index{programming model} by which we mean
the abstractions under which a program can be described (threads,
processes, data access patterns, synchronization and communication
mechanisms, \etc). A SIMD\index{SIMD} model, for instance, is one in which
there is a single ``logical'' instruction stream (that might, as in a
PRAM, be executed by many processors) that operates on some abstract
``vectors''~\cite{Blelloch90}.

The characterization \emph{Single-Program, Multiple-Data}
(\impi{SPMD}) is sometimes used to describe the situation where all
processors in a parallel system execute the same program, but each
processor may, at any time instant, be in a different part of the
program and thus operate on a different ``data stream'' than the
other processors. Our PRAM pseudo-code is SPMD as is typical
for most real parallel code, as we will see with \openmp\index{OpenMP}
and \mpi\index{MPI} later in the lecture notes.
There are relevant counter examples,
though, where the processor-cores\index{processor-core}
in a system actually do run
different programs, but nevertheless cooperate to solve a given,
computational problem. Complex simulations working at many levels at
the same time with different program packages and code could be such
an example. GPU\index{GPU} programming
models\index{programming model} sometimes
use the term \emph{Single-Instruction Multiple-Threads} (\impi{SIMT})
to emphasize that a single instruction
can be executed simultaneously, concurrently by multiple threads,
where batches of threads execute in lock-step as in the PRAM.

\section{Second block (1--2 lectures)}

The bar for \parco is high.  We judge parallel algorithms and
implementations by comparing them against the \emph{best possible}
sequential algorithm or implementation for solving the given
computational problem, and in cases where the best possible (lower
bound) is not known, against the \emph{best known} sequential
algorithm or implementation.  The reasoning is that we, by using the
dedicated parallel resources at hand, want to improve over what we can
already do with a sequential algorithm on our system. With our
parallel machine, we want to solve problems faster and/or better on
some account.

For now, our parallel model and system will be left unspecified. Some
number $p$ of processor-cores\index{processor-core}
interact to solve the problem at hand.

\subsection{Sequential and Parallel Time}
\label{sec:timecomplexity}

\parco is both a theoretical discipline and a
practical/experimental endeavor. As a theoretical discipline,
\parco is interested in the performance of algorithms in some models
(RAM\index{RAM}, PRAM\index{PRAM}, and more realistic settings), and
typically looks at the performance in the worst possible case (worst
possible inputs) when the input size is sufficiently large.
Let \aseq and \apar denote
sequential and parallel algorithms for a problem we are interested in
solving. The parallel algorithm, in contrast to the sequential
algorithm, additionally specifies how processors are to be employed in
the solution, how they interact and coordinate, and how they exchange
information. The sequential and parallel algorithms may be ``similar''
in idea and structure; they may also, as we have already seen
(Theorem~\ref{alg:fastmax}), be completely different.  This is fine as
long as we can argue or even prove that they both correctly solve the given
problem.

By $\tseq(n)$ and $\tpar{p}(n)$ we denote the running times (depending
on how our model accounts for time\index{parallel time},
for instance, number of steps
taken) of \aseq and \apar on worst-case inputs of size $n$ with one
processor for the sequential algorithm \aseq and with $p$
processor-cores\index{processor-core} for the parallel algorithm \apar.
The best possible
and best known algorithms for solving a given problem are those with
the best worst-case asymptotic complexities. For a given problem, the
best possible sequential running time is often denoted as $T^{*}(n)$,
a function of the input size $n$~\cite{JaJa92,RauberRunger13}, which
then defines the \impi{sequential complexity} of the given problem. In
the same way, we can define the \impi{parallel time complexity}
$\tinf(n)$ for a given parallel algorithm \apar as the smallest
running time that this algorithm can achieve using sufficiently many
processors. The number of processors to use to achieve this best
running time can then be turned into a function of the input size $n$.
If the parallel algorithm is the \emph{fastest possible} algorithm for
our given problem, $\tinf(n)$ is the parallel time complexity of the
problem.

As always, constants do matter(!), but they will often be ignored here
and hidden behind $O, \Omega, \Theta, o, \omega$. Recall the
definitions and rules for manipulating such expressions, see for
instance~\cite{CormenLeisersonRivestStein22} or any other algorithms
text, and note that, for parallel algorithms, the worst-case time
is a function of two variables, problem size $n$ and number
of processor-cores $p$.  Saying that some $\tpar{p}(n)$ is in
$O(f(p,n))$ then means that

\begin{displaymath}
  \exists C>0, \exists N,P>0: \forall n\geq N, p\geq P:
  0\leq\tpar{p}(n)\leq C f(p,n)
\end{displaymath}
and that some $\tpar{p}(n)$ is in $\Theta(f(p,n))$ that
\begin{displaymath}
  \exists C_0, C_1>0, \exists N,P>0: \forall n\geq N, p\geq P: 0\leq
  C_0 f(p,n)\leq\tpar{p}(n)\leq C_1 f(p,n) \quad .
\end{displaymath}

We may sometimes let the number of processors $p$ change as a function
of the problem size, $p=f(n)$ (``What is the best number of processors
for this problem size?'' as in the definition of parallel time
complexity\index{parallel time complexity}), or the problem size
change as a function of the number of processors, $n=g(p)$ (``What is
a good problem size for this number of processors?''), in which case
the asymptotics are of one variable.

Typical sequential, best known/best possible worst-case complexities
for some of our computational problems\index{computational problem}
are~\cite{CormenLeisersonRivestStein22}:
\begin{itemize}
\item
  $\Theta(\log n)$: Searching for an element in an ordered array of
  size $n$.
\item
  $\Theta(n)$: Maximum finding in an unordered $n$ element sequence,
  computing the sum of the elements in an array (reduction\index{reduction}),
  computing all prefix sums\index{prefix sums} over an array.
\item
  $\Theta(n\log n)$: Comparison-based sorting of an $n$ element array.
\item
  $\Theta(n^2)$: Matrix--vector multiplication with dense, square matrices
  of order $n$ (inputs of size $\Theta(n^2)$).
\item
  $O(n^3)$: Dense
  matrix--matrix multiplication\index{matrix--matrix multiplication},
  which we will take as the
  best bound known to us in this lecture (but far from best known,
  see, \eg,~\cite{Strassen69}).
\item
  $O(n+m)$: Breadth-First Search (BFS\index{BFS}) and
  Depth-First Search (DFS\index{DFS}) in graphs with $n$
  vertices and $m$ edges.
\item
  $\Theta(m+n)$: Merging\index{merging}
  two ordered sequences of length $n$ and $m$
  with a constant time comparison function, identifying the connected
  components of undirected graphs with $n$ vertices and $m$ edges.
\item
  $O(n\log n+m)$: Dijkstra's Single-Source Shortest Problem algorithm
  on real, non-negative weight, directed graphs with $n$ vertices and
  $m$ arcs using a best known priority queue.
\end{itemize}

Regardless of how time per processor-core\index{processor-core}
is accounted for, the time
of the parallel algorithm \apar when executed on $p$ processor-cores
is the time for the last processor-core to finish, assuming that all
cores started at the same time. Note that we here make a lot of implicit
assumptions, ``same time'' \etc, that will not be discussed further but
are worth thinking much more about. The rationale for this convention is
twofold: Our problem is solved when the last processor has finished
(and we know that this is the case), and since our parallel system is
dedicated, it has to be paid for until all processor-cores are again
free for something else.

In \parco as a practical, experimental endeavor, \aseq and
\apar denote concrete implementations of the algorithms, and
$\tseq(n)$ and $\tpar{p}(n)$ are measured running times for concrete,
precisely specified inputs of size $O(n)$ on concrete and precisely
specified systems.  Designing measuring procedures and selecting
inputs belong to experimental Computer Science and are
highly non-trivial tasks; they will not be treated in great
detail in these lectures.  Suffice it to say that time is measured by
starting the processor-cores at the same time as far as this is
possible, and accounting for the time $\tpar{p}(n)$ by the last
processor-core\index{processor-core} to finish.
Inputs may be either single, concrete
inputs or a whole larger set of inputs. Worst-case inputs may be
difficult (impossible) to construct and are often also not interesting,
so inputs are rather ``typical'' instances, ``average-case''
instances, randomly generated instances, inputs with particular
structure, \etc (for recent criticism of and alternatives to
worst-case analysis of algorithms, see~\cite{Roughgarden21}). The
important point for now is that inputs and generally the whole
experimental set-up be clearly described, so that claims and
observations can be objectively verified (reproducibility).

\subsection{Speed-Up}

We measure the gain of the parallel algorithm \apar over the best
known or possible sequential algorithm \aseq for inputs of size $O(n)$ by
relating the two running times. \parco aims to improve on the best
that we can already do with a single processor-core\index{processor-core}.
This is the fundamental notion of absolute \impi{speed-up} over a given
baseline:

\begin{definition}[Absolute Speed-up]  
The \emph{absolute speed-up}\index{speed-up!absolute} of parallel
algorithm \apar over best known or best possible sequential algorithm
\aseq (solving the same problem) for input of size $O(n)$ on a $p$
processor-core parallel system is the ratio of sequential to parallel
running time, \ie,
\begin{eqnarray*}
  \SU{p}{n} & = & \frac{\tseq(n)}{\tpar{p}(n)} \quad .
\end{eqnarray*}
\end{definition}

The notion of speed-up is meaningful in both theoretical (analyzed, in
some model) and practical (measured running times for specific inputs)
settings. Often, speed-up is analyzed by keeping the problem size $n$
fixed and varying the number of processor-cores\index{processor-core} $p$
(strong scaling\index{strong scaling}, see later). Sometimes (scaled
speed-up\index{speed-up!scaled}, see later) both input size $n$ and
number of processor-cores $p$ are varied. For the definition, it is
assumed that $\tpar{p}(n)$ is meaningful for any number of processors
$p$ (and any problem size $n$), which for concrete algorithms and
implementations is not always the case: Some algorithms assume $p=2^d$
for some $d$, a power-of-two number of processors, or $p=d^2, p=d^3$,
a square or cubic number of processors, \etc. The speed-up is
well-defined only for the cases for which the algorithms actually
work. For any input size $n$, there is obviously also a maximum
number of processors beyond which the parallel algorithm does not
become faster (or even work), namely when there is not enough
computational work in the input of size $n$ to keep any more
processors busy with anything useful. Beyond this number, speed-up
will decrease: Any additional processors are useless and wasted.

As an example, a parallel algorithm $\apar$ with $\tpar{p}(n)=O(n/p)$
would have an absolute speed-up of $O(p)$ for a best known sequential
algorithm with $\tseq(n)=O(n)$, assuming that $n\geq p$ ($p$ in $O(n)$
or, equivalently, $n$ in $\Omega(p)$). If $\tpar{p}(n)=O(n/\sqrt{p})$
the speed-up would be only $O(\sqrt{p})$.

A speed-up\index{speed-up}
of $\Theta(p)$, with upper bounding constant of at most one
and $n$ allowed to increase with $p$, is said to be
\emph{linear}\index{speed-up!linear}, and linear speed-up of $p$ where
both bounding constants are indeed close to one is said to be
\emph{perfect}\index{speed-up!perfect} (by measurement, or by analysis
of constants). Perfect speed-up is rare and hardly achievable
(sometimes provably not, an important example is given later in
these lecture notes, see Theorem~\ref{thm:prefix-tradeoff}).

According to the definitions of linear and perfect speed-up, a
parallel algorithm \apar with running time of at most
$\tpar{p}(n)=c(\frac{n}{p}+\log n)$ for some constant $c$ would have
perfect speed-up relative to a best possible sequential algorithm with
running time of at most $\tseq(n)=cn$ steps. We have
\begin{eqnarray*}
  \SU{p}{n} & = & \frac{cn}{c(n/p+\log n)} \\ & = & \frac{p}{1+(p\log
    n)/n}
\end{eqnarray*}
which is as close to $p$ as desired for $n/\log n>p$: For any
$\varepsilon, \varepsilon>0$, it holds that $(p\log n/n)<\varepsilon
\Leftrightarrow n/\log n>p/\varepsilon$. If the sequential and
parallel algorithms have different leading constants $c_0$ and
$c_1$, respectively (with $c_0<c_1$), the speed-up is linear
with upper bounding constant $\frac{c_0}{c_1}<1$.  In other words,
linear speed-up means that for any number of processors $p$, the
parallel running time multiplied by $p$ differs by a constant factor
from the best (possible or known) sequential running time (the
sequential time being lower) for sufficiently large $n$; perfect
speed-up means that this constant is practically one.

\subsection{``Linear Speed-Up is Best Possible''}
\label{sec:linearbest}

Linear speed-up\index{speed-up!linear}
is the best that is possible. The argument for this is
that a parallel algorithm running on $p$ dedicated cores can be
\emph{simulated} on a single core in time no worse than $p
\tpar{p}(n)$ time steps by simulating the steps of the $p$ processors one
after the other in a round-robin fashion.
If the speed-up would be more than linear,
then $\tseq(n)>p\tpar{p}(n)$, and the simulated execution would run
faster than the best known sequential algorithm for our problem, which
cannot be. Or: in that case, an even better algorithm would have been
constructed! Sometimes, indeed, a new parallel algorithm can by
a clever simulation lead to a better than previously known sequential
algorithm.

For the PRAM\index{PRAM} model,
the simulation argument\index{simulation argument}
can be worked out in detail, for instance, by writing a sequential
simulator for programs in our PRAM pseudo-code: Within each
\texttt{par}-construct, execute the instructions of the assigned
processors one after the other in a round-robin fashion, with some
care taken to resolve concurrent writing correctly.

Despite this argument, \emph{super-linear
speed-up}\index{speed-up!super-linear} larger than the number of
pro\-ces\-sor-co\-res $p$ is sometimes reported (mostly in practical
settings)~\cite{FaberLubeckWhite86,HelmboldMcDowell91}. If the reasons
for this are algorithmic, it can only be that the sequential and
parallel algorithms are, on specific inputs, not doing the same amount
of work (see below). Randomized algorithms, where more and different
coin tosses are possibly done by the parallel algorithm than by the
sequential algorithm, can likewise sometimes exhibit super-linear
speed-up. But also deterministic algorithms, like search
algorithms, can exhibit this behavior if the way the search space is
divided over the parallel processors
depends on the number of processor-cores causing the
parallel algorithm to complete the search more than proportionally
faster than the sequential algorithm. Finally, on ``real'' parallel
computing systems, the memory system and in particular the average
memory access times can differ between algorithms running on a single
processor-core\index{processor-core}
and on many processor-cores where memory is accessed in
a distributed fashion and faster memory ``closer to the core'' can be
used to a larger extent (see \Sec~\ref{sec:cachelocality}).

The argument that linear speed-up\index{speed-up!linear}
is best possible also tells us
that for any parallel algorithm it holds that $\tpar{p}(n) \geq
\frac{\tseq(n)}{p}$. In other words, the best possible parallel
algorithm \apar for the problem solved by \aseq cannot run faster than
$\tseq(n)/p$. This observation provides us with a first, useful
\impi{lower bound} on parallel running time.

For any parallel algorithm \apar on concrete input of size $O(n)$,
there is, of course a limit on the number of processor-cores that can
be sensibly employed. For instance, putting in more processor-cores
than there is actual work (operations) to be done makes no sense, and
some processors would sit idle for parts of the computation.  Specific
speed-up claims are therefore (or should be) qualified with the range
of processor-cores for which they apply\index{processor-core}.

\subsection{Cost and Work}
\label{sec:cost-work}

Our dedicated parallel system with $p$ processor-cores running \apar
is kept occupied for $\tpar{p}(n)$ units of time, and this is what we
have to ``pay'' for. The \emph{cost} of a parallel algorithm is,
accordingly, defined as the product $p\times\tpar{p}(n)$.
If we picture a parallel
computation as a rectangle with the processor-cores $i$ on one axis,
listed densely from $0$ to $p-1$ and the time spent by the
processor-cores on the other axis, the parallel time\index{parallel time}
$\tpar{p}(n)$ is
the largest time for some processor-core\index{processor-core} $i$,
and the cost is the area of the rectangle $p\times\tpar{p}(n)$. The parallel
algorithm \apar exploits the parallel system well if the parallel
cost invested for a given input is proportional to the cost of solving
the given problem sequentially by \aseq. This motivates the notion of
\impi{cost-optimality}.

\begin{definition}[Cost-optimal Parallel Algorithm]
  \label{def:costoptimality}
  A parallel algorithm \apar for a given problem is
  \emph{cost-optimal}\index{algorithm!cost-optimal} if its cost $p
  \tpar{p}(n)$ is in $O(\tseq(n))$ for a best known sequential
  algorithm \aseq for any number of processors $p$ up to some bound
  that is an increasing function of $n$.
\end{definition}

Cost-optimality requires that, for any given input size $n$, there is
a certain number of processors $p$ for which the cost $p'\tpar{p'}(n)$ for
any $p'\leq p$ is in $O(\tseq(n))$ and the bounding constant in
$O(\tseq(n))$ does not depend on $p'$ or $p$. The bound on the number of
processors must be an increasing function of the problem size $n$. The
intention is that the cost of \apar is in the ballpark of the
sequential running time of \aseq.  Almost per definition, cost-optimal
algorithms have linear speed-up, since $p \tpar{p}(n)\leq c\tseq(n))$
implies $\frac{\tseq(n)}{\tpar{p}(n)}\geq \frac{p}{c}$ which is the
speed-up. The requirement that the upper bound on the number of
processors $p$ increases with
$n$ makes it possible to find an increasing function of $p$ for which
the speed-up\index{speed-up} is in $\Theta(p)$.  Cost-optimality is a strong
property\index{cost-optimality}.

A different way of looking at cost-optimality is via
the parallel time complexity\index{parallel time complexity}
and the number of processors needed to reach this fastest
time. The product of this number of processors and this fastest
possible time should still be in the order of the effort required by a
best (known or possible) sequential algorithm.
This is captured in the following definition.

\begin{definition}[Asymptotically cost-optimal Parallel Algorithm]
  \label{def:costoptimality2}
  Let for some given problem \apar be a parallel algorithm with parallel
  time complexity $\tinf(n)$. Let $P(n)$ be the smallest number of processors
  needed to reach $\tinf(n)$. The cost of \apar with this number of
  processors is $P(n)\tinf(n)$ and \apar is cost-optimal if
  $P(n)\tinf(n)$ is in $O(\tseq(n))$ for a best known sequential
  algorithm \aseq for the given problem.
\end{definition}
  
We often use the term \emph{work}\index{work} to quantify the real
``effort'' that an algorithm puts into solving one of our
computational problems\index{computational problem}.  The work of a
sequential algorithm \aseq on input of size $O(n)$ is the number of
operations (of some kind) carried out by the algorithm.
Sequentially speaking, ``work is time''. The work of a parallel
algorithm \apar on a system with $p$ processor-cores is the total work
carried out by all of the $p$ cores, excluding time and operations
spent idling by some processors or by processors that are not
assigned to do anything (useful).
That is, anything that the cores might be doing that is not
strictly related to the algorithm does not count as work.
With a formal model like the
PRAM\index{PRAM}, this can be given a precise definition (``work is
the operations carried out by assigned processors''). In more realistic
settings, we have to be careful which idle times should count and which
not. The work of parallel algorithm \apar on input $n$ is denoted
$\wpar{p}(n)$. Ideally, work is independent of the number of processors
$p$ and we might write just $\wpar{}(n)$. This means that the work to be
done by the algorithm \apar has been separated from how the $p$
processors that will eventually perform this work share the work. This
is a very useful point of view which leads to a productive separation of
concerns between what has to be done (``the work'') and who does it (``which
processors''). This point of view motivates the next definition.

\begin{definition}[Work-optimal Parallel Algorithm]
  A parallel algorithm \apar with work $\wpar{}(n)$ is
  \emph{work-optimal}\index{algorithm!work-optimal} if $\wpar{}(n)$ is
  $O(\tseq(n))$ for a best known sequential algorithm \aseq.
\end{definition}

If an algorithm is work-optimal algorithms but not cost-optimal this
indicates either that the way the processors are used in the parallel
algorithms is not efficient (some processors sit idle for too long) or
that most of the work must necessarily be done sequentially, one piece
after the other (because of sequential dependencies).  From a
work-optimal algorithm that is not cost-optimal for the first reason,
a better, cost-optimal algorithm with the same amount of work that
runs on fewer processor-cores\index{processor-core}
can sometimes be constructed, but this may not be easy.

A cost-optimal parallel algorithm is per definition work-optimal but
not the other way around: A parallel algorithm that is not
work-optimal cannot be cost-optimal. Thus, a first step towards designing
a good parallel algorithm is to look for a solution that is (at least)
work-optimal.

Another useful observation following from the notion of parallel work
is that the best possible parallel running time of an algorithm with
work $\wpar{}(n)$ is at least
\begin{eqnarray*}
\tpar{p}(n) & \geq & \frac{\wpar{}(n)}{p} \quad .
\end{eqnarray*}
This is another useful lower bound\index{lower bound} which is sometimes
called the \emph{Work Law}\index{Law!Work Law} (See
\Sec~\ref{sec:taskgraphs}). The lower bound is met if the work
$\wpar{}(n)$ that has to be done has been perfectly distributed over the
$p$ processors and no extra costs have been incurred.

As an extreme example, consider a ``parallel'' algorithm that is just a (best)
sequential algorithm executed on one out of the $p$ processors.
This is a work-optimal parallel algorithm, but it is clearly not
cost-optimal since all but one processor are idle.
Its cost $O(p\tseq(n))$ is optimal when running it on
one or a small, constant number of processors $p$; but as long as the
number of processors that can be efficiently exploited cannot be
increased with increasing problem size, such an algorithm is not
cost-optimal according to our definition,
and speed-up beyond a limited, constant number of
processors cannot be achieved. This is not what is desired of a good
parallel algorithm.  Cost- and work-optimality are asymptotic notions
of properties that hold for large problems and large numbers of
processors.

Algorithms that are not cost-optimal do not have
linear speed-up\index{speed-up!linear}. The
PRAM\index{PRAM} maximum finding algorithm of Theorem~\ref{alg:fastmax}
takes $O(1)$ time with
$O(n^2)$ processors and therefore has cost $O(n^2)$, which is far from
$\tseq(n) = O(n)$.
To determine the speed-up of this algorithm, we first have to
observe that the algorithm can be simulated with $p\leq n^2$
processors in $O(n^2/p)$ parallel time\index{parallel time}
steps\index{simulation argument}.
The speed-up is $\SU{p}{n}=O(n/(n^2/p))= p/n$. The
speed-up is \emph{not} independent of $n$, and actually decreases with
$n$: The larger the input, the lower the speed-up\index{speed-up}.

The point of distinguishing work\index{work} and cost\index{cost} is
to separate the discovery of parallelism from an all too specific
assignment of the work to the actually available processors. A good,
parallel algorithm is work-optimal and can become fast when enough
processors are given. A next
design step is then to carefully assign the work to only as many
processors as allowed to keep the algorithm cost-optimal.
The PRAM\index{PRAM} abstraction supports this strategy well:
Processors can be assigned freely (with the \textbf{par}-construct),
and the analysis can focus on the number of operations actually done by
the assigned processors (the work).

More precisely, let us assume that a work-optimal PRAM algorithm with
work $\wpar{}(n)$ and parallel time complexity\index{parallel time complexity}
of $\tinf(n)$ has been found. Such an algorithm can (in
principle) be implemented to run on a $p$-processor PRAM (same variant)
in at most $\floor{\frac{\wpar{}(n)}{p}}+\tinf(n)$
parallel time\index{parallel time} steps.
This follows easily. In each of the $\tinf(n)$ parallel steps some
amount of work $\wpar{i}(n)$ has to be done. This work can be done in
parallel on the $p$ processors in $\ceiling{\frac{\wpar{i}(n)}{p}}$ time steps
by a straightforward round-robin execution\index{simulation argument} of the
work units over the $p$ processors.  Summing over the steps gives
\begin{eqnarray*}
  \sum_{i=0}^{\tinf(n)-1} \ceiling{\frac{\wpar{i}(n)}{p}}
  & \leq & \sum_{i=0}^{\tinf(n)-1} (\floor{\frac{\wpar{i}(n)}{p}}+1) \\
  & \leq & \floor{\frac{\wpar{}(n)}{p}}+\tinf(n)
\end{eqnarray*}

This observation is also known as \impi{Brent's Theorem}~\cite{Brent74}. The
observation only tells us that an efficient execution of the algorithm is
possible on a $p$-processor PRAM\index{PRAM},
but not how the work units for each step
can be identified. Sometimes this is obvious and sometimes not.

\subsection{Relative Speed-Up and Scalability}

While the absolute speed-up measures how well a parallel algorithm can
improve over its best known sequential counterpart, it does not
measure whether the parallel algorithm by itself is able to exploit
the $p$ processors well.  This notion of \impi{scalability} is
the \emph{relative speed-up}.

\begin{definition}[Relative Speed-up]\index{speed-up!relative}
  The \emph{relative speed-up} of a parallel algorithm \apar is the
  ratio of the parallel running time with one processor-core to the
  parallel running time with $p$ processor-cores, \ie,
\begin{eqnarray*}
  \SUR{p}{n} & = & \frac{\tpar{1}(n)}{\tpar{p}(n)} \quad .
\end{eqnarray*}
\end{definition}

Assume that an arbitrary number of processors is available. Any
parallel algorithm has, for any (fixed) input of size $O(n)$, a
fastest running time that it can achieve, denoted by $\tinf(n)$ which
is the time $\tpar{p'}(n)$ for some number of processors $p'$; this was
defined as the parallel time complexity\index{parallel time complexity}
(see \Sec~\ref{sec:timecomplexity}). Per definition, 
$\tpar{p}(n)\geq \tinf(n)$ for any number of processors $p$. It
thus holds that $\SUR{p}{n} = \frac{\tpar{1}(n)}{\tpar{p}(n)} \leq
\frac{\tpar{1}(n)}{\tinf(n)}$.

The ratio $\frac{\tpar{1}(n)}{\tinf(n)}$ which is a function of the
input size $n$ only is called the \impi{parallelism} of the parallel
algorithm. It is clearly both the largest, relative speed-up that can be
achieved, as well as an upper bound on the number of processors up to which
linear, relative speed-up\index{speed-up!relative}
can be achieved. If some number of
processors $p'$ larger than the parallelism is chosen, the definition
says that $\SUR{p'}{n} < p'$, that is, less than linear speed-up.
The parallelism is also the asymptotically smallest number of processor
needed to achieve the best possible running time  $\tinf(n)$.

It is important to clearly distinguish between absolute and relative
speed-up.  Relative speed-up compares a parallel algorithm or
implementation against itself, and expresses to what extent the
processors are exploited well (linear, relative speed-up). Absolute
speed-up compares the parallel algorithm against a (best known or
possible) baseline, and expresses how well it improves over the
baseline. A parallel algorithm may have excellent
relative speed-up\index{speed-up!relative},
but poor absolute speed-up. Is such a good algorithm? In any case,
reporting only the relative speed-up for a parallel algorithm or
implementation can be grossly misleading and should never be
done in serious \parco.
An absolute baseline always must be defined (that which we want to
improve over) and absolute running times also stated. There are
plenty of examples of basing claims
on relative speed-ups only also in the scientific literature.
For more on such pitfalls and
misrepresentations, see the now well-known and often paraphrased
``\ldots Ways to fool the masses\ldots''~\cite{Bailey92}, see
also~\url{https://blogs.fau.de/hager/archives/5299}.

The absolute speed-up compares the running time of the parallel
algorithm against the running time of a best known or possible
sequential algorithm.  For such an algorithm it holds that
$\tseq(n)\leq\tpar{1}(n)$ and therefore
\begin{eqnarray*}
  \SU{p}{n} & \leq & \SUR{p}{n} \quad .
\end{eqnarray*}
The absolute speed-up is at most as large as the relative speed-up and
also in that sense a tougher
measure\index{speed-up!absolute}\index{speed-up!relative}.

\subsection{Overhead and Load Balance}

A parallel algorithm for a computational problem usually performs more
work\index{work} than a corresponding best known sequential
algorithm. In summary, such work is termed
\emph{overhead}\index{overhead}; thus, overhead is work incurred by the
parallel algorithm that does not have to be done by the sequential
algorithm. Beware that this definition tacitly assumes that sequential
and parallel algorithms are somehow similar and can be compared
(``extra work''). This is not always the case. Sometimes, a
parallel algorithm is totally different from the best known sequential
algorithm.  Overheads can be caused by several factors, \eg,
\begin{itemize}
\item preparation of data for other processor-cores\index{processor-core},
\item communication between and coordination of processor-cores,
\item synchronization\index{synchronization}, and
\item algorithmic overheads: extra or redundant work
\end{itemize}
when compared to a corresponding, somehow similar sequential
algorithm. When a parallel algorithm \apar is derived from a
sequential algorithm \aseq, we can loosely speak of
\impi{parallelization} and say that \aseq has been
\emph{parallelized} into \apar. Parallel algorithms implemented with
\openmp\index{OpenMP} (see \Sec~\ref{sec:openmpframework}) are, for instance,
often very concrete parallelizations of corresponding
sequential algorithms. Again, it is
important to stress that many parallel algorithms are specifically not
parallelizations of some sequential algorithm.

Overheads are more or less inevitable, but if they are on the order of
(within the bounds of) the sequential work, $O(\tseq(n))$ the
parallel algorithm can still be work- and
cost-optimal\index{work-optimal}\index{cost-optimal}, and thus have
linear, although not perhaps perfect speed-up. Often, overheads
increase with the number of processors $p$, giving, for fixed problem
size $n$, a limit on the number of processors that can be used while
still giving linear speed-up. If the overheads are asymptotically
larger than the sequential work, the parallel algorithm will never
have linear speed-up\index{speed-up}.

The overheads caused by communication and
synchronization\index{synchronization} between pro\-ces\-sor-co\-res are
often significant. Later in these lecture notes, we will introduce
a simple model for accounting for communication operations. Suffice it
here to say that a simple synchronization between $p$ processors, which
means ascertaining that a processor cannot continue beyond a certain
point in its computation before all other processors have reached a
certain point in their computations (see \Sec~\ref{sec:barrier}),
may (and must) take $\Omega(\log
p)$ operations. An exchange of data will typically take time
proportional to the amount of the data (per processor) and an
additive term dependent on the number of processors $p$.

Between communication operations,
the processor-cores\index{processor-core} operate
independently on parts of the problem although they could interfere
indirectly through the memory and cache system (this will be discussed
in later parts of these lecture notes, see
\Sec~\ref{sec:cachelocality}). The length of the intervals between
communication and synchronization operations is sometimes referred to
as the \impi{granularity} of the parallel algorithm.  A parallel
computation in which communication and synchronization occur rarely is
called \emph{coarse grained}\index{granularity!coarse grained}. If
communication and synchronization occur frequently, the computation is
called \emph{fine grained}\index{granularity!fine rained}.  These are
relative (and vague) terms. Machine models that can support fine
grained algorithms, are also called fine grained. The PRAM\index{PRAM}
is an extreme example: The processors can (and often do) communicate
via the shared memory in every step, and they are lock-step
synchronized with no overhead for synchronization\index{synchronization}.

In some parallel algorithms, the processors may not perform the same
amount of work, and/or have different amounts of overhead.  If we, for
the moment, let $\tpar{i}(n)$ denote the time taken by some
processor-core $i,0\leq i<p$ from the time this
processor-core\index{processor-core} starts
until it terminates, the (absolute) \impi{load imbalance} is defined as
\begin{displaymath}
  \max_{0\leq i,j<p}{|\tpar{i}(n)-\tpar{j}(n)|}=
  \max_{0\leq i<p}\tpar{i}(n)-\min_{0\leq i<p}\tpar{i}(n) \quad .
\end{displaymath}
The relative load imbalance is the ratio of absolute load balance to
parallel time\index{parallel time} (completion time of slowest processor).
Too large load imbalance is another reason that a parallel algorithm
may have a too small (or non-linear) speed-up\index{speed-up}.
Too large load imbalance may likewise be a reason why an otherwise
work-optimal\index{work-optimal} parallel algorithm is not
cost-optimal\index{cost-optimal}: Too many processors take too
small a share of the total work.

Good load balance means that $\tpar{i}(n)\approx\tpar{j}(n)$ for all
pairs of 
processors $(i,j)$. Achieving good, even load balance over the
processors is called \impi{load balancing} and is always an issue in
designing a parallel algorithm, explicitly by the construction of the
algorithm or implicitly by taking steps later to ensure a good load
balance. We distinguish between \impi{static load-balancing}, where
the amount of work\index{work} to be done can be divided upfront
among the processors, and \impi{dynamic load balancing}, where the
processors have to communicate and exchange work during the execution
of the parallel algorithm.  Static load balancing can be further
subdivided into \emph{oblivious, static load-balancing}\index{oblivious},
where the problem can be divided over the processors based on the input size and
structure alone but regardless of the actual input, and
\emph{adaptive, problem-dependent, static load-balancing}, where the
input itself is needed in order to divide the work and 
preprocessing may be required.  Some aspects of the load balancing
problem (work-stealing\index{work-stealing}, loop
scheduling\index{loop scheduling}) will be discussed later in this
part of the lecture notes. However, load balancing \emph{per se} is too
large a subfield of \parco to be treated in much detail here.

Problems and algorithms where the input and work can be statically
distributed to the processors and where no further explicit
interaction is required are called either \emph{embarrassingly
parallel}\index{parallel!embarrassingly}, \emph{trivially
parallel}\index{parallel!trivially}, or \emph{pleasantly
parallel}\index{parallel!pleasantly}.  These are the best (but
uninteresting, in the sense of being unchallenging) cases of easily
parallelizable problems with linear or even perfect speed-up.  The
realization that the problem is trivially or embarrassingly parallel
can, of course, be highly non-trivial and the way to see this unpleasant.

\subsection{Amdahl's Law}

Gene Amdahl made a simple observation on how to speed up
programs~\cite{Amdahl67}, which when applied to \parco yields severe
bounds on the speed-up\index{speed-up} that certain parallel algorithms can
achieve. The observation assumes that the parallel algorithm is
somehow derived by parallelization\index{parallelization} of the
sequential algorithms.

\begin{theorem}[Amdahl's Law]\index{Law!Amdahl's Law}
Assume that the work performed by sequential algorithm \aseq can be
divided into a strictly sequential fraction $s, 0<s\leq 1$,
independent of $n$, that cannot be parallelized at all, and a fraction
$r=(1-s)$ that can be perfectly parallelized. The parallelized
algorithm is \apar.
Then, the maximum speed-up that can be achieved by \apar over \aseq is
bounded by $1/s$.
\end{theorem}

The proof is straightforward. With the assumption that
\begin{eqnarray*}
  \tpar{p}(n) & = & s\tseq(n)+\frac{(1-s)\tseq(n)}{p}
\end{eqnarray*}
we get
\begin{eqnarray*}
  \SU{p}{n} & = & \frac{\tseq(n)}{s\tseq(n)+\frac{(1-s)\tseq(n)}{p}} \\
  & = & \frac{1}{s+\frac{1-s}{p}} \\
  & \rightarrow & \frac{1}{s}\ \mbox{for $p\rightarrow\infty$} \quad .
\end{eqnarray*}

Amdahl's Law is devastating. Even the smallest, constant sequential
fraction of the algorithm to be parallelized will limit and eventually
kill speed-up\index{speed-up}. A sequential fraction of $10$\%, or $1$\%, sounds
reasonable and harmless but limits the speed-up to $10$, or $100$, no
matter what else is done, no matter how large the problem,
and no matter how many processors are
invested. Note that the parallelization considered is
work-optimal\index{work-optimal}; but it is surely not
cost-optimal\index{cost-optimal}. The running time of the parallel
algorithm is at least $s\tseq(n)$ and since $s, s<1$ is
constant, the cost is therefore $O(p\tseq(n))$ which is not in
$O(\tseq(n))$.

A sequential algorithm which falls under Amdahl's Law cannot be used
as the basis of a good, parallel algorithm: Its speed-up will be
severely limited and bounded by a constant.
Amdahl's Law is therefore rather an analysis tool: If it
turns out that a (large) fraction of the algorithm at hand cannot be
parallelized, we have to look for a different, better algorithm.  This
is what makes \parco a creative activity: Simple
parallelization\index{parallelization} of a sequential algorithm will
often not lead to a good, parallel counterpart. New ideas for old
problems are sometimes needed.

Typical victims of Amdahl's Law are:
\begin{itemize}
\item
  Input/output: For linear work algorithms, reading the input and
  possibly also writing the output will take $\Omega(n)$ time steps,
  and thus be a constant fraction of $O(n)$.
\item
  Sequential preprocessing: As above.
\item
  Maintaining sequential data structures, in particular sequential
  initialization, can easily turn out to be a constant fraction of the
  total work.
\item
  Hard-to-parallelize parts that are done sequentially (which might
  look innocent enough for just small parts): If such parts take a
  constant fraction of the total work, Amdahl's Law applies.
\item
  Long chains of dependent operations (operations that have to be
  performed one after the other and cannot be done in parallel), not
  necessarily on the same processor-core.
\end{itemize}

When analyzing and benchmarking parallel algorithms, input/output is
often disregarded when accounting for sequential and parallel
time\index{parallel time}.
The defensible reason for this is that we are interested in how
the core parallel algorithm performs (speeds up), under the
assumption that the input has already been read and properly
distributed to the processor-cores\index{processor-core}
according to the specification. In these lecture notes, our
algorithms are small parts (building blocks) of larger applications
and in this larger context would not need input/output: The data
are already where they should be. Also results do not have to be
output but should just stay and be available for the next building block to
use. We, therefore, analyze the building blocks in isolation without
the input/output parts that might fall victim to Amdahl's Law.

In a good parallel algorithm, not falling victim to Amdahl's Law, the
sequential part $s(n)$ will not be a constant fraction of the total
work but depend on and decrease with $n$. If such is the case,
Amdahl's Law does not apply. Instead, a good speed-up\index{speed-up}
can be achieved
with large enough inputs. \parco is about solving large,
work-intensive problems, and in good parallel algorithms the parts
doing the parallel work dominate the total work as the input gets
large enough.

\subsection{Efficiency and Weak Scaling}

As observed, there is, for any parallel algorithm on input of size
$O(n)$, always a fastest possible time, $\tinf(n)$, that the algorithm
can achieve (the parallel time complexity\index{parallel time complexity}).
Thus, the parallel running time of an algorithm with
good, linear speed-up\index{speed-up}
(up to the number of processor-cores\index{processor-core} determined
by the parallelism\index{parallelism}),
can be written as $\tpar{p}(n)=O(T(n)/p+t(n))$,
that is, as a parallelizable term $T(n)$ and a non-parallelizable term
$t(n)=\tinf(n)$.  If speed-up is not linear, the parallel running time
is instead something like $\tpar{p}(n)= O(T(n)/f(p)+t(n))$ 
strictly with $f(p)<p$ and $f(p)$ in $o(p)$, or $T(n)$ is not in $O(\tseq(n))$.

If we compare against a sequential algorithm with $\tseq(n)=O(T(n))=
O(T(n)+t(n))$, a parallel algorithm where $t(n)/T(n)
\rightarrow 0$ as $n\rightarrow\infty$ is also good and can have
linear speed-up for large enough $n$. The speed-up is namely
\begin{displaymath}
  \SU{p}{n} = \frac{\tseq(n)}{\tpar{p}(n)} =
  O(\frac{T(n)}{T(n)/p+t(n)}) = O(\frac{1}{1/p+t(n)/T(n)}) \rightarrow
  O(p)
\end{displaymath}
as $n$ increases. This is called \emph{scaled
speed-up}\index{speed-up!scaled}, and the faster $t(n)/T(n)$
converges, the faster the speed-up becomes linear. Against Amdahl's
Law\index{Law!Amdahl's Law}, the sequential part $t(n)$ should be as small
as possible and increase more slowly with $n$ than the parallelizable
part $T(n)$. Algorithms with this property are
cost-optimal\index{cost-optimal} according to
Definition~\ref{def:costoptimality}.

It is a good way which we use throughout these lecture notes to state
the performance of a (work-optimal) parallel algorithm as
$\tpar{p}(n)=O(T(n)/p+t(n,p))$ with the assumption that $t(n,p)$ is in
$O(T(n))$ for fixed $p$, and $\tseq(n)=O(T(n))$. That is, we allow the
non-parallelizable part to depend on both $n$ and $p$.  Often,
however, $t(n,p)$ is just $t(n)$ independent of $p$ or $t(p)$
depending on $p$ only (synchronization\index{synchronization}
costs). An iterative parallel algorithm with a convergence check
involving synchronization could, for instance, run in $O(n/p+\log
n\log p)$ parallel time\index{parallel time}
with $t(n,p)=O(\log n\log p)$. Such an
algorithm would perform total linear $O(n)$ work which has been well
distributed over the $p$ processors; the algorithm performs $O(\log
n)$ iterations each of which incurs a synchronization overhead of
$O(\log p)$ operations.

The \emph{parallel efficiency} of a parallel algorithm \apar is
measured by comparing \apar against a best possible parallelization
of \aseq as given by the Work Law\index{Law!Work Law}
(see \Sec~\ref{sec:taskgraphs}).

\begin{definition}[Parallel Efficiency]\index{parallel efficiency}\index{efficiency}
  The efficiency $\EFF{p}{n}$ for input of size $O(n)$ and $p$
  processors of parallel algorithm \apar compared to sequential
  algorithm \aseq is defined as
\begin{displaymath}
  \EFF{p}{n} = \frac{\tseq(n)}{p} \big/ \tpar{p}(n) =
  \frac{\tseq(n)}{p \tpar{p}(n)} = \frac{\SU{p}{n}}{p} \quad .
\end{displaymath}
\end{definition}

As worked out in the definition, the efficiency\index{efficiency}
is also the achieved
speed-up divided by $p$ as well as the sequential time divided by the cost
of the parallel algorithm.  It therefore holds that
\begin{itemize}
\item
  $\EFF{p}{n}\leq 1$.
\item
  If $\EFF{p}{n}=e$ for some constant $e, 0<e\leq 1$, the speed-up is
  linear.
\item
  Cost-optimal\index{cost-optimal} algorithms have constant
  efficiency.
\end{itemize}

Should it happen that the efficiency\index{efficiency} $\EFF{p}{n}$,
contrary to the
statement above, for some $n$ and number of processors $p$ is larger
than $1$, equivalently that the absolute
speed-up\index{speed-up!absolute} is larger than $p$, this tells us
that the sequential baseline is not the best (known) possible. It can
be replaced by some variation of the parallel algorithm. In such a
case, \parco has helped to discover a better sequential algorithm for
the given problem.

We note that this is a definition of
\impi{algorithmic efficiency}\index{efficiency}:
How close is the time of the parallel algorithm with $p$ processors to
that of a best possible parallelization of a best (known) sequential
algorithm? This definition does not say anything about how well the
parallel or sequential algorithm exploits the hardware capabilities
and how close the performance can come to the nominal
performance\index{nominal processor performance} of the parallel
processor system at hand. This notion of \impi{hardware efficiency}
plays a role in High-Performance Computing\index{HPC} (HPC),
understood here as the
discipline of getting the best out of the given system.

If an algorithm does not have constant efficiency and
linear speed-up\index{speed-up!linear}
for fixed, constant input sizes $n$, we can try to maintain a desired,
constant $e$ efficiency\index{efficiency}
by instead increasing the problem size $n$
with the number of processors $p$. This is the notion of
\impi{iso-efficiency}\cite{GramaGuptaKumar93,GramaKarypisKumarGupta03}
and can be achieved for cost-optimal algorithms.

\begin{definition}[Weak Scalability (constant efficiency)]
A parallel algorithm \apar is said to be \impi{weakly scaling}
relative to sequential algorithm \aseq if, for a desired, constant
efficiency $e$, there is a slowly growing function $f(p)$ such that the
efficiency is $\EFF{p}{n}=e$ for $n$ in $\Omega(f(p))$. The function
$f(p)$ is called the \impi{iso-efficiency function}.
\end{definition}

How slowly should $f(p)$ grow? A possible answer is found in another
definition of weak scaling\index{weakly scaling}.

\begin{definition}[Weak Scalability (constant work)]
A parallel algorithm \apar with work $\wpar{}(n)$ is said to be
\impi{weakly scaling} relative to sequential algorithm \aseq if, by
keeping the average work per processor $\wpar{}(n)/p$ constant at $w$,
the running time of the parallel algorithm $\tpar{p}(n)$ remains
constant. The input size scaling function is $g(p) = \tseq^{-1}(pw)$.
\end{definition}

Ideally, the iso-efficiency function $f(p)$, which tells how $n$ should grow as
a function of $p$ to maintain constant efficiency, should not grow
faster than the input size scaling function $g(p)$, which tells how
much $n$ can at most grow if the average work is to be kept constant:
$f(p)$ should be $O(g(p))$. The two notions may contradict. Constant
efficiency could require larger $n$ than permitted for maintaining
constant average work.  This happens if the sequential running time is
more than linear. Keeping constant efficiency requires $n$ to increase
faster than allowed by constant work weak scaling\index{weakly scaling}.
For such algorithms, constant work is maintained with decreasing
efficiency\index{efficiency}.

\subsection{Scalability Analysis}

How well does a parallel algorithm or implementation now perform
against a sequential counterpart for the problem that we are
interested in, in particular how well can it exploit the available
processor resources? \impi{Scalability analysis} examines this,
theoretically and practically by analyzing (measuring) the parallel
time\index{parallel time}
that can be reached for different number of processors $p$ and
possibly different problem sizes $n$.

\begin{itemize}
  \item
    Strong scaling analysis: Keep the input (size) $n$ constant.  The
    algorithm is \impi{strongly scalable} up to some maximum number
    of processors, as expressed by the parallelism\index{parallelism}
    of the algorithm if the parallel time\index{parallel time}
    decreases proportionally to $p$ (linear speed-up\index{speed-up!linear}).
  \item
    Weak scaling analysis: Keep the average work per processor
    constant by increasing $n$ with the number of processors $p$.  The
    algorithm is \emph{weakly scalable}\index{weakly scaling} if the
    parallel running time remains constant with increasing number of
    processors.
\end{itemize}

A strongly scaling algorithm, a strong property, is able to speed up
the solution of the given problem for some fixed size $n$ (large
enough for parallel execution to make sense) proportionally to the
number of employed processor-cores: our primary \parco goal.  A weakly
scaling algorithm in the sense of constant work per processor is able
to solve larger and larger instances of the problem within an allotted
time frame. Ideally, the time spent when the
processor-cores\index{processor-core} are
performing the same amount of work remains constant regardless of the
number of processors employed. If this is not the case, and the
parallel time\index{parallel time}
is increasing with the number of processors, this
indicates that the parallelization overhead (due to communication,
synchronization, unfavorable load balancing, or redundant
computation)\index{overhead} is increasing with $p$.

\subsection{Relativized Speed-Up and Efficiency}

For very large parallel systems with tens or hundred thousands of
processor-cores, measuring speed-up\index{speed-up}
relative to a sequential baseline
running on one processor may not make sense or even be possible.
The problem size needed
to keep the extreme number of processors busy may simply be too large
(and time consuming) to run on a single processor. Scalability
analysis may in such cases use as baseline the parallel algorithm
running on some number $p'$ of processors (say, $p'=2, p'=100,
p'=1000$ processor-cores). What happens if the number of processors is
doubled? What happens when going from $p'$ to $2p'$ to $10p'$ or to
some $p>p'$ processors? Does the problem size need to increase to
maintain a certain efficiency? The definitions of relative speed-up
and (relative) efficiency\index{efficiency}
can easily be modified to use a different processor baseline $p'$.

\subsection{Measuring Parallel Time and Speed-Up Empirically}

Running parallel programs on a parallel
multi-core processor\index{multi-core processor} or a
large parallel computing system requires quite considerable support
from the system's run-time system: Processor-cores\index{processor-core}
must be allocated
to the program, the program's active entities (processes, threads,
\ldots) must be started and so on, the execution monitored, the
program execution terminated, and the resources be given free for the
next program to use. The measured time for running a full, parallel
application is taken as the \emph{wall-clock time} from starting the
application until the system is free again, in accordance with our
definition of parallel time\index{parallel time}
and assuming that accurate timers are
available, and therefore includes all these surrounding
``overheads''. Benchmarking and assessing the performance (``is this
good enough?'') of an application in this context is done by varying
the inputs, the number of processors used, the system, and other
relevant factors in a systematic and well-documented way.

\parco is most often concerned with the algorithmic building blocks of
such larger applications and these building blocks are the
computational problems we are studying. Benchmarking and performance
assessment is therefore rather done by conducting dedicated
experiments, possibly using specific benchmarking tools, with our
developed kernels and building blocks. A benchmarking program or tool
will invoke the kernel to be benchmarked in a controlled manner. For
\parco with our definition of parallel time, it is thus common to
assume (and therefore ensure) that the available processor-cores to be
used in the assessment will start at the same time (as far as this
makes sense).  This will entail some form of \impi{temporal synchronization}
between the processor-cores, which is in itself a
non-trivial problem in \parco.  Also some means of detecting which
processor-core\index{processor-core} was the last to finish is needed,
possibly by again synchronizing the processor-cores.
As always in experimental science,
measurement and synchronization should be non-intrusive and not affect
or distort the experimental outcome, which is another highly
non-trivial issue. Since computer systems are effectively not
deterministic objects (with respect to timing) and measured run-times may
fluctuate from run to run, kernel benchmarks are repeated a certain
number of times, say $10$, $30$, $100$ times, or until results are
considered stable enough under some statistical measure, or until the
experimenter runs out of time. The time reported by the experiment as
the parallel running time of the algorithm in question may be based on
a statistical measure like average time of the slowest processor-core
over the repetitions or the median time of the measured
times. Sometimes, the fastest time over the repetitions of the
slowest processor-core in each repetition is taken as the parallel
running time. The argument for this is that this best time that the
system could produce can be reproducible and stable over repeated
experiments. A good experiment will clearly describe the experimental
setup and the statistics used in computing and reporting the
run-times. For others to reproduce an experiment and verify claims on
performance, a precise description of the parallel systems is likewise
required: Processor architecture, instruction set, number of
processor-cores, organization and grouping of the cores, clock
frequency, memory, cache sizes, \etc.

In these lecture notes, asymptotic worst-case analysis is used to
judge and compare algorithms, but most often worst-case inputs are not
known and may also not be interesting, common use-cases at
all. Experimental analysis aims at showing performance under many
different inputs, in particular those that are realistic and typical
for the uses of the algorithmic kernel under examination. Experiment
design deals with the construction of good experimental inputs. For
non-oblivious\index{oblivious}
algorithms that are sensitive to the actual input (and
not only the size of the input) it is good practice to always consider
extreme and otherwise special case inputs, such as are expected to
lead to either extremely good or extremely bad performance. Average
case and otherwise ``typical'' inputs are likewise probably of
interest and should be considered.

In \parco we are most often interested in aspects of scalability in
problem size and in particular in number of processor-cores. On both
accounts, it can be considered bad practice to focus only on input sizes
$n$ and especially number of processors $p$ that are powers of
two. The reason for this is that in many algorithms, powers-of-two
are special, and performance in these cases might be either extremely
good or extremely bad. In particular, parallel algorithms are
sometimes designed around communication structures or patterns where
the number of processors is first considered to be some
$p=2^q$. Likewise, some algorithms, for instance, dealing with
two-dimensional matrices, may be special for inputs and number of
processor-cores\index{processor-core} that are square numbers.
Benchmarking for only inputs
$n$ and $p$ that are squares can likewise be highly misleading.
Excluded from these considerations are of course algorithms and
kernels that only work for such special numbers.

\subsection{Examples}
\label{sec:scalabilityexamples}

It is illustrative(!) to strengthen intuition to visualize parallel
running time, (absolute) speed-up\index{speed-up!absolute},
efficiency\index{efficiency}, and iso-efficiency\index{iso-efficiency}
as functions of the number of processors put into solving a problem of
size $n$ (for different $n$).  Let some such problems be given with
best known sequential running times $O(n)\leq c n$, $O(n\log n)\leq c
(n\log n)$, and $O(n^2)\leq c n^2$ as seen many times now in these
lecture notes, for some bounding constant $c, c>0$ (the notation is
sloppy: We mean that the constant of the dominating term hidden within
the $O$ is $c$).

We first assume that the linear $O(n)$ algorithm has been parallelized
by algorithms running work-optimally in $O(n/p+1)\leq C(n/p+1)$,
$O(n/p+\log p)\leq C(n/p+\log p)$, $O(n/p+\log n)\leq C(n/p+\log n)$,
and $O(n/p+p)\leq C(n/p+p)$, respectively, for some bounding constant
$C, C>0$: Also many examples of such algorithms have been (and will
be) seen in the lecture notes.

We first assume that the bounding constants in our sequential and parallel
algorithms are ``in the same ballpark'', and
normalize both constants to $c=C=1$. We plot the parallel running time
as functions of the number of processors $p$ for $1\leq p\leq 128$,
and take $n=128, 128^2$, respectively; these are really small
problems for a linear time algorithm, $128^2=16K$ (and even $128^3=2M$).
The running times are shown in the following two plots.

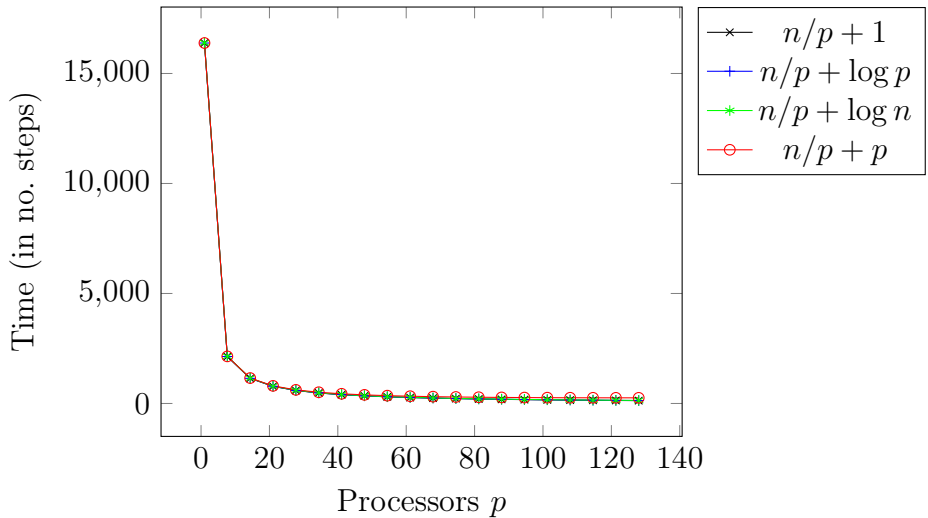
\begin{center}
\begin{tikzpicture}
\begin{axis}[title={Parallel time for $n=128$ and $C=1$.},
    legend entries={$\frac{n}{p}+1$,$\frac{n}{p}+\log p$,$\frac{n}{p}+\log n$,$\frac{n}{p}+p$},
    legend pos=outer north east,
    xlabel={Processors $p$},
    ylabel={Time (in no.\ steps)},
    scaled ticks=false]
  \newcommand{\n}{128}
  \newcommand{\maxp}{128}
  \addplot[mark=x,variable=p,domain=1:\maxp,samples=20,sharp plot] {\n/p+1};
  \addplot[mark=+,blue,variable=p,domain=1:\maxp,samples=20,sharp plot]
      {\n/p+ln(p)};
      \addplot
          [mark=asterisk,green,variable=p,domain=1:\maxp,samples=20,sharp plot]
          {\n/p+ln(\n)};
          \addplot
      [mark=o,red,variable=p,domain=1:\maxp,samples=20,sharp plot]
      {\n/p+p};
\end{axis} 
\end{tikzpicture}
\end{center}

\begin{center}
\begin{tikzpicture}
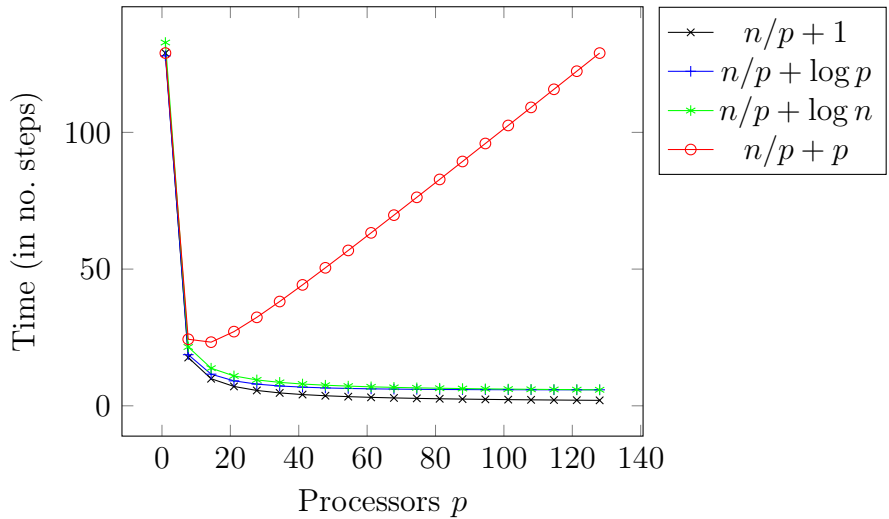
\begin{axis}[title={Parallel time for $n=128^2$ and $C=1$.},
    legend entries={$\frac{n}{p}+1$,$\frac{n}{p}+\log p$,$\frac{n}{p}+\log n$,$\frac{n}{p}+p$},
    legend pos=outer north east,
    xlabel={Processors $p$},
    ylabel={Time (in no.\ steps)},
    scaled ticks=false]
  \newcommand{\n}{128*128}
  \newcommand{\maxp}{128}
  \addplot
      [mark=x,variable=p,domain=1:\maxp,samples=20,sharp plot] {\n/p+1};
      \addplot
          [mark=+,blue,variable=p,domain=1:\maxp,samples=20,sharp plot]
          {\n/p+ln(p)}; \addplot
          [mark=asterisk,green,variable=p,domain=1:\maxp,samples=20,sharp plot]
          {\n/p+ln(\n)}; \addplot
          [mark=o,red,variable=p,domain=1:\maxp,samples=20,sharp plot]
          {\n/p+p};
\end{axis} 
\end{tikzpicture}
\end{center}

The running time (number of steps) plots do not very well
differentiate the four different parallel algorithms. For the larger
problem size, $n=128^2$, there is virtually no difference to be
seen. The shape of the curves for these linearly (perfect) scaling
algorithms is hyperbolic (like $1/p$). The parallel algorithm with
running time $O(n/p+p)$ is interesting: For the small input with
$n=128$, running time decreases until about $p=10$ processors, and
then increases. Indeed the best possible running time of this
algorithm is $\tinf(n)=\sqrt{n}$, and the parallelism\index{parallelism}
is also $n/\sqrt{n}=\sqrt{n}$. This can be seen by minimizing $C(n/p+p)$ for
$p$, which can be done by solving $Cn/p=Cp$ for $p$, giving
$p=\sqrt{n}$ (or more tediously, by calculus).

Plotting instead the absolute (unit-less)
speed-up\index{speed-up!absolute} against the linear (best known)
$O(n)$ algorithm (with $c=C=1$) can highlight the actually different
behavior of the four parallel algorithms. We plot for three problem
sizes $n=128,128^2,128^3$.

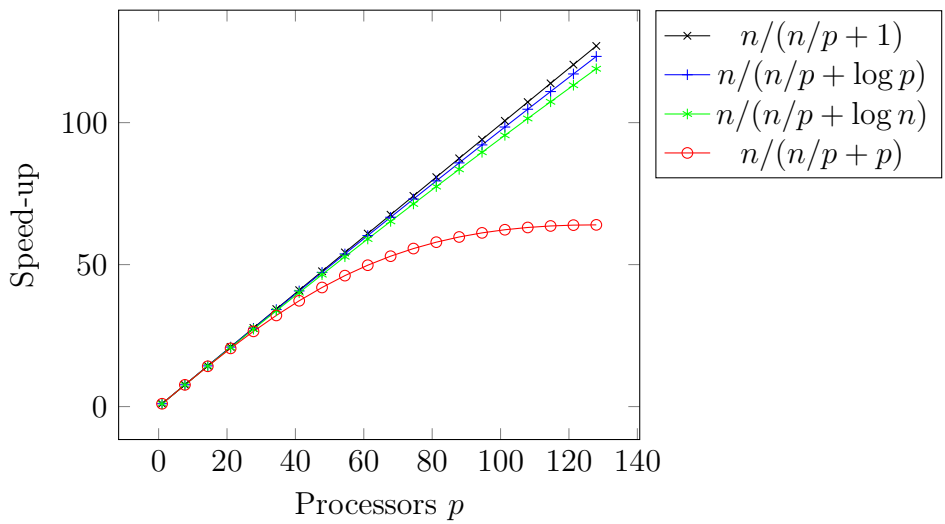
\begin{center}
\begin{tikzpicture}
\begin{axis}[title={Speed-up for $n=128$ and $c=C=1$.},
    legend entries={$\frac{n}{n/p+1}$,$\frac{n}{n/p+\log p}$,$\frac{n}{n/p+\log n}$,$\frac{n}{n/p+p}$},
    legend pos=outer north east,
    xlabel={Processors $p$},
    ylabel={Speed-up},
    scaled ticks=false]
  \newcommand{\n}{128}
  \newcommand{\maxp}{128}
  \addplot
      [mark=x,variable=p,domain=1:\maxp,samples=20,sharp plot]
      {\n/(\n/p+1)}; \addplot
      [mark=+,blue,variable=p,domain=1:\maxp,samples=20,sharp plot]
      {\n/(\n/p+ln(p))}; \addplot
      [mark=asterisk,green,variable=p,domain=1:\maxp,samples=20,sharp plot]
      {\n/(\n/p+ln(\n))}; \addplot
      [mark=o,red,variable=p,domain=1:\maxp,samples=20,sharp plot]
      {\n/(\n/p+p)};
\end{axis} 
\end{tikzpicture}
\end{center}

\begin{center}
\begin{tikzpicture}
\begin{axis}[title={Speed-up for $n=128^2$ and $c=C=1$.},
    legend entries={$\frac{n}{n/p+1}$,$\frac{n}{n/p+\log p}$,$\frac{n}{n/p+\log n}$,$\frac{n}{n/p+p}$},
    legend pos=outer north east,
    xlabel={Processors $p$},
    ylabel={Speed-up},
    scaled ticks=false]
  \newcommand{\n}{128*128}
  \newcommand{\maxp}{128}
  \addplot
      [mark=x,variable=p,domain=1:\maxp,samples=20,sharp plot]
      {\n/(\n/p+1)}; \addplot
      [mark=+,blue,variable=p,domain=1:\maxp,samples=20,sharp plot]
      {\n/(\n/p+ln(p))}; \addplot
      [mark=asterisk,green,variable=p,domain=1:\maxp,samples=20,sharp plot]
      {\n/(\n/p+ln(\n))}; \addplot
      [mark=o,red,variable=p,domain=1:\maxp,samples=20,sharp plot]
      {\n/(\n/p+p)};
\end{axis} 
\end{tikzpicture}
\end{center}

\begin{center}
\begin{tikzpicture}
\begin{axis}[title={Speed-up for $n=128^3$ and $c=C=1$.},
    legend entries={$\frac{n}{n/p+1}$,$\frac{n}{n/p+\log p}$,$\frac{n}{n/p+\log n}$,$\frac{n}{n/p+p}$},
    legend pos=outer north east,
    xlabel={Processors $p$},
    ylabel={Speed-up},
    scaled ticks=false]
  \newcommand{\n}{128*128*128}
  \newcommand{\maxp}{128}
  \addplot
      [mark=x,variable=p,domain=1:\maxp,samples=20,sharp plot]
      {\n/(\n/p+1)}; \addplot
      [mark=+,blue,variable=p,domain=1:\maxp,samples=20,sharp plot]
      {\n/(\n/p+ln(p))}; \addplot
      [mark=asterisk,green,variable=p,domain=1:\maxp,samples=20,sharp plot]
      {\n/(\n/p+ln(\n))}; \addplot
      [mark=o,red,variable=p,domain=1:\maxp,samples=20,sharp plot]
      {\n/(\n/p+p)};
\end{axis} 
\end{tikzpicture}
\end{center}

Speed-up\index{speed-up!absolute}
for the small problem size $n=128$ is not impressive and as
we would like, except for the first parallel algorithm. This
changes drastically and impressively as $n$ grows.  Indeed, for
the ``large'' $n=128^3$ problem, all four parallel algorithms show perfect
speed-up of almost $128$ for $p=128$.

If there is a difference in the bounding constants between sequential
and parallel algorithms, say $c=1$ and $C=10$, which means that the
parallel algorithm is a constant factor of $10$ slower than the
sequential one when executed with only one processor, speed-ups change
proportionally:

\begin{center}
\begin{tikzpicture}
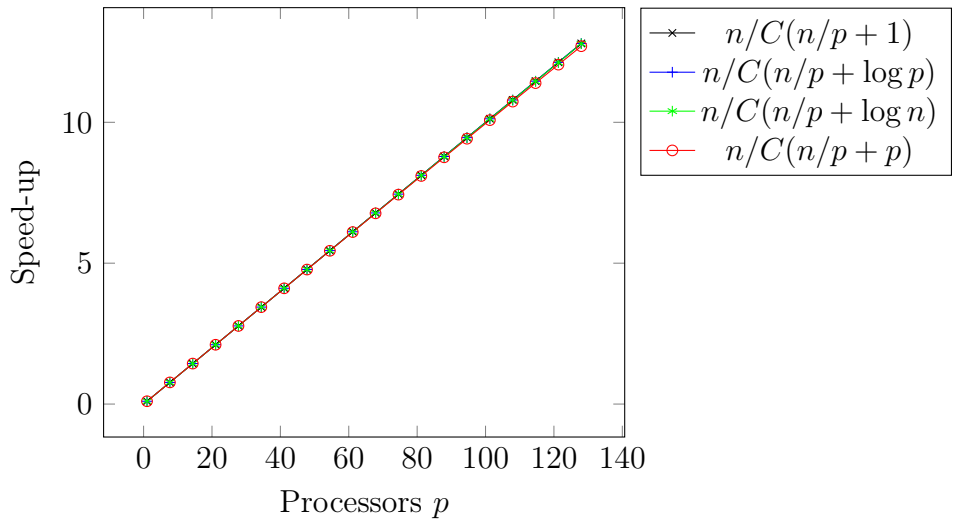
\begin{axis}[title={Speed-up for $n=128^3$ and $c=1, C=10$.},
    legend entries={$\frac{n}{C(n/p+1})$,$\frac{n}{C(n/p+\log p)}$,$\frac{n}{C(n/p+\log n)}$,$\frac{n}{C(n/p+p)}$},
    legend pos=outer north east,
    xlabel={Processors $p$},
    ylabel={Speed-up},
    scaled ticks=false]
  \newcommand{\n}{128*128*128}
  \newcommand{\maxp}{128}
  \newcommand{\CON}{10}
  \addplot
      [mark=x,variable=p,domain=1:\maxp,samples=20,sharp plot]
      {\n/((\n/p+1)*\CON)}; \addplot
      [mark=+,blue,variable=p,domain=1:\maxp,samples=20,sharp plot]
      {\n/((\n/p+ln(p))*\CON)}; \addplot
      [mark=asterisk,green,variable=p,domain=1:\maxp,samples=20,sharp plot]
      {\n/((\n/p+ln(\n))*\CON)}; \addplot
      [mark=o,red,variable=p,domain=1:\maxp,samples=20,sharp plot]
      {\n/((\n/p+p)*\CON)};
\end{axis} 
\end{tikzpicture}
\end{center}

Here, only $1/C$th of the processors are doing productive work in
comparison to the sequential algorithm. Constants \emph{do} matter,
and it is obviously important that sequential and parallel algorithms
have leading constants in the same ballpark. Otherwise, a proportional
part of the processors is somehow wasted.

The parallel efficiency\index{efficiency} indicates how well the
parallel algorithms behave in comparison to a best possible
parallelization with running time $cn/p$. The (unit-less) parallel
efficiencies for the four parallel algorithms are plotted for $n=128,
128^2, 128^3$.

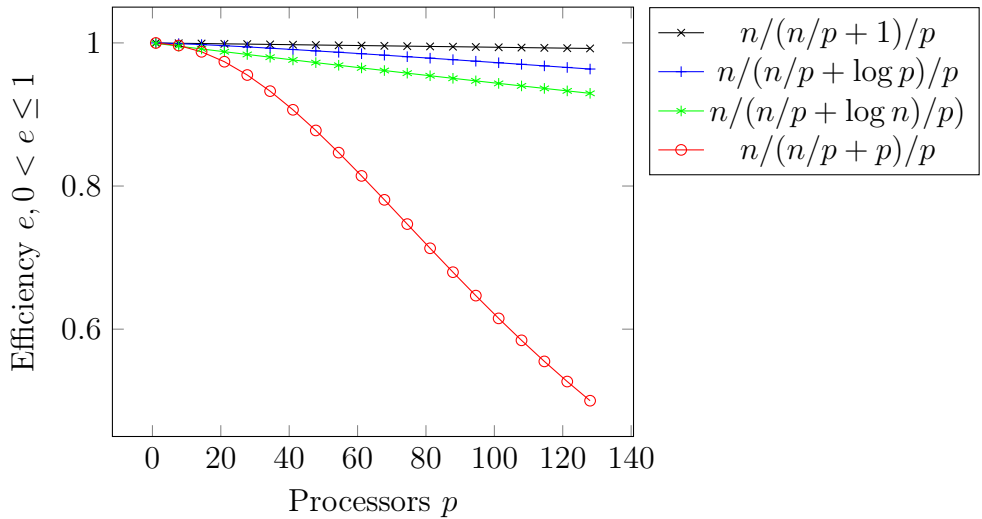
\begin{center}
\begin{tikzpicture}
\begin{axis}[title={Parallel efficiency for $n=128$ and $c=C=1$.},
    legend entries={$\frac{n}{p(n/p+1)}$,$\frac{n}{p(n/p+\log p)}$,$\frac{n}{p(n/p+\log n)}$,$\frac{n}{p(n/p+p)}$},
    legend pos=outer north east,
    xlabel={Processors $p$},
    ylabel={Efficiency $e, 0<e\leq 1$},
    scaled ticks=false]
  \newcommand{\n}{128}
  \newcommand{\maxp}{128}
  \addplot
      [mark=x,variable=p,domain=1:\maxp,samples=30,sharp plot]
      {\n/(\n/p+1)/p}; \addplot
      [mark=+,blue,variable=p,domain=1:\maxp,samples=30,sharp plot]
      {\n/(\n/p+ln(p))/p}; \addplot
      [mark=asterisk,green,variable=p,domain=1:\maxp,samples=30,sharp plot]
      {\n/(\n/p+ln(\n))/p}; \addplot
      [mark=o,red,variable=p,domain=1:\maxp,samples=30,sharp plot]
      {\n/(\n/p+p)/p};
\end{axis} 
\end{tikzpicture}
\end{center}

\begin{center}
\begin{tikzpicture}
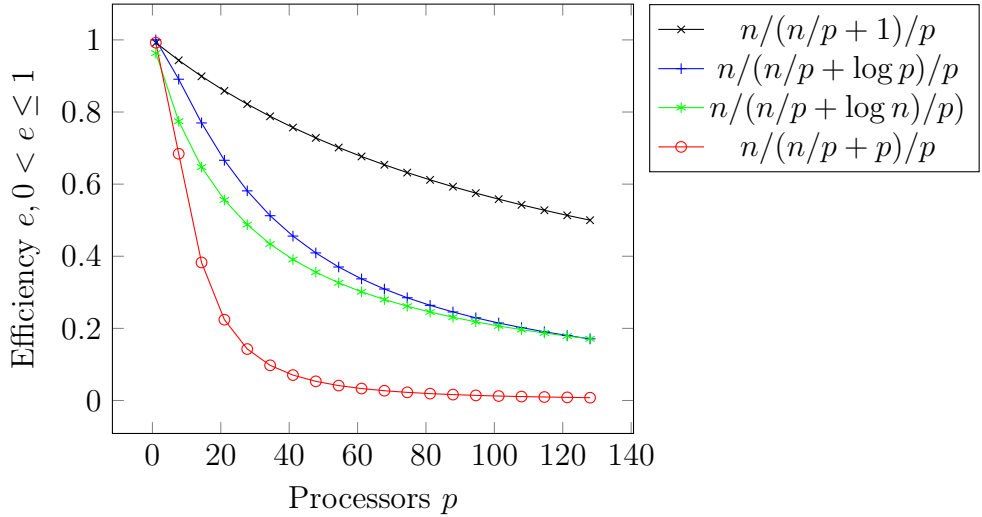
\begin{axis}[title={Parallel efficiency for $n=128^2$ and $c=C=1$.},
    legend entries={$\frac{n}{p(n/p+1)}$,$\frac{n}{p(n/p+\log p)}$,$\frac{n}{p(n/p+\log n)}$,$\frac{n}{p(n/p+p)}$},
    legend pos=outer north east,
    xlabel={Processors $p$},
    ylabel={Efficiency $e, 0<e\leq 1$},
    scaled ticks=false]
  \newcommand{\n}{128*128}
  \newcommand{\maxp}{128}
  \addplot
      [mark=x,variable=p,domain=1:\maxp,samples=30,sharp plot]
      {\n/(\n/p+1)/p}; \addplot
      [mark=+,blue,variable=p,domain=1:\maxp,samples=30,sharp plot]
      {\n/(\n/p+ln(p))/p}; \addplot
      [mark=asterisk,green,variable=p,domain=1:\maxp,samples=30,sharp plot]
      {\n/(\n/p+ln(\n))/p}; \addplot
      [mark=o,red,variable=p,domain=1:\maxp,samples=30,sharp plot]
      {\n/(\n/p+p)/p};
\end{axis} 
\end{tikzpicture}
\end{center}

\begin{center}
\begin{tikzpicture}
\begin{axis}[title={Parallel efficiency for $n=128^3$ and $c=C=1$.},
    legend entries={$\frac{n}{p(n/p+1)}$,$\frac{n}{p(n/p+\log p)}$,$\frac{n}{p(n/p+\log n)}$,$\frac{n}{p(n/p+p)}$},
    legend pos=outer north east,
    xlabel={Processors $p$},
    ylabel={Efficiency $e, 0<e\leq 1$},
    scaled ticks=false]
  \newcommand{\n}{128*128*128}
  \newcommand{\maxp}{128}
  \addplot
      [mark=x,variable=p,domain=1:\maxp,samples=30,sharp plot]
      {\n/(\n/p+1)/p}; \addplot
      [mark=+,blue,variable=p,domain=1:\maxp,samples=30,sharp plot]
      {\n/(\n/p+ln(p))/p}; \addplot
      [mark=asterisk,green,variable=p,domain=1:\maxp,samples=30,sharp plot]
      {\n/(\n/p+ln(\n))/p}; \addplot
      [mark=o,red,variable=p,domain=1:\maxp,samples=30,sharp plot]
      {\n/(\n/p+p)/p};
\end{axis} 
\end{tikzpicture}
\end{center}

Indeed, for work-optimal parallelizations, the
efficiency\index{efficiency} improves greatly with growing problem
size $n$ and is already for $n=128^3$ very close to $1$ for all of the
four parallelizations. The iso-efficiency
functions\index{iso-efficiency} more precisely tell how problem size
must increase with $p$ in order to maintain a given constant
efficiency $e$. We calculate the iso-efficiency functions for the
parallel algorithms as follows.

\begin{itemize}
  \item For parallel running time $n/p+1$ and desired efficiency $e$,
    we have $e=\frac{n}{p(n/p+1)}= n/(n+p)\Leftrightarrow e(n+p) =
    n\Leftrightarrow n = ep/(1-e)$.
  \item For parallel running time $n/p+\log p$ and desired efficiency
    $e$, we have $e=\frac{n}{p(n/p+\log p)} = n/(n+p\log p)\Leftrightarrow
    e(n+p\log p)=n\Leftrightarrow n=ep\log p/(1-e)$
  \item For parallel running time $n/p+p$ and desired efficiency $e$,
    we have $e=\frac{n}{p(n/p+p)}=n/(n+p^2)\Leftrightarrow
    e(n+p^2)=n\Leftrightarrow n=ep^2/(1-e)$
\end{itemize}

The case with parallel running time $n/p+\log n$ is more
difficult. The efficiency\index{efficiency}
calculation gives $e=\frac{n}{p(n/p+\log
n)}=n/(n+p\log n)$ and therefore $n/\log n = ep/(1-e)$, for which we
do not know an analytic, closed-form solution.

We plot the three analytic iso-efficiency functions below for
$p,1\leq p\leq 512$ and $e=90\%$.

\begin{center}
\begin{tikzpicture}
\begin{axis}[title={Iso-efficiency functions for desired efficiency $e=90\%$.},
    legend entries={$\frac{ep}{1-e}$, $\frac{ep\log p}{1-e}$, $\frac{ep^2}{1-e}$},
    legend pos=outer north east,
    xlabel={Processors $p$},
    ylabel={Required $n$},
    scaled ticks=false]
  \newcommand{\efficiency}{(9e-1)}
  \newcommand{\maxp}{512}
  \addplot
      [mark=x,variable=p,domain=1:\maxp,samples=26,sharp plot]
      {\efficiency*p/(1-\efficiency)}; \addplot
      [mark=+,blue,variable=p,domain=1:\maxp,samples=26,sharp plot]
      {\efficiency*p*ln(p)/(1-\efficiency)}; \addplot
      [mark=asterisk,green,variable=p,domain=1:\maxp,samples=26,sharp plot]
      {\efficiency*p*p/(1-\efficiency)};
\end{axis} 
\end{tikzpicture}
\end{center}

For the first two parallel algorithms, the iso-efficiency function is
indeed ``slowly growing'', and according to the first definition of
weak scalability, these algorithms are both strongly and weakly
scaling\index{strongly scaling}\index{weakly scaling}.  With the last
function, where the iso-efficiency function is in $O(p^2)$, it is a
matter of taste whether to still consider it slowly growing. In the speed-up
plots, we indeed let $n$ grow exponentially $n=128, 128^2, 128^3$, and
the speed-up\index{speed-up!absolute} for the latter algorithms was excellent.

We now look at non-linear time sequential algorithms. The $O(n\log n)$
algorithm could be a sorting algorithm (mergesort, say) which could
have been parallelized with running time $O(\frac{n\log n}{p}+\log^2
n)$. The second algorithm is perhaps matrix--vector multiplication,
which can easily be done work-optimally in parallel
time\index{parallel time} $O(\frac{n^2}{p}+n)$ (but also faster).

The corresponding speed-ups for $n=100,1\,000,10\,000,100\,000$ and
$p,1\leq p\leq 1000$ are shown below.

\begin{center}
\begin{tikzpicture}
\begin{axis}[title={Speed-up for $n=100$ and $c=C=1$.},
    legend entries={$\frac{n\log n}{(n\log n)/p+log^2 n}$,$\frac{n^2}{n^2/p+n}$},
    legend pos=outer north east,
    xlabel={Processors $p$},
    ylabel={Speed-up},
    scaled ticks=false]
  \newcommand{\n}{100}
  \newcommand{\maxp}{1000}
  \addplot
      [mark=x,variable=p,domain=1:\maxp,samples=40,sharp plot]
      {(\n*ln(\n))/((\n*ln(\n))/p+ln(\n)*ln(\n))};
      \addplot
      [mark=+,blue,variable=p,domain=1:\maxp,samples=40,sharp plot]
      {\n*\n/((\n*\n)/p+\n)};
\end{axis} 
\end{tikzpicture}
\end{center}

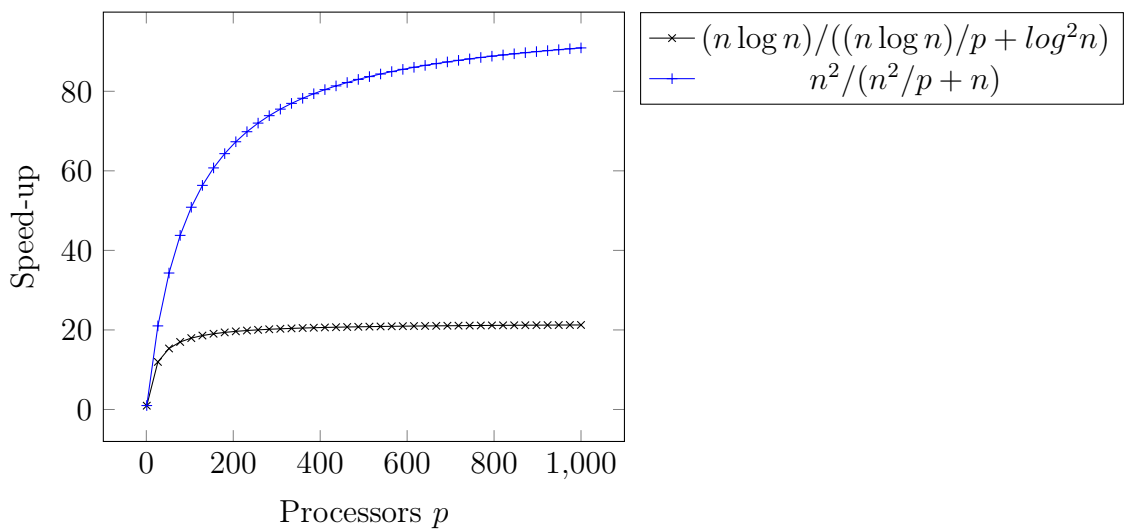
\begin{center}
\begin{tikzpicture}
\begin{axis}[title={Speed-up for $n=1\,000$ and $c=C=1$.},
    legend entries={$\frac{n\log n}{(n\log n)/p+log^2 n}$,$\frac{n^2}{n^2/p+n}$},
    legend pos=outer north east,
    xlabel={Processors $p$},
    ylabel={Speed-up},
    scaled ticks=false]
  \newcommand{\n}{1000}
  \newcommand{\maxp}{1000}
  \addplot
      [mark=x,variable=p,domain=1:\maxp,samples=40,sharp plot]
      {(\n*ln(\n))/((\n*ln(\n))/p+ln(\n)*ln(\n))};
      \addplot
          [mark=+,blue,variable=p,domain=1:\maxp,samples=40,sharp plot]
          {\n*\n/((\n*\n)/p+\n)};
\end{axis} 
\end{tikzpicture}
\end{center}

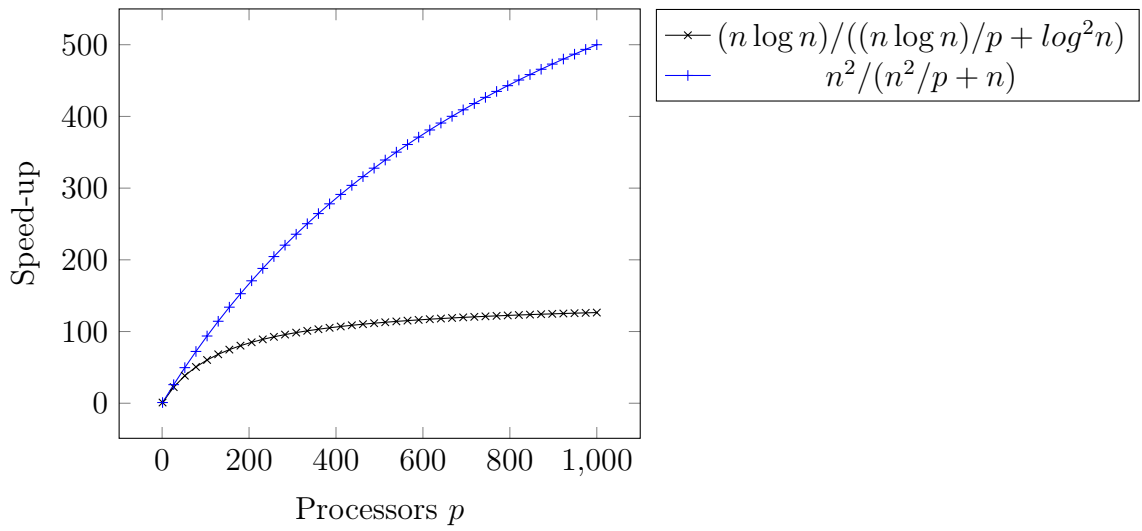
\begin{center}
\begin{tikzpicture}
\begin{axis}[title={Speed-up for $n=10\,000$ and $c=C=1$.},
    legend entries={$\frac{n\log n}{(n\log n)/p+log^2 n}$,$\frac{n^2}{n^2/p+n}$},
    legend pos=outer north east,
    xlabel={Processors $p$},
    ylabel={Speed-up},
    scaled ticks=false]
  \newcommand{\n}{10000}
  \newcommand{\maxp}{1000}
  \addplot
      [mark=x,variable=p,domain=1:\maxp,samples=40,sharp plot]
      {(\n*ln(\n))/((\n*ln(\n))/p+ln(\n)*ln(\n))};
      \addplot
      [mark=+,blue,variable=p,domain=1:\maxp,samples=40,sharp plot]
      {\n*\n/((\n*\n)/p+\n)};
\end{axis} 
\end{tikzpicture}
\end{center}

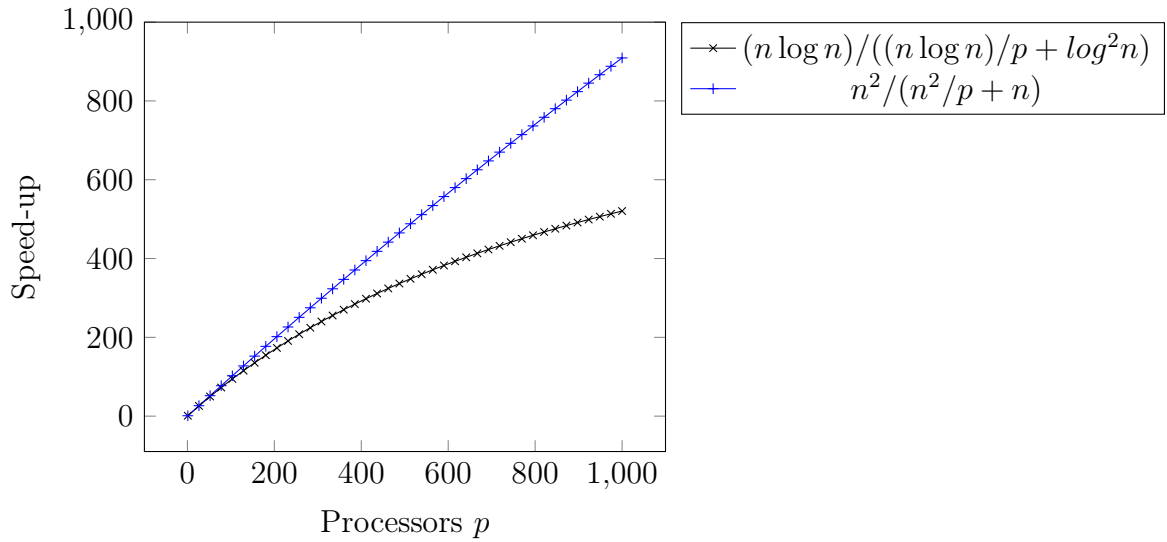
\begin{center}
\begin{tikzpicture}
\begin{axis}[title={Speed-up for $n=100\,000$ and $c=C=1$.},
    legend entries={$\frac{n\log n}{(n\log n)/p+log^2 n}$,$\frac{n^2}{n^2/p+n}$},
    legend pos=outer north east,
    xlabel={Processors $p$},
    ylabel={Speed-up},
    scaled ticks=false]
  \newcommand{\n}{100000}
  \newcommand{\maxp}{1000}
  \addplot
      [mark=x,variable=p,domain=1:\maxp,samples=40,sharp plot]
      {(\n*ln(\n))/((\n*ln(\n))/p+ln(\n)*ln(\n))};
      \addplot
      [mark=+,blue,variable=p,domain=1:\maxp,samples=40,sharp plot]
      {\n*\n/((\n*\n)/p+\n)};
\end{axis} 
\end{tikzpicture}
\end{center}

The parallelization of the low complexity algorithm with sequential
running time $O(n\log n)$ does not scale as well as the other
algorithm. For an $O(n^2)$ algorithm, an input of size $n=100\,000$
is already large, and we did not plot for this large $n$ here. However,
both algorithms clearly approach a perfect speed-up\index{speed-up!perfect}
with growing $n$.

Finally, we illustrate what happens with non work-optimal parallel
algorithms.  Assume we have a parallel algorithm with running times
$O(\frac{n\log n}{p}+1$ relative to a linear time sequential
algorithm, an $O(\frac{n^2}{p}+n)$ parallel algorithm relative to an
$O(n\log n)$ best possible sequential algorithm, and an Amdahl case
where the parallel algorithm has a sequential fraction $s, 0<s<1$ and
parallel running time $O(sn+\frac{(1-s)n}{p})$. Lastly, a parallel
algorithm with a running time of
$O(\frac{n}{\sqrt{p}}+\sqrt{p})=O(\frac{n\sqrt{p}}{p}+\sqrt{p})$
relative to an algorithm that solves an $O(n)$ problem.

\begin{center}
\begin{tikzpicture}
\begin{axis}[title={Speed-up for $n=128$ and $c=C=1$ and sequential fraction $s=0.1$.},
    legend entries={$\frac{n}{(n\log n)/p+1}$,$\frac{n\log n}{n^2/p+n}$,$\frac{n}{sn+(1-s)n/p}$,$\frac{n}{n/\sqrt{p}+\sqrt{p}}$},
    legend pos=outer north east,
    xlabel={Processors $p$},
    ylabel={Speed-up},
    scaled ticks=false]
  \newcommand{\n}{128}
  \newcommand{\maxp}{128}
  \newcommand{\seqfrac}{1e-1}
  \addplot
      [mark=x,variable=p,domain=1:\maxp,samples=20,sharp plot]
      {\n/((\n*ln(\n))/p+1)}; \addplot
      [mark=+,blue,variable=p,domain=1:\maxp,samples=20,sharp plot]
      {\n/(\n*\n/p+ln(\n))}; \addplot
      [mark=asterisk,green,variable=p,domain=1:\maxp,samples=20,sharp plot]
      {\n/(\seqfrac*\n+(1-\seqfrac)*\n/p)}; \addplot
      [mark=o,red,variable=p,domain=1:\maxp,samples=20,sharp plot]
      {\n/(\n/p+sqrt(p))};
\end{axis} 
\end{tikzpicture}
\end{center}

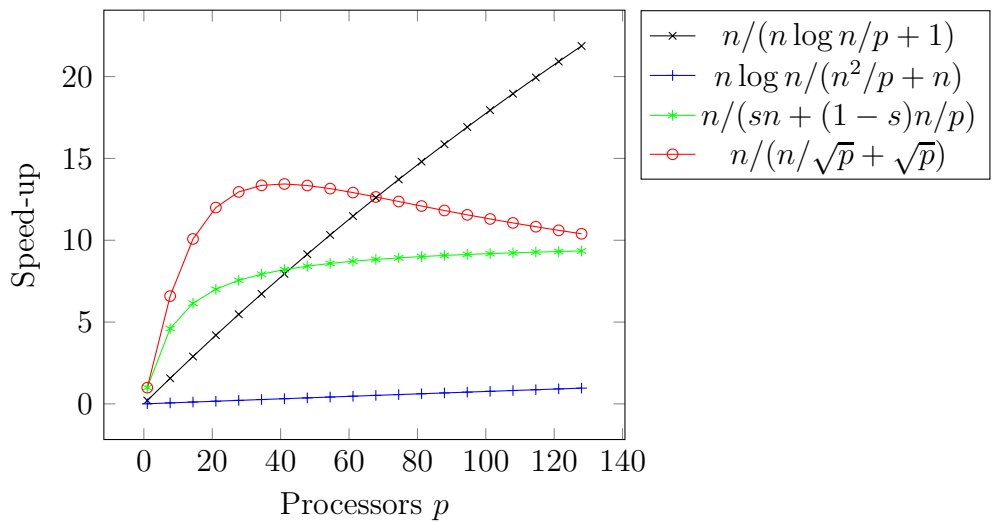
\begin{center}
\begin{tikzpicture}
\begin{axis}[title={Speed-up for $n=128^2$ and $c=C=1$ and sequential fraction $s=0.1$.},
    legend entries={$\frac{n}{(n\log n)/p+1}$,$\frac{n\log n}{n^2/p+n}$,$\frac{n}{sn+(1-s)n/p}$,$\frac{n}{n/\sqrt{p}+\sqrt{p}}$},
    legend pos=outer north east,
    xlabel={Processors $p$},
    ylabel={Speed-up},
    scaled ticks=false]
  \newcommand{\n}{128*128}
  \newcommand{\maxp}{128}
  \newcommand{\seqfrac}{1e-1}
  \addplot
      [mark=x,variable=p,domain=1:\maxp,samples=20,sharp plot] {\n/((\n *
        ln(\n))/p+1)}; \addplot
      [mark=+,blue,variable=p,domain=1:\maxp,samples=20,sharp plot]
      {\n/(\n*\n/p+ln(\n))}; \addplot
      [mark=asterisk,green,variable=p,domain=1:\maxp,samples=20,sharp plot]
      {\n/(\seqfrac*\n+(1-\seqfrac)*\n/p)}; \addplot
      [mark=o,red,variable=p,domain=1:\maxp,samples=20,sharp plot]
      {\n/(\n/sqrt(p)+sqrt(p))};
\end{axis} 
\end{tikzpicture}
\end{center}

The two plots illustrate the Amdahl case well: Speed-up is bounded by
$1/s$ (here $10$ for $s=10\%$) independently of $n$. The first two
algorithms have a diminishing speed-up\index{speed-up}
with increasing $n$. These two
algorithms have parallel work determined by the problem size which is
asymptotically larger than the sequential work. For the last
algorithm, the parallel work increases ``slowly'' by a factor of
$\sqrt{p}$ with $p$, and therefore the speed-up of this algorithm does
indeed improve with increasing problem size $n$, but is $o(p)$ and not
linear.

\section{Third block (1--2 Lectures)}
\label{sec:patterns}

In this part of the lecture notes, we take a closer look at the way
(parallel) work may be structured. The most important structures
discussed are work expressed as dependent tasks and work expressed as
loops of independent iterations. The latter can be seen as an
expression of recurring, similar computations in algorithms,
pseudo-code and actual programs:
A \emph{parallel programming pattern}\index{pattern!parallel} or
\emph{parallel design pattern}\index{pattern!design}.
The later part of this lecture block
gives further examples of parallel algorithmic design patterns for
which (good) parallelizations are known, including
pipeline, stencil, master-slave/master-worker, reductions, data
redistribution, and barrier synchronization.
Parallel design patterns can, explicitly and implicitly, provide
useful guidance for building
parallel applications, sometimes even as concrete building
blocks~\cite{MattsonSandersMassingill05,McCoolRobisonReinders12}.
We illustrate many of the patterns by sequential code snippets
using C~\cite{KernighanRitchie88} to specify the intended outcome
and semantics and use these descriptions to argue for lower
bounds on the parallel performance with given numbers of processors $p$.

\subsection{Directed Acyclic Task Graphs\marksec}
\label{sec:taskgraphs}

A \impi{Directed Acyclic (task) Graph (DAG)}, $G=(V,E)$, consists of a
set of \emph{tasks}, $ t_i\in V$, which are sequential computations
that will not be analyzed further (sometimes also called
\impi{strands}). Tasks are connected by directed \impi{dependency edges},
$(t_i,t_j)\in E$. An edge $(t_i,t_j)$ means that task $t_j$
is \emph{directly dependent} on task $t_i$ and cannot be executed
before task $t_i$ has completed, for instance, because the input data
for task $t_j$ are produced as output data by task $t_i$. In general,
a task $t_j$ is \emph{dependent} on a task $t_i$ if there is a
directed path from $t_i$ to $t_j$ in $G$. If there is neither a
directed path from $t_i$ to $t_j$ nor a directed path from $t_j$ to
$t_i$ in $G$, the two tasks $t_i$ and $t_j$ are said to be
\emph{independent}. Independent tasks could possibly be executed in
parallel, if enough processor-cores\index{processor-core} are available,
since neither task
needs input from nor produces output to the other. A task
$t_i$ may produce data for more than one other task, so there may
be several outgoing edges from $t_i$. Likewise, a task $t_j$ may need
immediate input from more than one task, so there may be several
incoming edges to $t_j$. Since $G$ is acyclic, there is at least one
task $t_r$ in $G$ with no incoming edges; such tasks are called
\emph{root}\index{root task} or \emph{start}\index{start task}
tasks. Likewise, there is at least one task $t_f$ with no outgoing
edges. Such tasks are called \emph{final}\index{final task}.

Many computations can be pictured as task graphs. Consider as a first
example the execution of the recursive Quicksort\index{Quicksort} algorithm.
The tasks may be the computations done in pivot selection and partitioning with
a dependency from a pivoting task to the ensuing partitioning
task. The root task will be the initial pivot selection in the input
array, followed by the first partitioning task of the whole
array. Dependent tasks will now be the pivot selection and partitioning of
the two independent parts of the partitioned array and so on and
so on. A final task will depend on all partitioning tasks to have
completed and will indicate that the array has been Quicksorted.  We will
see later in these lecture notes how such task graphs suitable for
parallel execution can be generated dynamically as \openmp
tasks\index{OpenMP} or with Cilk\index{Cilk}. Another often
encountered type of task DAG is the \impi{fork-join} DAG: A dependent
sequence of fork-join tasks, where each task has a number of
dependent, forked tasks that
are all connected to the next join task. A fork-join DAG is the standard
structure of \openmp programs corresponding to a sequence of loops of
independent
iterations\index{loop of independent iterations}, each of which
can be executed in parallel as a set of forked, independent tasks.

For computations structured as task graphs, there is normally a single
start task taking input of size $O(n)$ and a single, final task
producing the results of the computation. In a dynamic setting, the
task graph typically depends on the input, which can be emphasized by
writing $G(n)$. This $n$ is not to be confused with the number of tasks in
$G$.

Each task $t_i$ has an associated amount of work and takes sequential
time $T(t_i)$, typically also depending on $n$. The total amount of
work\index{DAG!work} of a given task graph $G=(V,E)$ with $k$ tasks
$t_0,t_1,\ldots,t_{k-1}$ is given by the total time of all tasks and
is denoted by
$T_1(n)=\sum_{i=0}^{k-1}T(t_i)$. We will again compare against a best
known sequential algorithm for the problem we are solving, so it holds
that $T_1(n)\geq\tseq(n)$.

Doing a computation as specified by a task graph $G$ sequentially on
a single processor-core amounts to the following: Pick a task $t$ with
no incoming edges and execute it. Remove all outgoing edges $(t,t')$
from $G$. Continue this process until there are no more tasks in
$G$. Since $G$ is acyclic, there is at least one root task from which
the execution can be started. After execution of this $t$, if $t$ is not
the last task, there will
be at least one task with no incoming edges, \etc (if not, $G$ would not be
acyclic).  Sequential execution of a task graph, therefore, amounts to
executing the tasks (nodes) in some \impi{topological order}. Any DAG
has a topological order (as can be determined sequentially in
$O(k)$ time steps~\cite{CormenLeisersonRivestStein22}).
A task that has become
eligible for execution by having no incoming edges is said to be
\emph{ready}. Since all tasks of $G$ are executed, each task exactly
once, and since there is at least one ready task after completion of
a task, the time taken for the sequential execution is $O(T_1(n))$.

Imagine that several processor-cores\index{processor-core} are available.
A parallel execution of a computation specified by a task graph $G$ could
proceed as follows: Pick a ready task. If there is a processor-core that is
not busy executing, assign the task to this core. When a task is
completed, remove all outgoing edges, possibly giving rise to further,
ready tasks (but also possibly not, tasks may have many incoming
edges). Continue this process until there are no more ready tasks.
The resulting order of tasks and assignment to processor-cores is
called a \impi{schedule}. The central property of a schedule is that
both dependencies and processor availability are respected: A task is not
executed before all incoming edges have been removed, which means that
dependencies have been resolved and data been made available to the task;
at no time, a processor-core is assigned more than one task; but at
times, cores may be unassigned and idle.

We are interested in the time taken to execute the work $T_1(n)$ with
some schedule with $p$ processors. This is given by the time for the
last task to finish. We denote the execution time of a (for now not
further specified) $p$ processor schedule by $T_p(n)$ and are, of
course, interested in finding fast schedules.

No matter how scheduling is done, the total amount of work $T_1(n)$
can never be completed faster than $T_1(n)/p$, the best possible
parallelization. Also, no matter how scheduling is done, tasks that
are dependent on each other must be executed in order. Consider a
heaviest path $(t_r,t_1,\ldots,t_f)$ from the start task $t_r$ to a
final task $t_f$ with the largest amount of total work over the tasks
$t_i$ on the path and define $\tinf(n)=T(t_r)+T(t_1)+\ldots+T(T_f)$ as
the work along such a heaviest path. With sufficiently many processor-cores
available (this number is suggested by $\infty$),
indeed a schedule exists that can achieve
running time $\tinf(n)$ (think about this). Clearly, for any schedule,
$T_p(n)\geq\tinf(n)$. These two observations are often summarized as
follows:

\begin{itemize}
\item
  \emph{Work Law}: $T_p(n)\geq T_1(n)/p\geq \tseq(n)/p$\index{Law!Work Law},
\item
  \emph{Depth Law}: $T_p(n) \geq \tinf(n)$\index{Law!Depth Law}.
\end{itemize}

The work on a heaviest path in a task graph $G$ is often also called
the \impi{span} or the \emph{depth}\index{DAG!depth}\index{DAG!span} of the
DAG. A heaviest path is commonly referred to as a \impi{critical path}
with \emph{length} or \emph{weight} $\tinf$.
It is also the parallel time complexity
of the DAG\index{parallel time complexity}.

As an example, consider a fork-join DAG with start and final tasks
$t_r$ and $t_f$, with $T(t_r)=1$ and $T(t_f)=1$. The start task forks
to a heavier task $t_1$ with $T(t_1)=4$, and to, say, $27$ light tasks
with one unit of work. All forked tasks join at the final task. Thus,
$T_1(n)=1+4+27+1=33$ and $\tinf(n)=1+4+1=6$. With $p=3$, the Work Law
says that $T_p(n)\geq 33/3=11$ with a (relative) speed-up of at most
$T_p(n)/T_1(n) = 3$ and the Depth Law that $T_p(n)\geq 6$.

With more than, say, $p=10$ processors, the Work law\index{Law!Work Law}
gives a running time of at least $T_1(n)/p\geq 33/11 = 3$ which is
less than $\tinf(n)=6$ and, therefore, not possible according to the
Depth Law. The maximum speed-up\index{speed-up!relative} achievable
is obviously given by $T_1(n)/\tinf(n)=33/6=5.5$.

For any schedule, the speed-up is bounded as follows:
\begin{displaymath}
  \SU{p}{n} = \frac{\tseq(n)}{\tpar{p}(n)} \leq \frac{T_1(n)}{T_p(n)}
  \leq \frac{T_1(n)}{\tinf(n)} \quad .
\end{displaymath}

The \impi{parallelism} $\frac{T_1(n)}{\tinf(n)}$ is, therefore, an upper
bound on the achievable relative speed-up and also gives the largest number of
processor-cores for which linear speed-up\index{speed-up!relative}
could be possible.
  
\emph{Critical path analysis} consisting in finding the longest chain
of dependent, sequential work over all tasks, as used in the Depth
Law, is an important tool to analyze the potential for parallelizing a
computation when thinking of the computation as a task graph. If, for
instance, the critical path $\tinf(n)$ is a constant
fraction of $T_1(n)$, Amdahl's Law applies\index{Law!Amdahl's Law}, which is a
sign that a better algorithm and a better DAG must be found.

We now consider a specific scheduling strategy, so-called
\impi{greedy scheduling}.  A greedy scheduler assigns a ready task to an
available processor as soon as possible (task ready and processor
available), meaning that a processor-core\index{processor-core}
is idle only in the case
when there is no ready task. Greedy schedules have a nice upper bound on
the achieved running time, which is captured in the following theorem.

\begin{theorem}[Two-optimality of greedy scheduling]
  \label{thm:greedy}
Let $T_p(n)$ be the execution time of a DAG $G(n)$ with any \emph{greedy
schedule} on $p$ processors, and let $T_p^{*}(n)$ be the execution time
with a best possible $p$ processor schedule. It holds that
\begin{eqnarray*}
  T_p(n) & \leq & \floor{T_1(n)/p} + \tinf(n) \\
  & \leq & 2T^{*}_p(n)
  \quad .
  \end{eqnarray*}
\end{theorem}

The proof can be sketched as follows: Divide the work of the scheduler
into discrete steps. A step is called \emph{complete} if all
processor-cores are busy on some tasks and \emph{incomplete} if some
cores are idle, which is the case when there are less ready tasks than
processor-cores in that step.  Then, the number of complete steps is
bounded by $\floor{T_1(n)/p}$; if there were more, more than the total
work $T_1(n)$ would have been executed. The number of incomplete steps
is bounded by $\tinf(n)$, since each incomplete step reduces the work
on a critical path. The Work and the Depth Law hold for any $p$
processor schedule, in particular for a best possible schedule,
so $T_1(n)/p\leq T^{*}(n)$ and $\tinf(n)\leq T^{*}(n)$ and
the last upper bound follows. The theorem, therefore, states that
the execution time that can be achieved by a greedy schedule is
bounded by two times what can be achieved by a best possible schedule,
a guaranteed two-approximation!

Neither the definition of greedy schedules nor the theorem says how a
greedy scheduler can or should be implemented. But if it can be shown
by some means that a proposed scheduling algorithm is greedy, the
greedy scheduling theorem says that the running time is within a
factor two of best possible. Greedy scheduling is sometimes called
\impi{list scheduling} and the argument for Theorem~\ref{thm:greedy} is
also known as Brent's Theorem\index{list scheduling}\index{Brent's Theorem}
as discussed in \Sec~\ref{sec:cost-work}.
Later in these lecture notes, we will briefly touch on
\impi{work-stealing} which is a decentralized, randomized, greedy
scheduling strategy for certain kinds of DAGs, like the one explained for
Quicksort\index{Quicksort}
(called strict, spawn-join DAG's)~\cite{AroraBlumofePlaxton01}.

Some parallel programming models\index{programming model}
and frameworks make it possible to
dynamically construct what effectively amounts to directed acyclic
task graphs, sometimes with additional structural properties, as the
parallel execution progresses. The run-time system for such frameworks
execute a (greedy) scheduling algorithm using the properties of the
task graph. With the help of Theorem~\ref{thm:greedy}, it is sometimes
possible to give provable time bounds for programs executed on such systems.
Examples are \openmp tasks, which will
be covered in detail later (see \Sec~\ref{sec:omptask}), and \cilk, \index{Cilk}
which we will briefly touch
upon~\cite{BlumofeJoergKuszmaulLeisersonRandallZhou96,Leiserson10,SchardlLee23}\footnote{See also \url{www.opencilk.org}}.

\subsection{Loops of Independent Iterations}
\label{sec:looppattern}

Computations are often expressed as loops, in algorithmic pseudo-code
and in real programs. A computation is to be performed for the
different values of the loop iteration variable in the range of this
variable, typically in increasing order of the loop variable:

\begin{lstlisting}[style=SnippetStyle]
for (i=0; i<n; i++) {
  c[i] = F(a[i],b[i]);
}
\end{lstlisting}

In this loop, the iterations (different values of the
iteration variable \texttt{i}) are \emph{independent} of each other
(provided the function \texttt{F} has no side effects): No computation
for iteration $i$ is affected by any computation for iteration $i'$
before $i$, $i'<i$, and no computation for a later iteration $i''$,
$i''>i$, could possibly affect the computation for iteration $i$. In
such a case, the loop could be trivially parallelized by dividing the
iteration space into $p$ roughly equal-sized blocks of about $n/p$
iterations and letting each block be executed by a chosen
\index{processor-core}processor-core.

The assignment of blocks, more generally individual iterations, to
processor-cores is called \impi{loop scheduling} and can be done
either fully explicitly (as sometimes needed when parallelizing with
\pthreads, see \Sec~\ref{sec:threadloop},
or with \mpi, see lecture Block~\ref{blk:mpi})\index{pthreads}\index{MPI}
or implicitly with the aid of
a suitable compiler and runtime system by marking the loop
(actually a bad name, since ``loop'' normally implies an order)
as consisting of independent iterations (another misnomer in this context,
``iteration'' implies sequential dependency) and, therefore,
parallelizable. An example, which we will see again in much detail later, is
the following \openmp\index{OpenMP} style parallelization of a loop:

\begin{lstlisting}[style=SnippetStyle]
#pragma omp parallel for
for (i=0; i<n; i++) {
  c[i] = F(a[i],b[i]);
}
\end{lstlisting}

With the PRAM\index{PRAM} model, independent loop computations were handled by
simply assigning a processor to each iteration with the
\textbf{par}-construct:

\begin{lstlisting}[style=SnippetStyle]
par (0<=i<n) {
  c[i] = F(a[i],b[i]);
}
\end{lstlisting}

The parallel time\index{parallel time} of this ``loop'' on a
PRAM would be $O(1)$ 
steps and the total number of operations $O(n)$ assuming that
each evaluation of the function $F$ also takes only a constant number
of time steps. On a parallel computer with $p$ processor-cores,
optimistically, the parallel loop can be executed in $\Omega(n/p+1)$ time
steps by splitting the $n$ iterations roughly evenly between the $p$
processors. The constant term is supposed to account for overheads in
splitting and assigning the iterations to the processors. This assumes
that also the number of iterations $n$ is known in advance and that
this $n$ is not changed during the iterations. On parallel computers
where the processors are not operating synchronously in lock-step like 
in the PRAM\index{PRAM}, a barrier synchronization (see \Sec~\ref{sec:barrier})
may be needed after the processor-cores have finished their iterations
in order to ensure that the results in the \texttt{c}-array are all available
to all processors. The parallel time of a ``parallel loop'' may, thus,
have to include the time needed for the barrier synchronization and will
be determined by the slowest processor-core\index{processor-core} to finish.
Load imbalance\index{load imbalance} could become an issue.

The loop of independent iterations pattern with the function
\texttt{F} being a simple, arithmetic-logic expression with
the same number of primitive instructions to be executed independently
of the actual argument values is a standard way of expressing a SIMD
parallel computation\index{SIMD}. One single stream of instructions
controls the computations on multiple data, namely for all the $n$
inputs of the iteration space. If the processor-architecture has
actual SIMD\index{SIMD} instructions, the loop of independent instructions
could be a way to instruct the compiler to use these instructions (see
\Sec~\ref{sec:simdloops}).

\subsection{Independence of Program Fragments\marksec}

Independent loop iterations, in general, independent program fragments
(which could be tasks as in \Sec~\ref{sec:taskgraphs}) could
possibly be executed concurrently, in parallel by different, available
processor-cores. Independence of program fragments is a
sufficient condition for allowing parallel execution.

Straightforward conditions for independence of program fragments are
the three \impi{Bernstein conditions}~\cite{Bernstein66}. Let $P_i$
and $P_j$ be two program fragments, with $P_j$ following after $P_i$
in the program flow. Each of $P_i$ and $P_j$ has a set of (potential)
input variables $I_i$ and a set of (potential) output variables $O_i$.
These sets can be determined statically by program analysis, but whether a
potential output variable will actually be assigned is, in general,
undecidable. The fragments $P_i$ and $P_j$ are \emph{dependent} if
either
\begin{enumerate}
\item
  $O_i\cap I_j\neq\emptyset$ (a \emph{true dependency}, or \emph{flow
dependency}), or
\item
  $I_i\cap O_j\neq\emptyset$ (an anti-dependency), or
\item
  $O_i\cap O_j\neq\emptyset$ (an output dependency).
\end{enumerate}

The conditions are obviously sufficient but not necessary: Either may
hold, but input or output may not be read or written by the program
fragment or read or written in some specific order such that the outcome of
the parallel execution is still correct.

Dependencies between the iterations of a loop are called \emph{loop
carried dependencies}\index{loop dependency!loop carried dependency},
and there are three types, corresponding to the three Bernstein conditions.

In a
\emph{loop carried flow dependency}\index{loop dependency!loop carried flow dependency},
the outcome of an earlier iteration affects the computation of a later
iteration:

\begin{lstlisting}[style=SnippetStyle]
for (i=k; i<n; i++) {
  a[i] = a[i]+a[i-k];
}
\end{lstlisting}

Here, the simple computation in iteration $i$ is dependent on output in
variable \texttt{a[i-k]} produced
in iteration $i-k$ (assuming that $k>0$; for $k=0$ there would be no
such dependency), an earlier iteration if the
iterations were executed in increasing, sequential order.
Such iterations can, therefore, not be done in parallel when expecting a
correct outcome.

In a
\emph{loop carried anti-dependency}\index{loop dependency!loop carried anti-dependency},
the outcome of a later iteration 
affects an earlier iteration, if the two iterations were
reversed or carried out simultaneously:

\begin{lstlisting}[style=SnippetStyle]
for (i=0; i<n-k; i++) {
  a[i] = a[i]+a[i+k];
}
\end{lstlisting}

The later iteration $i+k$ updates a variable that is used in iteration $i$,
so if iteration $i+k$ would have been executed before or concurrently with
iteration $i$, the output would be different than expected from a
sequential execution in increasing iteration order
and presumably not be correct.

Finally, in a
\emph{loop carried output dependency}\index{loop dependency!loop carried output dependency},
two or more iterations write to the same output variable(s).
If executed simultaneously, the output
would not be well-defined unless the same value is written for all
iterations $i$ (as allowed on the Common
CRCW PRAM\index{PRAM!Common CRCW}).

\begin{lstlisting}[style=SnippetStyle]
for (i=0; i<n-k; i++) {
  a[0] = a[i];
}
\end{lstlisting}

This is a first example of a \impi{race condition}, about which we will
learn more in later parts of the lecture notes.

Some loop carried dependencies can be removed by appropriate program
transformations. For instance, the loop carried anti-dependency can be
eliminated by introducing an auxiliary array \texttt{b} into which the
results from the computations on array \texttt{a} are written:

\begin{center}
\begin{minipage}{0.40\linewidth}
\begin{lstlisting}[style=SnippetStyle]
for (i=0; i<n-k; i++) {
  a[i] = a[i]+a[i+k];
}
\end{lstlisting}  
\end{minipage}
\hspace{0.05\linewidth}
$\longrightarrow$
\hspace{0.05\linewidth}
\begin{minipage}{0.40\linewidth}
\begin{lstlisting}[style=SnippetStyle]
for (i=0; i<n-k; i++) {
  b[i] = a[i]+a[i+k];
}
\end{lstlisting}
\end{minipage}
\end{center}

The transformed (rewritten) loop now consists of independent iterations,
and, therefore, the iterations can be executed in any order and concurrently,
in parallel. Depending on the surrounding program logic (where is the
result expected?),
this may have to be followed by a loop (of independent iterations) to copy
\texttt{b} back to \texttt{a}, taking $O(n)$ operations, or by a swapping of
the two arrays, taking $O(1)$ operations. By similar tricks, sometimes
other types of dependencies can be eliminated.

A \impi{parallelizing compiler} would analyze loops and other
constructs for dependencies and remove dependencies where possible by
appropriate transformations in order to generate code that can exploit a
large number of available processor-cores. Since the dependency
problem is in general undecidable, there is a limit to what such
compilers can do. Results may be modest~\cite{Midkiff12}.

\subsection{Pipeline\marksec}
\label{sec:pipeline}

Consider the following nested loop computation:

\begin{lstlisting}[style=SnippetStyle]
for (i=0; i<m; i++) {
  for (j=1; j<n; j++) {
    a[i][j] = a[i][j-1]+a[i][j];
  }
}
\end{lstlisting}

The inner loop on \texttt{j} clearly contains a loop carried
flow-dependency and, therefore, cannot be parallelized without
sacrificing correctness as defined by the sequential loop order. The
outer loop on \texttt{i} contains $O(n)$ work per iteration which could
be performed in parallel with up to $m$ processors. The parallel time
would be $\Omega(\frac{m}{p} n)$ with up to at most $m$ processors. We write
this argument compactly as $\Omega(\frac{m}{p}n+n)$
parallel time\index{parallel time}.

A different way of assigning processors to the doubly nested loop work
would be the following: Assume that up to $n$
\index{processor-core}processor-cores are
available.  A processor is assigned for each index $j$ in the inner
loop. The $j$th such processor first sits idle for $j-1$ rounds to wait for
\texttt{a[0][j-1]} to have been computed by processors
$0,1,\ldots, j-1$ before $j$. Now, processor $j$ can compute the value
\texttt{a[0][j]} followed by the values \texttt{a[i][j]} for
$i=1,\ldots,m-1$.  This latter viewpoint of the computation
is a \impi{linear pipeline}.
The parallel running time for computing all the values in
the two-dimensional array can be found by looking at the last processor
$n-1$. This processor will have to wait for $n-1$ rounds before it can
start computing values, after which it can compute a new value for the
remaining $m-1$ elements. This gives a running time of $O(n+m-1)=O(n+m)$
with $p=n$ processor-cores. For $p\leq n$ processors the running time
can be stated as $\Omega(\frac{n}{p}m+m+n)$.

The general, linear pipelining\index{pipelining} pattern
assumes that a number of $m$
work items are to be processed, each requiring work that can be
structured into a sequence of $n$ successive \emph{stages} that have
to be carried out one after the other and each take roughly the same
(not necessarily constant amount of) time.  The pipelining pattern allows
parallelization by assigning up to $n$ processors to the individual
stages.

Pipelining is a surprisingly versatile technique, which can lead to
highly efficient and fast parallel algorithms for some
problems. Pipelining\index{pipelining} is,
for instance, used in algorithms for data
exchange problems (see \Sec~\ref{sec:switching}).
Pipelines can be more complex, directed, acyclic dependency graphs like,
for instance, series-parallel graphs.
The essence is that work items pass through the stages of the pipeline,
perhaps being split or combined, and that the parallelism comes from
stages that work in parallel on different work items. The number
of processors that can be employed is thus determined by the number of
pipeline stages and the parallel time\index{parallel time}
by the number of work items.

\subsection{Stencil}
\label{sec:stencil}

Here is another frequently occurring nested loop computation. An
element of a two-dimensional $m\times n$ matrix \texttt{b[i][j]} is
updated with the result of a constant time computation on a small set
of elements of another matrix \texttt{a[i][j]}. In the example here,
each update is a simple average function \texttt{avg} on eight
elements neighboring \texttt{a[i][j]} and \texttt{a[i][j]} itself.
The updates in the \texttt{b} matrix depend on the \texttt{a} matrix,
but by using two matrices, the computation has no loop carried
dependencies. Therefore, both loops could possibly be perfectly
parallelized. Since each element update takes constant time, the total
amount of computation for updating all elements is in $O(mn)$ which
with $p$ processor-cores could ideally be done in $\Theta(mn/p+1)$
time steps. In a PRAM implementation\index{PRAM}, a processor is
assigned to each matrix element to do the update; with less than $mn$
processor-cores, the matrix is thought of as divided into $p$ parts,
typically block submatrices, and a
processor-core\index{processor-core} is assigned to each block to do
all the updates for the block (see \Sec~\ref{sec:semanticterms}
and~\ref{sec:organizingprocesses}).  After the update step, the two
matrices are swapped. The computation is repeated until some criteria
is met and the \texttt{done}-flag is set to \textbf{true}.  The
pattern is an example of a two-dimensional, so-called $9$-point
\impi{stencil computation}.  The matrices are assumed to have also
rows and columns indexed as \texttt{a[-1][j]}, \texttt{a[m][j]},
\texttt{a[i][-1]} and \texttt{a[i][n]}, respectively. This border of
\impi{ghost} rows and columns is sometimes called the \impi{halo}, and
in this example, the halo is of depth one.

As an aside, the code snippet illustrates a handy way of handling
two-dimensional arrays in C (for the best introduction to C,
see~\cite{KernighanRitchie88}).  Matrices are stored in row-major
order\index{row-major order}:
one row of consecutive elements (here of type \texttt{double})
after the other.  Each matrix is declared as a pointer to an array of
rows of $n+2$ elements, and $(m+2)(n+2)$ elements are allocated for
each matrix.  By pointer arithmetic, adding one full row and one
element, the matrix with halo can be conveniently addressed. The C
compiler can, since $n+2$ is known (although not static), compute the
starting address of each row in the allocated storage. This will also
work for higher-dimensional matrices as long as the sizes of the
lowest, faster changing dimensions are given in the declaration.

\begin{lstlisting}[style=SnippetStyle]
double (*a)[n+2];
double (*b)[n+2];
double (*c)[n+2];
double (*aa)[n+2];
double (*bb)[n+2];

a  = (double(*)[n+2])malloc((m+2)*(n+2)*sizeof(double));
aa = a; // save allocated address
// and shift address by one row and one column
a  = (double(*)[n+2])((char*)a+(n+2+1)*sizeof(double)); 

b  = (double(*)[n+2])malloc((m+2)*(n+2)*sizeof(double));
bb = b; // same for b
b  = (double(*)[n+2])((char*)b+(n+2+1)*sizeof(double));

int done = 0;
while (!done) {
  for (i=0; i<m; i++) {
    for (j=0; j<n; j++) {
      // 9-point stencil
      b[i][j] =
        avg(a[i][j-1],a[i+1][j-1],a[i+1][j],a[i+1][j+1],
            a[i][j+1],a[i-1][j+1],a[i-1][j],a[i-1][j-1],
            a[i][j]);
    }
  }
  c = a; a = b; b = c; // swap matrices a and b
  
  done = ... ; // set when done
}

free(aa); // free as allocated
free(bb);
\end{lstlisting}

A stencil computation\index{stencil computation}
on a $d$-dimensional matrix consists in updating
all matrix elements according to a (most often) constant-time
\impi{stencil rule} that depends on and
describes a small, bounded, constant-sized neighborhood of each matrix
element. The total amount of computation per stencil iteration
is then proportional to
the size of the $d$-dimensional matrix. The $9$-point stencil above
has as neighbors of matrix element \texttt{a[i][j]} the elements whose
distance is at most one in the maximum metric (Chebyshev distance), which is
sometimes called a Moore-neighborhood.  A two-dimensional, $5$-point
stencil would have as neighbors the elements whose Manhattan distance
(taxi cab metric) is at most one. This is sometimes called a von Neumann
neighborhood. Both are examples of first-order stencils; higher order
stencils include neighbors that are farther away in the chosen metric.
The stencil rule\index{stencil rule}
above is simply a computation of the average of the nine stencil elements
but could be any other constant-time function, for instance, the rules
of Conway's amazing \emph{Game of Life}~\cite{BerlekampConwayGuy04:4}.
In Conway's game, life evolves in a two-dimensional, but potentially
infinite universe. It is an example of a cellular
automaton~\cite{Codd68,ToffoliMargolus87} and, thus, not strictly a
stencil computation\index{stencil computation};
but a finite universe could easily be imagined
and perhaps still be interesting. The standard use of a $5$-point
stencil computation is Jacobi's method for solving the Poisson
differential equation, where the matrix updates are repeated until
convergence~\cite[Chapter 16]{Sourcebook03}.
The value in the ghost rows and columns define the
\emph{boundary conditions}. Other, higher-dimensional stencils, \eg,
$27$-point (Chebyshev) or $7$-point (Manhattan) in three dimensions are also
frequently used, as are many other, sometimes also asymmetric
stencils of higher order.  Accordingly, there are much terminology and
different notations for stencils in different application areas.

A single iteration of a one-dimensional, second-order
stencil computation\index{stencil computation}
is expressed by the following loop.

\begin{lstlisting}[style=SnippetStyle]
for (i=0; i<n; i++) {
  b[i] = a[i-2]+a[i-1]+a[i]+a[i+1]+a[i+2];    
}
\end{lstlisting}

It can be parallelized to run in $\Theta(n/p+1)$ parallel time with $p$
processor-cores\index{parallel time}.

\subsection{Work Pool}
\label{sec:workpool}

A \impi{work pool} maintains items of work to be performed to solve
our given computational problem in a data structure, the \emph{pool},
which makes it possible to insert and remove items in no particular
order.  A work item being solved may give rise to new work items that
are inserted into the pool. The process is repeated until all work
items have been processed and the work pool is empty. Work items can
be many things, like tasks ready for execution, parts of the input,
partial results, depending on the situation. The work pool pattern is
clearly attractive for parallelization. Since there are no
dependencies among items in the pool, several/many processors can
conceivably remove and work on work items from the pool
independently. A good parallelization in the sense of good load
balance\index{load balance} might be possible if there is at any time
during the computation a sufficient number of work items in the 
pool (compared to the number of processor-cores\index{processor-core}).
A non-trivial issue
is the \impi{parallel data structure} needed to allow many processors
to concurrently remove and insert work items into the pool.  A
\emph{centralized} work pool is maintained by a single processor, and
may be easier to design and reason about, but may also become a
sequential bottleneck for parallelization when a large number of
processors at the same time access the work pool. A centralized
design can easily fall victim to Amdahl's Law\index{Law!Amdahl's Law}.
In a \emph{distributed} work pool\index{work pool},
the pool consists in a number of
local pools maintained by the individual processors.  As long as there are
enough items in any of the pools, the processors can be kept busy and
good load balance guaranteed. Problems arise when work pools run
out of work. There are two strategies for alleviating the ensuing load
balancing problem. With \impi{work-dealing}, processors whose pools
are too full relative to some (static or dynamic) threshold
spontaneously deal out work item(s)
to other processors whose pools may have few(er) items. With
\impi{work-stealing}, processors whose pools have become empty
\emph{steal} work from other pools, and continue stealing until they
have either been successful or until it can be inferred that all pools
are empty and the computation has come to the end. Work-stealing is
currently a favored strategy for which appropriate parallel data
structures have been developed, and where sometimes strong bounds on
parallel running time can be proven.  Regardless of what strategy is
chosen, the work pool pattern eventually needs to solve the
(distributed) \impi{termination detection problem}: when is the pool
definitely empty?

\subsection{Master-Worker/Master-Slave}
\label{sec:masterworker}

The \impi{master-slave} or \impi{master-worker} pattern is sometimes
used to implement the work pool\index{work pool} pattern. A dedicated
master processor maintains a central data structure, from which the
slaves or workers are given work (data to work on, tasks to execute)
upon explicit request. The pattern is often simple to implement but
fully centralized, highly asymmetric, and, thus, easily subject to
Amdahl's Law\index{Law!Amdahl's Law} and similar serialization issues.

\subsection{Domain Decomposition\marksec}

The stencil computation\index{stencil computation}
employs a localized, constant time, mostly
position-oblivious\index{oblivious} update operation to each element
of a structured
domain, typically a $d$-dimensional matrix, which is iterated a
(large) number of times until a convergence criterion is met. It
appears easy to parallelize efficiently and can utilize a considerable
number of processor-cores.  In the more general
\impi{domain decomposition} pattern, a term which we use here very loosely to
characterize a computational pattern and not necessarily in accordance
with terminology from other domains, the situation is like this: A
more or less abstract domain in $d$-dimensional space over which
computations are to be performed is subdivided into subdomains (not
necessarily disjunct) which are assigned to the available $p$
processors.  The work to be done in the subdomains, say on moving
particles, may not be uniformly distributed over the domain and may
possibly move around in the domain. The computation per work item may
or may not be uniform and constant. The computation over the domain
is, like in the stencil pattern, typically to be repeated a large
number of times until convergence.

This pattern generalizes the
stencil computation\index{stencil computation} in several respects.
Thus, a static decomposition of
the domain may not perform well, since the subdomains can contain
different amounts of work items. Since the items may move, and since
the amount of required computation may change from iteration to
iteration, this pattern will typically need dynamic load
balancing\index{load balancing} to keep the $p$ processors equally
active throughout the computation.

\subsection{Iteration until Convergence, Bulk Synchronous Parallel}
\label{sec:bsppattern}

In both the stencil and the domain decomposition pattern, a parallel
computation is iterated a known or unknown number of times $k$ until
some convergence criterion is met. The parallel time\index{parallel time}
that can be achieved regardless of the number of
processor-cores\index{processor-core} employed is,
therefore, bounded from below by $\Omega(k)$. If $k$ is large compared to
the total work to be carried out, the achievable speed-up\index{speed-up}
will be limited by Amdahl's Law\index{Law!Amdahl's Law}.

Another way of looking at the pattern is as follows. Let some
number $p$ of processors be given. In each iteration, each processor
is assigned a part of the computation to be done, ideally in such a
way that the work load is evenly balanced over the
processors\index{load balancing}.  The processors perform their work
and in cooperation decide whether the termination condition has been
met or not. If not, work for the next iteration is redistributed over
the processors. When the work per iteration is large compared to the
coordination at the end of the iteration consisting in communication
(data exchange) and synchronization, this is a typical \emph{coarse
grained} parallel computation\index{granularity!coarse grained},
often referred to as a
\emph{Bulk Synchronous Parallel} (BSP)\index{BSP} computation.
The term was probably coined by Les Valiant~\cite{Valiant90,Bisseling04}.

An interesting example that can be cast in the bulk synchronous
parallel pattern is level-wise Breadth-First Search (BFS)\index{BFS}
in a(n un)directed graph from some starting vertex.  Let $G=(V,E)$ be
the given graph with $n=|V|$ vertices and $m=|E|$ arcs, and $s\in V$ a
given source vertex. The problem is to find the distance from $s$ to
all other vertices $u\in V$ defined as the number of arcs on a
shortest path from $s$ to $u$. A standard, sequential BFS algorithm
(see any algorithms textbook~\cite{CormenLeisersonRivestStein22})
maintains a queue of vertices being explored in the current iteration
and a queue of new vertices to be explored in the next iteration. It
maintains a distance label for each vertex which is the length of
a shortest path from $s$ in the part of the graph that has been
explored so far. Initially, all vertices have distance label $\infty$,
except $s$ which has distance label $0$, since no part of $G$ has been
explored. The invariant\index{invariant} to be maintained for iteration $k,
k=0,1,\ldots$ is that all vertices in the queue of vertices to be
explored have correct distance label $k$. In iteration $k$, all vertices $u$
in this queue are explored by examining the outgoing arcs $(u,v)\in
E$. If $v$ has a finite distance label already, there is nothing to be
done. If the distance label of $v$ is $\infty$, it is updated to $k+1$
and $v$ is put into the queue of vertices for the next iteration. At the
end of iteration $k$, the two queues are swapped.

It is clear that the algorithm terminates after $K$ iterations where
$K$ is the largest finite distance from $s$ of some vertex in $G$. It
is also clear that all arcs are examined at most once. Thus, assuming
that all vertices are reachable from $s$, the complexity of this
algorithm is $\Theta(n+m)$ if the queue operations are in $O(1)$.

There is much potential for parallelization. Vertices in the queue of
vertices to be explored in iteration $k$ can be processed in parallel
since order is not important and all arcs out of such vertices can
also be examined in parallel; provided that vertices and arcs are
available to the processor-cores\index{processor-core}
and that conflicts, for instance,
when inserting vertices into the queue of vertices for the next
iteration can be handled. By the end of an iteration, arcs and
vertices may have to be exchanged between processor-cores and queues
consolidated for the next iteration.  In the best possible case, a
parallel running time in $\Omega(\frac{n+m}{p}+K)$ could be
possible. If $m$ is large compared to $K$ and perhaps $n$, that is if
$G$ is not sparse and has low diameter, there might be enough ``bulk''
work for the processor-cores so that reasonable speed-up\index{speed-up}
can be achieved in practice.

\subsection{Data Distribution}
\label{sec:datadistributions}

Parallel algorithms working on structured data often seek to split the
data into (disjoint) parts on which processor-cores can work
independently in an embarrassingly parallel
fashion\index{parallel!embarrassingly}.  We have seen this approach
with the stencil pattern\index{stencil computation}
and will see it again many times. The
splitting of the data can be explicit by reorganizing the data into
disjoint parts accessible to the available processors; or implicit by
providing naming schemes and transformations to conveniently access
the different parts.  Structured data are here thought of as arrays,
vectors, matrices, higher-dimensional matrices, \etc, of objects that
can themselves be structured. We refer to the splitting of such
objects as \impi{data distribution}, which may be an active operation
to be performed (repeatedly) during the execution of a parallel
algorithm or a matter of providing means to refer to the parts of the
data in the required fashion.

Let some linear data structure of $n$ elements be given, \eg, an
array, and let $b, b\geq 1$ be a chosen \emph{block size}.  Let $p$ be
the number of processors, numbered from $0$ to $p-1$.  In a
\emph{block cyclic}
data distribution\index{data distribution!block cyclic}
the $n$-element array is split into \emph{blocks} of consecutive
elements of $b$ elements each; one last piece may have fewer than $b$
elements, depending on whether $b$ divides $n$. Number these blocks
consecutively starting from $0$. Then blocks $0,p,2p,\ldots$ can or
will be accessed by processor $0$, blocks $1,p+1,2p+1,\ldots$ by
processor $1$, blocks $2,p+2,2p+2,\ldots$ by processor $2$, in general
blocks $i,p+i,2p+i,\ldots$ by processor $i, 0\leq i<p$.

A \emph{cyclic} data distribution\index{data distribution!cyclic}
is the special case where $b=1$.
A \emph{blockwise} data distribution\index{data distribution!blockwise} is
the special case where roughly $b=n/p$ and where rounding is done such
that, as far as possible, each processor has a block of at least one element.

A higher-dimensional matrix may likewise be divided into smaller blocks (many
possibilities) and distributed in a block cyclic way. Special cases
for two-dimensional matrices are the
\emph{row-wise} distribution\index{data distribution!row-wise}
where each processor is assigned to work on a consecutive number of
full rows of the matrix, and the
\emph{column-wise} distribution\index{data distribution!column-wise}
where each processor
is assigned to work on a consecutive number of full columns of the matrix.
We will see examples of the use of such distribution in \Sec~\ref{sec:la}.

\subsection{Compaction, Gather and Scatter}
\label{sec:arraycompaction}

Consider the loop below. A marker array \texttt{mark[i]} is given,
which for each element of the array \texttt{a[i]}, tells whether
the element is to be kept or not for some later computation. The marked
elements are copied into the \texttt{b[j]} array in the loop order
of appearance.

\begin{lstlisting}[style=SnippetStyle]
j = 0;
for (i=0; i<n; i++) {
  if (mark[i]) b[j++] = a[i];
}
\end{lstlisting}

As we will see in the rest of these lectures, this pattern, which is
called \impi{array compaction}, is important and surprisingly
versatile. Unfortunately, the loop has obvious dependencies,
\eg, the increment of \texttt{j}, and so far we have no means of
parallelizing it. \Sec~\ref{sec:prefixsums} will be devoted to this
problem.

The (dense) \emph{gather} and \emph{scatter} patterns rearrange array elements
and are illustrated below. Given an index array \texttt{ix[i]} with
values $0\leq\texttt{ix[i]}<n$ which is not necessarily required to be
a permutation (it may be that \texttt{ix[i]==ix[j]} for some,
different \texttt{i} and \texttt{j}), the gather pattern copies
elements from \texttt{b} in the order given by the index array into
\texttt{a}. The scatter pattern is the opposite and copies into the
\texttt{a} array in index order from the \texttt{b} array in sequential
loop order.

\begin{lstlisting}[style=SnippetStyle]
// gather
for (i=0; i<n; i++) {
  a[i] = b[ix[i]];
}
// scatter
for (i=0; i<n; i++) {
  b[ix[i]] = a[i];
}
\end{lstlisting}

Ideally, with $p$ processor cores, both of the patterns can be
parallelized to run in $\Theta(n/p+1)$ parallel time steps. Dependent
on the index array, there may be concurrent reading in the gather
pattern. If implemented on a PRAM\index{PRAM}, this pattern requires
either concurrent read capabilities\index{PRAM!CREW} or prior
knowledge that the index array is indeed a permutation. Likewise, the
scatter pattern may incur concurrent writing. If implemented on a
PRAM, sufficiently strong concurrent write capabilities are
required\index{PRAM!CRCW} depending on which values may be written.
More liberal gather and scatter patterns would allow the \texttt{b}-array
to be an $m$-element array with $m\geq n$ and indices in this array.

\subsection{Data Exchange, Collective Communication}
\label{sec:exchangepatterns}

Different parts of the data being processed at different stages of the
execution of a parallel algorithm may be managed by or have special affinity
to different processor-cores; indeed, this was the case for many of
the parallel patterns discussed above. It can, therefore, be convenient or
even necessary (as will be seen in Chapter~\ref{chp:distributedmemory})
to explicitly exchange or reorganize data between processor-cores
at different stages of the computation. Such \emph{exchange patterns}
and operations are frequent in \parco and are also often referred to as
\impi{collective communication} or just \impi{collectives} because all
affected processor-cores jointly take part in and jointly, by appropriate
underlying algorithms, effect the exchange.

The following exchange and reduction patterns are traditionally considered.

\begin{itemize}
\item
  \emph{Broadcast} and all-to-all broadcast, in which one, or all,
  processors have data to be distributed to all other processors.
\item
  \emph{Gather}, in which a specific processor collects individual
  data from all other processors.
\item
  \emph{Scatter}, in which a specific processor has individual data to
  be transmitted to each of the other processors.
\item
  \emph{All-to-all}, in which all processors have specific, individual
  data to each of the other processors.
\item
  \emph{Reduction}, reduction-broadcast (all-reduce) and
  reduction-scatter, in which data are combined together under an
  associative operator, with the results stored either at a specific
  processor, at all processors, or distributed in parts over all
  processors.
\end{itemize}

All these patterns are explicitly found in, for instance,
\mpi\index{MPI} and will be discussed in great detail in
\Sec~\ref{sec:collective}; but they do appear explicitly and
implicitly in many other \parco contexts as well.

\subsection{Reduction, Map-Reduce and Scan}

Surprisingly many problems can be viewed as reduction problems: A
(large) number of input values which can be numbers, vectors,
matrices, complex objects (texts, pictures, data bases) \etc are
combined together using an associative, functional rule to arrive at
the solution. Subsets of elements can be assigned to processor-cores,
and by associativity, the reduction can be performed by repeated
reduction of disjoint pairs of sets of values. The reduction pattern
is a well parallelizable design pattern.  The pattern can be made
more powerful and flexible by allowing, for instance, a precomputation
on the input values before reduction; this operation is often, for
instance, in functional programming, called a \emph{map operation}, and
the combined pattern has been popularized as
\impi{map-reduce}~\cite{DeanGhemawat08,DeanGhemawat10}. Many
variations are possible and have been proposed.

A related pattern is the \impi{scan} or prefix sums\index{prefix sums}
where the associative rule is applied on the input values in sequence: The
result for the $i$th input is the associative reduction of all inputs
before, up to and possibly including input $i$. A scan computes these
prefix sums for all $i$~\cite{Blelloch89}.  We will see many
applications of prefix sums and the scan operation throughout these
lectures, see \Sec~\ref{sec:prefixsums} also for efficient algorithms
for computing the scan.

\subsection{Barrier Synchronization}
\label{sec:barrier}

Some of the patterns above divide a computation into separate stages
that are executed one after the other, for instance, until some
convergence criterion is met. Other patterns and computations assume
that computations done by other processors have been completed before
a processor can continue to its next phase of computation. Ensuring
such requires some form of \emph{synchronization} between
processor-cores and is the task of what we here call the
\impi{barrier} parallel pattern.  Semantic barrier synchronization
means that a processor that has reached a certain point in its
computation, called the \emph{barrier} (point),
is not allowed to proceed before all
other processors have reached their barrier point.  After the barrier
synchronization, updates and computations performed by the other
processors shall be available to the processor to the extent that this
is required.

Barrier synchronization can be implicit or explicit; for arguing for the
correctness of a particular parallel algorithms it is often required,
though, to know at which points the processors are synchronized in the
sense of all having reached a certain point and having a consistent view
of the computation.

In a lock-step, synchronized model like the PRAM\index{PRAM}, explicit
barrier synchronization is not (or rarely) needed. In asynchronous,
shared memory models and systems, various forms of barriers are needed
to ensure correctness, and they are typically provided.
The \openmp\index{OpenMP}
thread model, see \Sec~\ref{sec:openmpframework}, provides implicit
barrier points as well as explicit barrier synchronization constructs.
Unlike for the PRAM, barriers are typically not for free.  In
asynchronous models, barrier synchronization of $p$ processors takes
$\Theta(\log p)$ parallel time\index{parallel time} steps.
Many interesting standard
algorithms for barrier synchronization on non-synchronous (non-PRAM)
shared memory systems can be found in~\cite{MellorCrummeyScott91}.

In distributed memory models, the required synchronization is
sometimes guaranteed by the semantics of the provided communication
operations, whether implicit or explicit. Explicit semantic barrier
constructs may be provided as well, although they may be needed less
often (to preview: in \mpi, an explicit barrier is almost never
needed!\index{MPI}). Also here, $\Theta(\log p)$ dependent parallel operations
go into enforcing a semantic barrier.

\section{Fourth block (1 lecture)}

As examples of not quite trivial parallel algorithms for computational
problems that are not obviously parallelizable,
we now look at two concrete problems, namely merging\index{merging}
of two ordered sequences and computing the prefix sums\index{prefix sums}
of elements in an array. The
aim is to derive good, parallel algorithms that can actually be
implemented on real, parallel systems with both shared- and distributed
memory.  While the usefulness of the merging problem is obvious, this part of
the lecture also motivates why computing prefix sums is such an important
\parco problem. We also state the so-called ``Master
Theorem''\index{Master Theorem}, a useful tool that will immediately
solve (most of) the recurrence relations of these lectures.

\subsection{Merging Ordered Sequences in Arrays}
\label{sec:merging}

The \emph{merging problem}\index{merging} is the following: Given two ordered
sequences stored in arrays $A$ and $B$ with $n$ and $m$ elements,
respectively, from some universe with a total order $\leq$, construct
an ordered $n+m$ element array $C$ containing exactly the elements
from $A$ and $B$.

The standard, straightforward sequential algorithm for merging steps
through the arrays $A$ and $B$ hand in hand and in each iteration
writes out the smaller element to the $C$ array. This is captured by
the following \texttt{seq\_merge} function (for arrays of C
\texttt{double}s).

\begin{lstlisting}[style=SnippetStyle]
void seq_merge(double A[], int n, double B[], int m,
               double C[]) {
  int i, j, k;

  i = 0; j = 0; k = 0;
  while (i<n&&j<m) {
    C[k++] = (A[i] <= B[j]) ? A[i++] : B[j++];
  }
  while (i<n) C[k++] = A[i++];
  while (j<m) C[k++] = B[j++];
}
\end{lstlisting}

This algorithm (which is not the best possible in terms of
constants~\cite{Knuth73:taocp3}) unfortunately seems strictly
sequential: The output at position $i$ of $C$ depends on the relative
order of all the previous elements in $A$ and $B$, and there is not
much that can be done in parallel. Possibly the last two copy
loops of independent iterations could be parallelized, but it is not
known in advance how many elements of the input will be handled by these
loops and, therefore, the observation does not help much. The
complexity of the standard algorithm is $\tseq(n)=\Theta(n+m)$.  A
new, different idea is required for a good parallel solution.

Recall that merging\index{merging} and sorting algorithms are
called \impi{stable} if the relative order of equal elements
in the input is preserved. For
the merging problem, this means that the relative order of equal
elements in the inputs arrays $A$ and $B$ is preserved in the output
and also that elements in array $A$ that are equal to an element of $B$ occur
before the $B$ element in the output array $C$. Stability is often a
useful or even desired property. Some merging and sorting algorithms
are naturally stable, like the standard, sequential merging algorithm
listed above, some are not.

For some of the merging algorithms\index{merging} in the following,
it is convenient
to assume that all elements are distinct. Distinctness can be assumed
without loss of generality, because elements can always be made
distinct: Instead of merging elements, we merge triples $(x,F,i)$
where $x$ is an element from either $A$ or $B$, $F$ marks whether the
element comes from $A$ or from $B$, and $i$ is the index of the
element in its array, whether in $A$ or in $B$. We use a lexicographic order,
defined by $(x,F,i)<(x',F',i')$ if either $x<x'$, or if $x=x'$ and
$F\neq F'$ and $F=A$, or if $x=x'$, $F=F'$ and $i<i'$.

Using this order will ensure stability of any merging or sorting
algorithm. The cost is extra space and a more expensive comparison operation
(which should not be neglected. Try it!). It is, therefore, most often
better if the merging\index{merging} or sorting algorithm is stable by design,
without resorting to the ``make-distinct trick''.

\subsection{Merging by Ranking}
\label{sec:ranking}

A different approach to merging\index{merging} is the following:
For each element
$A[i]$ in $A$, find the position $j$ in $B$ such that
$B[j-1]<A[i]<B[j]$; here we assume element distinctness and for
convenience that $B[-1]=-\infty$ and $B[m]=\infty$. The position $j$
is called the \impi{rank} of $A[i]$ in $B$, denoted by
$\rank(A[i],B)$.  The rank of $A[i]$ in $B$, thus, counts the number of
$B$ elements that are strictly smaller than $A[i]$.  By knowing the
rank of element $A[i]$ in $B$, we also know the position of $A[i]$ in
the output array $C$: It is $i+\rank(A[i],B)$.

We can now merge the elements of $A$ and $B$ into $C$ by computing the
ranks for all elements in $A$ and in $B$ in the respective other array.
The rank of any element of $A$ in $B$ can be computed by binary search
in $O(\log m)$ time steps.
The sequential complexity\index{sequential complexity}
of \impi{merging by ranking} is, therefore, $O(n\log m+m\log
n)=O((n+m)\log\max(n,m))$, far worse than the standard, sequential
merging algorithm.

However, merging by ranking can be performed in parallel: Assign a
processor to each element of $A$ and of $B$, let it compute the rank
of the element in the other array and write the element to its
position in the output array $C$. With $n+m$ processors, the algorithm
takes $O(\log \max(n,m))$ time steps, so it is fast, but it is clearly
not work-optimal: The work is the same as sequential merging by the
ranking algorithm, namely $O((m+n)\log\max(n,m))$. We note also that when
ranking is done concurrently by many processors, concurrent read
capabilities (as in the CREW PRAM\index{PRAM!CREW}) are required of our system.

To reduce the work, a new idea is needed. We want to design an
algorithm using $p$ processors. The idea is to rank only some of the
elements, more precisely $O(p)$ of them. The input array $A$ is
divided into disjoint, consecutive \emph{blocks} of size roughly
$n/p$, and the first element of each $A$ block is ranked in $B$ (it is
helpful to graphically visualize this).  Now, the $A$ block can be
merged with the corresponding part of $B$ determined by the rank of
the first element of the $A$ block in $B$ and the first element of
the next $A$ block in $B$, using our best known sequential merging
algorithm. These pairs of blocks are disjoint and can all be merged in
parallel. We now have a work-optimal, parallel merging\index{merging}
algorithm. There are $p$ processors, which together spend
$O(p\log\max(n,m))$ work on ranking the $p$ elements from $A$ and
$O(n+m)$ time for merging pairs of blocks. It should
also be obvious that the algorithm is correct (given the distinctness
assumption; use pictures to see this).

Unfortunately, we cannot give a good bound on the time. Desired is
$O(\frac{n+m}{p}+\log\max(n,m)$. Since we do not know the inputs, and
the arrays $A$ and $B$ can be arbitrarily interleaved in $C$, it can
happen for one of the $A$ blocks that the first element has a rank in $B$
close to $0$ and the first element of the next $A$ block a rank close
to $m-1$. Merging\index{merging} this pair which is done by one
processor would, therefore, take $\Omega(n/p+m)$ sequential
time steps, and there would be no speed-up over the sequential
algorithm\index{speed-up}.
This is a classical \emph{load balancing problem}\index{load balancing}:
One processor is doing almost all of the work.

There are at least two possible solutions to this problem. Assume that
for some block in $A$ the ranks in $B$ of the first element and the
rank of the first element of the next $A$ block in $B$ are far apart
(close to $m$ elements). Such a \emph{bad segment} in $B$ could be
divided roughly evenly into $p$ blocks of size about $m/p$ elements
and the rank in $A$ for the first elements of each of these blocks 
computed (in parallel). It can easily be seen (use a picture) that all
these ranks in $A$ will lie within the $A$ block which gave rise to
the bad segment in the first place. Therefore, the pairs of the blocks
of the bad $B$ segment and the blocks now found in the $A$ block will all
have size at most $n/p+m/p$, and can, therefore, be merged
sequentially within the desired bound of $O(\frac{n+m}{p})$ time
steps. This would lead to a fast and work-optimal parallel
algorithm. The only problem remaining is to be able to identify the
bad $B$ segments (there could be more than one) and to re-allocate the
processors to work on these segments. This problem can be solved with
use of prefix sums\index{prefix sums}
(see later)~\cite{JaJa92,ShiloachVishkin81}.

The other solution is to divide from the outset both
the $A$ and $B$ arrays into blocks of roughly equal size $n/p$ and
$m/p$ elements and rank the first elements of these blocks in the
other sequence. This gives rise to $2p$ pairs of blocks of size at
most $n/p+m/p$ that can be merged in
parallel with $p$ (or $2p$) processor-cores. However, seeing that
the blocks are indeed disjoint and cover the $A$ and $B$ arrays takes
some care~\cite{HagerupRub89,Traff12:merge}.
Nevertheless, we claim the following theorem.

\begin{theorem}
  On a $p$ processor system where binary search can be performed,
  two ordered arrays $A$ and $B$ can be merged cost-optimally in
  $O(\frac{n+m}{p}+\log\max(n,m))$ time steps.
\end{theorem}

Concurrent, simultaneous reading of the same location in either $A$ or
$B$ array can happen during the binary searches, dependent on both the
timing of the processors and the input, so if implemented on a PRAM,
CREW capabilities are required\index{PRAM!CREW}.

\subsection{Merging by Co-Ranking}
\label{sec:corankmerge}

A completely different approach turns the parallel
merging problem\index{merging}
upside-down and focuses on what will be written into the $C$ array. The
idea is to find for each position $i$ in the output array $C$ the
unique positions $j$ and $k$ in the input arrays $A$ and $B$, such
that by (stably!) merging $A[0,\ldots,j-1]$ and $B[0,\ldots,k-1]$, we
get exactly the $i$-element prefix $C[0,\ldots,i-1]$ of $C$. The
positions $j$ and $k$ are called the \emph{co-ranks}\index{co-rank}
for $i$ and the
approach itself \impi{merging by co-ranking}~\cite{Traff14:perfectmerge}.
Note that it holds that $j+k=i$ which will be an essential
invariant\index{invariant} in the algorithm for finding the co-ranks.
If a processor can determine the co-ranks for the first element of a
block of $(n+m)/p$ elements of $C$ and the co-ranks for the first
element of the next block of $C$, the $(n+m)/p$ element block of $C$
can be constructed by (sequentially) merging the blocks of $A$ and $B$
determined by the respective co-ranks.

By this approach, we can ensure that all of the $p$ processors have
blocks of exactly the same size by diving the $C$ array into blocks of
size $(n+m)/p$ (plus/minus one element, if $p$ does not divide
$(n+m)$), and in that sense arrive at a perfectly load-balanced
merging\index{merging} algorithm.

The observation of the following lemma tells how
co-ranks\index{co-rank} can be computed.
\begin{lemma}
  \label{lem:coranks}
  For any index $i, 0\leq i<n+m$, there are unique co-ranks $j$ and
  $k$ with $j+k=i$ such that
  \begin{enumerate}
  \item
    either $j=0$ or $A[j-1]\leq B[k]$, and
  \item
    either $k=0$ or $B[k-1]<A[j]$.
  \end{enumerate}
\end{lemma}

To see this, consider the element $C[i-1]$ of the output array that
corresponds to the co-ranks $j$ and $k$. Since each $C$-element comes
from either $A$ or $B$, either $C[i-1]=A[j-1]$ or $C[i-1]=B[k-1]$.
Consider first the case where $C[i-1]=A[j-1]$ and $j>0$. Then $B[k]$
is the first element of $B$ that is not in $C[0,\ldots,i-1]$, and
since the merge is stable, it follows that $A[j-1]\leq B[k]$. Also
$B[k-1]<A[j-1]$, and therefore, since $A$ is ordered,
$B[k-1]<A[j-1]\leq A[j]$. For the other case, $C[i-1]=B[k-1]$ and
$k>0$, it similarly follows that $B[k-1]<A[j]$ (since the merge is stable,
equal elements of $A$ are before elements of $B$), and also that
$A[j-1]\leq B[k-1]\leq B[k]$.

To find the co-ranks $j$ and $k$ for a given index $i$ in $C$,
a binary-search like
procedure in both $A$ and $B$ can be applied, halving
intervals in $A$ and $B$ until the conditions of
Lemma~\ref{lem:coranks} are both fulfilled while maintaining throughout
the invariant\index{invariant} that $i=j+k$.
This will take $O(\log (n+m))$ iterations.
The co-ranking code\index{co-rank}
is shown below, and a full merge algorithm can (for parallel systems with
shared memory) readily be implemented.

\begin{lstlisting}[style=SnippetStyle]
j = min(i,m); k = i-j;
jlow = max(0,i-n);
klow = 0;

done = 0;
do { // invariant: i = j+k
  if (j>0&&k<n&&A[j-1]>B[k]) {
    // condition 1 violated
    d = (1+j-jlow)/2;
    klow = k;
    j -= d; k += d;
  } else if (k>0&&j<m&&B[k-1]>=A[j]) {
    // condition 2 violated
    d = (1+k-klow)/2;
    jlow = j;
    k -= d; j += d;
  } else done = 1;
  assert(i==j+k);
} while (!done);
\end{lstlisting}

We summarize in the following theorem.
\begin{theorem}
  On a $p$ processor system where co-ranking can be performed, the
  merging problem can be solved cost-optimally in
  $O(\frac{n+m}{p}+\log (n+m))$ time steps with $p$
  processor-cores. The algorithm is perfectly load balanced and
  stable.
\end{theorem}

Like for binary-search based merging\index{merging}, concurrent,
simultaneous reading of the same location in either $A$ or $B$ array
can also happen with co-ranking\index{co-rank},
dependent both on the timing of the processors and on the input, so if
implemented on a PRAM, CREW capabilities are required\index{PRAM!CREW}.

Ranking and co-ranking are examples of static, problem-dependent load
balancing\index{load balancing}: Eventually, the blocks of the $A$ and
$B$ arrays assigned to the processors to be merged sequentially have
approximately the same total size, for the co-ranking approach 
exactly so ($\pm 1$),
but how exactly the blocks are cut depends on the input. The
preprocessing needed for the load balancing step, after which the
sequential block merging is done, takes $O(\log\max(n,m))$, which is
not a constant fraction of the total work $O(n+m)$, so Amdahl's
Law\index{Law!Amdahl's Law} does not apply.

\subsection{Bitonic Merge\marksec}

Bitonic merging is an example of an \impi{oblivious} merging\index{merging}
algorithm: The indices that are compared against each other depend
only on $n$ and $m$, the size of the input, and not the input
itself. Bitonic merging does not require concurrent read capabilities
of the system and can be implemented on an EREW PRAM\index{PRAM!EREW}.
Bitonic merging is an important example algorithm and
can in some situations have practical advantages over the merging
algorithms in the previous sections. Bitonic merging and Bitonic
Merge sort were invented by Kenneth Batcher~\cite{Batcher68}.

Let $a_0,a_1,\ldots a_{n-1}$ be a sequence of $n, n>1$ comparable
elements, $a_i\leq a_j$ or $a_j\leq a_i$.  The sequence is a
\impi{Bitonic sequence} if either
\begin{enumerate}
\item
  there is an $i, 0\leq i<n$ such that $a_0\leq a_1\leq\ldots\leq a_i$
  and $a_{i+1}\geq a_{i+2}\geq\ldots\geq a_{n-1}$, or
\item
  there is a cyclic shift of the sequence, such that the first
  condition holds.
  \end{enumerate}
For convenience, a sequence of $n=1$ elements is also Bitonic.

It is not so difficult to see that the following lemma holds.
\begin{lemma}
  \label{lem:bitonicsplit}
  Let $a_0,a_1,\ldots a_{n-1}$ be a Bitonic sequence of even
  length. The two sequences
  \begin{itemize}
  \item
    $\min(a_0,a_{n/2}),\min(a_1,a_{n/2+1}),\ldots,\min(a_{n/2-1},a_{n-1})$
    and
  \item
    $\max(a_0,a_{n/2}),\max(a_1,a_{n/2+1}),\ldots,\max(a_{n/2-1},a_{n-1})$
  \end{itemize}
  of length $n/2$ are Bitonic and all elements of the first sequence
  are smaller than or equal to the elements of the second sequence.
\end{lemma}

A Bitonic sequence of length $n=2^d$ can recursively be put into
non-decreasing order as follows: By Lemma~\ref{lem:bitonicsplit},
split the sequence into two Bitonic halves with all elements of the
first half smaller than the elements of the second half
and recursively order the two Bitonic halves.
In each recursive call, the number of
elements to split is halved, so the number of calls in any successive
sequence of calls needed to arrive at a single element is $d=\log_2 n$. The
total number of comparisons performed and thus the total work
measured as the number of operations as a function of $n$ is given by
the recurrence relation
\begin{eqnarray*}
  W(1) & = & 0 \\
  W(n) & = & 2W(n/2)+n/2
\end{eqnarray*}
which has the solution $W(n) = (n/2)\log_2 n$. This can be seen by
induction or estimated by the Master Theorem~\ref{thm:master}
(Case 2).  It is plausible that this can be turned into a parallel
algorithm with $\log_2 n$ parallel time steps, in each of which $n/2$
comparisons are performed by recursive calls being carried out in
parallel by the available processors.

Bitonic ordering can be used to merge two ordered sequences. From the
two ordered sequences in arrays $A$ and $B$ of length $n$ and $m$, a
Bitonic sequence is constructed by listing the $n$ elements from $A$
in increasing order, followed by listing the $m$ elements of $B$ in
reverse, that is in decreasing order.  Bitonic merging can be extended
to sequences of any length by padding with virtual $-\infty$ elements
in front of the first sequence to get a virtual sequence of length
some power of two. With some care, this can be made to work without
doing any comparisons with the virtual $-\infty$ elements (outcome is
always known).  Compared to our sequential merge algorithm, this
approach is not work-optimal\index{work-optimal}.  Bitonic merge can
elegantly be employed to sort a given $n$-element sequence in
$O(\log^2 n)$ parallel time steps and $O(n\log^2n)$ work (total number
of operations).

Parallel Bitonic merging and sorting is commonly analyzed using another model
of parallel computation: \impi{comparator networks}. Bitonic ordering
can be implemented on such a network of depth $\log_2 n$ and
$(n/2)\log_2 n$ comparators. Bitonic Merge sort, which can also be
implemented on such a \impi{sorting network}, is not work-optimal, and
it was a long standing open question of theoretical importance whether
sorting networks of depth $O(\log n)$ and size $O(n\log n)$ (number of
comparators) exist~\cite[\Sec~5.3.4, Exercise~51]{Knuth73:taocp3}.
The question was answered affirmatively in a
famous paper by Ajtai, Koml\'os, and
Szemer\'edi~\cite{AjtaiKomlosSzemeredi83}. Another important result is
``Cole's parallel Merge sort'', which shows that sorting by
merging\index{merging}
can be done in $O(\log n)$ parallel time steps with $n$
processors on a(n EREW) PRAM~\cite{Cole88,Cole93}\index{PRAM}\index{PRAM!EREW}.
Both results have very large
constants hidden in the $O$s and are in their original forms not
practically relevant~\cite{Paterson90,BaddarBatcher11}.

\subsection{The Prefix Sums Problem}
\label{sec:prefixsums}

We now turn our attention to another immensely important problem whose
usefulness may not be obvious at first glance.  Let an input array $A$
of $n$ elements from a set with a binary, associative operator $\oplus$ be
given.  The $i$th \emph{inclusive prefix sum} for $0\leq i<n$ is 
\begin{eqnarray*}
  B[i] & = & \bigoplus_{j=0}^{i}A[j]
\end{eqnarray*}
and the $i$th \emph{exclusive prefix sum} for $0<i<n$ is
\begin{eqnarray*}
  B[i] = \bigoplus_{j=0}^{i-1}A[j] \quad .
\end{eqnarray*}
where the exclusive prefix sum $B[0]$ is left undefined.

The \impi{prefix sums problem}\index{prefix sums}
is to compute the (exclusive or
inclusive) prefix sums for all indices $i$.  Computing all prefix sums
over an array is sometimes also called \impi{scan} which mostly, but
not always, denotes the inclusive prefix sums, with \impi{exscan} for
the exclusive prefix sums~\cite{Blelloch89}.
Note that the $i$th inclusive prefix sum
can be computed from the $i$th exclusive prefix sum by just adding the
$i$th input element with $\oplus$ to the $i$th exclusive prefix sum.  The
converse does not hold, unless an inverse of the $\oplus$ operator is
given, and that may not always be the case.

The prefix sums problem\index{prefix sums problem} is a generalization of the
\impi{reduction problem} which is to compute only the last,
inclusive prefix sum
\begin{eqnarray*}
  B[n-1] & = & \bigoplus_{j=0}^{n-1}A[j] \quad .
\end{eqnarray*}

Since the operator $\oplus$ is associative, the sums are well-defined
with $A[i]\oplus A[i+1]\oplus A[i+2] = (A[i]\oplus A[i+1])\oplus
A[i+2] = A[i]\oplus (A[i+1]\oplus A[i+2])$. If the $\oplus$ operator is
in addition commutative, any two summands can be swapped and still yield
the same result. Commutativity can give more freedom to algorithms to apply
$\oplus$ in a convenient order but is normally not assumed.

Both problems are trivial to solve sequentially by a scan through the
$A$ array (thus the term), keeping a running sum in a register and
writing it to $B[i]$.

\begin{lstlisting}[style=SnippetStyle]
sum = A[0]; // running sum in register
B[0] = sum;  
for (i=1; i<n; i++) {
  sum = sum+A[i];
  B[i] = sum;
}
\end{lstlisting}

Improvements are possible by exploiting vector (SIMD\index{SIMD})
capabilities of the processor which is not quite trivial but can to
some extent be accomplished by compiler loop unrolling. The
sequential complexity is $\Theta(n)$ steps\index{sequential complexity},
since $n-1$ sum computations are necessary because the result can depend on
any element.

Both reduction\index{reduction} and prefix sums\index{prefix sums}
are examples of parallel
patterns\index{pattern!parallel} or \impi{collective operations}
(see \Sec~\ref{sec:exchangepatterns}):
Each of the $p$ processors contributes some of the $n, n\geq p$
elements, and the processors together perform a reduction or compute
the prefix sums with results stored at the processors (prefix sums) or
some selected processor (reduction).

\subsection{Load Balancing with Prefix Sums}
\label{sec:prefixsumloadbalance}

The reduction operation is clearly useful. A frequently occurring
book-keeping task in parallel computations is for the processor-cores
to agree on some common value (could be a flag, telling whether the
computation is done). This common value is computed by a parallel
reduction. A \impi{broadcast operation} may be needed to
provide the outcome to all processors or, even better, a combined
reduce-broadcast, which is commonly called an \impi{allreduce operation}
(a somewhat unfortunate name choice).

Applications of the prefix sums problem are perhaps less
obvious. Consider the following situation: Some expensive computation
is to be done on some elements of a large array of $n$ elements. It is
not known a priori where these elements are. Instead, there is an
associated marker array, also of size $n$, that for each index tells
whether the associated element is to be worked on or not. All
computations are independent of each other. Thus, there is potential
for doing the work in parallel. We want to assign the element
computations to $p$ processors.  The strategies for parallelizing
loops that we have seen before (splitting the iteration range into $p$
disjoint blocks, one for each of the $p$ processor-cores) will not
work well. Since it is not known which element indices are marked, it
can easily happen that some blocks have many marked elements, while
other blocks have no marked elements at all and, therefore, little to
do apart from checking $n/p$ indices and finding them unmarked. This
is a typical load-balancing problem; the blocked merging\index{merging}
by ranking
algorithm had a similar problem. One processor may end up with all the
work with no speed-up\index{speed-up} possible. Prefix sums solve this load
balancing problem. This application is one of the most important
uses of prefix sums\index{prefix sums} in \parco and one reason why
the problem is so important.

The solution is as follows: In a marker-index array $M$ of size
$n$, put a $1$ for each marked element and a $0$ for each non-marked
element.  This can be done in $O(n/p)$ parallel time steps by a
parallelized loop of independent iterations. Perform an exclusive
prefix sums computation on $M$.  Now for each marked element, $M[i]$
is the number of marked elements up to (but not including) element
$i$. It can, therefore, be used as index into another array which stores
the marked elements consecutively.  Assume that there are $m$ marked
elements; $m$ can be computed by a reduction over the array of marks
or directly from $M[n-1]$.  Since these are now stored consecutively,
the array of the marked elements can be partitioned into $p$ blocks of
about $m/p$ elements, on all of which the expensive computation has to
be performed. All $p$ processors now have about the same amount of
non-trivial work to do, and much better load balance is achieved,
especially if the element computations all take about the same time.

This pattern, often called
\emph{parallel array compaction}\index{array compaction} (see
\Sec~\ref{sec:arraycompaction}), occurs in many guises. One is
parallelizing the sequential, linear-time partitioning step of the
Quicksort\index{Quicksort} algorithm.
We do three mark-and-compact steps. First, the
elements strictly smaller than the pivot are marked and compacted into
an array for the recursive call on the smaller elements. Second,
the elements equal to the pivot (no recursive call needed) are
compacted, and third, the elements strictly larger than the pivot are
compacted into an array of the larger elements. The total work
is $O(n)$, although the constants are larger than in standard,
sequential partition implementations. How fast this is, depends on how
fast the prefix sums problem can be solved. The two Quicksort calls
(on smaller and larger elements) are independent of each other and
can possibly be done in parallel as will be discussed in later
parts of these lectures.

If the partitioning step is not parallelized, it will become a severe
bottleneck for a parallel Quicksort\index{Quicksort} implementation,
consuming $O(n)$ time steps for the first Quicksort recursion level out
of $O(n\log n)$ work in total, resulting in parallelism\index{parallelism}
in the best case of only
$O(\frac{n\log n}{n}) = O(\log n)$. The scan\index{scan} operation (parallel
pattern) is similarly useful for sorting by counting and bucket
sorting (see \Sec~\ref{sec:mpisorting}).

We now discuss three different solutions to the
inclusive prefix sums\index{prefix sums problem}.

\subsection{Recursive Prefix Sums}
\label{sec:recursiveprefix}

The first algorithm is a recursive, divide-and-conquer approach. Let
$A$ be an array of $n$ elements for which to compute the inclusive
prefix sums. We reduce the problem to a prefix sums\index{prefix sums problem}
problem of only $\floor{n/2}$ elements by computing into an array $B$
the sums of pairs of immediately consecutive elements of $A$:
$B[i] = A[2i]\oplus A[2i+1]$,
and recursively solve the prefix sums problem on $B$.
The prefix sums of the $A$ array can be constructed from
$B$: $A[2i] = B[i-1] \oplus A[2i]$ and $A[2i+1] = B[i]$ with some
care for the first and for the last element when $n$ is odd. This can be
implemented as shown below by a recursive function \texttt{Scan} that
computes the prefix sums of the \texttt{n}-element \texttt{A}.

\begin{lstlisting}[style=SnippetStyle]
void Scan(int A[], int n) {
  if (n==1) return;

  int B[n/2]; // careful with stack allocation for large n
  int i;
  for (i=0; i<n/2; i++) B[i] = A[2*i]+A[2*i+1];
  
  Scan(B,n/2);

  A[1] = B[0]; // A[0] is per definition correct
  for (i=1; i<n/2; i++) {
    A[2*i]   = B[i-1]+A[2*i];
    A[2*i+1] = B[i];
  }
  if (n%2==1) A[n-1] = B[n/2-1]+A[n-1];
}
\end{lstlisting}

It is easy to see by an inductive argument that the recursive
algorithm and program correctly compute the inclusive prefix sums of
$A$.  If there is only one element in $A$ ($n=1$), $A[0]$ is indeed
the prefix sum. Now, assume that the function correctly computes the
prefix sums of an array $B$ of $\floor{n/2}$ elements.  For $i>0$, the
$i$th prefix sum of $A$ can be written as
$\oplus_{j=0}^{i}A[i]=\oplus_{j=0}^{\floor{i/2}}(A[2j]\oplus A[2j+1])$
when $i$ is odd and as
$\oplus_{j=0}^{i}A[i]=\oplus_{j=0}^{\floor{i/2}}(A[2j]\oplus
A[2j+1])\oplus A[i]$ when $i$ is even.  By the initialization of $B$
with $B[i] = A[2i]\oplus A[2i+1], 0\leq i<\floor{n/2}$, it will then
hold by the induction hypothesis that $B[i]=\oplus_{j=0}^i(A[2j]\oplus
A[2j+1])$ after the recursive call. Then,
$\oplus_{j=0}^{i}A[i]=B[\floor{i/2}]$ when $i$ is odd, and
$\oplus_{j=0}^{i}A[i]=B[\floor{i/2}-1]\oplus A[i]$ when $i$ is
even. This is what the program computes after the recursive call.

At each level of the recursion, there is $O(n)$ work to be done for
computing the pairwise sums of the input array.
Thus, the total work can be expressed by
the following \impi{recurrence relation}
\begin{eqnarray*}
  W(n) & = & W(n/2)+O(n) \\ W(1) & = & 1
\end{eqnarray*}
which can be solved by induction to give $W(n) = O(n)$. On each level
of the recursion, the pairwise sums can be computed in parallel by a
loop of independent iterations over the intermediate $B$ array of size
$\floor{n/2}$) in $O(n/p)$ time steps. Using, say, $\floor{n/2}$
processors, this is $O(1)$, and the parallel time\index{parallel time}
over all recursion levels is, therefore, expressed by
\begin{eqnarray*}
  T(n) & = & T(n/2)+O(1) \\ T(1) & = & 1
\end{eqnarray*}
which by induction gives $T(n) = O(\log n)$. The parallel running time
with $p$ processors is, therefore, in the best case $O(n/p+\log n)$.
The Master Theorem~\ref{thm:master} applies to both recurrences.

To implement the algorithm with some fixed number $p$ of processors,
the pairwise summing (loop) must be parallelized. The recursive call
is done by all processors, but before each call, the processors must
wait for each other to have completed their part of the summing loop for
which a \impi{barrier synchronization} operation is needed. Likewise, after
the recursive call, the processors must again wait for each other
before they compute the results. Two barrier synchronizations are
needed at each level for the recursion, for a total of $2\floor{\log
  n}$.

\begin{theorem}
  \label{thm:optiprefix}
  The inclusive prefix sums problem can be solved in parallel time
  $\Omega(n/p+\log n)$.
\end{theorem}

In the theorem, we have tacitly assumed that barrier
synchronization is done in $O(1)$
parallel time, which would be the case on a PRAM\index{PRAM}. On other
parallel computing systems, barrier synchronization takes $\Omega(\log p)$
parallel time steps. Therefore, a more realistic estimate of the algorithm is
$O(n/p+\log n\log p)$ parallel time\index{parallel time} steps
with $p$ processor-cores.

The recursive prefix sums algorithm needs to allocate an intermediate
array of size $\floor{n/2}$ elements at each recursive call (for a
total of $n$ elements). The pairwise summing has optimal spatial
locality (see the next lecture) and can exploit the cache system
well. It does about $2n$ summations with the $\oplus$ operator in
the two parallel loops, about twice as many as the sequential
algorithm.

\subsection{Solving Recurrences with the Master Theorem}

Recurrence relations, similar to the expression of work and time in
the previous section, will often occur in the remainder of these lectures
and many recursive algorithms give rise to this kind of very regular recurrence
relations. Instead of doing an induction proof for each new
recurrence, the solution to recurrences of this form can be summarized
in a general theorem. This is often called the ``Master Theorem'' (for
simple, regular divide-and-conquer recurrences), which exist in
different versions\index{Master Theorem}. Here is one which covers
most of the recurrences that will come up in these lectures:

\begin{theorem}
  \label{thm:master}
  Given a recurrence of the form
  \begin{eqnarray*}
    T(n) & = & a T(n/b) + \Theta(n^d\log^e n)
  \end{eqnarray*}
for constants $a\geq 1$, $b>1,$ $d\geq 0$, $e\geq 0$, and $T(1)$ some
constant.  The recurrence has the following closed-form solution
\begin{enumerate}
\item
  $T(n)=\Theta(n^d\log^e n)$ if $a/b^d < 1$ (equivalently $b^d/a>1$),
\item
  $T(n) = \Theta(n^d\log^{e+1} n)$ if $a/b^d = 1$ (equivalently
  $b^d/a=1$), and
\item
  $T(n) = \Theta(n^{\log_b a})$ if $a/b^d > 1$ (equivalently
  $b^d/a<1$).
\end{enumerate}
\end{theorem}

When the recurrence relation models a recursive procedure, $b$ is the
shrinkage or reduction factor by which the subproblems get smaller,
and $a$ is the proliferation or expansion factor, roughly the
``number'' (not necessarily integer) of subproblems to be solved at
each recursion level. It is clear that the number of levels of the
recursion is $\ceiling{\log_b n}$. A proof analyzes such recursion
trees and can be found in any good algorithms' textbook, see for
instance~\cite{ahu74,ahu87,CormenLeisersonRivestStein22,SedgewickFlajolet13,Roughgarden17}
and also a recent paper by Kuszmaul and
Leiserson~\cite{KuszmaulLeiserson21}.  A proof can be found in the
appendix to these lecture notes and is much recommended to study.

We can immediately apply the Master Theorem to the simple parallel
prefix sums recurrences.  For the $W(n)$ recurrence,
$W(n)=W(n/2)+O(n)$, Case 1 applies (with $a=1, b=2, d=1, e=0$) which
gives $W(n)=O(n)$.  For the $T(n)$ recurrence, $T(n/2)+O(1)$, Case 2
applies (with $a=1, b=2, d=0, e=0$) and gives $T(n)=O(\log n)$.

\subsection{Iterative Prefix Sums}
\label{sec:iterativeprefix}

Theorem~\ref{thm:optiprefix} can be achieved by a different looking,
iterative algorithm.  In fact, the iterative algorithm can be found
by unfolding the recursions of the recursive algorithm. An advantage of
the iterative prefix sums algorithm is that no intermediate array has
to be allocated.

The algorithm has two phases, an up-phase, corresponding to the
pairwise sum computations before the recursive call and a
down-phase, corresponding to the sum computations on return from
the recursive call. Both up- and down-phases take $\floor{\log_2 n}$
iterations.

In the first up-phase iteration, sums of even-odd pairs are
computed. In the next iteration, sums of pairs of every second
element are computed, in the third iteration, sums of pairs of every
fourth element, and so on. The down-phase reverses this pattern.  The
following code illustrates the algorithm.

\begin{lstlisting}[style=SnippetStyle]
int k, kk;
int i;

// up-phase
for (k=1; k<n; k=kk) {
  kk = k<<1; // double the loop increment
  for (i=kk-1; i<n; i+=kk) A[i] = A[i-k]+A[i];
}
// down-phase
for (k=k>>1; k>1; k=kk) {
  kk = k>>1; // halve the loop increment
  for (i=k-1; i<n-kk; i+=k) A[i+kk] = A[i]+A[i+kk];
}
\end{lstlisting}

The correctness of the up-down-phase inclusive prefix sums\index{prefix sums}
algorithm
(and implementation) can be proven by showing that certain invariant
properties are maintained for each iteration and, in the end, imply the
desired end result. To formulate the invariants\index{invariant}, let
$a_i, 0\leq i<n$ be the input sequence for which the inclusive
prefix sums are to be computed in $A[i]$, that is, $A[i] =
\oplus_{j=0}^{i}a_i$.

For the up-phase, the following invariant\index{invariant}
will hold before iteration
$k, k=0,1,\ldots,\floor{\log_2 n}$: For each $i, i<n$ of the form
$i=j2^k-1$ for some $j>0$,
\begin{eqnarray*}
  A[i] = \oplus_{j=i+1-2^k}^{i}a_j
\end{eqnarray*}

That is, every $2^k$th $A[i]$ will store the sum of the $2^k$ previous
elements up to and including the $i$th element itself. This clearly
holds before the first iteration ($k=0$), since the input array is
$A[i]=a_i=\sum_{j=i}^{i}a_j$. Assuming that the property holds before
iteration $k, k>0$, we have for that iteration which computes
$A[i-2^k]\oplus A[i]$ into $A[i]$ for elements $i=j2^{k+1}$ that
\begin{displaymath}
A[i]=(\oplus_{j=i-2^k+1-2^k}^{i-2^k}a_j)\oplus
(\oplus_{j=i+1-2^k}^{i}a_j)=\oplus_{j=i+1-2^{k+1}}^{i}a_j
\end{displaymath}
for all $i$ of the form $i=j2^{k+1}+1$.
Thus, the invariant\index{invariant} holds before the start
of iteration $k+1$. We can, by the way, observe that all $A[i]$ with
$i=2^k-1$ for $k=0,\ldots, \floor{\log_2 n}$ are ``good'' in the sense
of correctly containing the $i$th prefix sum. The task of the
down-phase is to make all other elements in $A$ ``good'' as well. Also
note here that the variables \texttt{k} and \texttt{kk} in the program
are $2^k$ and $2^{k+1}$, respectively, for the iteration count $k$ in
the proof.

The down-phase starts with the results computed in the $A$ array by
the up-phase. The invariant\index{invariant}
for the $k$th iteration for $k=\floor{\log_2
  n}, \floor{\log_2 n}-1,\ldots,0$ is that each $2^k$th element is
``good'' in the sense that $A[i]=\oplus_{j=0}^{i}a_j$ for $i$ of the form
$i=j2^k-1$. From the up-phase, this holds before the first
iteration. In the iteration, the program computes $A[i]\oplus A[i+2^{k-1}]$
into $A[i+2^{k-1}]$. So, assuming the invariant to hold, we have that
\begin{displaymath}
A[i+2^{k-1}]=(\oplus_{j=0}^{i}a_j)\oplus
(\oplus_{j=i+2^{k-1}+1-2^{k-1}}^{i+2^{k-1}}a_j)=\oplus_{j=0}^{i+2^{k-1}}a_j
\end{displaymath}
by the ``goodness'' of $A[i]$ and the invariant\index{invariant}
from the up-phase for
$A[i+2^{k-1}]$. The iteration, therefore, makes $A[i+2^{k-1}]$ ``good'',
and $i+2^{k-1}$ is of the form $j2^{k-1}-1$ for the next
iteration. After the last iteration when $k=1$, this implies that
$A[i]=\oplus_{j=0}^{i}a_i$ for all $i$. Thus, the prefix sums for
all indices are correctly computed in the $A$ array.

The algorithm consists of the $\floor{\log_2 n}$ iterations of
loops of independent iterations with successively half the number of
elements in the up-phase and double the number of elements in the down-phase.
After each loop, the processor-cores employed have to be synchronized with
some form of barrier synchronization, each of which may take $\Omega(\log p)$
parallel time\index{parallel time} steps.

The algorithm achieves the bound stated in
Theorem~\ref{thm:optiprefix}.  It also does about $2n$ summations with
the $\oplus$ operator in the up- and down-phase parallel loops,
about twice as many as the sequential algorithm. A drawback is that
the pairs being summed are farther and farther apart ($1,2,4,\ldots$).
Thus, the iterative algorithm has worse \impi{spatial locality}
than the recursive algorithm (more on spatial locality in the next
lecture).

It is an important theoretical result that any logarithmically fast
parallel prefix sums algorithm\index{prefix sums} has to do twice the
number of sequentially required $\oplus$ operations. Paraphrasing,
something like the following result has been proved (using yet another
model of parallel computation: the \impi{arithmetical circuit}).

\begin{theorem}
  \label{thm:prefix-tradeoff}
  For computing the inclusive prefix sums of an $n$-element input
  sequence, the following trade-off holds between \emph{size} $s$
  (roughly number of $\oplus$ operations done by gates) and
  \emph{depth} $t$ (parallel time, longest path from an input to an
  output): $s+t\geq 2n-2$.
\end{theorem}
This was proved by Snir~\cite{Snir86}, a more intuitive proof can be
found in~\cite{ZhuChengGraham06}.

The theorem says that for any fast (sublinear, logarithmic) parallel
prefix sums\index{prefix sums} algorithm,
the speed-up\index{speed-up} (when counting the possibly
expensive $\oplus$ operations) is at most $p/2$. This is bad news for
highly parallel algorithms running on a large number of processors
which may use prefix sums for array compaction\index{array compaction}
and other important computations.  The trade-off also tells us how
many operations a best possible parallel prefix sums algorithm is
allowed to perform.

\subsection{Non Work-Optimal, Faster Prefix Sums}
\label{sec:hillissteele}

The two previous algorithms executed the loops summing pairs of
elements $2\floor{\log n}$ times. The next algorithm will reduce this
to about $\ceiling{\log n}$ loops, but the price is that it is no
longer work-optimal. The algorithm has been discovered many times, and
in these lectures we use the name Hillis--Steele after some of the
discoverers~\cite{HillisSteele86}. The algorithm computes the
inclusive prefix sums in-place\index{prefix sums} in an array $A$.

In the Hillis--Steele algorithm\index{Hillis--Steele algorithm},
a $\oplus$ operation is done for (almost) all of the
$n$ array elements in each iteration. In the first iteration, for each
element $i$, except the first, $A[i]$ is updated by summing with its
adjacent element, $A[i] = A[i-1] \oplus A[i]$. In the next iterations,
the update is $A[i] = A[i-2]\oplus A[i]$, in the third iteration $A[i]
= A[i-4]\oplus A[i]$, and so on, in iteration $k$, $A[i] =
A[i-2^k]\oplus A[i]$ when $i-2^k\geq 0$.
Each iteration can be written as a loop of,
unfortunately, flow (forward) dependent iterations. The dependencies
can easily be eliminated by performing the updates into a result
array $B$ and swapping $A$ and $B$ after the iteration. The following
code snippet shows how.

\begin{lstlisting}[style=SnippetStyle]
int *a, *b, *t;

a = A; b = B;
  
k = 1;
while (k<n) {
  // update into B
  for (i=0; i<k; i++) b[i] = a[i];
  for (i=k; i<n; i++) b[i] = a[i-k]+a[i];

  t = a; a = b; b = t; // swap
  k <<= 1;             // and double
}
if (a!=A) {
  for (i=0; i<n; i++) A[i] = B[i]; // copy back when necessary
}
\end{lstlisting}

It is easy to prove by invariants\index{invariant} that the
Hillis--Steele algorithm\index{Hillis--Steele algorithm}
correctly computes all inclusive
prefix sums. Assuming that $a[i]=a_i$ for the input sequence $a_i$,
one invariant is clearly that before iteration $k$ it holds that
$a[i]=\oplus_{\max(i-2^k+1,0)}^{i}a_i$ for each $i>0$, which implies
the claim when $2^k\geq n-1$.  As in the iterative prefix sums
program, the variable \texttt{k} is $2^k$ for iteration count $k,
k\geq 0$.  The number of iterations is clearly $\ceiling{\log n}$.
The work of the algorithm is $O(n\log n)$ and it is therefore not
work-optimal. This is summarized in the theorem below.

\begin{theorem}
  The inclusive prefix sums problem can be solved in parallel time
  $O(\frac{n\log n}{p}+\log n)$.
\end{theorem}

Also, for this algorithm to be correct, the processor-cores must all
have completed their part of the parallel loop before moving on to the
next iteration, so a barrier synchronization is needed in each
iteration of the \texttt{while}-loop. This would increase the parallel
time\index{parallel time} by an $\Theta(\log p)$ factor.

\subsection{Blocking}
\label{sec:blocking}

What is the use of a prefix sums algorithm\index{prefix sums}
that is not work-optimal?
In itself, for solving the problem on input of size $n$, it is not
useful, as the larger the $n$, the smaller the
absolute speed-up\index{speed-up!absolute}.

Algorithms that are not work-optimal can, however, be useful in
context, as building blocks, where some of their advantages (like
being fast) may pay off without being hurt by the extra work they
perform. The situation is like this: If $p$ processors have already
been allocated, we may as well use them to reduce the parallel
time\index{parallel time}. There is no point in rescheduling the work to fewer
processors. The processors are there anyway and have to be paid for.

The general idea is to reduce the problem at hand to a smaller
(possibly different) problem that can be solved on $p$ processors, and
then use this solution to compute the solution to the original
problem, both steps by the use of work-optimal algorithms.  For the
whole algorithm to be work-optimal, the problem reduction and
computation of the final solution must be done by work-optimal
algorithms, but for the step in the middle, where a smaller problem is solved
on the available $p$ processors, there may be ``room enough'' to
employ a faster, but not work-optimal algorithm. Overall, this can pay
off to arrive at an algorithm that is work-optimal and possibly
faster.

When applied to the prefix sums problem, the idea is sometimes called
\impi{blocking}. An $n$-element input array $A$ is given, as are $p$
processors to solve the problem. The input array is divided into $p$
\emph{blocks} of about $n/p$ elements each, one for each of the
processors. Each processor performs a sequential reduction on its
block of elements and puts the results into an array $B$ of $p$
elements, one for each processor. Now, all prefix sums of $B$ are
computed (by either of the parallel prefix sums algorithm). After
this, each processor $i$ adds $B[i]$ to the first element of its $A$
block and computes the prefix sums over its $A$ block. This completes
the computation of the prefix sums of $A$.

The time complexity of this blocked prefix sums\index{prefix sums} algorithm,
using Hillis--Steele\index{Hillis--Steele algorithm}
as building block is $O(n/p+(p\log p)/p+n/p)=O(n/p+\log
p)$, since Hillis--Steele is applied to an array of $p$ elements
only. In contrast to Theorem~\ref{thm:optiprefix}, the
non-parallelizable term is $\log p$, not $\log n$, and with
Hillis--Steele, the constant is $1$, and not $2$, as would have been
the case with the recursive or iterative prefix sums algorithm.

\begin{theorem}
  \label{thm:optiprefix2}
  The inclusive prefix sums problem can be solved in parallel time
  $\Omega(n/p+\log p)$.
\end{theorem}

The saving of a factor of $2$ in the $\log p$ term does not sound like
much. However, if pairwise summing involves expensive communication as
is the case when the algorithm is used for distributed memory systems
and implemented with \mpi\index{MPI} (see \Sec~\ref{blk:mpi}),
such a factor can
be worthwhile. There are more dramatic applications of the blocking
technique in the literature. For instance, the fast, but not
work-optimal Common CRCW PRAM\index{PRAM!Common CRCW} maximum
finding algorithm of Theorem~\ref{alg:fastmax} can be used to devise a
work-optimal Common CRCW PRAM maximum finding algorithm running
in $O(\log\log n)$ time steps (see \Sec~\ref{sec:veryfastmax}).

\subsection{Related Problems}

In the prefix sums\index{prefix sums}
and reduction problems, the elements were given in
an array, and the array order determined the order of the application
of the associative function $\oplus$. In this sense, the prefix sums
problems are \emph{oblivious}\index{oblivious},
\emph{data-independent} problems.  It
is a natural and, it has turned out, extremely useful generalization
to consider also \emph{data-dependent} prefix sums problems where the
order in which to apply the associative function is determined by an
additional list structure (array of pointers) given as part of the
input. To solve the data-dependent prefix sums problem it would
suffice to traverse the list structure and for each element count the
distance to the last (tail) element of the list (or from the first, head,
element of the list). Based on this, the input
elements can easily be put into an array in list-order on which
the prefix sums can be computed by either of our efficient algorithms.
Doing the traversal efficiently in parallel is the notorious
\impi{list ranking problem}.

Although similar to the prefix sums problem\index{prefix sums problem},
the list ranking problem
turns out to be much more intricate and much more difficult to
solve. For instance, although there are list ranking algorithms
similar to the Hillis--Steele algorithm\index{Hillis--Steele algorithm},
the simple blocking technique
does not work here.  It was a long standing problem to devise a fast,
work-optimal, deterministic list ranking\index{list ranking} algorithm.
The best deterministic result on an EREW PRAM\index{PRAM!EREW}
is the $O(n/p+\log n)$
time algorithm of Anderson and Miller~\cite{AndersonMiller91}.

\subsection{A Careful Application of Blocking\marksec}
\label{sec:carefulblocking}

By a more careful application of blocking as described in
\Sec~\ref{sec:blocking}, we can arrive at a parallel inclusive
prefix sums algorithm\index{prefix sums}
that achieves the best possible combination of
time and work, $t$ and $s$ both measured as the
number of $\oplus$ operations, captured in
Theorem~\ref{thm:prefix-tradeoff}, namely $s+t=2n-2$ (equality). The
algorithm is reasonably fast when $n$ is large compared to $p$.  The
trick is to divide the input sequence of $n$ elements into $p+1$
blocks (for $p$ processors) of $n/(p+1)$ elements, instead of just $p$
blocks as was done above. Assume now that $(p+1)$ divides $n$; this
assumption can with a little care easily be lifted by dealing with
some blocks of $\ceiling{n/(p+1)}$ elements and some blocks with
$\floor{n/(p+1)}$ elements. The blocks are ordered, the first block
contains the first $n/(p+1)$ input elements (block $0$), the second
block the next $n/(p+1)$ elements (block $1$), and so on; the last
block (block $p$) contains the last $n/(p+1)$ elements.

We measure the time $t$ in the number of $\oplus$ computations that
have to be carried out in sequence (the depth\index{depth})
and work (or size) $s$ as the
total number of $\oplus$ operations carried out by the $p$
processors. The prefix sums algorithm consists of three steps.

\begin{enumerate}
\item Compute, for each of the first $p$ blocks, the inclusive
  prefix sums for the $n/(p+1)$ elements in the block. This takes
  $t_1=\frac{n}{p+1}-1$ operations (time) and requires a total work
  of $s_1=p\left(\frac{n}{p+1}-1\right)$ operations.
\item Compute the inclusive prefix sums for a sequence consisting of
  the $p$ sums of the elements in each of the first $p$ blocks: This
  is for each block the prefix sum for the last element computed in Step 1.
  This takes time $t_2=p-1$ and work $s_2=p-1$ operations.
\item For the $p-1$ blocks $1,2,\ldots,p-1$, excluding the first block
  $0$, which is done (all prefix sums computed by the first step), and
  the last block $p$, which is special, add the prefix sum for the last
  block to the first $\frac{n}{p+1}-1$ elements of the block. This
  results in the correct prefix sums for all elements, since the
  prefix sum for the last element of each block is the prefix sum for
  the block that was computed in Step $2$.  This takes time
  $t_3=\frac{n}{p+1}-1$ and work of $(p-1)\left(\frac{n}{p+1}-1\right)$
  operations. For the last block (block $p$), instead the prefix sum
  for block $p-1$ is added to the first element of the block, and the
  inclusive prefix sums for the $n/(p+1)$ elements of the block are
  computed. This takes time of $n/(p+1)=t_3+1$ operations and
  another $n/(p+1)$ operations of work. The total work (number of
  operations) for the last step is therefore
  $s_3=1+p\left(\frac{n}{p+1}-1\right)$.
\end{enumerate}

The total time for this algorithm is
\begin{eqnarray*}
  t & = & t_1+t_2+t_3 \\ & = & \left(\frac{n}{p+1}-1\right) + p-1 +
  \left(\frac{n}{p+1}-1\right)+1 \\ & = &
  2\left(\frac{n}{p+1}-1\right)+p
\end{eqnarray*}
    
The total work for this algorithm is
\begin{eqnarray*}
  s & = & s_1+s_2+s_3 \\ & = &
  p\left(\frac{n}{p+1}-1\right)+(p-1)+p\left(\frac{n}{p+1}-1\right)+1
  \\ & = & 2p\left(\frac{n}{p+1}-1\right)+p
\end{eqnarray*}

The sum of work and time is
\begin{eqnarray*}
  s+t & = &
  2p\left(\frac{n}{p+1}-1\right)+p+2\left(\frac{n}{p+1}-1\right)+p
  \\ & = & 2(p+1)\frac{n}{p+1}-2(p+1)+2p \\ & = & 2n-2
\end{eqnarray*}
which meets the trade-off of Theorem~\ref{thm:prefix-tradeoff}. When
carefully implemented, the algorithm could run in $O(n/p+p)$ parallel time
steps not counting the two barrier synchronization operations
that may be needed.

The same trick of dividing an input sequence into $p+1$ blocks was
used by Snir~\cite{Snir85} to speed up binary search from $\log_2 n$
to $\log_{p+1} n$ comparison steps. It was also shown that this is the
best possible. Note that the number of steps is constant if $n$ is in
$O(p^k)$ for some constant $k\geq 1$.

\subsection{A very Fast, Work-Optimal Maximum Algorithm}
\label{sec:veryfastmax}

Can the maximally fast, $O(1)$ time step
Common CRCW PRAM\index{PRAM!Common CRCW} algorithm of
Theorem~\ref{alg:fastmax} be made work-optimal or more efficient? By
itself not, but with the blocking technique, it can be put to
use to achieve a very fast and work-optimal algorithm for finding the
maximum of a sequence of $n$ numbers. We prove the following theorem
constructively by outlining the corresponding algorithm.

\begin{theorem}
  \label{alg:veryfastmax}
The maximum of $n$ numbers stored in an array can be found in
$O(\log\log n)$ parallel time steps, using $n/\log\log n$ processors
and performing $O(n)$ operations on a Common CRCW PRAM.
\end{theorem}

Divide the array into blocks of roughly $\sqrt{n}$ numbers. Assume
(recursively) that the maximum has been found for each of these
$\sqrt{n}$ blocks. Now, we can employ the optimally fast
maximum finding algorithm to find the maximum among these $\sqrt{n}$
block maxima in $O(1)$ time steps and
$O((\sqrt{n})^2)=O((n^{\frac{1}{2}})^2)=O(n)$ work.  The time and work,
including the recursive solution to the $\sqrt{n}$ subproblems of roughly
$\sqrt{n}$ numbers each, is given by the following recurrence relations.

For the time, we have
\begin{eqnarray*}
  T(n) & = & T(\sqrt{n})+1\\ T(1) & = & 1
\end{eqnarray*}
and for the work
\begin{eqnarray*}
  W(n) & = & \sqrt{n}W(\sqrt{n})+n\\ W(1) & = & 1
\end{eqnarray*}

Neither of these recursions are covered by the Master
Theorem~\ref{thm:master}. Its is, however, easy to guess a closed form
and verify the guess by induction. For the time recurrence $T(n)$,
we see that we have to repeat taking the square root of
$n$ until we get down to some constant. We conjecture that
$T(n)=\log\log n$.  With this as induction hypothesis, we get $T(n) =
T(\sqrt{n})+1 = \log\log\sqrt{n}+1 = \log(\frac{1}{2}\log n)+1 =
\log\frac{1}{2}+\log\log n+1 = -1+\log\log n+1 = \log\log n$.
Similarly, we can find that $W(n)=n\log\log n$.

This recursive algorithm gives the time claimed in
Theorem~\ref{alg:veryfastmax}, but the work of $O(n\log\log n)$
operations is still too much. Precomputation, in parallel, by
blocking, with the right number of processors, decreases the work to
the desired $O(n)$ operations. Let $n$ be the size of the given input
array. The work-optimal algorithm does the following.
\begin{enumerate}
\item
  Divide the input into $n/\log\log n$ blocks of roughly $\log\log n$
  elements. Assign a processor to each of the blocks to find a maximum
  for each block. This preprocessing has reduced the problem size to
  $n/\log\log n$ block maxima and takes $O(\log\log n)$ parallel
  time\index{parallel time} steps and $O(n)$ work using
  the $n/\log\log n$ processors.
\item
  Apply the fast, recursive algorithm with $n/\log\log n$ processors
  to the reduced problem to find the maximum (of the original input)
  in
  \begin{displaymath}
    O(\log\log(n/\log\log n)) = O(\log\log n)
  \end{displaymath}
  parallel time steps. The parallel work is
  \begin{displaymath}
    O((n/\log\log n)\log\log(n/\log\log n))=O(n)
  \end{displaymath}
  as desired.
\end{enumerate}

This very fast maximum finding algorithm dates back to early work on
fast and efficient PRAM\index{PRAM}
algorithms~\cite{ColeVishkin86:micromacro,ShiloachVishkin81}.

\subsection{Do Fast Parallel Algorithms always Exist?\marksec}

Cost-optimal algorithms have linear speed-up\index{speed-up!linear},
and are especially attractive if they are efficient with a slowly growing
iso-efficiency function\index{iso-efficiency function}. For a fixed
input size, they can possibly achieve a solid, linear speed-up up to
some large number of processors. Many examples of work- and
cost-optimal algorithms were discussed (for merging\index{merging} and
prefix sums\index{prefix sums},
for instance) that can even achieve a logarithmic running time in the
size of the input given that enough processors are available.
It is a central question,
not only for complexity theory and theoretical computer science, but
also for the practitioner, whether such fast
parallel algorithms exist for all problems.

A qualified, ``most likely no'' answer is given by standard complexity
theory~\cite{Papadimitriou94}. In outline, the answer is as follows.

The complexity class $\cal{P}$\index{complexity class!$\cal{P}$} is
the class of \impi{tractable problems}, computational problems that
can be solved in polynomial time in the worst case in the size of the
input on a Random Access Machine (RAM)\index{RAM}
or other reasonable model of sequential computation.  The question is now
whether all tractable problems can be solved fast in parallel.  The
complexity class $\cal{NC}$ (``Nick's class'', after Nicholas
Pippenger)\index{complexity class!$\cal{NC}$} is the class of problems
than can be solved in poly-logarithmic parallel time $O(\log^c n)$ in
the size of the input $n$ for some constant $c$ with a polynomial
number of processors in the size of the input $n$ on a Parallel Random
Access Machine (PRAM)\index{PRAM} or other reasonable model of parallel
computation. Put differently, $\cal{NC}$ is the class of parallel
algorithms with tractable (polynomial) costs and poly-logarithmic
parallel time complexity\index{parallel time complexity}.  By a
simulation argument\index{simulation argument}, it is clear that
$\cal{NC}\subseteq\cal{P}$. So, the question is whether
$\cal{NC}=\cal{P}$ or $\cal{NC}\neq\cal{P}$. That is, whether all
tractable problems can be solved by an algorithm that runs in
poly-logarithmic time with a polynomial number of processors on a
suitable PRAM variant.  A complete problem in $\cal{P}$ with respect
to $\cal{NC}$-reduction is a problem to which all other problems in
$\cal{P}$ can be reduced by a reduction (an algorithm) that is also in
$\cal{NC}$. If such a complete problem is in $\cal{NC}$, then all
problems in $\cal{P}$ would be in $\cal{NC}$ and admit fast
parallelization.

Interestingly, it can be shown that there are indeed such complete
problems in $\cal{P}$ and that many important, classic, and
practically relevant problems are $\cal{P}$-complete. Furthermore, for
none of these problems, an $\cal{NC}$-algorithm has been found
despite much effort. It may, therefore, be that $\cal{NC}\subset\cal{P}$
and that there are problems (namely the $\cal{P}$-complete problems)
that do not admit fast parallel solution with ``only'' polynomial
resources (number of processors). Some problems that are
$\cal{P}$-complete under $\cal{NC}$-reduction are (ordered) depth
first search (ODFS)\index{ordered depth first search}, maximum
flow\index{maximum flow problem}, and linear
programming\index{linear programming}~\cite{Cook85,GreenlawHooverRuzzo95,JaJa92,JonesLaaser77}.
These problems may, thus, turn out to be, in a sense
\impi{inherently sequential}.

The emphasis in these lectures has been on work- and cost-optimal
algorithms with provable, linear speed-up\index{speed-up!linear}.
The emphasis in the small
part of parallel complexity theory outlined in this section is on
tractable problems with poly-logarithmically fast running time
(parallel time complexity\index{parallel time complexity})
and not on work- and cost-optimality.
A problem in $\cal{NC}$ may or may not be cost- or
work-optimal and indeed very far removed from that. Establishing
membership in $\cal{NC}$ is, therefore, only one aspect of parallel
algorithmics and parallel complexity theory.

\section{Exercises}

\begin{enumerate}
\item
  Is the PRAM a NUMA or a UMA model? Is the PRAM a SIMD or a MIMD
  model?  Does the SPMD characterization apply to the PRAM?
  Anticipating the programming frameworks to come, what can the
  advantages of adhering to an SPMD style possibly be? Anticipating
  even further, is the PRAM a PGAS model?
\item
  Let $A$ be a two-dimensional $m\times n$ element matrix stored as
  \texttt{A[i,j]} and $x$ an $n$-element vector stored as
  \texttt{x[j]}.  Consider the following PRAM algorithm:

\begin{lstlisting}[style=SnippetStyle]
par (0<=i<m) b[i] = 0;
par (0<=i<m) {
  for (j=0; j<n; j++) b[i] = b[i]+A[i,j]*x[j];
}
\end{lstlisting}

Explain what this PRAM algorithm accomplishes. What is the number of
parallel steps of the algorithm? What is the parallel time taken
for the algorithm to finish? What is the total number of operations
performed by the processors of the algorithm (parallel work)? Which
PRAM variant is needed for the algorithm to work correctly? Which
PRAM variant is sufficient? Explain your answers.
\item
  Modify the parallel $O(\log n)$ time algorithm of
  Theorem~\ref{alg:logmax} for finding a maximum among $n$ elements
  stored in an array to perform a reduction, \eg, compute the sum,
  over the $n$ elements for a given, associative operator
  $\oplus$. You may assume that the operator is commutative, so that the
  summands may be used in any order. What is the total number of
  operations performed by the assigned processors? Which PRAM variant
  is needed?
\item
  Give a different (or modify the above) algorithm for performing
  reductions over the elements in an $n$-element array \texttt{a} that
  works for not
  necessarily commutative but still associative operators $\odot$. This
  means that the sum must be computed in a fixed order as
  $\texttt{a}[0]+\texttt{a}[1]+\ldots +\texttt{a}[n-1]$ (summations
  can, of course, be grouped into smaller parts, but the order of the
  summands must not be changed).  The algorithm must run in $O(\log
  n)$ parallel time steps. What is the total number of operations? Is
  the algorithm work-optimal? Is the algorithm cost-optimal?
\item
  Modify the parallel $O(\log n)$ time algorithm of
  Theorem~\ref{alg:logmax} for finding a maximum among $n$ elements
  stored in an array \texttt{a} to instead copy a specific element
  \texttt{a[r]} for some given index \texttt{r} between $0$ and $n-1$
  to all positions of \texttt{a}. The algorithm must work on an EREW
  PRAM. What is the total number of operations performed? Are any
  further assumptions needed to guarantee that the EREW PRAM
  capabilities suffice?  What is the time and work of the copy
  operation on a CREW PRAM?
\item
  Show how to make $p-1$ additional copies of a large array of
  $n$ elements on an EREW PRAM by actually writing a program in PRAM
  pseudo-code. More precisely, given an $n\times p$ matrix
  \texttt{a[n][p]}, copy a specific column \texttt{a[][r]} to all
  other columns \texttt{a[][i]} for $0\leq i<p,i\neq r$ for any given
  $r$ as input.  It may be assumed that $n$ is (much) larger than $p$
  and that $p$ divides $n$, $p|n$.  Note that the total number of
  operations required is $\Omega(n(p-1))$ so that the best possible
  number of parallel time steps is $\Omega(n-n/p)$.  What is the
  number of parallel time steps of your algorithm? What if $p$ does
  not divide $n$, $p\not|n$.
\item
  Give a PRAM algorithm for matrix--vector multiplication that runs in
  $O(\log n)$ time steps for vectors of $n$ elements. Hint: Use and
  modify the $O(\log n)$ time algorithm for finding the maximum of $n$
  numbers. Which PRAM variant is needed? Can the algorithm be made
  to work on an EREW PRAM?
\item
  Give an EREW PRAM algorithm for adding two $m\times n$ matrices $A$ and
  $B$, that is, for computing $C =A+B$. What is the number of parallel
  time steps of your algorithm? What is the total number of operations?
\item
  Give a work-optimal $n\times n$ matrix--matrix multiplication
  algorithm running in $O(\log n)$ time steps, first on a CREW PRAM, then
  on an EREW PRAM.  You may assume that the optimal work of sequential
  matrix--matrix multiplication is in $O(n^3)$.
\item
  A collection of $n$ list elements is stored in an array with
  an additional $n$-element array \texttt{next} that for each list element 
  gives the index of a next following element.
  The indices in this array must fulfill that $0\leq
  \texttt{next}[i]<n$ and that, for each $i, 0\leq i<n$, there is at
  most one $j,0\leq j<n$ such that $\texttt{next}[j]=i$.  A
  \emph{tail} (final, last element of the list) is an element $i$ with
  $\texttt{next}[i]=i$, that is, an element indexing itself.  A
  \emph{head} (initial, first, start element of the list) is an element
  $i$ to which no other element points; that is, with no $j$ such that
  $\texttt{next}[j]=i$. The \texttt{next}-array must have at least one tail
  and at least one head element. If there is one head and one tail element
  in the \texttt{next}-array, the collection is a single list, if there are
  more the collection consists of several shorter sublists.

  Let now an $n$-element index array \texttt{next} be given.  Devise a
  fast $O(\log n)$ time step PRAM algorithm to verify that the
  \texttt{next} array fulfils the conditions described above.  Attach
  flags, stored in $n$-element arrays, which for each list element
  tell whether the element is a head or a tail element. How many
  operations does your algorithm perform? How does it compare in terms
  of number of operations to a sequential algorithm that analyzes and
  traverses the \texttt{next}-array? Are there interesting trade-offs
  between different PRAM variants? The Arbitrary CRCW PRAM may be
  relevant to consider. Is is possible to easily decide whether the
  \texttt{next} array represents exactly one list?
\item
  Given a collection of $n$ list elements represented as described in
  the previous exercise: Devise a super fast and efficient (in number
  of operations performed) EREW PRAM algorithm to make this singly
  linked list a doubly linked list. The algorithm should compute an
  additional index array \texttt{prev} that for each list element $i$
  gives the previous (preceding) element. That is, it must hold for
  all $i,0\leq i<n$ that $\texttt{next}[\texttt{prev}[i]]=i$ (the
  preceding element of a head element shall be the element itself).
\item
  Consider the following PRAM program. It is intended to work on a
  list defined by an $n$-element array of \texttt{next} indices, as
  described in the two previous exercises. The $n$-element arrays
  \texttt{tail}, \texttt{dist}, and \texttt{sum} store new
  information for the list elements and can be assumed to already 
  have been allocated. The \texttt{sum} array stores results that have
  to be computed, and this array has been initialized with an input value
  for each list element.
\begin{lstlisting}[style=SnippetStyle]
par (0<=i<n) {
  tail[i] = next[i];
  if (tail[i]!=i) dist[i] = 1; else dist[i] = 0;
}
for (k=1; k<n; k<<=1) {
  par (0<=i<n) {
    if (tail[tail[i]]!=tail[i]) {
      dist[i] = dist[i]+dist[tail[i]];
      sum[i]  = sum[i]+sum[tail[i]];
      tail[i] = tail[tail[i]];
    }
  }
}
\end{lstlisting}

What does the algorithm accomplish, in particular, what will be the
contents of the \texttt{dist} and \texttt{sum} arrays after it has finished?
What is the number of parallel steps taken by the algorithm? What is
the number of operations performed?  Which PRAM variant does it need?
Can you make the algorithm work on an EREW PRAM?  Devise a
sequential algorithm achieving the same result. What is the complexity
of your sequential algorithm?  Is the PRAM algorithm
work-optimal when compared to your best possible sequential algorithm?
Note: This algorithm is Wyllie's list ranking algorithm and
illustrates an important technique (for achieving logarithmic parallel
time complexity) called \impi{pointer jumping}.
\item
  A directed graph $G=(V,E)$ with $n=|V|$ vertices numbered
  consecutively $0,\ldots,n-1$ is represented by an $n\times n$
  adjacency (incidence) matrix $\texttt{A}[n,n]$.  In the adjacency
  matrix, $\texttt{A}[i,j]=1$ iff there is a directed edge in $G$ from
  vertex $i$ to vertex $j$ and $\texttt{A}[i,j]=0$ if there is no
  such edge. This is the input to the program you have to devise.  It
  is not known from the input how many edges $G$ has and neither are
  the out-degree and the in-degree of the vertices.

  Write a (slow, \ie, not necessarily $O(\log n)$ steps) EREW PRAM
  program for computing the in-degree and the out-degree of all
  vertices $V$ in $G$. The out-degree and the in-degree of vertex $i$
  shall be stored as $\texttt{outdeg}[i]$ and $\texttt{indeg}[i]$,
  respectively.  What is the running time (number of time steps) and
  the work (total number of operations) of your program?  Write a fast
  $O(\log n)$ PRAM algorithm for computing $m$, the number of edges in
  $G$ (hint: see the previous exercises). Which PRAM model is needed?
  What is the number of operations performed by your algorithm? How
  does it compare to a best sequential algorithm operating on the
  same representation of $G$?
\item
  A directed graph $G=(V,E)$ with $n=|V|$ vertices numbered
  consecutively $0,\ldots,n-1$ is represented as a set of $n$
  adjacency lists: For each vertex $i$, there is a list of adjacent
  vertices $j$, stored as a consecutive array with
  $\texttt{outdeg}[i]$ elements. The \texttt{outdeg}-array is given as
  part of the input. It may be assumed that all adjacency lists are
  stored consecutively in a larger array with $m$ elements in total, where $m$
  is the number of edges of $G$. In this array, the list of adjacent
  vertices for vertex $i$ start at index $\texttt{adj}[i]$.  Devise a
  sequential algorithm to compute the in-degree for each vertex $i$.
  What is the complexity of this best possible sequential algorithm?
  Devise a fast PRAM algorithm to accomplish the same task. Which PRAM
  variant does your solution require? Is the algorithm efficient in
  comparison to the sequential algorithm (in number of operations
  performed by the PRAM processors)?

  Now, extend the algorithm to compute for each
  vertex $i$ an array storing the vertices that are adjacent to $i$
  (that is, the list of incoming edges of $i$). Hint: You will
  probably have to use extra space, $n^2$ instead of $m$, and perform
  asymptotically more operations than the sequential algorithm.
\item
  Consider the SPMD PRAM program execution of a conditional statement
  in which some processors execute one (the \textbf{true}) branch and
  some other processors execute the other (the \textbf{false}) branch.
  If the two branches consist of different numbers of instructions (as
  was disallowed for PRAM programs discussed in the text), the
  processors will not reach the end of the conditional statement in
  the same step (clock cycle) and in that sense they will not be
  synchronized anymore (at the algorithmic level) even though the
  individual instructions are executed in lock-step. Devise a PRAM
  barrier synchronization algorithm that will ensure that processors
  reach a specified synchronization point in the same
  instruction. Your algorithm should use $O(\log p)$ instructions on a
  $p$-processor PRAM. What is the smallest constant you can achieve?
  Your algorithm should preferably run on an EREW PRAM.
\item
  What is the parallel time complexity of finding the maximum of $n$
  numbers stored in an array?
\item
  Consider and give an example of a sequential algorithm running in
  $O(mn)$ operations: Which problem could be solved by such an
  algorithm? Is the algorithm of your choice in $\Theta(mn)$?
  Assume that different parallel algorithms for the problem have been
  developed running in parallel time $O(mn^2/p+n^2)$, $O((mn\log
  n)/p+n)$, $O(mn/p+n)$, $O(mn/p+\log n)$ and $O(mn/p+\log n\log p)$,
  respectively.  Explain how these running times could possibly
  have been achieved, say, on a PRAM with different algorithmic approaches.
  Do these parallelizations have linear
  speed-up? Can they have perfect speed-up? Are thy work-optimal?
  Are they cost-optimal?
\item
  Repeat the previous exercise with a sequential algorithm running in
  $O(n+m)$ time steps and with parallel algorithms running in
  $O(((n+m)\log n)/p+n)$, $O((n+m)/p+n)$, $O((n+m)/p+n\log n)$,
  $O(n+m/p)$ and $O((n+(m\log n)/p+\log n)$, respectively.
\item
  Explain why standard BFS (Breadth-First Search)
  algorithms on graphs $G=(V,E)$
  with $n$ vertices and $m$ edges starting from a given source vertex
  $s\in V)$ are \emph{not} in $\Theta(n+m)$. Give an example where the running
  time of the standard sequential algorithm is $o(n+m)$.
\item
  Some parallel algorithms with running times $O(n/p+\log n)$,
  $O((m+n)/p+\log n)$ and $O(n^2/p+\log n)$ are given. What are their
  parallel time complexities? Give expressions that characterize the number
  of processors needed to reach the claimed parallel time.
\item
  Different parallel algorithms for a computational problem that can
  be solved sequentially in $O(n)$ time have been given with running
  times $O(n/p+\log n)$, $O(\frac{n}{p/\log p}+\log n)$ and
  $O(\frac{n}{\sqrt{p}}+\log n)$, respectively.  What is the parallel
  time complexity of the three parallel algorithms?  Give expressions
  that characterize the number of processors needed to reach the
  minimum parallel running time. Which of the algorithms are
  cost-optimal?  Which algorithm would be preferable for solving the
  problem and why?
\item
  Let $\tseq(n)$ for four computational problems be in $O(n)$, $O(n\log
  n)$, $O(n\sqrt{n})$, $O(n^2)$, respectively. For each, there is a parallel
  algorithm with fastest possible running time (parallel time complexity) in
  $O(\log^2 n)$. Give corresponding expressions for the parallel running
  times with $p$ processors for four parallel algorithms that can give
  linear speed-up in the four cases.
\item
  What is the parallelism of an algorithm that is susceptible to Amdahl's Law?
  Can such an algorithm be cost-optimal? Is it work-optimal?
\item
  A program works on square matrices of order $n$ and performs a large
  number of matrix--vector multiplications. The number of such
  multiplications is some constant $k$. Each iteration takes $O(n^2)$
  time steps and has been perfectly parallelized (up to some number
  $p, p<n^2$ of processors). A sequential preprocessing of the input
  matrix is necessary and takes $cn^2$ operations for some (medium
  large) $c$.
  
  What is the maximum speed-up that this program can achieve as a
  function of $c,k,n$?
  Calculate the concrete speed-up with $c=100,k=10000,n=1000$, and
  $p=10$ and $p=100$ processors, respectively.
  What is the maximum speed-up that can be achieved for
  $c=1000,k=100000,n=10000$?
\item
  Two computational problems have best sequential algorithms with
  running times in $O(n)$ and $O(n^3)$, respectively. Assume they can
  be parallelized, however, with a fraction of the work being strictly
  sequential (non-parallelizable).  With a multi-core processor with
  $64$ cores, we want to achieve a speed-up of $60$.  How large can
  the sequential (non-parallelizable) fraction of the work be in the
  two cases? Are there differences between the two cases?
\item
  A simple, parallel, work-optimal matrix--vector multiplication
  algorithm is running in $O(n^2/p+n)$ time steps on input of $n^2 +
  n$ elements (matrix and vector), whereas sequential matrix--vector
  multiplication can be done (optimally) in $O(n^2)$ time steps.  How
  large must $n$ be in order to achieve a speed-up of 64 on 128
  processor-cores? Which assumptions on the constants in the parallel
  and sequential algorithms are needed for the calculation?
\item
  Consider two parallel matrix--matrix multiplication algorithms with
  running times $O(n^3/p+n)$ and $O(n^3/p+n^2)$, respectively.  You
  are asked to perform a weak-scaling analysis of the algorithms. In
  this analysis, the average work over the $p$ processors should stay
  fixed at some given number of operations $w$.  Assuming that the
  sequential algorithm used as (best known) baseline has work
  $O(n^3)$, you have to determine up to how many processors our
  algorithms can work efficiently if the average work per processor is
  to be kept at $w=n^3/p$.  How does $n$ have to grow as a function of
  $p$ to keep constant average work $w$?  What are the asymptotic
  running times for the two algorithms as a function of $w$ and $p$?
  Up to how many processors will the two algorithms be weakly scaling
  (that is, have constant running time independent of $p$)?  When do
  the second terms in the parallel running times start to dominate?
\item
  We have a number of parallel algorithms for matrix--matrix
  multiplication at our disposal, running in time 
  $O(n^3/p+\log n)$, $O(n^3/p+\sqrt{n})$, $O(n^3/p+n)$,
  $O(n^3/p+n\sqrt{n})$, $O(n^3/p+n^2)$, respectively. The best
  sequential algorithm known to us is, for now, running in $O(n^3)$
  time.

  For each of the five cases, assuming the asymptotic constants have
  been normalized to $1$, state the maximum number of processors that
  can sensibly be used, \ie, the maximum number of processors for
  which a linear speed-up can be achieved.  For each of the five cases,
  state $T\infty(n)$ and state the parallelism. State the
  iso-efficiency functions for the five cases, that is, the
  smallest input size $n$ as a function of $p$ that is required to
  achieve a given, fixed parallel efficiency $e$.  It may not be
  possible to give a closed-form formula in each case; if not state
  that: $n$ must be at least\ldots Compute a required (integer) input
  size $n$ to maintain efficiency $e=0.5$ and efficiency $e=0.95$
  for $p=10$, $p=100$ and $p=1000$ for the parallel algorithm with
  running time $O(n^3/p+n)$.
\item
  Consider algorithms with $\tseq(n)$ in $O(n)$, in $O(n^2)$ and in
  $O(n^3)$. Assume we have found parallel algorithms with running
  times in $O(n/p+\log p)$, in $O(n^2/p+\log p)$ and in $O(n^3/p+\log
  p)$, respectively. Consider the two different definitions of weak
  scaling. Either we want to maintain a constant, given parallel efficiency
  $e$ or we want to maintain
  constant average work $w$ ($=n/p$, $=n^2/p$ or $=n^3/p$, respectively)
  by increasing $n$ as a function of $p$. Compute the iso-efficiency
  function for the three algorithms. Compute the input size scaling function
  for constant average work for the three algorithms. Can the running
  times be kept constant under constant efficiency?

  Repeat the exercise with three different parallel algorithms now running
  $O(n/p+\sqrt{n})$, in $O(n^2/p+\sqrt{n})$ and in $O(n^3/p+\sqrt{n})$,
  respectively.
\item
  A (best known) sequential algorithm for some interesting problem
  runs in $\tseq(n)=O(n\log\log n)$ time steps for input of size $n$
  (for an example, see \Sec~\ref{sec:primesieve}).  A parallel
  algorithm for the same problem running in $\tpar{p}(n)=O((n\log\log
  n)/p+\sqrt{n})$ time steps has been found. Is this parallel
  algorithm work-optimal? Does the algorithm give linear speed-up and
  if so, up to which number of processors $p$? Derive the
  iso-efficiency function for the parallel algorithm relative to the
  best known sequential algorithm. Is the parallel algorithm weakly
  scalable?
\item
  Implement the $\rank(x,A,n)$ operation for computing the number of
  elements in an ordered $n$-element array $A$ that are smaller than
  the element $x$. Assume first that elements in $A$ are distinct and
  different from $x$. What if this is not the case? Modify the
  definition of rank accordingly to either count elements $A[i]<x$ or
  elements $A[i]\leq x<A[i+1]$.
\item
  Implement the merging by binary search algorithm as a sequential
  program working on input arrays $A$ and $B$ of $n$ and $m$ elements,
  respectively.  For each element in input array $A$, compute the rank
  of $A[i]$ in the other input array $B$. For each element in input
  array $B$, compute the rank of $B[j]$ in the other input array $A$.
  Use indices and ranks to put each element from $A$ and $B$ to its
  correct position in the output array $C$. What is the sequential
  complexity of this algorithm? Compare it (experimentally) against
  the standard \texttt{seq\_merge()} function from
  \Sec~\ref{sec:merging}. What are the assumptions on the inputs in
  $A$ and $B$ for your program to be correct?  Find a way to make the
  algorithm \emph{stable} (Hint: consider the previous exercise).
\item
  Implement a function
\begin{lstlisting}[style=SnippetStyle]
void corank(int A[], const int n, int B[], const int m,
            const int i, int *j, int *k);   
\end{lstlisting}
for computing the co-ranks $j,k$ for $i, 0\leq i< n+m$ in arrays $A$ and $B$.
\item
  Assume that $p$ processors numbered from $0$ to $p-1$ are available that can
  all access input arrays $A$ and $B$ from memory. Write out pseudo-code for
  a parallel merging by co-ranking algorithm describing what each processor
  $i, 0\leq i<p$ has to do. Use the \texttt{seq\_merge()} and
  \texttt{corank()} functions and indicate where barrier synchronization is
  required in order to guarantee correct output in the $C$ array.
  You may look ahead and implement your parallel algorithm with \openmp.
\item
  Devise a synchronization free merging by co-ranking algorithm, \ie,
  an implementation where no internal barrier synchronization is
  required. Hint: Use two \texttt{corank()} calls. Challenge:
  Can you do with only one call and still be synchronization free?
  Hint: By stability, this is possible, but requires a slight change
  in the sequential merge function.
\item
  Describe how to do mergesort (sorting by merging) in parallel by doing
  $\ceiling{\log n}$ iterations of parallel merge operations. Here, $n$ is
  the size of the input array $A$ to be sorted. What is the running time
  with $p$ processors of your algorithm? Is this cost-optimal? Hint: It
  is possible to achieve $O(\frac{n\log n}{p}+\log^2 n)$ by this approach.
\item
  Write a sequential (recursive) program using Bitonic merging to merge
  any two ordered
  sequences of $n$ and $m$ elements, respectively, for any $n, n>0$ and
  $m, m>0$ (not necessarily powers of two). Make sure that the implementation
  remains oblivious, meaning that the splitting of sequences depends only
  on length and position and never on actual values of elements.
\item
  Consider the following two, semantically equivalent and
  correct implementations of a sequential, inclusive prefix sums computation.
\begin{lstlisting}[style=SnippetStyle]
for (i=1; i<n; i++) {
  a[i] = a[i-1]+a[i];
}      
\end{lstlisting}
and
\begin{lstlisting}[style=SnippetStyle]
register s = a[0]; // running sum
for (i=1; i<n; i++) {
  s += a[i];
  a[i] = s;
}
\end{lstlisting}
Create a benchmark to compare the performance of the two
implementations (you can use the \openmp framework and the \ompwtime
timing function; alternatively, implement with a C timing library like
\texttt{time.h}) with large, preinitialized arrays. Experiment with
various compiler optimization options, including no optimization. Are
there notable differences? Use different element types for the array
\texttt{a} (\texttt{int}, \texttt{double}, \ldots). Think of a model
that can explain the expected and observed differences between the two
implementations assuming the compiler does not transform one into the
other. Study the assembly output (\texttt{gcc -s}).
\item
  What is the exact number of recursive calls performed by
  the \texttt{Scan} algorithm as a function of the input array size $n$?
  What is the exact number of applications of the \texttt{+} operator as
  a function of $n$? The function $\texttt{popcount}(n)$ which counts the
  number of ones (set bits) in the binary representation of $n$ will be helpful
  to express these numbers.
  Verify your solution by implementing the recursive \texttt{Scan()} function
  and instrumenting it with a count of the number of element \texttt{+}
  operations performed (not the \texttt{i++} loop index increments). How
  much extra space for the intermediate \texttt{B}-arrays is allocated? Can
  you modify the program such that allocation is done only once and for all?
\item
  Devise an algorithm for recursively solving the exclusive
  prefix sums problem by modifying the \texttt{Scan} algorithm that
  motivated Theorem~\ref{thm:optiprefix}. What is the exact number of
  recursive calls as a function of the array size $n$? What is the
  exact number of applications of the \texttt{+} operator? Express as
  recurrence relations and solve by induction; be as general as
  possible (in the sense of exact solutions for as many $n$ as
  possible). As above, the $\texttt{popcount}(n)$ function will be helpful.
\item
  Give an algorithm for finding, in a not necessarily ordered array
  $a[n]$ of $n$ elements, the first and the last occurence of some
  element $x$ in $a$ (indices $i$ and $j$ with $a[i]=x$ and $a[j]=x$,
  $i\leq j$, as well as the total number of occurences of $x$ in $a[n]$.
\item
  Implement the iterative inclusive prefix sums algorithm (up- and down-phases)
  as a sequential function \texttt{inclusive\_prefix(int A[], int n)}.
  Modify the algorithm and your implementation to compute the exclusive
  prefix sums by a function \texttt{exclusive\_prefix(int A[], int n)}.
\item
  Prove that $a[i]=\oplus_{\max(i-2^k+1,0)}^{i}a_i$ is an invariant for
  the non work-optimal inclusive prefix sums algorithm of
  \Sec~\ref{sec:hillissteele}.
\item
  Prove the claim that $W(n)=n\log\log n$ for the recurrence
  $W(n)=W(\sqrt{n})+n$ for the very fast maximum finding algorithm in
  \Sec~\ref{sec:veryfastmax}.
\item
  Implement the optimal trade-off inclusive prefix sums algorithm
  outlined in \Sec~\ref{sec:carefulblocking}. The implementation
  should be entirely in-place, that is computation done on the input
  (and output) array with no extra arrays and only some constant
  number of additional variables (loop indices, running sums).
\item
  Give an algorithm for the exclusive prefix sums problem similar to
  the blocking algorithm of \Sec~\ref{sec:carefulblocking}.  Count the
  number of element \texttt{+} operations performed and the longest
  chain of dependent such operations.  What does this imply for the
  trade-off between the total number of additions (size) and the
  longest chain of dependent additions (depth) of
  Theorem~\ref{thm:prefix-tradeoff}?
\item
  Give an algorithm for performing $p+1$-ary (instead of binary)
  search in ordered arrays of $n$ elements with $p\geq 1$ processors.
  Show that the running time of your algorithm is $O(\log_{p+1} n)$
  (as claimed in \Sec~\ref{sec:carefulblocking}).
\item
  Explain why the following \emph{Work Law} argument is incorrect and
  does not improve the work and depth lower bounds: With $p$
  processor-cores, assign one core permanently to the work on a
  critical path. This leaves $p-1$ processor-cores to work on the
  remaining work, which can in the best case be sped up by a factor of
  $p-1$. That is, for any $p$ processor schedule it holds that $T_p(n)
  \geq \frac{T_1(n)-\tinf(n)}{p-1}$.
\item
  You are tasked with inventing a parallel algorithm for Depth First
  Search (DFS) that can provide provable speed-up for graphs that are
  not too sparse. Apparently and in contrast to Breadth-First Search (BFS)
  as discussed in \Sec~\ref{sec:bsppattern}, processing all
  the arcs directed from a found vertex in parallel will not help
  much. Another idea is needed. Describe your algorithm and state the
  parallel running time with $p$ processors for graphs with $n$ vertices and
  $m$ arcs assuming a PRAM model of computation.
  Compare the complexity to a parallel BFS algorithm.
  Hint: See~\cite{Traff13:arcelim}.
\item
  Which of the following parallel algorithms with running times $O(n/p+1)$,
  $O(n/\sqrt{p}+\log^3 p)$, $O(n/p+\sqrt{n})$, and $O(n^3/p+\log n\log p)$
  would belong to the complexity class $\cal{NC}$? Defend your answers.
\item
  Write out an iterative Common CRCW PRAM implementation of the very
  fast $O(\log\log n)$ maximum finding algorithm of
  Theorem~\ref{alg:veryfastmax} in detail. Use additional arrays for
  bookkeeping to make it possible in each iteration to look up which
  part of the input array the processor is assigned and with how many
  processors it shares work (comparisons). Use $n$ processors at first
  resulting in $O(n\log\log n)$ work and then improve by blocking to
  make the implementation work-optimal.
\end{enumerate}


\chapter{Shared Memory Parallel Systems and \openmp}
\label{chp:sharedmemory}

The middle third of the lectures on \parco is concerned with
efficient, practical use of real, parallel computing systems with a
shared memory through which processor(-core)s can exchange
information. It deals first with structure and properties of and
inevitable constraints on shared memory systems that must be
understood and taken into account in the development of algorithms and
implementations for such systems and distinguish real shared memory
systems from idealized constructs like the PRAM. As concrete
programming frameworks for such systems that illustrate many
fundamental ideas, the library \pthreads and the programming language
extension \openmp (and briefly touched upon: \cilk) are treated in
detail.

\section{Fifth block (1 lecture)}

This block is an introduction to performance-relevant aspects of real,
parallel, shared memory systems. The practical questions are how to
deal with these aspects in order to get the best possible
speed-up\index{speed-up} out
of our parallel algorithms and whether there are architectural
obstacles to achieving the linear-speed that our algorithm analysis
might suggest. Concrete results are often modest and contradictory to
first expectations. Practical \parco is challenging.

A na\"ive, parallel shared memory \emph{system model} consists of a
(fixed) number of processor-cores\index{processor-core} $p$ connected
to a large (but
finite) shared memory. Every processor-core can read/write every
location in memory, but memory access is significantly more expensive
than performing operations in the processor-core. Furthermore, memory
accesses are not uniform: From each processor-core's point of view,
some locations can be accessed (much) faster than others.  Processors
are not synchronized. All these assumptions are in stark contrast to
those made for the idealized PRAM\index{PRAM} model.

In a corresponding, shared memory \impi{programming model}, processes
or threads (being executed by the processor-cores) can likewise access
objects in a shared memory space. Processes or threads also have their
own, private memory spaces that cannot be directly accessed by other
processes or threads. There may be more processes or threads than
processor-cores. These are scheduled to run by the operating (runtime)
system (OS). Processes or threads are not synchronized, but the
programming model defines means for synchronization and exchange of
information via shared objects. In the next lectures, concrete
shared memory programming interfaces will be covered, namely the
thread programming models \pthreads\index{pthreads} and \openmp\index{OpenMP}.
A programming model\index{programming model} in which threads or processes
can be executed by any of the processor-cores, as chosen by the OS, is called
\emph{Symmetric MultiProcessing} (SMP\index{SMP}).  We here define SMP as a
property of the programming model; there are other uses of the term,
as in \emph{Symmetric MultiProcessor} where SMP\index{SMP} is rather an
architectural property. It can have advantages to leave it to the OS
to exploit the processor-cores well, but it can also have drawbacks
(for instance related to the cache system, see below).
In \parco, where our system is dedicated (check again
Definition~\ref{def:parco}), we often program with only as many
threads or processes as there are processor-cores\index{processor-core}
(dedicated to us for exclusive use) and make sure that each thread
or process is executed
by one specific core. Ensuring this binding is sometimes called
\impi{pinning} and will be discussed briefly in this lecture.

\subsection{On Caches and Locality}
\label{sec:cachelocality}

The first difference between real shared memory systems and the
na\"ive model is the existence of caches. A (hardware) \impi{cache} is
a small, fast memory close to the processor-core\index{processor-core}
that is used to store
frequently used values and, thus, to \emph{amortize} the slow access
times to the main memory. For instance, if a value that is read from
memory can be reused $10$ times, the effective main memory access time
is one tenth of what it would have been if the value had to be read at
every use.  On the other hand, with no reuse, a cache might even
introduce overhead in the memory access time.  Note that \emph{reuse}
is an \emph{algorithmic property}. Indeed, since many algorithms have
locality of access properties (see next section), caches help
immensely toward sustaining the illusion of fast, uniform memory
access (the RAM\index{RAM} model). However, some algorithms are truly
``random access'' and have no locality of accesses. For such
algorithms, caches do not help. Instead, the speed of the main memory
accesses determines the performance of such algorithms. Examples are
graph search problems (DFS,
BFS\index{DFS}\index{BFS}) on very large graphs, where the access
pattern is determined by the input graph and the next graph vertex to
be accessed would in most cases not be in the cache.

The ratio of the access times between data values fetched from main
memory and from cache memory has increased over time.  The ratio of
accessing data in main memory and accessing data in the fastest cache
(lowest level of the cache hierarchy) can easily be a factor of $10$
or more. Also nominal processor performance has (up to the early
$2000$ years) increased dramatically\index{Law!Moore's Law}.
Effectively, improvements in memory performance have not kept
up with improvements in nominal processor
performance\index{nominal processor performance}.
As a consequence, caches have grown
significantly and now typically take up a substantial amount of space
and transistors of the multi-core processor-chip\index{multi-core processor}.
Also, the cache
system itself has become more and more elaborate. The behavior of the
cache system of a standard processor can normally not be changed. The
development in caches accounted for much of the ``free lunch''.

\subsection{Cache System Recap}

The cache system of a standard processor does not work on the
granularity of single values or words in memory but on larger blocks
of memory addresses. Also, caches map addresses (locations) of words
in memory to addresses in the cache. The memory can be thought of as
being segmented into small \emph{blocks} (a typical block size could
be $64$ Bytes). Each block can be mapped to some cache line. A
\impi{cache line}, thus, stores a memory block but also some
additional meta information (bits and flags) needed by the cache
system. The terms block and cache line are sometimes used
interchangeably.

A cache in which each memory block is mapped to one, predetermined
cache line is called
\emph{directly mapped}\index{cache!directly mapped}.
The other extreme, a cache in which each memory block can
be mapped to any cache line, is called
\emph{fully associative}\index{cache!fully associative}.
A cache where each memory
can be mapped to some predetermined, small set of cache lines is
called \emph{set associative}\index{cache!set associative}. Modern
processors have set associative caches with small $k$-set sizes with
$k=2,4,8,\ldots$ and are called \emph{$k$-way set
associative}\index{cache!$k$-way set associative}.  A directly mapped
cache is a $1$-way set associative cache. Direct cache mapping schemes
can be easily implemented by means of a few integer division and
modulo operations. Associative caches need additional search logic and
are more involved.

When a processor reads a word, the memory block to which the word
belongs is calculated, and it is checked whether this block is already
in the cache. If so, the reference is a \emph{cache
hit}\index{cache!cache hit}, and the word can be read fast from the
cache. If not, the reference is a
\emph{cache miss}\index{cache!cache miss},
and the block has to be read from slow memory into a
corresponding cache line.  In an application, the cache \emph{miss/hit
rate}\index{cache!miss rate}\index{cache!hit rate} is the ratio of
cache hits/misses over a longer sequence of memory references.

On a cache miss, a new block has to be read into a corresponding cache
line. Since the cache is finite and much smaller than the main memory,
it can easily happen that the cache or cache line is full, in which
case there is a conflict and some cache line has to be \emph{evicted}.

There are three types of cache misses. A \emph{compulsory
(cold)}\index{cache!compulsory miss}\index{cache!cold miss} cache miss
happens when there are no address blocks in the cache. In this case
every first reference to some block address will lead to a cache
miss. A \emph{capacity miss}\index{cache!capacity miss} happens when
the cache (every cache line) is full; it is inevitable that some line
is evicted.  Finally, a
\emph{conflict miss}\index{cache!conflict miss}
happens when all cache lines in the set in which the block
being read can fit are occupied. Thus, a conflict miss can happen even
when the cache as a whole is not full. Conflict misses can be
particularly frequent for directly mapped caches, where it is normally
easy (if the mapping function is known) to construct cases where every
memory access will be a conflict miss. Examples include strided memory
accesses with a bad stride determined by the size of the cache.
Conflict misses can happen only for directly mapped or set-associative
caches. A fully associative cache would have only capacity misses; in
general, a capacity miss is also a conflict miss. In a $k$-way set
associative cache, either of the $k$ cache lines can be evicted upon a
conflict miss. The choice which cache line to evict is called the
\emph{eviction}\index{cache!eviction policy} or \emph{replacement
policy}\index{cache!replacement policy}.  Typically used replacement
policies are \emph{least recently used (LRU)} and \emph{least
frequently used (LFU)}. Such concrete details of the processor and
memory system may be difficult to find out.

On a write to a memory address, the workings of the cache system are a
little more involved.  If the block of the address written is already
in the cache, it is (must be) overwritten; otherwise, a subsequent
read from the cache would deliver an outdated value.  If it is not in
the cache, either a cache line for that block is \emph{allocated}
(thus, possibly resulting in a conflict miss), or the address is
updated directly in memory. The former policy is called \emph{write
allocate}\index{cache!write allocate}, the latter \emph{write
non-allocate}\index{cache!write non-allocate}.  On an update to a
block already in a cache line, the value written may nevertheless be
written to memory, which is called
\emph{write-through}\index{cache!write-through} cache. The other
possibility, that the cache line is not written to memory but kept
until it is eventually evicted, is called \emph{write
back}\index{cache!write back}.

The \emph{granularity}\index{granularity} of the cache system is in
units of memory blocks, which each hold several words (in todays
processors, typically $64$ bytes, \ie, $8$ \texttt{double} floating
point numbers). When an address is read into the cache, the whole
memory block to which the address belongs is read. Thus, at the cost
of one long read, a whole block of addresses will be in cache and some
cache misses can be avoided. Such a cache system can benefit
applications with two types of \emph{locality of access}.

An application is said to have \emph{temporal
locality}\index{cache!temporal locality} if the content of a memory
address is reused several time in brief succession with no or few
other uses in between, so that eviction will not happen.  An
application is said to have \emph{spatial
locality}\index{cache!spatial locality} if addresses in the same block
are also used (before the cache line is evicted). Again, we stress
that access locality is a property of applications and algorithms, and
only applications that have this property benefit from the cache
system. It is a lucky incident that many applications have either or
both temporal and spatial access locality, which is the reason why
hardware caching is such a successful idea.

A good computer architecture textbook can provide additional detail on
the cache system, some of which may be important for exploiting a
given system efficiently, see for instance~\cite{BryantOHallaron15}.

\subsection{Cache System and Performance: Matrix--Matrix Multiplication}

Access locality matters: A standard, and highly illustrative example
application is
matrix--matrix multiplication\index{matrix--matrix multiplication}
following the definition of the matrix--matrix product.

The matrix--matrix multiplication problem is to compute for an $m\times
l$ input matrix $A$, and an $l\times n$ input matrix $B$, in an
$m\times n$ output matrix $C$ all product-sums
$C[i,j]=\sum_{k=0}^{l-1}A[i,k]B[k,j]$.  The straightforward sequential
implementation takes three nested loops to do this, assuming that the
$C$ matrix has been initialized to all zeros (neutral element for
addition). In C, the programming language~\cite{KernighanRitchie88},
matrices are stored in row-major order\index{row-major order}, one row
after the other, as in one-dimensional arrays. Thus, the three
matrices are given by three one-dimensional arrays \texttt{a},
\texttt{b} and \texttt{c}, which we can cast into matrices (pointers
to rows) and address in matrix-notation.

\begin{lstlisting}[style=SnippetStyle]
double (*A)[l] = (double(*)[l])a; // indexing in mxl matrix a
double (*B)[n] = (double(*)[n])b; // indexing in lxn matrix b
double (*C)[n] = (double(*)[n])c; // indexing in mxn matrix c

... // allocate, initialize

for (i=0; i<m; i++) {
  for (j=0; j<n; j++) {
    for (k=0; k<l; k++) {
      C[i][j] += A[i][k]*B[k][j];
    }
  }
}
\end{lstlisting}

The work (sequential time) of this algorithm is clearly $O(mnl)$, and
$O(n^3)$ for square matrices of order $n$. How well does this
implementation perform (and compared to what)? In
Theorem~\ref{thm:mmslow}, we observed that, in this implementation,
two of the loops have independent iterations and can be parallelized.
A further observation is that the three loops can be interchanged and
essentially be done in any order.

There are six $3!=3\cdot 2\cdot 1=6$ permutations of the three
loops. We ran them all on a few standard (Intel, AMD) processors, on
medium-large, square matrices of order $n=1,000$, with and without
compiler optimizations (\texttt{gcc -O3}) and for both C \texttt{int}
and \texttt{double} matrix elements. The results are surprising, and
illustrative (do try this at home)!  Briefly, we observed a factor of
about $20-40$ between the worst and the best loop orders. The worst
are the versions where the $i$ loop is the innermost; the best when
the $j$ loop is innermost.

The differences can be grossly explained by looking at the cache miss
rate.  Matrices in C are conventionally stored in row-major
order\index{row-major order} with the elements of each row in
consecutive memory addresses and the rows one after the other. We
assume that the cache is large enough to hold a single row of each of
the three matrices, but no more. In that case, for the worst variants
(\texttt{i}-loop innermost), each load of \texttt{A[i][k]} and each
write to \texttt{C[i][j]} would result in a cache miss. For the best
variants (\texttt{j}-loop innermost), \texttt{B[k][j]} and
\texttt{C[i][j]} are both accessed in row-order (best possible spatial
locality): the miss rate is determined by the cache line size.

\subsection{Recursive, Divide-and-Conquer Matrix--Matrix Multiplication}

Other approaches to
matrix--matrix multiplication\index{matrix--matrix multiplication}
solve the problem by
doing the multiplications and additions not on individual elements,
but instead on smaller submatrices that may fit better in the cache. A
recursive formulation of such an approach splits the input matrices
$A$ and $B$ roughly in half along both dimensions, recursively
multiplies the submatrices, and computes the corresponding submatrices
of $C$ by adding the resulting submatrices.

Concretely, write the input matrices $A$ and $B$ as matrices of four
submatrices.
\begin{displaymath}
  A = \left(\begin{array}{cc} A_{00} & A_{01} \\ A_{10} & A_{11} \\
  \end{array}\right) \mbox{and}\
  B = \left(\begin{array}{cc} B_{00} & B_{01} \\ B_{10} & B_{11} \\
  \end{array}\right) \quad .
\end{displaymath}
Then
\begin{displaymath}
  C = \left(\begin{array}{cc} C_{00} & C_{01} \\ C_{10} & C_{11} \\
  \end{array}\right)
  = \left(\begin{array}{cc} A_{00}B_{00}+A_{01}B_{10} &
    A_{00}B_{01}+A_{01}B_{11} \\ A_{10}B_{00}+A_{11}B_{10} &
    A_{10}B_{01}+A_{11}B_{11} \\
  \end{array}\right) \quad .
\end{displaymath}
where the submatrix products $A_{00}B_{00}$ \etc are all computed
recursively. It is a good exercise to complete and implement this in C
(and compare the performance to the loop-based
implementations). Dealing with matrices in C is still cumbersome (see
the code snippets for how to declare and allocate efficiently) and
care is needed when allocating (and freeing) space for intermediate
submatrices. Submatrices are given implicitly by the start and end row
and column indices of the original input and output matrices. For
performance reasons, we usually look for a good cutoff value; that is,
the size of the matrix at which the recursive algorithm stops and the
remaining subproblem (a submatrix--matrix multiplication) is solved
iteratively.  Our implementation performs similarly to the second best
iterative implementation (see above), but can still be improved by
careful attention to cutoff and memory allocation.

The recursive formulation does $8$ (recursive) matrix--matrix
multiplications\index{matrix--matrix multiplication}
and $4$ matrix additions. The total amount of work
performed by the algorithm can be estimated by the following
recurrence relation:
\begin{eqnarray*}
  W(n) & = & 8W(n/2) + O(n^2)\\ W(1) & = & O(1) \quad .
\end{eqnarray*}

The recursion depth can be estimated by the following recurrence
relation.  Here, we are assuming that matrix addition is also done
recursively and has depth $O(\log n)$:
\begin{eqnarray*}
  T(n) & = & T(n/2) + O(\log n), \\ T(1) & = & O(1) \quad .
\end{eqnarray*}

The recurrences are readily solved by the Master
Theorem~\ref{thm:master}\index{Master Theorem} which gives $W(n) =
O(n^3)$ (Case 3 with $a=8,b=2,d=2,e=0$), and $T(n) = O(\log^2 n)$
(Case 2 with $a=1,b=1,d=0,e=1$). Thus, the work is of the same order
as the straightforward implementation, and the length of the critical
path(s) if the computation is viewed as a task graph is $O(\log^2 n)$.
The parallel time complexity\index{parallel time complexity}
can be improved by doing the matrix
additions with more processors in $O(1)$ time steps (depth).

Volker Strassen brilliantly discovered that it is possible to do with
only $7$ matrix--matrix multiplications\index{matrix--matrix multiplication}
and $18$ matrix
additions~\cite{Strassen69} which gives rise to an algorithm with
$W(n) = O(n^{2.81})$ (Master Theorem\index{Master Theorem} again).

\subsection{Blocked Matrix--Matrix Multiplication}

Instead of splitting the matrices recursively, the matrices can be
split into submatrices of size $k'\times k''$ up front for some
$k',k''$ and the
matrix--matrix multiplication\index{matrix--matrix multiplication}
performed as the
three-loop iterative algorithm on these submatrices.  This gives rise
to an implementation with the same work but with $6$ nested loops. If
the submatrices are small enough to fit in cache, this implementation
can perform better than the straightforward implementation. The choice
of best $k',k''$ depends on the size of the cache. Such an algorithm
which needs to know the sizes and other properties of the caches 
is called \emph{cache-aware}\index{cache-aware algorithm}, in contrast
to a \impi{cache-oblivious algorithm}\index{oblivious}.
Cache-oblivious algorithms can have good or even optimal cache
performance, regardless of the concrete size of the cache, which does
not have to be known by the
algorithm~\cite{FrigoLeisersonProkopRamachandran99,FrigoLeisersonProkopRamachandran12}.

\subsection{Multi-Core Caches}
\label{sec:multicorecaches}

The cache system in modern multi-core processor\index{multi-core processor}
systems is structured
in several dimensions. First, there is a hierarchy of caches of
increasing size, L1, L2, L3 (perhaps more), with L1 the lowest level,
closest to the\index{processor-core} processor-core,
smallest, but fastest cache (typically
$16$ KBytes), and L3 the \impi{last level cache} (LLC\index{LLC}), of typically
several MBytes. The L1 cache is often divided into a data cache
and an instruction cache. The memory management system has another
cache, the virtual page cache or \impi{translation look-aside buffer}
(TLB). The L1 and sometimes also the L2 caches are \emph{private} to one
processor-core (and, therefore, replicated for the number of cores of the
multi-core processor), whereas from some level in
the hierarchy, the caches are shared among more and more cores
For example, the L2 cache might be shared among the cores on a single CPU
``socket'', the L3 among all cores in the parallel, multi-CPU
``socket'' system. Processors differ in the way the cache system is
structured.

Caches in parallel multi-core systems pose new problems that do not
manifest when a single processor-core\index{processor-core} works
in isolation (doing, for
instance, matrix--matrix multiplication\index{matrix--matrix multiplication}),
related to both semantics and performance.

The first is the \impi{cache coherence problem} among private
caches~\cite{NagarajanSorinHillWood20}.
Assume that a memory block is in the private L1 caches of two
different cores. What should happen if one core updates an address in
the cache line where the block is kept? If the cache line will
\emph{eventually} be updated in the other core's cache to reflect the
change, the cache system is said to be
\emph{coherent}\index{cache!coherent}. If the cache line is
\emph{never} updated as a response to the update of the other core,
the cache system is
\emph{non-coherent}\index{cache!non-coherent}. Updated as a response
can mean that either the cache line is indeed modified with the new
value or that it is \emph{invalidated} such that the next reference
from the other core to the block in the cache line will result in a
cache miss. Keeping caches coherent is a non-trivial task that
requires a complex algorithm in the processor hardware, a
\impi{cache coherence protocol}. This protocol can affect performance by
\impi{cache coherence traffic}. The cache coherence protocol
of a processor can normally not be influenced by the application programmer
(or only with difficulty or to some
extent). Cache coherence is a strong property, that guarantees that
the processor-cores\index{processor-core} have a consistent
view of individual memory
addresses. Let $a$ be an address (location) in memory. A cache
coherent system fulfills:

\begin{enumerate}
\item
  If core $c$ writes to $a$ at time $t_1$ and reads $a$ at a later
  time $t_2, t_2>t_1$, and there are no other writes (by $c$ or any
  other core) to $a$ between $t_1$ and $t_2$, then $c$ reads the value
  written at $t_1$ (local consistency).
\item
  If core $c_1$ writes to $a$ at time $t_1$ and another core $c_2$
  reads $a$ at a later time $t_2, t_2>t_1$ and no other core writes to
  $a$ between $t_1$ and $t_2$, then $c_2$ reads the value written by
  $c_1$ at $t_1$ (update transfer).
\item
  If core $c_1$ and core $c_2$ write to $a$ at the same time, then
  either the value written by $c_1$ or the value written by $c_2$ is
  stored at $a$ (write consistency, order).
\end{enumerate}

The terms \emph{eventually}, \emph{later}, \emph{at the same time} are
modalities: Something will happen. When something will happen is not
said. Also, note that the term \emph{later} assumes that the read and
write \emph{events} corresponding to the memory instructions performed by
the core can be ordered relative to some (virtual) global
time. It is possible to formulate the cache coherency axioms without
any reference to such a virtual, global time.

Current, shared memory multi-core systems are cache coherent, but
there have been exceptions (often in the
HPC area\index{HPC}) and it is
frequently debated whether cache coherence is a reasonable expectation
for many-core parallel systems with very large numbers of
cores~\cite{MartinHillSorin12}.

The second problem is a phenomenon called \emph{false
sharing}\index{cache!false sharing} which is caused by the granularity
of the cache system. Recall that cache lines map blocks of consecutive
addresses, say $8$ double words for cache blocks of $64$ Bytes.  If
some block is in the private caches of two or more cores, any update
that one core performs to an address of that block will affect the
other core's cache, either by an update or by an invalidation of the
cache line. In particular, updates to two different addresses
\texttt{\&c0} and \texttt{\&c1} in the block mapped by the two cores,
will create coherence traffic, even if the two variables \texttt{c0}
and \texttt{c1} are not in any way related. This can degrade the
expected performance significantly~\cite{TorrellasLamHennessy94}. Here
are two classical examples of false sharing with \openmp\index{OpenMP} (see
\Sec~\ref{sec:openmpframework}).  In the first example, the elements
in a C structure of integers filling a memory block of $64$ Bytes ($16$
integers) are updated by individual threads.  In the second example,
the $16$ integers are stored in an array, and again updated by the
individual threads.

\begin{lstlisting}[style=SnippetStyle]
struct { // 16 int in consecutive block
  int c0, c1, c2, c3, c4, c5, c6, c7, c8, c9, c10,
      c11, c12, c13, c14, c15;
} cl;
int cs[16]; // 16 int in array
  
// false sharing in struct
#pragma omp parallel
{
  int t = omp_get_thread_num();
  for (i=0; i<r; i++) {
    switch (t) {
    case 0:
      if (i==0) cl.c0 = 0; else cl.c0 += t;
      break;
    case 1:
      if (i==0) cl.c1 = 0; else cl.c1 += t;
      break;
      ...
    }
  }
}
// false sharing in array
#pragma omp parallel
{
  int t = omp_get_thread_num();
  for (i=0; i<r; i++) {
    switch (t) {
    case 0:
      if (i==0) cs[t] = 0; else cs[t] += t;
      break;
    case 1:
      if (i==0) cs[t] = 0; else cs[t] += t;
      break;
      ...
    }
  }
}
// no false sharing, local variable
#pragma omp parallel
{
  int t = omp_get_thread_num();
  int c; // local variable, hopefully each on own cache line
  for (i=0; i<r; i++) {
    switch (t) {
    case 0:
      if (i==0) c = 0; else c += t;
      break;
    case 1:
      if (i==0) c = 0; else c += t;
      break;
      ...
    }
  }
}
\end{lstlisting}

In both cases, the updates to the different integer elements are in no
way related, but by being on the same cache line, each update will
lead to activity in the cache system. The different running times
compared to when the updates are performed on a local variable for the
threads that (presumably) will not share a cache line can be dramatic
and indeed a large factor. It is extremely illustrative to try the
example on a multi-core processor\index{multi-core processor}
using at most 16 \openmp\index{OpenMP} threads
with and without compiler optimizations with a sufficiently large
number of iterations (the variable \texttt{r}).

Avoiding false sharing requires attention to allocation and use of
variables, attempting to ensure that independent and frequently used
and updated variables will always reside on different cache lines. The
strategy called \impi{padding} which ensures that there is only one
critical variable per cache line by allocating variables at the
granularity of the cache block size will work, but is obviously
wasteful in memory; in the integer example above by a factor of $16$.
Using local variables for the threads and updating the global
structure or array only once or rarely is often an effective solution
as also illustrated in the example. The example shows that variables that
are updated frequently by individual threads should not be put too
spatially close to each other in arrays. Performance counters introduced
in an application to may be an example of such variables.

\subsection{The Memory System}

The cache system is part of the \impi{memory hierarchy} which, for our
purposes, will mainly be the large \emph{main memory}, beyond which
there are disks and other types of \impi{external memory}. The
characteristic of the memory hierarchy is that as memory up (from L1
to L2 to L3 caches to main memory, \etc) in the hierarchy becomes larger and
larger, the access times (and often also the granularity of access)
also get larger and larger. Any textbook on computer architecture will give
approximate ratios of access times and details on
granularity~\cite{BryantOHallaron15,HennessyPatterson17}.

A final, important part of the memory system not mentioned so far, is
the \impi{write buffer} in which writes to the main memory are
buffered and written to the memory in the pace that the memory system can
process updates. The write buffer, as long as it has capacity, makes writes to
memory appear fast. Write buffers may be simple FIFO buffers but can
also be sorted and usually coalesce writes to the same address. The
interaction with the cache system is highly non-trivial, but for
single-core processors, write buffers (and caches) were part of the
``free lunch'' in that they transparently made (most) memory writes (and reads)
appear much faster than the actual main memory access times. For multi-core
processors, the existence of write buffers is no longer transparent,
as will be explained below\index{multi-core processor}.

In a hierarchical memory system, memory access times are not
uniform. The first time an address or block is accessed, access time
depends on where in the hierarchy the address is located, and later
accesses may be less expensive due to the cache system. Different
addresses, residing in different parts of the hierarchy likewise have
different access times. Modern memory systems are highly
NUMA\index{NUMA}.

Memory systems for multi-core parallel systems have additional
structure and additional restrictions. In a multi-core CPU, not every
core has a direct connection to the main memory. Instead, the cores
share a small(er than the number of cores) number of
\impi{memory controllers}.
The memory is divided into separate banks over the memory controllers. The
memory access times for a particular core depend on the ``closeness''
to the memory controller for the bank in which the accessed address is
contained. Access times to different addresses are again
non-uniform. The non-uniformity becomes even more prominent for
parallel systems consisting of several multi-core CPUs. Access to
memory that is controlled by a different CPU than the core issuing the
access requires communication between the CPUs and can take
significantly longer than access to memory controlled by the CPU of
the core.

Not taking the NUMA architecture and behavior of the memory system
into account can become a serious performance issue.  To some extent,
NUMA effects can be alleviated by paying attention to the placement of
data used by an application. Partly, this is done automatically by
the virtual memory system. An often used virtual memory page
allocation policy is the so-called ``first touch''\index{first touch}
policy, by which a virtual memory page will be put physically in the
memory bank closest to the core that does the first access to the
page. An application can attempt a good placement of virtual memory
pages by first ``touching'' pages (addresses) by the cores that will later
most heavily use the pages.

\subsection{Super-Linear Speed-Up caused by the Memory System}

Although super-linear (absolute) speed-up\index{speed-up!super-linear}
was claimed to be impossible, it can nevertheless happen and be observed
on real, parallel systems. What is wrong with the argument presented in
\Sec~\ref{sec:linearbest}?

The argument that linear (perfect)
speed-up\index{speed-up!linear}\index{speed-up!perfect} is best possible
assumes that the sequential and parallel system behave identically, in
particular that memory accesses behave identically and take the same
time for the two systems. Due to the memory hierarchy with large
caches, exactly this may not be the case. Assume for simplicity an
algorithm that can be parallelized well in the sense that the working
set with $p$ processors is $1/p$ of the working set on just one
processor. As $p$ grows, the smaller and smaller working set will fit
in faster and faster caches in the memory hierarchy, effectively
leading the memory accesses of the parallel algorithm to be much
faster than for the sequential algorithm. The speed-up can exceed $p$
by a factor equal to the ratio between effective, average sequential
memory access time and effective, average parallel memory access
time. As a consequence, super-linear speed-up of the form $kp$ with
some constant $k>1$ can indeed be possible and is indeed sometimes
observed\index{speed-up!super-linear}.

\subsection{Application Performance and the Memory Hierarchy}

The nominal performance of the CPU and processor-cores\index{processor-core}
do not alone determine what the performance of some given application
on a system will be\index{nominal processor performance}.
If the memory system is not able to supply data fast enough
to the processor-cores, the performance of the memory system (access
times) will eventually determine the performance. What ``fast enough''
is, is determined by the application.

We say that an application is
\begin{itemize}
\item
  \impi{memory-bound}, if the operations to be performed per unit read
  from or written to the memory take less time than reading/writing a
  unit from/to memory, and
\item
  \impi{compute-bound}, if the operations to be performed per unit
  read from or written to the memory take more time than
  reading/writing a unit from/to memory.
\end{itemize}

In a memory-bound application, the memory system and memory access
times will determine the application's performance including
its speed-up\index{speed-up}, and in a compute bound application the
nominal processor performance\index{nominal processor performance}
will determine the application performance.  Thus, the application is
the determining factor, whether a fast memory or on a fast processor
would be the better investment.  This trade-off is worked out
quantitatively in the so-called
\impi{roofline performance model}~\cite{WilliamsWatermanPatterson09}.

\subsection{Memory Consistency}

While the memory hierarchy, cache system, and write buffer are all
functionally and semantically transparent for a single core,
this is no longer the case when multiple cores are doing \parco together.

When a program is executing sequentially, reads and writes to memory
addresses (appear to) take place in the execution order of the
program's instructions (a read instruction of an address written by an
already executed write to that address, will return the value that was
written). This is called the \impi{program order} which is assumed in
order to prove properties of the program by state
invariants\index{invariant}. When two programs are being executed
concurrently by our asynchronous, parallel, multi-core system, it is
(probably) a natural expectation that the outcome will be as if some
\impi{interleaving} of the two program order executions has taken place;
that is, that memory order follows program order. This is a particular kind of
memory consistency which is called
\impi{sequential consistency}~\cite{Lamport79:sc} which would allow us
to prove properties of parallel programs much like we do for sequential
programs. Only the possibility of different interleavings has to be
considered.

Unfortunately, often due to the existence of per-core write buffers
and the complex, banked structure of the memory system, modern
multi-core systems are \emph{not} sequentially consistent. The
consequences of this can be seen by considering the simple example given
below. Two cores execute the respective pieces of code. The idea is to
protect the code which is in the body of the \texttt{if}-statement
such that at most one of the cores will be executing this code body. The
two flags \texttt{f0} and \texttt{f1} are in shared memory and can be
read and written by both cores. The question is whether we can prove
the property that ``at most one of the two cores can execute the
\texttt{if}-body''?

\begin{lstlisting}[style=SnippetStyle]
int f0, f1; // shared flags initialized to 0
  
// (thread) code for core 0
f0 = 0; 

... // some code
  
f0 = 1; // core 0 now wants to enter
if (f1==0) {
  ... // protect: core 0 alone?
}
\end{lstlisting}  

\begin{lstlisting}[style=SnippetStyle]
// (thread) code for core 1
f1 = 0; 

... // some code
  
f1 = 1; // core 1 now wants to enter
if (f0==0) {
  ... // protect: core 1 alone?
}
\end{lstlisting}

We can try to argue by contradiction. Assume that one of the cores,
say core $0$, has entered the \texttt{if}-body. In that case, it has set
its flag \texttt{f0} to $1$, and read the other flag \texttt{f1} and
found it to be $0$. This means that core $1$ cannot have reached the
instruction where it sets its flag \texttt{f1} to $1$. Therefore, it is
not in the \texttt{if}-body and will also not be able to enter, since
\texttt{f0} is still $1$. If one of the cores is in the
\texttt{if}-body, the other cannot be, and the desired property
holds. There is no interleaving of the two pieces of code that will
lead to both cores being in the \texttt{if}-body, and the parallel
program has the desired effect under sequential consistency.  As can
easily be seen, though, it can of course happen that none of the cores
enter, but that was not the claim.

The crucial observation is that the argument holds only under the
assumption that reads and writes to memory happen in program order. If
the memory system is not sequentially consistent, this might not be
the case. For instance, with write buffers for the two cores, the
following could happen. Both cores execute the initialization of the
flags and the $0$ values are written to memory. Now the cores proceed,
execute their flag updates to $1$, but these updates end up in the
write buffers. Both cores execute the read of the flag in the
\texttt{if}-expression, both return $0$, and both enter the body,
exactly what should not happen. What happened was that the outcome
of the write and the read instructions in memory did not follow program
order. This is a major problem: How can we reason about parallel
programs running on such systems? How can we prove fundamental
correctness properties?

Answering these questions is beyond these lectures. The programming
interfaces that we will see in the next lectures
(\pthreads, \openmp)\index{pthreads}\index{OpenMP}
will help us in that they give constructs to ensure 
guarantees that, at certain points in the execution, the memory is in a
well-defined state. The guarantees are typically
of the form that updates performed
by one thread are, at this specific point in the execution after a
particular construct visible to other threads.
If used correctly, it will ideally not be necessary to
pay attention to the exact behavior of the memory system. To do so, it
is important that the hardware provides mechanisms to ensure that
operations on memory (read and writes) have indeed been
performed. Such mechanisms are operations to \emph{flush} the write
(and other) buffers. They are often called \emph{memory fences}. Also
so-called \emph{atomic operations}\index{atomic operation}
can serve as memory fences. Another means of ensuring
some (total) order between codes executed by different threads is to have
special, privileged hardware instructions or mechanisms that always execute in
order: If instruction \texttt{IA} happened before instruction \texttt{IB}
as observed by some thread, the same order can be observed by (any other)
thread(s). Often, \emph{atomic operations}\index{atomic operation}
provide such ordering guarantees. In C, ordering between atomic (and other)
instructions by different threads can be controlled and enforced, but this
is beyond these lectures.

Memory and cache behavior for parallel multi-core systems is intriguingly
and painfully intricate.
Being aware of the issues is essential for writing correct
programs and for getting the best possible performance out of the
system at hand. We summarize the two kinds of issues we have
discussed:
\begin{itemize}
\item
The \impi{cache coherence problem}: What happens when different cores
read/write the same address?
\item
The \impi{memory consistency problem}: What happens when different
cores read/write different addresses?
\end{itemize}

\section{Sixth block (1--2 lectures)}

\pthreads\index{pthreads} is our first example of a concrete
programming interface in the form of a library that implements a
shared memory programming model\index{programming model} and is
intended for running on parallel shared memory systems. \pthreads is
an early example of a thread programming interface for the C
programming language~\cite{KernighanRitchie88}, is still widely used,
and has been taken as a blueprint for many subsequent thread
interfaces(despite issues with correctly realizing a thread interface
as a library~\cite{Boehm05}).  Native threads in C are defined since
C11 and essentially follow the \pthreads interface but are often not
supported and seem little used. \pthreads is standardized in POSIX
(Portable Operating Systems Interface for uniX) as an IEEE standard
(IEEE POSIX 1003.1c).

From now on, the lectures will use C as programming
language, and the practical projects given in the exercises are
intended to be implemented in C.  The
standard reference text is the book by Kernighan and
Ritchie~\cite{KernighanRitchie88}. For learning good programming style in C,
the book by Kernighan and Pike~\cite{KernighanPike99} is likewise
valuable.

\subsection{\pthreads Programming Model}

A \impi{thread} is the \emph{smallest unit of execution} that can be
scheduled and preempted by the operating system (OS). In C and
Unix/Linux, threads live inside \emph{processes} and different threads
share information that is global to their process. Threads in C are
functions, and shared information are, for instance, global variables,
static variables, file pointers, and the heap used for dynamic memory
allocation. Threads maintain their own stack. Also, the registers
can be thought of as private to a thread. It is also possible to
allocate thread-local storage: special memory that is bound to the
allocating thread.

The main characteristics of the \pthreads\index{pthreads}
programming model\index{programming model} are:
\begin{enumerate}
\item
  Fork-join parallelism. A thread can \emph{spawn} any number of new
  threads (up to system limitations) and wait for completion of any
  other thread. Threads are referenced by \emph{thread
  identifiers}. Initially, a single (master) thread is running.
\item
  Threads are symmetric \emph{peers}. Any thread can wait for
  the completion of any other thread via the thread identifier.
\item
  Threads execute functions in the same program (SPMD
  model\index{SPMD}) but possibly different functions for different
  threads (MIMD model\index{MIMD}). Initially, only one main function
  thread is active.
\item
  Threads are scheduled by the operating system (OS) and may or may
  not run simultaneously on the different cores of the parallel
  system.
\item
  There is no implicit synchronization among threads. Threads progress
  independently of each other.
\item
  Threads share global objects and information.
\item
  Coordination constructs for synchronization and updates to
  shared objects are provided: mutexes, readers-writer locks,
  condition variables. All updates to shared information must be
  protected by coordination constructs. Otherwise, the program is
  illegal and the outcome undefined.
\end{enumerate}

\pthreads\index{pthreads} does not come with a performance
model (for analyzing the
performance of \pthreads programs) and does not come with (much of) a
memory model, either (for writing correct programs on hardware memory
that is not sequentially consistent\index{sequential consistency}).
It just requires that updates to shared information are done via
the coordination constructs of \pthreads.

\pthreads allows any number of threads to be spawned (subject to
system limitations). Spawning more threads than the number of
available cores in the parallel system at hand is called
\impi{oversubscription}. It is up to the operating system (OS) how
and when threads are scheduled to run (even when there are fewer
threads than cores). Threads can also be preempted or suspend
themselves, which can, to some extent, be influenced by (non-standard)
\pthreads functionality that we will not go into in these lectures.

Oversubscription can have advantages (hiding latencies, giving freedom
to the OS), but the \emph{pragmatics} of \parco is mostly to have only
as many threads as there are processor-cores\index{processor-core}
and to assume that these threads all run simultaneously.

\subsection{\pthreads in C}

\pthreads\index{pthreads} is a library and the thread functionality
can be used by
linking the code against the \pthreads library. C code using \pthreads
must include the function prototype header with the
\texttt{\#include <pthread.h>} preprocessor directive. All functions 
and predefined objects relevant to \pthreads
are prefixed with \texttt{pthread\_}
which identifies the \pthreads ``name space''.  With \gcc, code can be
compiled using the \texttt{-pthread} option which enables linking
against the library.

Most \pthreads functions return an error code, and it is good practice
to check the error code (which is often not done). The error code $0$
indicates success.

\subsection{Creating Threads}
\label{sec:createpthreads}

When a C program with \pthreads\index{pthreads} is started, the \texttt{main()}
function is the only (``master'') thread running. The master thread
and any other thread can start new threads and wait for termination of
any other thread. A thread is identified by an \emph{opaque} object of type
\texttt{pthread\_t} which is set by the creation call and used to
reference the now started thread. Such objects can be used and manipulated
exclusively through defined operations; their implementation and
structures is neither defined nor accessible.
Thread identifiers can be compared
for equality but otherwise not manipulated.

\begin{lstlisting}[style=SnippetStyle]
int pthread_create(pthread_t *thread,
                   const pthread_attr_t *attr,
                   void *(*start_routine)(void *), void *arg);
void pthread_exit(void *retval);
int pthread_join(pthread_t thread, void **retval);

pthread_t pthread_self(void); // return own thread identifier
int pthread_equal(pthread_t t1, pthread_t t2);
\end{lstlisting}

Code that is to run as a thread must be written as a C function with a
single \texttt{void*} pointer argument. This pointer is used to point
to a structure holding the actual arguments to the thread. The thread
function will, therefore, often cast this void pointer to something
more meaningful.  The pointer to the function together with a pointer
to the actual arguments are given as arguments to the thread creation
call.  Attributes will not be covered in these lectures; but they can
be used to control the way the thread is to run. In most cases,
\texttt{NULL} can just be given as the attribute argument. C
programming is brittle: It is easy to make mistakes with function and
argument pointers and such mistakes have grave consequences by
leading to memory corruption and program crashes often much later than
the call where the mistake was made.

Once a thread has been created, the corresponding function runs on its
own, asynchronously and concurrently with other activities, possibly
on its own\index{processor-core} processor-core.
When a thread function comes to an end, it
should terminate itself by making the exit call. This call also takes a
pointer that can point to information to be given back to the thread
that intercepts the terminating thread. If return information is used,
it must be allocated on the heap and definitely not on the stack where it
will sooner or later disappear. Waiting for a thread to exit is done
by a join call, which will update its \texttt{void**} pointer
argument to point to the structure returned by the exiting thread.
Thread identifiers can be exchanged freely between threads, and any
thread can wait for any other thread to finish. In that sense, threads
are ``peers''.

The following simple, almost full-fledged \pthreads\index{pthreads} program
shows how to start \texttt{p} threads one after the other
and assign each a ``rank'' (a unique identifier between $0$ and \texttt{p-1})
by passing a corresponding argument.

\begin{lstlisting}[style=SnippetStyle]
#include <pthread.h>

typedef struct { // the real arguments to thread functions
  int rank;
} realargs;

void *hello(void *arguments) {
  realargs *args = (realargs*)arguments;

  // a classic race; try it, and see later
  printf("Thread %d starting\n",args->rank);
  pthread_exit(NULL);
}

int main(int argc, char *argv[])
{
  int p = ...; // (small) number of threads
  int i;

  pthread_t thread[p];
  realargs threadargs[p];

  // create and start the threads
  for (i=0; i<p; i++) {
    threadargs[i].rank = i;
    pthread_create(&thread[i],NULL,hello,&threadargs[i]);
  }

  // wait for termination and intercept return values (none)
  for (i=0; i<p; i++) {
    pthread_join(thread[i],NULL);
  }

  return 0;
}
\end{lstlisting}

The program snippet illustrates how threads are created and started,
but is technically wrong. One problem is the call to the C
\texttt{printf()} library function in the thread function
\texttt{hello()} which will possibly lead to the threads competing for a
resource: the printing device. This is a first example of a classic
\impi{race condition} (for more, see later).  The general problem
is that a function called from a thread may not work properly
when other threads can also be calling the function; the calls to such
functions are \impi{unsafe}. Conversely, a function that can be called
concurrently by any number of threads is called \impi{thread safe}.
Pure functions without side effects (for instance, not
updating shared state in the form of global or static variables) are thread
safe. System functions may or may not be thread safe, and one should always
check; a notoriously thread unsafe function is the ``old'' random number
generator \texttt{rand()}.

The binding of threads to processor cores can be controlled by the
following (non-standard) \pthreads\index{pthreads} functions.
A \texttt{cpuset} is a
set data structure (bit vector) representing a set of possible
physical cores, numbered consecutively and corresponding to the
numbering of the cores on the shared memory system. They should be
manipulated through predefined macros.

\begin{lstlisting}[style=SnippetStyle]
int pthread_setaffinity_np(pthread_t thread,
                           size_t cpusetsize,
                           const cpu_set_t *cpuset);
int pthread_getaffinity_np(pthread_t thread,
                           size_t cpusetsize,
                           cpu_set_t *cpuset);
\end{lstlisting}

\subsection{Loops of Independent Iterations in \pthreads}
\label{sec:threadloop}

The parallel patterns we have seen in the previous lectures
(\Sec~\ref{sec:patterns}) can all be implemented with
\pthreads\index{pthreads}.
A loop of independent iterations, for instance, can be
parallelized by assigning each thread a consecutive range of iterations.
The thread function performs the iterations, taking the
arguments for the loop from a suitable argument data structure. This is
shown in the code snippet below which reuses the linear thread creation
and argument structure from the previous example.

\begin{lstlisting}[style=SnippetStyle]
#include <pthread.h>

typedef struct {
  int rank;
  int size;
  int n;
  int *a; // pointer to shared array
} realargs; // argument structure

void *loop(void *arguments)
{
  realargs *args = (realargs*)arguments;

  int i;
  int s, nn; // start index and number of iterations
  int *a = args->a;
  
  nn = (args->n)/args->size;
  s  = nn*args->rank;
  if (args->rank==args->size-1) nn = args->n-s;

  // part of loop for thread
  for (i=s; i<s+nn; i++) a[i] = i; // some loop body
  
  pthread_exit(NULL);
}
\end{lstlisting}

The thread function uses the total number of iterations to be
parallelized to compute the iteration start and the local number of
consecutive iterations that the thread will perform. This is taken as
$\floor{n/p}$ where $n$ is the number of iterations and $p$ the number
of threads in the field \texttt{size}. The last thread will perform
the remaining $n-(p-1)\floor{n/p}$ iterations.

\subsection{Race Conditions and Data Races}

In a thread model with shared memory, executed on a shared memory
multi-core system, it is possible for different threads to access and
update shared variables. Since threads may execute concurrently, such
updates may happen ``at the same time''. In such a situation the
outcome is (for almost all systems, and we will assume this behavior)
the update by one of the threads and not something in between
(also not: no update). But which thread
succeeds with its update is undetermined. We say that the outcome of a
concurrent update to a shared variable is \emph{non-deterministic},
and such non-determinism may affect the final result of the whole
program, an often undesirable situation. Since threads execute
asynchronously (our thread model makes this assumption: no synchronicity
among threads, unlike the PRAM model\index{PRAM}), the order of
updates to shared objects is not defined, and either thread can be the
``last'' thread to perform an update. Depending on the memory
system behavior, updates may or may not become visible to the other
threads in the order in which they were performed.
Thread programs are inherently non-deterministic. In
order to write correct programs that give a determinate, final output,
we need to be able to deal with and restrict the non-determinism in
updates and accesses to shared variables and objects.

Such non-deterministic updates to shared objects and variables in a program
which can lead to different results of the program,
some of which are not correct, are commonly called \emph{race
conditions}\index{race condition}.
It is important to keep in mind that asynchronous parallel programs
are inherently non-deterministic. Non-determinism is the price for the
potential performance benefits of asynchronous parallelism. Also,
concurrent updates may not always lead to different, or wrong, final
results.

Any thread programming model\index{programming model}
needs either means to reason about
non-deterministic executions and updates to shared objects or means
to restrict and control non-determinism wherever it is crucial that
updates are done in a certain, specific order, or both.

A particular kind of race condition is the \emph{data race}.
Technically, a \impi{data race} is a situation where two or
more threads access a shared object, and at least one of the accesses
is an update (write). It is undecidable to determine
whether a program will have a data race, so automatically finding
\emph{all} race conditions (by a compiler) is algorithmically
impossible.

Thread models like \pthreads\index{pthreads} and \openmp\index{OpenMP} 
forbid uncontrolled, concurrent updates to shared variables and
objects. In particular, they forbid data races. Instead, they have
constructs for threads to access and update shared objects that
make concurrent updates well-behaved and technically eliminate
data races. Such programs are informally called \impi{data race free} here.
A way to look at such constructs is that they restrict the possible
interleavings of asynchronous thread executions. We will see the main
\pthreads construct in the next section.

The following simple example shows why data races can be harmful
and lead to race conditions\index{race condition}. Let \texttt{a} be
a variable that is shared among many threads. The threads all execute the
following simple (but composite) update:

\begin{lstlisting}[style=SnippetStyle]
a = a+27;
\end{lstlisting}

With typical processors and instruction sets, this simple expression
evaluation and assignment translates into at least three instructions,
namely (1) a load of \texttt{a} into a register, (2) an
addition with a constant (here \texttt{27}),
and (3) a write back to the location of
\texttt{a}. The sequential semantics of the statement is that \texttt{a} is
incremented by the constant $27$.
When several threads execute this code, the following can easily
happen: The threads all read the old value of \texttt{a}, all perform
the addition in their respective (private, non-shared) registers and
then race on the update to \texttt{a}. Instead of each thread
incrementing by $27$, only one increment will effectively have happened.
With many threads, many outputs of this sort are possible
(any increment by some multiple $k$ of $27$ with $k<p$),
most of which are probably not what was intended (intended was probably
an increment of $27p$ when all threads have finished).

Not all updates that are technically data races may be harmful.  For
instance, it might be unproblematic if all threads write the same
value to the shared variable, as allowed by the Common CRCW
PRAM\index{PRAM!Common CRCW}, for instance.  In the above
example, it was harmful and leading to very unintended outcome.

\subsection{Critical Sections, Mutual Exclusion, Locks}
\label{sec:melocks}

\pthreads\index{pthreads} and \openmp\index{OpenMP} programs (see later)
with data races
are technically not correct, and programs with updates to shared
variables by several threads that could happen concurrently are
illegal. \pthreads provides constructs to control accesses and updates
to shared variables and shared objects.

The problem in the example above is not so much the individual data
races on the shared variable \texttt{a} but rather the whole sequence
of read-modify-write instructions involved in the update.
When two threads at the same time come to this little piece of code,
what is required for the
intended outcome is that either of the threads runs entirely before the
other. In order to get the (presumably) desired behavior, we
need to exclude the interleavings of read and modify and write sketched above
from the possible interleavings of the two thread executions.

A piece of code that should not be executed concurrently by several
threads is commonly called a \impi{critical section}. A thread running
code in a critical section must exclude other threads from doing so.
Threads need to cooperate to ensure this, and guaranteeing that
a critical section is indeed being executed by at most one thread is commonly
called \impi{mutual exclusion}.  The \impi{mutual exclusion problem}
is to guarantee mutual exclusion and is not a trivial problem. It
is not the purpose of this lecture to go into solutions or algorithms
for the mutual exclusion problem which has a long and
ongoing history~\cite{HerlihyShavit12,Raynal13a}.
Note that the code in a thread's critical section must not necessarily
be the same for all threads. Rather, a critical section is a piece of
a thread's code that should not be executed concurrently, in parallel
with certain other pieces of code of other threads. The mutual
exclusion problem is to ensure that this is the case.

A programming model\index{programming model} mechanism
that guarantees mutual exclusion is
commonly called a \emph{lock}\index{Lock}.  Locks provide mutual
exclusion as follows:  A thread that wants to enter a critical section
tries to \emph{acquire}\index{Lock!acquire} the corresponding lock. If
it succeeds, the thread is on its own in the critical section and does
what it needs to do, typically reading and writing shared variables. No
other thread can acquire the same lock as long as it is being held.
Therefore, there can be no data races on objects updated by the thread
having the lock as long as they are not referenced by threads not being in
the critical section. When finished, the thread exits the
critical section by \emph{releasing}\index{Lock!release} the lock. Then,
other threads can again enter the critical section by trying to acquire
the lock. If a thread tries to but cannot acquire the lock,
it cannot progress and is \emph{blocked} waiting for the lock to become
available.  The lock
acquire and release operations are often also called just
\emph{lock}\index{Lock!lock} and \emph{unlock}\index{Lock!unlock}. A
lock is a brittle construct:  If the unlock is forgotten
or does not happen, either due to program logic or because the
corresponding thread is not progressing (suspended by the operating
system, sleeping or entirely gone), other threads wanting
to acquire the lock will become blocked and eventually the whole
program execution will grind to a stand-still (allusion intended).  This
is the dreaded \impi{deadlock} situation. In general, a deadlock
in \parco or \conco is the following: A thread, a process or a
processor-core\index{processor-core} needs a resource from another
thread before it can proceed which in turn needs a resource from another
thread and so on infinitely or cyclically, preventing the thread, process
or core from ever getting the resource. All waiting entities are stuck forever
or until the deadlock is dissolved from the outside by an arbiter that
breaks the dependencies.

Apart from guaranteeing mutual exclusion (at most one thread at a time
can hold a given lock), the fundamental property of a lock is that it
must itself be \emph{deadlock free}\index{Lock!deadlock free}. This means
that whenever a number of threads (one, some, many, all) is trying to acquire
the lock, \emph{eventually} one thread \emph{must} succeed and get the
lock. A perhaps desirable property is that whenever a specific thread is
trying to acquire the lock it will \emph{eventually} acquire the lock,
no matter which other threads are also trying to acquire the lock.  A lock is
said to be \emph{starvation free}\index{Lock!starvation free} if it
has this property that no thread can be \emph{starved} forever. Locks
are said to be \emph{fair}\index{Lock!fair} if they provide stronger
starvation freedom guarantees, like that when a thread is trying to
acquire a lock ``before'' some other thread it will also get the lock
before that other thread (whatever ``before'' means).

In \pthreads\index{pthreads} terminology,
a lock is called a \impi{mutex} (for mutual
exclusion\index{mutual exclusion}) and shared objects are only allowed
to be updated by acquiring a mutex before doing so (concurrent reading
alone is allowed).  A mutex is identified by
an opaque \texttt{pthread\_mutex\_t} type. Mutexes must be initialized
either statically (by assigning
\texttt{PTHREAD\_\-MUTEX\_\-INITIALIZER}) or dynamically before they
can be used.

\pthreads mutexes guarantee mutual exclusion and are deadlock free,
but they are \emph{not} starvation free. In addition, they guarantee that all
memory updates performed by a thread in the critical section before
the release of the mutex will be visible to any other thread
acquiring the lock afterwards. This is the \pthreads memory model.

\begin{lstlisting}[style=SnippetStyle]
int pthread_mutex_init(pthread_mutex_t *restrict mutex,
                       const pthread_mutexattr_t
                       *restrict attr);
int pthread_mutex_destroy(pthread_mutex_t *mutex);
                                  
int pthread_mutex_lock(pthread_mutex_t *mutex);
int pthread_mutex_trylock(pthread_mutex_t *mutex);
int pthread_mutex_unlock(pthread_mutex_t *mutex);

int pthread_rwlock_init(pthread_rwlock_t *restrict rwlock,
                        const pthread_rwlockattr_t
                        *restrict attr);
int pthread_rwlock_destroy(pthread_rwlock_t *rwlock);

int pthread_rwlock_rdlock(pthread_rwlock_t *rwlock);
int pthread_rwlock_tryrdlock(pthread_rwlock_t *rwlock);

int pthread_rwlock_trywrlock(pthread_rwlock_t *rwlock);
int pthread_rwlock_wrlock(pthread_rwlock_t *rwlock);

int pthread_rwlock_unlock(pthread_rwlock_t *rwlock);
\end{lstlisting}

The data race on the shared, global variable \texttt{a}
in the \texttt{a = a+27;} example from above is correctly
avoided by protecting this critical section by a mutex.

\begin{lstlisting}[style=SnippetStyle]
// A lock shared by all threads
pthread_mutex_t lock = PTHREAD_MUTEX_INITIALIZER;

pthread_mutex_lock(&lock); // acquire lock
a = a+27; // thread alone in critical section, no race
pthread_mutex_unlock(&lock); // release lock
\end{lstlisting}

Threads that try to execute the update concurrently will
\impi{serialize}: One thread after the other will be allowed to enter
the critical section. If there is repeated competition for acquiring
the lock, it may even happen (if the mutex is not starvation free)
that some thread will never enter the critical section.
If this happens, such a thread does not contribute
to the parallel computation any more and the speed-up\index{speed-up}
that might be
possible is reduced accordingly.  A lock for which many threads are
competing is said to be \emph{contended}\index{Lock!contention}.
Contention is always a source of slowdown, since threads are waiting
for their critical section instead of doing useful work.

To allow threads to do something useful in case of contention, many
lock models offer a \emph{try-lock}\index{Lock!try-lock} operation.
Try-lock tries to acquire the lock, and if the lock is not already
held by some other thread, it immediately acquires the lock. If the
lock is held by another thread, try-lock returns with a condition
code (\textbf{false}). It is, clearly, essential that try-lock acquires the lock
immediately when the lock is free instead of returning with a
condition code. This would be useless, since trying to acquire the lock after
checking the condition code could well fail because of some other
thread having taken the lock in-between. A great application of
try-lock to the implementation of a concurrent priority queue with
certain guarantees can be found in~\cite{WilliamsSandersDementiev21}.

A means of alleviating lock serialization effects and slowdown takes
advantage of the situation that accesses and updates to shared objects
are often asymmetric.  In some (many) critical sections, shared
variables are only read, while in other (fewer) also actual updates
(writes) have to be performed. All the threads that only need to read
some shared object can do this concurrently, in parallel. For the
writes, full mutual exclusion is needed, and all other reading as well
as writing threads must be excluded from the critical section.
\emph{Readers-writer} locks\index{Lock!readers-writer} that are found
in many thread programming models\index{programming model}, provide
this functionality. Readers-writer locks have a lock acquire operation
for threads that want to read (concurrently), and another lock acquire
operation for threads that want to write under strict mutual
exclusion.  It is the programmer's responsibility to make sure that no
updates (to shared variables) are performed in the critical sections
when the lock is acquired for reading.

There are many ideas and algorithms for implementing locks (not
treated in this lecture). An important pragmatic issue is how waiting
for a lock is implemented and how waiting (blocking) interacts with
the operating system (OS). In a
\emph{spin lock}\index{Lock!spin lock},
the processor-core\index{processor-core} executing the blocked thread actively
keeps testing (spinning) for the lock to become free. That is, the
processor-core is kept busy for as long as the thread is blocked on
the lock acquire operation. Acquiring the lock is fast for spin locks,
and this implementation is typically advantageous when the critical
sections are short and there is no thread oversubscription. With a
\emph{blocking lock}\index{Lock!blocking}, the thread that is waiting
for the lock to become free is suspended by the OS, and the
processor-core that was executing the thread is free to do something
else. It could, for instance, wake up and run another thread.
Blocking locks may be advantageous when the shared memory system is
oversubscribed and the lock waiting times can be productively spent by
the core doing something else. In \pthreads,
spinning behavior can be requested explicitly by using spin
locks. This (strange) \pthreads\index{pthreads} design decision
means that code has to be rewritten, if spin locks are desired.

\begin{lstlisting}[style=SnippetStyle]
int pthread_spin_destroy(pthread_spinlock_t *lock);
int pthread_spin_init(pthread_spinlock_t *lock, int pshared);

int pthread_spin_lock(pthread_spinlock_t *lock);
int pthread_spin_trylock(pthread_spinlock_t *lock);
int pthread_spin_unlock(pthread_spinlock_t *lock);
\end{lstlisting}

\subsection{Flexibility in Critical Sections with Condition Variables}

Since \pthreads\index{pthreads} programs must be data
race free\index{data race free},
locks (or other constructs, see the following)
must be used when transferring information between threads. For instance,
a value updated by a writing thread may be needed by several reading
threads. The following first solution is obviously wrong since it
easily leads to a deadlock: A reading thread entering its critical
section before the writer thread will stay in the while-loop and prevent
the writer thread from ever setting the \texttt{written} flag.

\begin{lstlisting}[style=SnippetStyle]
// reader threads
pthread_rwlock_rdlock(&lock);
while (!written);
a = b; // information transfer
pthread_rwlock_unlock(&lock);
\end{lstlisting}  

\begin{lstlisting}[style=SnippetStyle]
// writer thread
pthread_rwlock_wrlock(&lock);
b = ... ; // update
written = 1;
pthread_rwlock_unlock(&lock);
\end{lstlisting}

This situation is quite common. A thread having entered its critical
section cannot proceed before some condition that involves other
threads to enter their critical section is fulfilled. A solution
that sometimes works is for the thread to leave the critical section and
try again later, hoping for the condition to have been fulfilled. A
more elegant solution involves so-called condition variables. A
\impi{condition variable} is an object associated with a mutex\index{mutex}
variable. A thread can perform a \emph{wait}\index{condition variable!wait} on
a condition variable, meaning that the thread will be suspended and
effectively out of the critical section (the lock is released) until
some other thread performs a
\emph{signal}\index{condition variable!signal}
operation on the condition variable. When a waiting
thread receives the signal and is woken up, the signalling thread will
have left the critical section, such that mutual exclusion in the critical
section is always maintained. More threads, for instance,
readers as in the example above, can wait on the same condition
variable. A single signal operation will wake up either of the
threads:
\pthreads\index{pthreads} provides no fairness guarantee and no guarantee
that a thread is not starved. To wake up all waiting threads, one
after the other, a
\emph{broadcast}\index{condition variable!broadcast} is also provided.
Note that mutual exclusion is always maintained, one thread after the other
will be in the critical section.
A signal operation on a condition variable where no thread is
suspended is lost. This is different from the \impi{semaphore}, another
well-known primitive synchronization mechanism, that is not natively supported
with \pthreads, though. The standard usage pattern for locks with
condition variables is called a \impi{monitor}~\cite{Hoare74}. Some
thread models and interfaces support monitors directly, \pthreads
only indirectly via the condition variable mechanism.

\begin{lstlisting}[style=SnippetStyle]
int pthread_cond_destroy(pthread_cond_t *cond);
int pthread_cond_init(pthread_cond_t *restrict cond,
                      const pthread_condattr_t *restrict attr);

int pthread_cond_wait(pthread_cond_t *restrict cond,
                      pthread_mutex_t *restrict mutex);
int pthread_cond_signal(pthread_cond_t *cond);
int pthread_cond_broadcast(pthread_cond_t *cond);
\end{lstlisting}

The problem with the readers-writer lock transfer of information that
we saw above can now be solved with condition variables.

\begin{lstlisting}[style=SnippetStyle]
pthread_cond_t data = PTHREAD_COND_INITIALIZER; // shared
  
// reader threads
pthread_lock(&lock);
while (!written) {
  pthread_cond_wait(&data,&lock);
}
a = b;
pthread_unlock(&lock);
\end{lstlisting}  

\begin{lstlisting}[style=SnippetStyle]
// writer thread
pthread_lock(&lock);
b = ... ;
written = 1;
pthread_cond_broadcast(&data);
pthread_unlock(&lock);
\end{lstlisting}

Often the condition variable mechanism permits so-called
\emph{spurious signals} or \emph{spurious wakeups}. These are false or
outdated signals being sent to a waiting thread (this could be for
implementation reasons). With \pthreads this can indeed be the
case. It is, therefore, good and common practice to always recheck the
desired condition (in the example the \texttt{written} flag) when a
sleeping thread is woken up. Condition variables can easily lead to
deadlocks if used wrongly; there must always be a thread and a
condition that signals and wakes up waiting threads.

\subsection{Versatile Locks from Simpler Ones}

The condition variable mechanism (monitor) is a powerful addition to
the simple mutexes\index{mutex}. For instance, the more versatile
readers-writer locks can be constructed from simple locks using
condition variables. Also, different priority schemes (writer or
readers preferred) can be implemented. The following code example
gives one such implementation.

\begin{lstlisting}[style=SnippetStyle]
typedef struct {
  int readers; // count number of readers
  int waiting, writer; // writers waiting
  pthread_cond_t read_ok, write_ok;
  pthread_mutex_t gatekeeper;
} rwlock_t;

void lock_read(rwlock_t *rwlock)
{
  pthread_mutex_lock(&rwlock->gatekeeper);
  while (rwlock->waiting>0||rwlock->writer) {
    pthread_cond_wait(&rwlock->read_ok,&rwlock->gatekeeper);
  }
  // acquired for read (possibly more than one)
  rwlock->readers++;
  pthread_mutex_unlock(&rwlock->gatekeeper);
  assert(rwlock->writer==0); // at any time before unlock
}

void lock_write(rwlock_t *rwlock)
{
  pthread_mutex_lock(&rwlock->gatekeeper);
  rwlock->waiting++;
  while (rwlock->writer||rwlock->readers>0) {
    pthread_cond_wait(&rwlock->write_ok,&rwlock->gatekeeper);
  }
  // acquired for write (exactly one)
  rwlock->waiting--;
  rwlock->writer = 1; 
  pthread_mutex_unlock(&rwlock->gatekeeper);
  assert(rwlock->readers==0); // at any time before unlock
}

void unlock_readwrite(rwlock_t *rwlock)
{
  pthread_mutex_lock(&rwlock->gatekeeper);
  if (rwlock->writer) rwlock->writer = 0; // done writing
  else rwlock->readers--; // one less reading
  pthread_mutex_unlock(&rwlock->gatekeeper);

  // resume possibly waiting threads
  if (rwlock->readers==0&&rwlock->waiting>0) {
    // wake up writer
    pthread_cond_signal(&rwlock->write_ok);   
  } else {
    // wake up readers
    pthread_cond_broadcast(&rwlock->read_ok); 
  }
}
\end{lstlisting}

The functions implement the functionality of the \pthreads\index{pthreads}
readers-writer lock\index{Lock!readers-writer}. A simple
\texttt{gatekeeper} lock is used to provide mutual exclusion when
updating the variables that control the behavior of the readers-writer
lock: A number of readers are allowed to acquire the lock for reading,
but only one writing thread for writing.  The number of threads
waiting to acquire the lock for writing also needs to be kept track
of. By unlock, if there are no readers and at least one thread waiting
to acquire the lock for writing, one waiting thread is
signalled. Otherwise, all possibly waiting reader threads are notified
by a \texttt{pthread\_cond\_broadcast()}.  When used in a
multi-threaded application, our readers-writer lock is declared as a
shared variable and the fields determining the waiting conditions
initialized to zero. Also, the \texttt{gatekeeper} mutex\index{mutex} must be
initialized.

\begin{lstlisting}[style=SnippetStyle]
// declaration and initialization of readers-writer lock
rwlock_t lock; 
lock.readers = 0;
lock.waiting = 0;
lock.writer  = 0;
  
pthread_mutex_init(&lock.gatekeeper,NULL);
pthread_cond_init(&lock.read_ok,NULL);
pthread_cond_init(&lock.write_ok,NULL);
\end{lstlisting}

A \emph{thread barrier} is a construct which makes it possible for a thread
to define a point in the execution beyond which it cannot progress before a
certain number of other threads have reached the barrier
synchronization point (see \Sec~\ref{sec:barrier}).
\pthreads\index{pthreads} defines function interfaces for such
barriers; the count is the number of threads required to reach the
barrier point. Each barrier (there can be several) is identified by an
opaque \texttt{pthread\_barrier\_t} object, which needs to be shared
among the threads.

\begin{lstlisting}[style=SnippetStyle]
int pthread_barrier_init(pthread_barrier_t *restrict barrier,
                         const pthread_barrierattr_t
                         *restrict attr,
                         unsigned count);
int pthread_barrier_destroy(pthread_barrier_t *barrier);

int pthread_barrier_wait(pthread_barrier_t *barrier);
\end{lstlisting}

Barriers can also trivially be constructed from mutexes with
condition variables. Implementing efficient shared memory barriers is
non-trivial, however, see for instance~\cite{MellorCrummeyScott91}.

A final, common pattern is concurrent initialization, where one of the
threads (the ``first'') should carry out some initialization code
(function). This pattern can easily be implemented with mutexes, but
\pthreads\index{pthreads} provides a shorthand.

\begin{lstlisting}[style=SnippetStyle]
int pthread_once(pthread_once_t *once_control,
                 void (*init_routine)(void));
\end{lstlisting}

\subsection{Locks in Data Structures}

Sequential data structures with their particular semantics and
operations are often used in a parallel setting and this can make a
lot of sense. Threads might want to share a linked list, for instance,
used as the implementation of a set data structure with search,
insert, and delete operations, or a stack, or a queue, or a hash map,
or a priority queue, \etc, and use the data structure operations as
the means for communication and synchronization between threads. As
long as the data structure does not become a sequential bottleneck
(Amdahl's Law\index{Law!Amdahl's Law}) by being too slow or by leading
to thread serialization, shared data structures with sequential
semantics can be helpful in the implementation of parallel
algorithms.

The trivial way of making a(ny) sequential data structure useful in a
parallel algorithm, is to use a global lock to protect all data
structure operations. Mutual exclusion will ensure that the operations
on the data structure are done one after the other. A sequence of
concurrent operations will thus behave according to the sequential
semantics. The already available sequential implementation, perhaps
complex and highly tuned, can be used right away, but the price is
that all concurrent operations on the data structure will
serialize. This can limit the possible speed-up\index{speed-up} of the
algorithm. Thus, this solution is often not good enough. For data
structures with read and write operations, like the set, which
supports search (read) and insert/delete (write) operations, the more
versatile readers-writer locks\index{Lock!readers-writer} can
alleviate some of the drawbacks. Concurrent, perhaps frequent read
operations will have real parallelism, and only the write
operations may be bottleneck operations.

When this is, for performance reasons or others, not acceptable, data
structures and algorithms have to be rethought into more
\impi{concurrent data structures}. Some data structures, like
linked lists, easily allow for implementations with more
``fine-grained'', hands-over locking. The idea is to use a
lock for each list element. As the list is being traversed, only
the locks for the current element and its successor are acquired.
Having locks on two successive elements makes it possible to link out an
element or insert a new element between the two under mutual exclusion
and thus without interference from other threads. For
long lists, this makes it possible for many threads to perform
operations on different parts of the lists. But since a thread having
acquired the locks on elements at the front of the list will prevent
other threads from traversing the list past this point,
the improvement of this locking scheme is modest.

Developing data structures, even with the use of locks, that allow for
a large amount of concurrent uses by many threads, is highly non-trivial and
beyond the scope of these lectures. The point we make here is that locks
can still be useful, but need to be used carefully (localized, short
critical sections), and that in such cases a large number of locks
will have to be used. Therefore, the (space) efficiency of the lock
implementations provided by \pthreads\index{pthreads},
\openmp\index{OpenMP} and other thread models
is highly important.

\subsection{Problems with Locks}

Locks, semaphores, and similar constructs are
\conco\index{Concurrent Computing} constructs that were not
designed for \parco with large numbers of active processing elements
(threads, processes, processor-cores). The typically
(inherently) limited scalability is a reason to use them
sparingly. Locks have other problems:

\begin{itemize}
\item
  Deadlocks can easily be introduced by design (errors).  For
  instance, in a program with two or more locks $L_1$ and $L_2$ (like
  the linked list with hands-over locking), one thread may try to
  acquire the locks in the order $L_1, L_2$ and some other thread in
  the order $L_2,L_1$. If the two threads execute roughly at the same
  time, they will both come to a point where they cannot proceed,
  because the lock that each is trying to acquire is already taken by
  the other thread. This sounds trivial to avoid, but it is not.  The
  code for the two different threads may be in different parts of a
  large software package, may perhaps not written by the same people
  \etc. Each of the code pieces can in itself be correct so that when
  tested in isolation, the deadlock situation does not occur. When the
  pieces of code are run together, the program deadlocks. In that
  sense, locks are not a mechanism that supports modular software
  development.  A deadlock\index{deadlock} is always deadly, it
  proliferates and eventually the whole application cannot complete,
  because the deadlocked threads will not complete.  To avoid
  deadlocks when using multiple locks, locks can be acquired in an
  agreed upon order (stratified locking) or release-temporary
  backoff-acquire techniques be used. With multiple locks, the
  try-lock\index{Lock!try-lock} operation can often be useful.
\item
  A special case of deadlock can occur when a thread
  having acquired lock $L$ tries to acquire $L$ again. This may
  deadlock. So-called \emph{recursive locks}\index{Lock!recursive}
  (or \emph{nested locks}\index{Lock!nested})
  explicitly allow a thread having a lock to acquire the lock again.
  The number of unlock calls may have to match the
  number of lock calls. \pthreads makes it possible to initialize
  recursive locks by the use of a \texttt{PTHREAD\_MUTEX\_RECURSIVE}
  attribute.
\item
  Locks that protect long critical sections lead to possibly harmful
  serialization which can severely degrade performance (Amdahl's
  Law\index{Law!Amdahl's Law}).
\item
  Infinitely long critical sections, for instance, a thread crashing
  in the critical section, lead to deadlocks. Locks are not
  fault-tolerant.
\item
  Since locks are often not fair, threads can be starved and
  actually not be contributing to the progress of the parallel
  algorithm.
\item
  When threads have priorities (possible with \pthreads, but not
  covered in these lectures) locks can lead to the effect that a lower
  prioritized thread prevents a thread with high priority from
  running, even when this would have been possible. The phenomenon is
  called \impi{priority inversion}.
\end{itemize}

\subsection{Atomic Operations}
\label{sec:atomics}

The problem with the \texttt{a = a+27;} example was that the sequence
of instructions in one thread's complex assignment operation
(load, compute, store) could be
interleaved with instructions executed by another thread as well as
the data race on the final update to \texttt{a} (all threads writing).
To prevent such interleavings, the assignment should be executed as an
\emph{atomic}, that is as an indivisible, unit. Mutual exclusion with locks
is one way of guaranteeing atomic execution of the sequence of
instructions. The drawback is that during the execution of the lock code,
no other threads can do anything with the variables that are protected by
the lock.

Another way of ensuring atomic execution of compound operations is
offered by hardware
implemented \emph{atomic operations}\index{atomic operation}. An atomic
operation carries out a complex (but still relatively simple) compound
instruction as a unit that cannot be interfered with by other threads
or\index{processor-core} processor-cores.
One kind of atomic operation is, for instance,
the Fetch-And-Add (FAA) instruction\index{Fetch-And-Add (FAA)}
which can implement the particular \texttt{a = a+27;} assignment as
a single, indivisible instruction.

Special, \emph{atomic instructions} for performing atomic
operations\index{atomic operation}
are offered by all modern multi-core processors\index{multi-core processor}
and systems.  They
operate on one or more memory words given by their memory addresses,
sometimes with an additional value operand, and produce a result.
Memory words that are operated on by atomic instructions are called
and must be \impi{atomic}. Typical atomic instructions are for instance:

\begin{enumerate}
\item
  \emph{Test-And-Set} (TAS)\index{Test-And-Set (TAS)}: On an atomic
  memory word, returns the contents of the address and updates the
  contents to $1$ (\textbf{true}).
\item
  \emph{Fetch-And-Add} (FAA)\index{Fetch-And-Add (FAA)},
  \emph{Fetch-And-Increment} (FAI)\index{Fetch-And-Increment (FAI)}:
  On an atomic memory word, returns the contents of the address and
  updates the memory word by either adding a given value (FAA) or
  incrementing it by one (FAI). More generally,
  \emph{Fetch-And-Operate} (FAO)\index{Fetch-And-Operate (FAO)}
  updates the memory word by a simple operation (logical, for
  instance) while returning the original contents of the address.
\item
  \emph{Exchange} (EXCHG)\index{Exchange (EXCHG)}: On an atomic memory
  word, returns the contents of the address and replaces the contents
  with the given value.
\item
  \emph{Swap} (SWAP)\index{Swap (SWAP)}: Swaps the contents of two
  atomic memory addresses.
\item
  \emph{Compare-And-Swap} (CAS)\index{Compare-And-Swap (CAS)}: On an
  atomic memory word, checks whether the contents equals a given
  expected value and if so replaces the contents with a new update value, and
  returns \textbf{true}. If the contents are not equal to the expected
  value, \textbf{false} is returned and the contents are not changed.
\item
  \emph{Compare-Exchange} (CEX)\index{Compare-Exchange (CEX)}: On an
  atomic memory word, checks whether the contents equals a given
  expected value and if so replaces the contents with a new update value, and
  returns \textbf{true}. If the contents are not equal to the expected
  value, \textbf{false} is returned and the contents of the atomic memory
  word copied back to the expected value (given as a reference).
\end{enumerate}

Beyond this lecture: These atomic operations\index{atomic operation}
form a hierarchy (hence the numbering, with CAS and CEX being what is called
universal and the most powerful) characterized by the power of what
they can do~\cite{HerlihyShavit12}, more precisely for how many
threads they can solve the so-called \impi{consensus problem}. All
these operations are quite natural and helpful in many contexts. For
instance, the atomic Test-And-Set (TAS) instruction is
exactly what is needed to implement a lock.

Atomic operations are indeed instructions like all other processor
instructions, meaning that they complete in some finite, bounded
number of clock cycles, regardless of what other
processor-cores\index{processor-core} might
be doing (even executing atomic operations). This essential
property is called \impi{wait-freeness}. This does not mean that
atomic operations are always fast. Mostly, they are not when compared to
other operations offered by the processor instruction set.  On the
contrary: Atomic operations are expensive, since they need to interact
with the cache and memory system (locking and/or invalidating cache lines,
flushing the write buffer). So, like locks, they
should be used sparingly. But in contrast to locks, use of atomic
operations cannot lead to deadlocks. A crashed (failed) thread
will not affect the ability of the other threads to continue and make
progress. Optimistically, we might assume that atomic operations are
constant time $O(1)$ operations with relatively small constants, but
bounded does not always mean constant\index{atomic operation}.

In the \texttt{stdatomic.h} header for C, the following atomic
operations\index{atomic operation} are standardized for C;
however, there is more to the C
atomics than explained here (ordering guarantees and memory model,
for instance). These operations work on atomic
integer types. There is such an atomic integer type defined in
the header for all C integer types, \eg, \texttt{atomic\_\-bool,
  atomic\_\-char, atomic\_\-short, atomic\_\-int, atomic\_\-long}, \etc. There
is also a special, atomic flag type, \texttt{atomic\_flag}. We
list the operations as defined for atomic integers (they are also defined
for other C word datatypes).

\begin{lstlisting}[style=SnippetStyle]
atomic_init(atomic_int *object, int value);
int atomic_load(atomic_int *object); 
void atomic_store(atomic_int *object, int desired);

// TAS
_Bool atomic_flag_test_and_set(volatile atomic_flag* obj); 
void atomic_flag_clear(volatile atomic_flag* obj);

// FAA, FAO
int atomic_fetch_add(atomic_int *object, int operand);
int atomic_fetch_and(atomic_int *object, int operand);
int atomic_fetch_or(atomic_int *object, int operand);
int atomic_fetch_sub(atomic_int *object, int operand);
int atomic_fetch_xor(atomic_int *object, int operand);
// EXCHG
int atomic_exchange(atomic_int *object, int desired);

// CEX
_Bool atomic_compare_exchange_strong(atomic_int *object,
                                     int *expected,
                                     int desired);
_Bool atomic_compare_exchange_weak(atomic_int *object,
                                   int *expected,
                                   int desired);

// Are operations lock-free?
_Bool atomic_is_lock_free(const volatile A* obj); 

void atomic_thread_fence(memory_order order); // memory fence
\end{lstlisting}

Here is an interesting example: A number of threads update three
counters stored in a global C structure. One counter is updated
non-atomically, the two others with the \texttt{atomic\_fetch\_add}
instruction. After execution, it will not necessarily hold that, for
instance, \texttt{cnt0==cnt1} or \texttt{cnt0==cnt2}. And even if each
of the two counters \texttt{cnt1} and \texttt{cnt2} are updated
atomically, the compound update of both is not, therefore neither of
the stated assertions will (always) hold.

\begin{lstlisting}[style=SnippetStyle]
typedef struct {
  int cnt0;
  atomic_int cnt1, cnt2;
} count3;

void *updates(void *arguments) {
  count3 *counters = (count3*)arguments;
  
  int i;
  int c1, c2;

  for (i=0; i<1000; i++) {
    counters->cnt0++;
    c1 = atomic_fetch_add(&(counters->cnt1), 1);
    c2 = atomic_fetch_add(&(counters->cnt2), 1);
    //assert(c1==c2); ?
    //assert(counters->cnt1==counters->cnt2); ?
  }
  
  pthread_exit(NULL);
}
\end{lstlisting}

It is a good exercise to try this example with varying numbers of
threads.

Our final \pthreads example is concerned with finding and listing all
primes up to some upper \texttt{limit}. An obvious parallelization of
this problem would be to use a loop of independent iterations and to
check in each iteration whether the corresponding index is a prime by
calling \texttt{isprime(i)}. This function more or less naively tries
to find out whether \texttt{i} is divisible by some smaller
number. Since it is called by all the threads, it must be thread
safe\index{thread safe}.  A little thought shows that this is not an
efficient parallelization approach.  First, the time for checking
whether index \texttt{i} is prime depends strongly on \texttt{i} and
is fast for numbers with small prime factors and slow for large
primes. And second, primes are not uniformly distributed (Prime Number
Theorem). For these two reasons, parallelization of the loop will have
poor load balance\index{load balance}. Some threads will finish their
iterations fast and have to wait for other threads with many expensive
checks. Load balancing\index{load balancing} by array
compaction\index{array compaction} cannot help here: We neither know
which indices are primes nor which indices are either fast or slow to
check. Instead, we employ a shared counter which the threads can query
and increment to get the next index to check for primality.  A
corresponding thread function is shown below.

\begin{lstlisting}[style=SnippetStyle]
typedef struct {
  int rank;
  int limit;
  int *next; // shared counter
  int found;
} realargs;
  
void *primes_race(void *arguments)
{
  int i;
  realargs *next = (realargs*)arguments;

  do {
    i = (*(next->next))++;
    if (i<next->limit) {
      if (isprime(i)) { // prime found, take action
	next->found++;
      }
    } else break;
  } while (1);
  
  pthread_exit(NULL);
}
\end{lstlisting}

It is illustrative to try this code and check how many primes it
reports per thread in field \texttt{found}. The problem is the
non-atomic update of the shared counter \texttt{*next}, similarly to
what we saw in the \texttt{a = a+27;} example. Since increments can easily
be lost, the effect is that much
primality checking is repeated by the threads of which there may be many.
The solution is to use an atomic counter with a FAI instruction.

\begin{lstlisting}[style=SnippetStyle]
void *primes_atomic(void *arguments)
{
  int i;
  realargs *next = (realargs*)arguments;

  do {
    i = atomic_fetch_add(next->next,1);
    if (i<next->limit) {
      if (isprime(i)) { // prime found, take action
	next->found++;
      }
    } else break;
  } while (1);

  pthread_exit(NULL);
}
\end{lstlisting}

This is a simple example of a work pool\index{work pool} with
work-stealing\index{work-stealing} (see \Sec~\ref{sec:workpool}).  The work
items are the indices to be checked and are maintained as a single,
shared counter. Threads steal work atomically by reading the current
value of the counter and incrementing it.

In general, an operation on a data structure is said to be
\impi{wait-free} if a thread executing the operation can always
complete in a bounded amount of time, regardless of what the other
threads are doing (including also performing the operation).  An
operation is said to be \impi{lock-free}, if, when several threads are
performing the operation, some thread will be able to complete in a
bounded amount of time. Wait-freeness\index{wait-freeness} is the
nonblocking analogy of starvation-freeness and \impi{lock-freeness}
the nonblocking analogy of deadlock-freeness. Like starvation freedom
implies deadlock freedom, wait-freeness implies lock-freeness.

It can be shown that, with sufficiently
strong atomic operations\index{atomic operation}\index{Compare-And-Swap (CAS)}
(CAS), it is possible to give a wait-free implementation of
any sequential data structure~\cite{HerlihyShavit12}. This is a
theoretically strong result, but does not mean that wait- and
lock-free data structures also perform well in practical contexts. We
have seen that a wait-free counter can be useful. Other lock- and
wait-free algorithms and data structures are beyond these lectures.

\section{Seventh block (3 lectures)}
\label{sec:openmpframework}

\openmp\index{OpenMP} (``Open Multi Processing''),
a standard for C and Fortran dating
back to around 1997, is our next example of a concrete programming
interface that implements a shared memory
programming model\index{programming model} and is
intended for running on parallel shared memory systems. Like
\pthreads, \openmp is thread based, but offers much more and much
stronger support for \parco. Historically, the main unit of
parallelization in \openmp was the loop of independent iterations, see
\Sec~\ref{sec:looppattern}. Around \openmp 3.0, support for task
parallelism was introduced, see \Sec~\ref{sec:taskgraphs}. This
lecture and the following ones give an introduction to parallel programming
with \openmp and cover the main features and constructs needed in
\parco. There is more to \openmp than we will cover here, though. In
particular, thread teams and groups will be silently circumvented and also the
recent support for accelerators like GPUs will not be
treated\index{GPU}\index{accelerator}.
Some recommended and sometimes revealing books for \openmp\index{OpenMP}
programmers
are~\cite{OpenMP2008,OpenMP2017,OpenMP2019}.

\openmp\index{OpenMP} is maintained and developed further by an
\impi{Architecture Review Board} (ARB) which includes academic
institutions and industry in various roles.
The \openmp specification and additional
information are freely available via \url{www.openmp.org}, including
very helpful cheat-sheets, see for
instance~\url{https://www.openmp.org/wp-content/uploads/OpenMPRefCard-5.1-web.pdf}.

\subsection{The \openmp Programming Model}

Like \pthreads, \openmp\index{OpenMP}\index{pthreads}
is a fork-join thread model but threads are
less explicit than in \pthreads. There is no specific object
identifying a thread.
A \emph{master thread} can fork (activate) a consecutively numbered set
of working threads that includes the master thread itself. The threads
share in executing \emph{work} as specified by a
\impi{work sharing construct}, \eg, a loop of independent iterations
or a task graph. Upon
completion, threads join, leaving the master thread to fork a next
set of threads. An \openmp program is a single program and all forked
threads execute the same code (SPMD\index{SPMD}).

The main characteristics of the \openmp\index{OpenMP}
programming model\index{programming model} are:
\begin{enumerate}
\item
  Parallelism is (mostly) implicit through work sharing. All threads
  execute the same program (SPMD).
\item
  Fork-join parallelism: A master thread implicitly spawns threads
  through a parallel region construct. Threads join at the end of
  the parallel region.
\item
  Each thread in a parallel region has a unique integer thread identifier (id),
  and threads are consecutively numbered from $0$ to the number of
  threads minus one in the region.
\item
  The number of threads can exceed the number of available processors/cores.
  Threads are intended to be executed in parallel by available processor-cores.
\item
  Constructs for sharing work across threads are provided.
  Work is expressed as loops of independent iterations and task graphs.
\item
  Threads can share variables; shared variables are shared among all
  threads. Threads can have private variables.
\item
  Unprotected, parallel updates of shared variables lead to data
  races and are illegal.
\item
  Synchronization constructs for preventing
  race conditions\index{race condition} are provided.
\item
  Memory is in a consistent state after synchronization operations.
\end{enumerate}

As for \pthreads, \openmp does not come with any performance model and
gives neither guarantees nor prescriptions for the behavior and performance
of compiler and runtime system. Different compilers and runtime
systems for \openmp sometimes deliver very different performance for
the same code.

\subsection{\openmp in C}

\openmp\index{OpenMP} requires compiler, library and runtime system support
and must, therefore, be compiled with an \openmp-capable compiler and
linked against library and runtime system. Most C compilers are nowadays
\openmp-capable. For instance, \openmp programs can be compiled with
\gcc by giving the \texttt{-fopenmp} option.  C code using \openmp
must include the function prototype header file with the
\texttt{\#include <omp.h>} preprocessor directive.
All \openmp relevant functions and
predefined objects are prefixed with \texttt{omp\_}, which identifies
the \openmp ``name space''. Special \openmp environment variables are
prefixed with \texttt{OMP\_}. \openmp is not a language extension per
se, but requires extensive compiler and runtime support for parsing
and translating and efficiently executing
the \texttt{\#pragma omp}-directives. \openmp programs
are C programs, but constructs like \texttt{for}-loops and compound
statements are given their \openmp meaning by \texttt{\#pragma omp}
compiler directives. Pragmas are designations in the code that direct
the compiler to handle the following (compound) statement in a certain way.

For the concrete explanations in the following sections, we use
\texttt{<...>} as meta-language designation for statements and
non-empty lists of names, \texttt{[...]} to denote zero or more
(optionally comma-separated) repetitions of some pragma element
(clause), and \texttt{|} for exclusive choice.

\subsection{Fork-Join Parallelism with the Parallel Region}

Threads are activated when the master thread reaches
an \openmp\index{OpenMP} \impi{parallel region}
construct which is a structured C
statement (simple statement or compound statement in curly brackets
\texttt{\{...\}}) designated by the \texttt{omp parallel} pragma.
Actual threads and their binding to processor-cores\index{processor-core}
are managed by
the \openmp runtime system; threads may be created once and for all
and put to sleep for later reactivation or may be created afresh at
each parallel region. In the parallel region, a defined number of
threads $p$ will be active, all
executing the structured statement (SPMD style\index{SPMD}). Variables
declared in the parallel region exist per thread and are local to the
threads while variables declared before and outside of the parallel region are
per default shared by all the threads. Once
started, the number of threads in the parallel region cannot be
changed. The threads can, by suitable library function calls, look up
their thread identifier (id) and the number of threads executing in the
parallel region. The thread id is a C integer between $0$ and $p-1$.
That is, thread ids are consecutive $0,1,\ldots, p-1$.
Threads coming to the end of the parallel region \emph{join} with the
other threads by performing a
barrier synchronization\index{barrier synchronization},
leaving only the master thread active after the parallel region.
The barrier synchronization operation is implicit with the end of a
parallel region, and it is essential for the \openmp\index{OpenMP}
fork-join model
that this cannot be changed.  The thread id of the master thread is
always $0$.

\begin{lstlisting}[style=SnippetStyle]
#pragma omp parallel [clauses]
<structured statement>
\end{lstlisting}

Activation and deactivation of threads takes place at entry and (syntactic)
exit of the parallel region and entails a barrier synchronization where
all threads of the region have to be involved. It is, therefore, not allowed
to break into or out of a parallel region with \texttt{break;} or \texttt{goto}
statements. Sometimes such requirements can be checked by the \openmp compiler.

The number $p$ of threads in a region can be controlled either by the
runtime environment, by a library call, or by a
\texttt{num\_threads()} clause for the \texttt{omp parallel}
pragma. The last takes priority over the library call, which takes
priority over the environment setting.  When controlled by the
environment, either a default number of threads is used
or $p$ is determined by the environment
variable \ompnumthreads. The \ompnumthreads variable can be set to a
number of threads larger than the number of processor-cores; that is,
it is possible to run \openmp\index{OpenMP} programs
with \impi{oversubscription}. This is
often useful for debugging but rarely for performance.
The default number of threads is typically the number of
processor-cores\index{processor-core}
of the system where the program is running but can also be the number of
hardware supported threads: The CPU may support \impi{hardware multi-threading}
where each core can execute a small number of unrelated instruction streams
which makes it look as if a correspondingly larger number of
cores are available.

Regardless of the number of threads that might be available for a
parallel region, it can sometimes be useful and efficient not to employ
any additional threads at all -- the problem at hand could be so small
that parallelization with any number of threads larger than one is
actually detrimental. An \texttt{if (<expr>)} clause to a parallel
region will only activate threads if the expression evaluates to
\textbf{true}. Otherwise, the parallel region will be executed by the master
thread alone.

An \openmp\index{OpenMP} program
consists of a sequence of parallel regions and can
be depicted as a fork-join task graph (see \Sec~\ref{sec:taskgraphs}).
The work\index{work} of a parallel region executed with $p$ threads is
the sum of the sequential work done by the $p$ threads, and the time
of a parallel region is the time for the last thread to finish. Since
threads are activated at the beginning of the region and the region is
finished by a barrier synchronization, the time of a region can be
defined as the time that can be measured by the master thread $0$ from
start to end. Note that this always entails at least $\Omega(\log p)$
time for the barrier synchronization, with hopefully small constants
depending on the quality of the \openmp implementation.  The work of a
complete \openmp program is the total amount of work done by the
threads over all parallel regions. Different parallel regions may use
different numbers of threads. The parallel execution time of an
\openmp program is the time that can be measured by the master thread
from start to the end of the computation, that is, the sum of the
running times of the successive regions plus the time taken by the
master thread when no parallel region is active.  A good \openmp
program will have work proportional to the work of a best known
sequential program for the given problem and has a small number of
regions in each of which the work is well balanced over the threads
executing in the regions. In particular, the number and total time of
the regions will correspond to $\tinf(n)$, the longest path in the
program and the dependent part of the work that has not been
parallelized.

\subsection{\openmp Library Calls}

By suitable \openmp\index{OpenMP} library calls, a thread can look up its
non-negative integer thread number, determine the number of threads in
a parallel region, get the maximum number of threads allowed by the
environment, and set the number of threads for a parallel region.

\begin{lstlisting}[style=SnippetStyle]
int omp_get_thread_num(void);
int omp_get_num_threads(void);
int omp_get_max_threads(void);
void omp_set_num_threads(int num_threads);
\end{lstlisting}

These \openmp library calls are all \impi{thread safe}, that is, they can be
called concurrently, in parallel, without any risk of interference.

For measuring the time taken by the execution of a (sequence of)
parallel region(s), \openmp provides standardized access to a (stable,
high precision) timer.

\begin{lstlisting}[style=SnippetStyle]
double omp_get_wtime(void);
double omp_get_wtick(void);
\end{lstlisting}

The library function \ompwtime returns the
\impi{wall clock time} in seconds since some point in the past. To report the
time in milliseconds or microseconds of a piece \openmp code, read the
time before and after the piece of code and  multiply the difference by
$1000.0$ or $1.000.000$, respectively. The \texttt{omp\_get\_wtick()}
call returns the resolution (in seconds) of the timer.

\subsection{Sharing Variables}

Per default, all variables declared before a parallel region by the
master thread are shared by the threads in the region. Variables
declared in the structured statement (block) of the parallel region
are \emph{private} (local) to each thread which means that a private,
local copy for each thread will be created by the \openmp\index{OpenMP}
compiler.

Sharing of variables can be controlled through sharing clauses to the
\texttt{omp parallel} pragma directive.

\begin{lstlisting}[style=SnippetStyle]
private(<comma separated list of variables>)
firstprivate(<comma separated list of variables>)
shared(<comma separated list of variables>)
default(shared|none)
\end{lstlisting}

A list of variables declared by the master thread (before the parallel
region), that will per default be shared in the parallel region, can be
made private, which means that the compiler will generate a local copy
for each thread. Variables declared private by the \texttt{private()}
clause are \emph{not} initialized. The \texttt{firstprivate()} clause
additionally initializes each local copy to the value the variable had
before the parallel region. Often, this is the desired and perhaps
implicitly assumed behavior. Note, that this can be expensive if the
variable denotes a large, statically (compiler) allocated array as in
\texttt{int a[1000];}.  In contrast, for pointers the value of the
pointer is copied and not the object to which it points. There are many
possibilities for making non-sharing mistakes with \openmp.

It is good practice (many say) to explicitly not share any variables
declared by the master thread before a parallel region by using the
\texttt{default(none)} clause and to then explicitly list the variables
to be shared in a \texttt{shared()} clause. Such discipline forces
one to think about which variables need to be shared and which not.

Shared variables can be read concurrently by the threads in the
parallel region, but an \openmp program in which it can happen that a
thread updates a shared variable concurrently with other threads
reading (or writing) the shared variable is \emph{incorrect}. This  is a
\impi{data race}\index{data race} that may lead to a
race condition\index{race condition}
and correct \openmp programs must not exhibit data races.
\openmp provides different means for avoiding data races while
still allowing to
exchange information between threads via shared variables (see the following).

\subsection{Work Sharing: Master and Single}

The simplest work sharing \openmp\index{OpenMP}
constructs\index{work sharing construct} designate work that is
\emph{not} to be shared among the threads but rather to be executed by only
one thread.

\begin{lstlisting}[style=SnippetStyle]
#pragma omp master
<structured statement>
\end{lstlisting}

Here, the work of the structured statement is done by the master
thread alone (the thread with \texttt{omp\_get\_thread\_num()==0}). The other
threads will skip the structured statement code and just continue execution.
There is \emph{no} barrier synchronization implied following the master thread
code. Also, the code of the master thread is \emph{not} executed under mutual
exclusion\index{mutual exclusion}.
That is, the master thread must not update shared variables that
can potentially be read or updated concurrently by the other threads that
are not in the master statement.

\begin{lstlisting}[style=SnippetStyle]
#pragma omp single [clauses]
<structured statement>
\end{lstlisting}

With the \texttt{single} construct,
the work of the structured statement is done by either one of
the parallel, running threads, but it is not determined which of the
threads; the \openmp\index{OpenMP} runtime system (or compiler)
makes the decision.
A parallel region can, of course, have several \texttt{single} statement
blocks and each of the blocks may be executed by a different
thread. The code executed by the chosen, single thread is, like for
the \texttt{master} construct, not executed under mutual exclusion. So
updates to shared variables possibly read or written by other threads are
illegal. In contrast to the \texttt{master} construct, the
\texttt{single} construct has an implied barrier at the end of the
structured statement. A thread reaching this point, regardless of
whether it was the thread executing the \texttt{single}-designated
statement or one of the other threads, cannot proceed until all threads have
reached this point. This implies that the number of
encountered \texttt{single} statement blocks must be the same for all
threads and so one must be careful with branches and loops in parallel
regions.

The implied barrier at the end of the \texttt{single} block can be
eliminated with the \texttt{nowait} clause. This can sometimes lead to
better performance: A barrier can be expensive, and an \openmp program
should have no more barriers than absolutely necessary. On the other
hand, a \texttt{nowait} clause can as easily make a correct program
incorrect by introducing race conditions\index{race condition} 
(data races\index{data race}). The
\texttt{single} construct allows to make variables \texttt{private()}
or \texttt{firstprivate()}; the \texttt{master} construct does not.

In the following example, the master thread reads input for a
parallel computation to be done by private (non-parallelized)
\texttt{for}-loops by the threads. Since there is no implied
synchronization between the threads after the master has completed,
an explicit \openmp barrier (see next section) has been introduced,
after which all threads can safely work on the input in the array
\texttt{a}. The result in array \texttt{b} is written by some single
thread. In order to ensure that all threads have completed their
work before the array is written, again an explicit barrier is
needed. The implicit barrier of this \texttt{single} construct is not
needed here and is, therefore, eliminated
by a \texttt{nowait} clause.
Since there is always a barrier at the end of the parallel region,
the extra barrier implied by the \texttt{single} work sharing construct
would have been redundant here.

\begin{lstlisting}[style=SnippetStyle]
int n; // shared size
int *a, *b; // allocate somewhere
#pragma omp parallel if (n>10000) // only for large n
{
  int i; // private i for each thread    
...
#pragma omp master
  readdata(a,n);
#pragma omp barrier
  // compute
  int i0, n0; // loop range for thread
  i0 = ...; n0 = ...n;
  for (i=i0; i<n0; i++) {
    b[i] = ...; // per thread computation from a into b
  }
#pragma omp barrier
#pragma omp single nowait
  writedata(b,n);  
}
\end{lstlisting}

If the explicit barriers were omitted, correctness could be
guaranteed: There would be possible race conditions on both \texttt{a} and
\texttt{b} arrays\index{race condition}.

Code for single and master threads should be kept short, unless the
other threads have sufficient other work to do. All threads in a
parallel region should perform more or less the same amount of work.

\subsection{The Explicit Barrier}

An explicit barrier, a point in the code of a parallel region beyond
which no thread shall continue before all other threads have reached
this point\index{barrier}, can be designated with the \texttt{barrier}
construct as we saw in the previous section.

\begin{lstlisting}[style=SnippetStyle]
#pragma omp barrier
\end{lstlisting}

An explicit barrier is sometimes necessary, for instance, after a
\texttt{master} construct or in situations where threads read values
computed by other threads. Here, an explicit (or implicit) barrier can
be necessary to ensure that the other threads have indeed completed the
computation of the required values.

\subsection{Work Sharing: Sections}

The work to be done by some (part of an) algorithm can sometimes be
statically expressed as some finite set of independent pieces that can
potentially be executed in parallel by a set of available threads. In
\openmp\index{OpenMP}, such work can be identified and the independent pieces
can be designated as such.
This work sharing construct\index{work sharing construct}
is called \texttt{sections}
with each independent piece forming a \texttt{section} of code.

\begin{lstlisting}[style=SnippetStyle]
#pragma omp sections [clauses]
<section block>
\end{lstlisting}

Each independent section of code in the section block (enclosed in
\texttt{\{...\}}) is marked as such.

\begin{lstlisting}[style=SnippetStyle]
#pragma omp section
<structured statement>
\end{lstlisting}

A block of sections also ends with an implicit barrier synchronization
point: No thread can continue beyond the sections code before all
sections have been completed. This implicit barrier can be
circumvented with the \texttt{nowait} clause. Before the block of
sections, the sharing of variables can be restricted to either
\texttt{private()} or \texttt{firstprivate()}.

In a parallel region with sections, the individual sections are
assigned to the threads according to some schedule\index{schedule} chosen
by the \openmp\index{OpenMP} runtime system.
Ideally, each thread will execute a section,
and the threads will all run in parallel. If there are more sections
than threads in the parallel region, some thread(s) will necessarily
execute more than one section. Good \openmp code will aim to make the
amount of work in the sections balanced and, in particular, avoid having
(too) few, very large sections that could lead to harmful load
imbalance\index{load imbalance} by many threads sitting idle at the barrier
waiting for the other threads. \openmp sections is a static division of work
and normally the number of sections is small. We will see constructs for
dynamic division of work in the following sections.

\subsection{Work Sharing: Loops of Independent Iterations}

Work is very often expressed as loops of independent
iterations (see \Sec~\ref{sec:looppattern}).
This was and is still the basic, fundamental premise of
\openmp\index{OpenMP}
and the parallelized loop one of the basic,
work sharing constructs\index{work sharing construct}.
As we have seen, loops of independent iterations provide
ample opportunity for keeping threads\index{processor-core}
(processor-cores) busy by
assigning (blocks of consecutive) loop iterations to threads. The
assignment of particular iterations to threads is called
\impi{loop scheduling} in \openmp and is expressed by a \texttt{for}
work sharing pragma with clauses as part of a parallel region.
Loop scheduling must at least fulfill
that each iteration is executed exactly once by some thread as the
sequential semantics of the loop require. By the independence condition,
the iterations can be executed in any order and concurrently by the threads.
Loops must take a specific, syntactic form called the \impi{canonical form}.

\begin{lstlisting}[style=SnippetStyle]
#pragma omp for [clauses]
for (<canonical form loop range>)
<loop body>
\end{lstlisting}

In order that threads can independently of each other
(perhaps supported by data structures in the \openmp runtime system) schedule 
the iterations, the loop range must confirm to certain rules. The
most important such rule is that all threads in the parallel region will
be able to determine the \emph{same} loop range. Thus, in a standard
C for-loop
\begin{lstlisting}[style=SnippetStyle]
for (i=start; i<end; i+=inc)
<loop body>
\end{lstlisting}
all threads must compute the same values for the start and end
iteration and must use the same increment (here, \texttt{i},
\texttt{start}, \etc are arbitrary variable names and
expressions). These values must \emph{not} change in any way during
the execution of the loop. Also, loop ranges must be finite
and determined; that is, the for loop must \emph{not} be a camouflaged,
open-ended while loop. Such a range can easily be split into
blocks of iterations by the compiler. 

Finally, \openmp poses restrictions on the form of the loop 
bound condition, which must be of the form \texttt{i<n},
\texttt{i<=n}, \texttt{i>n}, \texttt{i>=n}, or \texttt{i!=n} only
(\texttt{i} is an arbitrary variable and \texttt{n} an arbitrary
expression).  Also, increments must take either of the forms
\texttt{i++}, \texttt{i+=inc}, or \texttt{i=i+inc} and similarly for
decrements. Loops fulfilling such restrictions are said to be in
\impi{canonical form}. The loop variable, here \texttt{i}, is 
automatically made private for the loop body; otherwise, each iteration
would be a race condition\index{race condition} on \texttt{i}.

There is a composite, shorthand directive that combines the parallel region
with one parallel loop. This is one of the most frequent directives
in \openmp programs.

\begin{lstlisting}[style=SnippetStyle]
#pragma omp parallel for [clauses]
for (<canonical form loop range>)
<loop body>
\end{lstlisting}

Inherited from the \texttt{omp parallel} construct, the composite loop
directive does not allow the \texttt{nowait} clause since only the master
thread $0$ is to be active after the parallel for loop. For this reason,
breaking into or out of a parallel loop is illegal. Such violations
may sometimes be caught by the compiler. Try compiling the following loop.

\begin{lstlisting}[style=SnippetStyle]
#pragma omp parallel for
for (i=0; i<n; i++) {
  if (i==10) continue; // this is ok
  if (i==11) break;    // but not this!
}
\end{lstlisting}

The sharing clauses from the \texttt{omp parallel} region also apply
to the \texttt{omp parallel for} loop. For parallel loop constructs
there is a further sharing option which allows to transfer the value
of a private variable to its shared counterpart, namely by capturing
the value of the variable at the sequentially last iteration of the
loop. Here is a handy use-case for capturing the value of the loop
index variable after the last iteration as often used in sequential code.

\begin{lstlisting}[style=SnippetStyle]
int i; // shared i
#pragma omp parallel for lastprivate(i) // now private
for (i=0; i<n; i++) {
  a[i] = b[i];
}
assert(i==n);
\end{lstlisting}

The \texttt{lastprivate()} clause can also be used with parallel
sections and will in that case capture the value of the variable in
the syntactically last section.

In order for an \openmp program to be correct, loop iterations,
regardless of the order in which they are executed by the threads,
must not cause data races by concurrent reads and writes to
shared variables: The loop iterations must be independent and have neither
forward, anti-, nor output dependencies\index{loop dependency}.
A simple, sufficient rule for independence of loops is the following.
The loop does array updates only, each iteration updates at most one
array element, and no iteration refers to an element updated by another
iteration.

Some loop carried dependencies in simple, array only loops can be
eliminated by transforming the loops. A loop like
\begin{lstlisting}[style=SnippetStyle]
for (i=k; i<n; i++) a[i] = a[i]+a[i+k];
\end{lstlisting}
where, sequentially, \texttt{a[$i$]} is updated in iteration $i$ with
the (sequentially) not yet updated, and, therefore, ``old'' value
\texttt{a[$i+k$]}, can equivalently be written as
\begin{lstlisting}[style=SnippetStyle]
for (i=k; i<n; i++) aa[i] = a[i]+a[i+k];
// swap
tmp = a; a = aa; aa = tmp;
\end{lstlisting}
by introducing a new array \texttt{aa} into which the updates are
computed from the ``old'' values in array \texttt{a} and
swapping the two arrays after the loop. A little care is required to allocate
and free the extra arrays correctly.

The transformed loop is now a loop of independent iterations (also
according to the simple rules for independent loops) and can, therefore,
readily be parallelized with \texttt{\#pragma omp parallel for}.

\subsection{Loop Scheduling}
\label{sec:ompschedule}

Loop scheduling denotes the assignment of loop iterations to
threads: How exactly is the work expressed by the loop of independent
iterations shared across the threads of the parallel region? 

For loop scheduling, the number of iterations in the loop range is divided into
not necessarily same-sized \emph{chunks} of consecutive
iterations. Like the iterations, chunks are numbered consecutively
such that they can be referred to by their number. The chunk
numbering is for reference only and not something that has to be
computed or maintained explicitly by the \openmp\index{OpenMP} runtime system.

\openmp provides three basic types of loop schedules\index{schedule}.  In a
\emph{static} schedule\index{Loop schedule!static}, all chunks have
(almost) the same size, and chunks are assigned in a round-robin
fashion to the threads. For a loop range of $n$ iterations, with
chunksize $c$, and $p$ threads, there are $k=\ceiling{n/c}$ chunks,
and the iterations of chunk $i, 0\leq i<\ceiling{n/c}$ are executed
(one after another) by thread $i\bmod p$. That is, thread $0$ executes
the iterations of chunk $0$, thread $1$ the iterations of chunk $1$,
thread $2$ the iterations of chunk $2$, and so on. If there are more
than $p$ chunks, again thread $0$ executes the iterations of chunk
$p$, thread $1$ the iterations of chunk $p+1$, and so on, until all
iterations of all chunks have been executed. If the loop range has
been divided into at least $p$ chunks, all threads can be kept busy,
but not necessarily all of the time: That depends on the exact number
of chunks and on the time that each iteration takes, which may be
different for different iteration indices. For instance, if the work
per iteration in chunks $0$, $p$, $2p$ \etc is very small, which could
be the case if a condition on the loop iteration index fails for these
chunks, there is nothing to do for these chunks except for going
through the iterations and checking the condition. So, thread $0$ might
be able to finish much faster than the other threads.

In a \emph{dynamic} schedule\index{Loop schedule!dynamic} all
chunks also have the same size $c$, but the chunks are not assigned to the
threads in any predetermined, static fashion. Chunks are executed by
the threads in increasing order. Instead of a fixed assignment, each thread
\emph{dynamically} grabs the next not yet assigned chunk as soon as it
has finished its previous chunk.
With a dynamic schedule, the situation sketched above will not
happen. As soon as thread $0$ finishes (fast) with chunk $0$ it will
grab the next unassigned chunk and, thus, help with finishing the loop
iterations faster than the static schedule could.

Like in a dynamic schedule, a \emph{guided}
schedule\index{Loop schedule!guided}
assigns chunks to threads dynamically as the
threads become available. Unlike both static and dynamic
schedules, the chunk size is no longer fixed. Instead, when a thread
has finished executing an earlier, smaller numbered chunk, it grabs a
chunk for the next iteration that has not yet been executed. Instead,
the size (number of iterations) of the chunk is computed dynamically
as the number of remaining, not yet
executed or assigned iterations divided by the number of threads $p$.

The advantage of the static schedule is that computation of chunk
numbers and assignment to threads can be done very fast and
efficiently. Essentially, each thread decides for itself which
chunks it will have to execute, which is possible due to the
restrictions on parallelizable loops by the canonical
form\index{canonical form}. Thus,
static schedules have low scheduling overhead. A static schedule can be
expected to give good performance when the work per iteration or per chunk
is more or less the same for all iterations or chunks.
Many, but certainly not all loops have this
property. The time per iteration can be influenced heavily
by the memory and cache access patterns even for code where the
iterations incur the same number of instructions to be
executed. Dynamic and guided schedules can be preferable for loops
with conditions depending on the iteration index or otherwise
varying amounts of work per iteration. The guided schedule is
motivated by the assumption that, when a thread becomes ready to execute the
next chunk, the work in the remaining iterations will be more or less the
same per iteration, in which case it makes sense to divide these remaining
iterations evenly into $p$ chunks. This assumption may or may not hold,
and a guided schedule may or may not perform better than a static or dynamic
schedule. Dynamic and guided schedules both
have a higher scheduling overhead than static schedules. For instance,
dynamic scheduling could be implemented by the \openmp runtime system
with a simple \emph{work pool}\index{work pool} (see \Sec~\ref{sec:workpool})
that maintains the next,
not yet executed loop iteration index. Implementing such a work pool would
require just one \emph{atomic counter}\index{atomic operation}:

\begin{lstlisting}[style=SnippetStyle]
do {
  start = atomic_fetch_and_add(&i,chunksize);
  if (start>=n) break;
  end = min(start+chunksize,n);
  for (j = start; j<end; j++) {
    // execute iterations in chunk
  }
} while (1);
\end{lstlisting}

Here, \texttt{chunksize} is the fixed chunksize $c$. It was
tacitly assumed that the loop increment was $1$.

In \openmp\index{OpenMP},
the particular schedule\index{schedule} type for a parallel for loop can be
determined by an explicit \emph{schedule clause} that takes an
optional, explicit chunk size parameter. For static and dynamic
schedules this optional chunk size is the exact size in number of
iterations of the chunk (except possibly for the last chunk),
whereas for guided schedules, the chunk size parameter
is a lower limit on the gradually decreasing number of iterations for
the chunks.

\begin{lstlisting}[style=SnippetStyle]
schedule(static[,chunksize])
schedule(dynamic[,chunksize])
schedule(guided[,chunksize])
\end{lstlisting}

If a schedule clause is not given, a static schedule with default chunk
size is used. If no chunk size is given, a default chunk size is used.
For a static schedule for a loop range with $n$ iterations,
this is approximately $n/p$, such that there are exactly $p$ chunks,
one for each thread. If $p$ does not divide $n$, the smallest chunk size
is $\floor{n/p}$ and one or more chunks will have one or more extra iterations.
The \openmp\index{OpenMP} specification deliberately does not
specify which chunks will get the extra iterations. For dynamic
and guided schedules, the default chunk size is $1$.

For simple loops over arrays, it can sometimes make sense to let the chunksize
$c$ be some multiple of the \impi{cache line} (block) size in order to
avoid \emph{false sharing}\index{cache!false sharing}.

There are two additional schedule types that can be given with the
schedule clause.

\begin{lstlisting}[style=SnippetStyle]
schedule(runtime)
schedule(auto)
\end{lstlisting}

With the \texttt{runtime} type schedule, the schedule can be set
externally through the \ompschedule environment variable, which can be very
useful for tuning and experimenting with different schedules. Here are
some examples of \ompschedule settings.

\begin{lstlisting}[style=SnippetStyle]
"static,1"
"static,8"
"dynamic"
"guided,100"
\end{lstlisting}

With the \texttt{auto} type schedule, the choice of ``best'' schedule
is left to the \openmp compiler and runtime system.

To understand and memorize the \openmp\index{OpenMP} loop schedules,
it is a good
exercise to implement code that for a (large) iteration range $n$ in a
parallel for loop records for each iteration which thread was responsible
for that iteration and as well makes sure that all iterations were
indeed executed (and no more).

\begin{lstlisting}[style=SnippetStyle]
int t = omp_get_max_threads();
int iter[t]; // iterations per thread
int loop[n]; // who did iteration i; careful for large n
  
for (i=0; i<t; i++) iter[i] = 0;
#pragma omp parallel for schedule(runtime)
for (i=0; i<n; i++) {
  loop[i] = omp_get_thread_num();
  iter[omp_get_thread_num()]++; // problematic for large n
}
int nn = 0; 
for (i=0; i<t; i++) nn += iter[i];
assert(nn==n); // all iterations done
\end{lstlisting}

Running the code with $n=20$ (small) and printing out
the values \texttt{loop[i]} and
\texttt{iter[i]} with \ompschedule set to \texttt{"static,2"}
and seven threads (\ompnumthreads set to $7$) gives the following output.

\begin{center}
  \begin{tabular}{rrrrrrrrrrrrrrrrrrrr}
    \multicolumn{20}{c}{Thread for iteration $i, 0\leq i<n$.} \\
    0 & 1 & 2 & 3 & 4 & 5 & 6 & 7 & 8 & 9 & 10 & 11 & 12 & 13 & 14 & 15 & 16 & 17 & 18 & 19 \\
    \midrule
0 & 0 & 1 & 1 & 2 & 2 & 3 & 3 & 4 & 4 & 5 & 5 & 6 & 6 & 0 & 0 & 1 & 1 & 2 & 2 \\
  \end{tabular}
\end{center}

\begin{center}
  \begin{tabular}{rrrrrrr}
    \multicolumn{7}{c}{Iterations per thread.} \\
    0 & 1 & 2 & 3 & 4 & 5 & 6 \\
    \midrule
    4 & 4 & 4 & 2 & 2 & 2 & 2 \\
  \end{tabular}
\end{center}

\subsection{Collapsing Nested Loops}

Many computations, for instance, computations involving matrices, are
often expressed with (doubly, triply) nested, parallelizable loops.
If all the loops
in the loop nests are loops of independent iterations, either of them
can be parallelized with the parallel for directive. Deciding which
one to parallelize may not be obvious and depends (among other
things) on the amount of work per iteration and the number of
iterations per loop. Often it makes sense to parallelize the loop with
the largest number of iterations, but since that could be either of
the loop nests, attempting this could lead to code blow up by having
to maintain parallelizations for all the possibilities.
A sometimes good solution is to treat the nested loops as
one single larger loop; that is, to transform code of the form

\begin{lstlisting}[style=SnippetStyle]
for (i=0; i<m; i++) {   // parallelize this loop?
  for (j=0; j<n; j++) { // or this loop?
    x[i][j] = f(i,j); 
  }
}
\end{lstlisting}
into
\begin{lstlisting}[style=SnippetStyle]
for (ij=0; ij<m*n; ij++) {
  i = ij/n; j = ij%n;
  x[i][j] = f(i,j);
}
\end{lstlisting}

This transformation is valid in the sense that each iteration of the
nested loop is performed exactly once by the transformed loop.  This
is possible under the condition that all loop bounds can be computed
before the two loops and do not change during the iterations.

By adding the \texttt{collapse(<nesting depth>)} clause to the
\texttt{parallel for} directive, the outlined transformation can be
performed automatically (to any nesting depth) by the \openmp
compiler. The loops must be perfectly nested. This means that the body
of an outer loop must consist of only the next inner loop (no extra
statements).  As for all \openmp parallelizable loops, the iteration
ranges must be in canonical form\index{canonical form} prescribed
by \openmp. The two nested loops can then be parallelized as follows:

\begin{lstlisting}[style=SnippetStyle]
#pragma omp parallel for collapse(2)
for (i=0; i<m; i++) {   // OpenMP makes one loop out of two
  for (j=0; j<n; j++) {
    x[i][j] = f(i,j); 
  }
}
\end{lstlisting}

If the loop has this form instead,
\begin{lstlisting}[style=SnippetStyle]
#pragma omp parallel for collapse(2)
for (i=0; i<m; i++) {
  for (j=i; j<n; j++) {
    x[i][j] = f(i,j); // upper matrix triangle 
  }
}
\end{lstlisting}
where the start index of the inner loop depends on the outer loop,
the loops cannot be collapsed automatically
and the \openmp compiler will (most likely) complain.

The \texttt{schedule()} clause and all other clauses allowed for parallel for
loops can be used and will be interpreted as if the loop had been
transformed (collapsed, flattened) as outlined. According to the
\openmp specification, the sequential execution order of the iterations in
uncollapsed loops determines the order of the iterations of the
collapsed iteration range.

\subsection{Reductions}

Two frequently occurring loop patterns are the following:
\begin{itemize}
  \item
    Prefix sums
\begin{lstlisting}[style=SnippetStyle]
for (i=1; i<n; i++) { 
  a[i] = a[i-1]+a[i]; // the classic flow dependency
}
\end{lstlisting}
\item
  Reduction
\begin{lstlisting}[style=SnippetStyle]
sum = a[0];
for (i=1; i<n; i++) { 
  sum += a[i]; // data race on sum
}
\end{lstlisting}
\end{itemize}

Both of these loop patterns are loops of dependent
iterations. Therefore, they cannot be correctly parallelized with the
\openmp constructs for loop parallelization seen so far. The
computations expressed by the two loops (parallel
prefix sums\index{prefix sums} and simple\index{reduction} reductions)
require different, parallel algorithms in order
to be performed with any speed-up\index{speed-up} by a set of threads
working together (see \Sec~\ref{sec:recursiveprefix} and
\Sec~\ref{sec:iterativeprefix}).
Thus, either non-trivial compiler transformations of the
loop patterns into better, parallel algorithms
or the execution of preimplemented
algorithms at runtime is required to handle such loop patterns
well. Good parallel algorithms require the binary operator used in the
patterns (here: \texttt{+}) to be at least associative.

The reduction pattern loop can be handled, \ie,
parallelized efficiently, with \openmp by using the \texttt{reduction()}
clause with the parallel \texttt{for} directive. How well the
parallelization works will depend on the \openmp compiler and runtime
system among other things.

The \texttt{reduction()} clause is quite flexible. It takes an
associative, binary reduction operator and a list of reduction
variables on which a reduction with this operator is to be performed
in the loop. The order of the reductions follow the loop iteration
order, but it is not defined where brackets are put: The associativity
of the reduction operator is exploited to allow efficient reduction in
parallel.  Different reduction operators can be used in the same loop
by giving a reduction clause for each.

\begin{lstlisting}[style=SnippetStyle]
reduction(<reduction operator>:<reduction variables>)
\end{lstlisting}

The allowed operators in C are \texttt{+}, \texttt{-}, \texttt{*},
\texttt{\&}, \texttt{|}, \texttt{\^}, \texttt{\&\&}, \texttt{||}, as
well as special \texttt{min} and \texttt{max} operators. Minimum and
maximum operations are expressed either with special operators or
with code patterns like 
\begin{lstlisting}[style=SnippetStyle]
mi = (x<mi) ? x : mi;
if (x>ma) ma = x;
\end{lstlisting}
that will be recognized by the compiler as minimum or maximum computations,
respectively. Here \texttt{mi} and \texttt{ma} are global
variables declared by the programmer.

The reduction clause can also be used with parallel regions and the
sections work sharing construct\index{work sharing construct}.
In such cases, the reduction will be
performed in thread or section order.

Since \openmp $5.0$ the scan/prefix sum\index{prefix sums} pattern can be also
handled\index{scan}.  This is expressed by modifying a reduction in a
parallelizable loop to ``capture'' the reduced result for the current
iteration, \ie, the prefix sum for that iteration. Prefix sums reductions
are called \texttt{inscan} reductions and are expressed as
follows\index{prefix sums}:
\begin{lstlisting}[style=SnippetStyle]
reduction(inscan,<reduction operator>:<reduction variables>)
\end{lstlisting}

A reduction is performed with the reduction operator on the reduction
variables in the sequential loop order. The corresponding prefix sum
is captured with either a
\begin{lstlisting}[style=SnippetStyle]
#pragma omp scan exclusive(<reduction variables>)
<structured statement>  
\end{lstlisting}
directive for a structured statement (for the exclusive prefix sums), or a
\begin{lstlisting}[style=SnippetStyle]
#pragma omp scan inclusive(<reduction variables>) 
<structured statement>  
\end{lstlisting}
directive for a structured statement (for the inclusive prefix sums). For
the inclusive prefix sums computation\index{prefix sums},
the reduction variables can be
used in the block of the \texttt{scan} directive and will contain the
result of applying the reduction operator up to and including the
current iteration of the parallel loop; conversely, the result of the
reduction for the current iteration used before the \texttt{scan}
directive will be the exclusive prefix sum up to but not including the
current iteration. There can be only one \texttt{scan} directive in a
parallel loop, and in such a loop scheduling clauses cannot be
used. The following example shows how to compute inclusive and
exclusive prefix sums for an input array \texttt{a} with the result
stored in \texttt{b}.

\begin{lstlisting}[style=SnippetStyle]
x = 0;
#pragma omp parallel for reduction(inscan,+:x)
for (i=0; i<n; i++) {
  x += a[i]; // reduce
#pragma omp scan inclusive(x) 
  b[i] = x;  // and save the prefix (current value)
}

x = 10;
#pragma omp parallel for reduction(inscan,+:x)
for (i=0; i<n; i++) {
  b[i] = x;  // save the prefix
#pragma omp scan exclusive(x) 
  x += a[i]; // and reduce for next iteration
}
\end{lstlisting}

A natural application of reduction with the scan directive is array
compaction\index{array compaction} as discussed in
\Sec~\ref{sec:arraycompaction} and \Sec~\ref{sec:prefixsumloadbalance}.
The marked elements of an input array \texttt{a} have to be
compacted into a shorter output array \texttt{b}. To do this,
a running index for each marked element is needed. Exactly this
can be captured by the running, exclusive prefix sum.

\begin{lstlisting}[style=SnippetStyle]
int mark[n];
// mark[i] == 0/1 for input elements a[i]

int j = 0;
#pragma omp parallel for reduction(inscan,+:j)
for (i=0; i<n; i++) {
  if (mark[i]) b[j] = a[i];
#pragma omp scan exclusive(j) 
  j += mark[i];
}
\end{lstlisting}

\subsection{Work Sharing: Tasks and Task Graphs}
\label{sec:omptask}

Dynamically evolving work can often be expressed as
a Directed Acyclic task Graph (DAG)\index{DAG} as discussed
in \Sec~\ref{sec:taskgraphs}. The \openmp \emph{task} work
sharing constructs\index{work sharing construct}
make it possible to express such computations.

Consider the recursive Quicksort\index{Quicksort} algorithm as discussed in
\Sec~\ref{sec:taskgraphs}. In each Quicksort invocation, the input
array is partitioned into two roughly equally large parts, each of
which can be Quicksorted independently of the other. With several
threads activated like in an \openmp parallel region, each Quicksort
call can be wrapped as a \emph{task} to be executed by a thread that
may happen to be available and has no other work to do. In an \openmp
parallel region, any piece of code, like a procedure call (Quicksort),
a function call, or even a structured block, can be marked as a
\emph{task} with the following work sharing construct.

\begin{lstlisting}[style=SnippetStyle]
#pragma omp task [clauses]
<structured statement>
\end{lstlisting}

During execution, the code designated as a task will be prepared and
wrapped by the thread executing the \texttt{omp task} pragma (with compile
time help from the \openmp compiler). A created task may be
executed by any (other) thread in the parallel region, possibly at a later
time. All created tasks will be executed and completed at the latest at
a point in the code where
completion is requested. One such point of completion is the implicit
barrier at the end of the parallel region. All generated tasks can
also be completed by an explicit \texttt{\#pragma omp barrier}
construct.  In the terminology of \Sec~\ref{sec:taskgraphs}, the
tasks being wrapped by a thread are \emph{ready}, but they can have
dependencies on (private and shared) variables of the thread that
generated the task. Thus, in a correct \openmp task program,
the generating task shall not update any
variable that can be referred to by the generated tasks. If it does,
data races, which are illegal in \openmp, may arise.

A thread that generates one or more tasks may depend on these tasks to
be executed and completed before it can continue its computation. A
thread can have many \texttt{omp task} directives, for instance,
through recursive calls, and it may need results from these tasks in
order to continue.  Waiting for completion of the tasks directly
generated by the thread can be enforced by the \texttt{taskwait}
construct.

\begin{lstlisting}[style=SnippetStyle]
#pragma omp taskwait [clauses]
\end{lstlisting}

The only allowed clauses are \texttt{depend()} clauses. They express
input and output dependencies on other tasks in the same lexical scope.
Dependencies are not treated in this
lecture. It is also possible to generate tasks in such a way that waiting
is done not only for the directly generated (children) tasks, but also for all
tasks descending from these tasks (for instance, from recursive calls). This is
done by generating the tasks in a \texttt{taskgroup} region which we shall
not describe further in these lectures.

Here is a standard example of an algorithm that can be parallelized
with tasks.  The problem is to count the number of occurrences of some
value $x$ in an unordered array \texttt{a} of $n$ elements. The
algorithm is recursive. If $n=1$ the problem is trivial: There is an
occurrence if $\texttt{a[0]}=x$, otherwise not. If $n>0$ the array is
split into two halves, the number of occurrences in both halves
counted and added together. This idea can obviously be formulated as a
computation on a task graph and be implemented in \openmp as shown
below.

\begin{lstlisting}[style=SnippetStyle]
int occurs(int x, int a[], int n)
{
  if (n==1) {
    return (a[0]==x) ? 1 : 0;
  } else {
    int s0, s1; // private variables for executing thread 
#pragma omp task shared(s0,a)
    s0 = occurs(x,a,n/2);
#pragma omp task shared(s1,a)
    s1 = occurs(x,a+n/2,n-n/2);
#pragma omp taskwait

    return s0+s1;
  }
}

int main(...)
{
  int a[n];
  int x;
  int s;
  
#pragma omp parallel shared(x) shared(s) shared(a)
  {
#pragma omp single
    s = occurs(x,a,n);
  }
}
\end{lstlisting}

Each recursive call is marked as an \texttt{omp task}. In order
to sum the number of occurrences for each half of the array, an
explicit \texttt{omp taskwait} is necessary. The computed
results (and the array pointer) must be classified as \texttt{shared}.
This is crucial, since each task can be executed by any 
thread of the parallel region,
in particular by a thread that is different from the one that
allocated the variable. The thread that executes a task must be able to
update the variable that was possibly allocated by another thread. This
is possible only if the variable is shared among the two
different threads.

In the main program, the threads are activated by the \texttt{parallel}
region. One of the threads, arbitrarily chosen by the \texttt{single}
work sharing construct\index{work sharing construct}, shall initiate the search.
If \texttt{single} (or \texttt{master}) is
forgotten, all threads will start performing the search operation, which
leads to superfluous work (by a factor of the number of threads $p$) and
(in this example) definitely to data races.

In the example, the recursion is done all the way down to the bottom
$n=1$ condition. This is rarely a good choice, neither sequentially,
nor in parallel. Generally, finding a good cut-off for recursive algorithms
is a difficult problem, which we will not solve here. In order to
prevent too many, too small tasks, a task can be
designated as \texttt{final}, meaning that the task will generate no
additional tasks. \texttt{untied} tasks are tasks that may be suspended
and are allowed to be resumed by any other thread.
Together with a conditional \texttt{if}-clause
this can possibly be used as a substitute for an explicit cut-off
programmed into the recursive task.

The \texttt{omp task} work sharing construct\index{work sharing construct}
offers further possibilities for controlling when a task will be ready for
execution. Input-output dependencies can be expressed with
\texttt{depend()} clauses. With the \texttt{priority()} clause,
tasks can be prioritized as hints to the \openmp runtime system in which
order the tasks should preferably be executed.

Here is a(n almost) complete parallelization of the sequential
Quicksort\index{Quicksort} algorithm with \openmp tasks.

\begin{lstlisting}[style=SnippetStyle]
void Quicksort(int a[], int n)
{
  int i, j;
  int aa;

  if (n<2) return; // recursion all the way down

  // partition
  int pivot = a[0]; // choose an element (non-randomly...)
  i = 0; j = n;
  while (1) {
    while (++i<j&&a[i]<pivot); 
    while (a[--j]>pivot);
    if (i>=j) break;
    aa = a[i]; a[i] = a[j]; a[j] = aa;
  }
  // swap pivot
  aa = a[0]; a[0] = a[j]; a[j] = aa;

#pragma omp task shared(a) 
  Quicksort(a,j);
#pragma omp task shared(a) 
  Quicksort(a+j+1,n-j-1);
  // #pragma omp taskwait - not needed
}

int main(int argc, char *argv[])
{
  ...
  
  start = omp_get_wtime();
#pragma omp parallel
  {
#pragma omp single nowait
    Quicksort(a,n);
    //#pragma omp taskwait
  }
  stop = omp_get_wtime();
}
\end{lstlisting}

Assuming that a good pivot can be selected that divides the input array
into two approximately equally large halves, the recurrence relations for
work and parallel time complexity are as follows.

\begin{eqnarray*}
  W(n) & = & 2W(n/2) + O(n) \\
  W(1) & = & O(1) \\
  T(n) & = & T(n/2) + O(n) \\
  T(1) & = & O(1)
\end{eqnarray*}

Using the Master Theorem~\ref{thm:master}, the solutions are
$W(n) = O(n\log n)$ and $T(n) = O(n)$, respectively (Case 2 and Case 1).
This tells us that the parallelism of this implementation is low,
in $(n\log n/n) = O(\log n)$ and that only up to that number of
processors can be employed with linear speed-up\index{speed-up!linear}.
The culprit is obviously the linear-time partition step which we have
to pay for in full at the first recursive call.
In order to improve, a parallel algorithm, for instance by
using prefix sums and compaction as discussed, would have to be employed.
Unfortunately, in this lecture, we will have no means for using task and
loop parallelism (with scan-reduction) together. We could use our own
parallel prefix sums algorithms\index{prefix sums} parallelized
with \texttt{omp taskloop} (see \Sec~\ref{sec:simdloops}).

\subsection{Mutual Exclusion Constructs and Atomic Operations}

In order to prevent data races in parallel regions, 
\openmp\index{OpenMP} provides direct support
for \impi{mutual exclusion} through named critical sections.

\begin{lstlisting}[style=SnippetStyle]
#pragma omp critical [(name)]
<structured statement>
\end{lstlisting}

Threads that encounter a (named) \impi{critical section} will all
execute the code in the critical section, but under mutual exclusion;
that is, at most one thread at a time can execute the code for its
critical section. In a critical section, one or more shared variables
can be updated, shared variables can be read, and the thread can make
decisions based on the read values. Since no other threads will be
executing code for the named critical section at the same time, such
updates are technically not data races, and it is possible to ensure a
definite outcome of the parallel execution of the threads. The order
in which the threads enter the critical section is
nondeterministic. It depends on the relative speeds of the threads, on
when they encounter the critical section, on how many threads
arrive ``at the same time'' and compete for the critical section,
and on how the mutual exclusion (locking)
algorithms of the runtime system resolve the conflicts.
Thus, relying on some specific
behavior of the critical section construct will lead to incorrect
programs. A concrete case is the implementation of reduction like
operations: Implementations with critical sections will be correct
only when the reduction operators being used are commutative.

Critical sections are always (relatively) expensive constructs and will,
therefore, have an impact on the overall performance of a parallel
program. In particular, they lead to serialization between the
threads and should, therefore, be used sparingly and with care.

Here is a technically incorrect, but defensible use of mutual
exclusion to find the maximum of values computed by threads in a
parallel region which can, dependent on the input, alleviate
some serialization penalties.

\begin{lstlisting}[style=SnippetStyle]
int max = 0;
#pragma omp parallel shared(max)
{
  int x;
  x = ...;

  if (x>max) {
#pragma omp critical
    if (x>max) max = x;
  }
}
\end{lstlisting}

With this kind of speculative, \impi{optimistic locking} the critical section
is entered only by threads whose \texttt{x} value could potentially be
the maximum.  Under mutual exclusion in the critical section where the
thread is alone, it must then be rechecked that the \texttt{x} value
is still a candidate in which case the shared \texttt{max} variable is
updated.  The recheck is necessary since another thread could have
updated \texttt{max} before the thread enters the critical section
such that \texttt{x} is after all not the largest value seen ``so
far'' (try this).  The pattern is also known as
test-and-test-and-set (TATAS)\index{test-and-test-and-set}.
The advantage is that values that definitely cannot
be the maximum do not cause serialization. Technically, such code is not
data race free\index{data race free};
but as argued it leads to correct results: not all
race conditions\index{race condition} are harmful.

In case the update and work to be done in a critical section has
a particularly simple form, it may be possible to use a hardware assisted
atomic operation instead. \openmp provides access to certain types of
\emph{atomic operations}\index{atomic operation} by the following construct.

\begin{lstlisting}[style=SnippetStyle]
#pragma omp atomic [read|write|update|capture|compare]
<atomic statement>
\end{lstlisting}

The atomic operations ensure that operations by different threads are
performed in order without interruption (atomically). Atomic
operations also lead to serialization\index{atomic operation},
but since they are performed
directly by the hardware they can be considerably less costly than
explicit critical sections.  Atomic \texttt{update} and
\texttt{capture} operations allow the use of Fetch-And-Add (FAA) type
atomic operations\index{Fetch-And-Add (FAA)}.
For the \texttt{atomic update} construct, the atomic statement is
restricted to be of the form \texttt{x++;}, \texttt{++x;},
\verb|x--;|, \verb|--x;|, and \texttt{x = x binop expr;} and
similar syntactically equivalent forms that will result in a
simple arithmetic-logical update to \texttt{x}.
For the \texttt{atomic capture} construct, the atomic statement
must have the form \texttt{y = x++;} or a similar syntactically
equivalent form that will read the previous content of \texttt{x} and
cause an update. Here \texttt{x} and \texttt{y} are variables where
the C operators apply and \texttt{binop} one of the word wise C
operations \texttt{+}, \texttt{*}, \verb|-|, \texttt{/},
\texttt{\&}, \texttt{\^}, \texttt{|}, \texttt{<<} or \texttt{>>}.

A frequent, convenient use case for captured
atomic operations\index{atomic operation} is the
computation of unique indices for threads that execute a parallel loop.

\begin{lstlisting}[style=SnippetStyle]
int ixc = 0; // shared index counter
#pragma omp parallel for shared(ixc)
for (i=0; i<n; i++) {
  int ix; // next index

  if (need) { // index needed
#pragma omp atomic capture
    ix = ixc++;
  }
}
\end{lstlisting}

Each time a thread performing some iteration of the loop needs an
index, it performs the atomic Fetch-And-Increment
(FAI)\index{Fetch-And-Increment (FAI)} to retrieve the current
value of the shared counter and atomically increments this to the next
index.

After \openmp $5.0$, also the (hardware)
Compare-And-Swap (CAS)\index{Compare-And-Swap (CAS)} operation
is supported as an \openmp atomic operation\index{atomic operation}.
For this, \texttt{compare}
or \texttt{capture compare} clauses have to be given. A conditional
statement like \texttt{if (x==e) x = d;} is executed atomically, preferably by
a corresponding hardware atomic operation. This makes it possible to
implement certain lock- and wait-free algorithms directly in
\openmp, but this
is beyond this lecture, see for instance~\cite{HerlihyShavit12}.
The atomic operations defined for C in the \texttt{stdatomic.h} library
can also be used with \openmp.

\subsection{Locks}

Sometimes named critical sections are insufficient, for instance, for
implementing list-based algorithms with hands-over locking, where a
lock (critical section) is needed for each element of the list. For
that reason, \openmp provides locks that can be allocated dynamically
similarly to the \pthreads locks. Locking and unlocking a lock is called
set and unset in \openmp\index{OpenMP}.

\begin{lstlisting}[style=SnippetStyle]
void omp_init_lock(omp_lock_t *lock);
void omp_init_nest_lock(omp_nest_lock_t *lock);
void omp_destroy_lock(omp_lock_t *lock);
void omp_destroy_nest_lock(omp_nest_lock_t *lock);
void omp_set_lock(omp_lock_t *lock);
void omp_set_nest_lock(omp_nest_lock_t *lock);
void omp_unset_lock(omp_lock_t *lock);
void omp_unset_nest_lock(omp_nest_lock_t *lock);
int omp_test_lock(omp_lock_t * lock);
int omp_test_nest_lock(omp_nest_lock_t * lock);
\end{lstlisting}

Locks in \openmp do not have condition variables\index{condition variable}.
\openmp provides nested (recursive\index{Lock!recursive})
locks\index{Lock!nested}. Nested (recursive) locks in \openmp must be unlocked
as many times as locked by the threads having successfully acquired the lock.
Locks (now) also have a
try-lock\index{Lock!try-lock} operation called \texttt{omp\_test\_lock}.
\openmp does not provide
readers-and-writer locks\index{Lock!readers-writer}. Thus,
\openmp is not intended for involved programming with locks the same
way that \pthreads and other thread interfaces are.

\subsection{Special Loops}
\label{sec:simdloops}

Loops of independent iterations where the operation(s) per iteration
have a particularly simple form, for instance, expressing a simple
$n$-element vector addition like:

\begin{lstlisting}[style=SnippetStyle]
for (i=0; i<n; i++)
  c[i] = a[i]+b[i];
\end{lstlisting}
can benefit from hardware capabilities for operating on small vectors
with a single instruction. Modern processors typically have such
capabilities in the form of extended vector instructions for operating
on $2, 4, 8, 16$ float or double elements with one single instruction (SSE,
AVX instructions).
Such instructions are called SIMD instructions\index{SIMD}. With
\openmp\index{OpenMP},
the compiler can be instructed to try to use SIMD instructions with
the following three loop parallelization constructs.
A sequential loop to be executed by one thread with SIMD instructions is
designated with the \texttt{simd} pragma.

\begin{lstlisting}[style=SnippetStyle]
#pragma omp simd [clauses]
for (<canonical form loop range>)
<loop body>
\end{lstlisting}

A loop within a parallel region to be shared among the threads of the
region with each chunk executed with SIMD\index{SIMD}
instructions is designated with the \texttt{for simd} pragma.

\begin{lstlisting}[style=SnippetStyle]
#pragma omp for simd [clauses]
for (<canonical form loop range>)
<loop body>
\end{lstlisting}

A parallel region with SIMD loop sharing can be
written with a shorthand, composite construct.
\begin{lstlisting}[style=SnippetStyle]
#pragma omp parallel for simd [clauses]
for (<canonical form loop range>)
<loop body>
\end{lstlisting}

For the compiler to be able to exploit SIMD\index{SIMD} instructions, often
certain preconditions must be fulfilled, for instance, on alignment
and size of loop ranges. Hints and assertions that this is the
case can be expressed by additional clauses. Such hints and conditions
are beyond these \openmp lectures.

A different way of parallelizing a loop of independent iterations is
to (recursively) break the iteration range into smaller ranges that
are executed as tasks. While such a loop parallelization can easily be
done by hand (give a recipe for this), \openmp provides a construct
for automatically performing the transformation into and execution of
a loop as tasks.

\begin{lstlisting}[style=SnippetStyle]
#pragma omp taskloop [clauses]
for (<canonical form loop range>)
<loop body>
\end{lstlisting}

A taskloop is initiated by a single thread in a parallel region.
A taskloop does not take a \texttt{schedule()} clause (scheduling is
done by the task scheduling algorithm); instead, the size
of the parts of the iteration range can be controlled by a
\texttt{grainsize()} clause. Alternatively, the number of tasks across
which the loop is split can be set with the \texttt{num\_tasks()}
clause. Nested loops can be collapsed with the \texttt{collapse()} clause.
Also, reductions can be performed over task loops by adding a
\texttt{reduction()} clause. Prefix sums \texttt{inscan}-reductions are
not allowed, though (think about reasons why).

\subsection{Parallelizing Loops with Hopeless Dependencies}

Instead of completely giving up on (not parallelizing) loops with
dependency patterns that cannot be handled by reductions, scans or any other
of the means that we have seen, \openmp\index{OpenMP} makes it possible
as a last resort to mark a part of the loop code as having to be
executed in the sequential iteration order. This is done by 
the \texttt{ordered} parallel for loop clause and by marking the
section of code that has to be done in the sequential iteration order
with the corresponding \openmp construct.
\begin{lstlisting}[style=SnippetStyle]
#pragma omp ordered
<structured statement>
\end{lstlisting}

There can be only one ordered statement block of code in a parallel loop.
While the ordered blocks are performed in sequential order, other code
for the iterations could possibly be performed in parallel. However,
the construct usually brings only overhead. Its use is not recommended.

\subsection{Example: Parallelizing a Sequential Algorithm with Dependencies}
\label{sec:primesieve}

The prime sieve of Erathostenes is an amazing recipe for listing all prime
numbers in increasing order starting at $2$ (the first prime) up to some given
$n$. The idea is this.
Write all (optimization: odd) numbers in a list. Start going
through the list. The first number ($2$) is a prime, write this down, and
cross out all multiples of this prime on the list.
Go to the next number still on
the list ($3$). This must be a prime, otherwise it would have been crossed
out by being a multiple of some earlier found prime.
Write it down, and cross out all multiples of
this prime. Continue like this ($5, 7, 11, \ldots$) until all numbers
have been considered.

The function \texttt{primesieve()} below implements Erathostenes prime
sieve with a few clever, well-known optimizations worth pondering.

\begin{lstlisting}[style=SnippetStyle]
int primesieve(int n, int primes[])
{
  int i, j, k;
  unsigned char *mark;
  mark = (unsigned char*)malloc(n*sizeof(unsigned char));
  
  for (i=2; i<n; i++) mark[i] = 1; // possibly prime
  k = 0;
  for (i=2; i*i<n; i++) {
    if (mark[i]) {
      primes[k++] = i; // list prime
      for (j=i*i; j<n; j+=i) mark[j] = 0; // j not prime
    }
  }
  for (; i<n; i++) {
    if (mark[i]) primes[k++] = i; // list remaining primes
  }
  free(mark);
  
  return k;
}
\end{lstlisting}

First, if some $i$ is composite, $i=pq$, then either $p\leq\sqrt{i}$ or
$q\leq\sqrt{i}$. Therefore, it suffices to examine the list only up to
$\sqrt{n}$. At iteration $i$, multiples of all $j<i$ have been crossed
out, therefore if $i$ is found to be prime and is still on the list
(\texttt{mark[i]} is \textbf{true}), then $2i, 3i,
\ldots (i-1)i$ have already been crossed out. It is, therefore, correct
to start the crossing out from index $i^2$.
The algorithm performs $O((n-i)/i)$ operations for the crossing out for
each found prime $i, 2\leq i<n$. The complexity is thus bounded by
$O(\sum_{i=2, i\mathrm{prime}}^{n}n/i)$ which is in $O(n\log\log n)$ by a
result from number theory, see for instance~\cite[Theorem 427]{HardyWright79}.

\begin{proposition}
  The prime sieve algorithm lists all primes in the range from $2$ to $n$
  in $O(n\log\log n)$ operations.
\end{proposition}

The idea and the program above have obvious room for parallelization
and some obstacles in the form of dependencies.  The inner loop for
crossing out multiples is a standard loop of independent iterations
and straightforwardly parallelizable with \openmp\index{OpenMP}. It is performed
$\sqrt{n}$ times. However, the final loop for compacting the marked
primes has our now well-known dependency on \texttt{k}.  Still, we
might just give up and mark these iterations as to be performed in the
sequential order.

\begin{lstlisting}[style=SnippetStyle]
#pragma omp parallel for private(i) schedule(static,1024)
for (i=2; i<n; i++) mark[i] = 1; // possibly prime

k = 0;
for (i=2; i*i<n; i++) {
  if (mark[i]) {
    primes[k++] = i;
#pragma omp parallel for private(j) schedule(static)
    for (j=i*i; j<n; j+=i) mark[j] = 0; // j not prime
  }
}
j = i;
#pragma omp parallel for ordered
for (i=j; i<n; i++) {
  if (mark[i]) 
#pragma omp ordered
  primes[k++] = i;
}
\end{lstlisting}

Performance will likely not be any better than just performing the loop
sequentially: there is no other work in the loop that could benefit
from being executed in parallel.
It is instructive to try it out. However, by now, 
the array compaction pattern is familiar
(see \Sec~\ref{sec:arraycompaction})\index{array compaction}.
We can parallelize either manually by using any one of our
algorithms for computing the prefix sums\index{prefix sums} of the \texttt{mark}
array or by using the \openmp\index{OpenMP} construct for
capturing the prefix sums in a loop reduction.

\begin{lstlisting}[style=SnippetStyle]
#pragma omp parallel for private(i) schedule(static,1024)
for (i=2; i<n; i++) mark[i] = 1; // possibly prime

k = 0;
for (i=2; i*i<n; i++) {
  if (mark[i]) {
    primes[k++] = i;
#pragma omp parallel for private(j) schedule(static)
    for (j=i*i; j<n; j+=i) mark[j] = 0; // j not prime
  }
}
j = i;
#pragma omp parallel for reduction(inscan,+:k)
for (i=j; i<n; i++) {
  if (mark[i]) primes[k] = i;
#pragma omp scan exclusive(k)
  if (mark[i]) k = k+1;
}
\end{lstlisting}

Again, it is instructive to compare this parallelization against
the version with the \texttt{omp ordered} clause and, of
course, against the sequential implementation to estimate the achievable
(relative) speed-up\index{speed-up}.
The parallel time complexity\index{parallel time complexity}
of the parallel solution
with prefix sums compaction\index{prefix sums} is clearly $O(\sqrt{n})$ (why?).

\subsection{Cilk: A Task Parallel C Extension}
\label{sec:cilk}

\impi{Cilk} (alluding to ``silk'', C and ``ilk'') is (was) a C language
extension for task parallel programming originally developed at MIT in
the mid-$90$ties.
It focuses on provably efficient execution by the runtime system
of dynamically generated acyclic task graphs~\cite{AroraBlumofePlaxton01,BlumofeLeiserson99,BlumofeJoergKuszmaulLeisersonRandallZhou96,Leiserson10}.
\cilk was supported by \gcc and other compilers for a number of years, but is
unfortunately being deprecated since 2018. However, there is
recently an open version called OpenCilk,
see \url{www.opencilk.org}~\cite{Leiserson10,SchardlLee23}.
The \openmp task model has surely been inspired by \cilk, and
\cilk programs can easily be reimplemented with \openmp. \cilk
provides three new keywords to C.

\begin{lstlisting}[style=SnippetStyle]
cilk_spawn <function call>
cilk_sync
cilk_for (<canonical form iteration space>) <loop body>
\end{lstlisting}

Generation of tasks is called \impi{spawning} in \cilk, \index{Cilk} and the
\cilkspawn keyword indicates to compiler and run-time system
that a function or procedure call can be executed as a task, thus,
concurrently with other tasks on processor-cores\index{processor-core}
that may be available. This corresponds to the
\texttt{omp task} construct, which is, however, more general: With
\openmp\index{OpenMP}
an arbitrary structured statement can be wrapped as a task.
Directly spawned children tasks will be waited for at the end of the statement
block doing the task spawns. If waiting for the directly spawned
tasks to complete is required (as in the search program discussed in
\Sec~\ref{sec:omptask}) the keyword \cilksync can be
used, much in the same way as \texttt{omp taskwait}. Finally, the
\cilkfor keyword is used as a shorthand for parallelizing
loops as collections of tasks, much in the same way as
\texttt{omp taskloop}.

\cilk\index{Cilk} has no explicit concept of threads. The \cilkspawn construct
indicates that a function or procedure call \emph{may} be executed
concurrently with the code following the spawn (called the
\impi{continuation}); but not \emph{how} or by which processor-core or
thread this is done. The \cilksync construct introduces a dependency
point where the execution must wait for the spawned calls to
complete. By removing the \cilkspawn and \cilksync keywords and
replacing \cilkfor with a C \texttt{for} a \cilk program should be a
correct, sequential C program. This is sometimes helpful for
understanding the \cilk semantics (a similar observation, \btw, should
hold for \openmp\index{OpenMP} programs).
The \cilk\index{Cilk} runtime system executes
spawned threads with a clever \impi{work-stealing} algorithm. In the
multi-threaded runtime system, threads execute spawned tasks from a
local task-queue and, when running out of local tasks, \emph{steal}
tasks from other runtime threads. They continue this until there are
no more tasks to be executed. The \cilk constructs give rise to highly
structured, acyclic tasks graphs, so-called
\impi{(fully) strict computations}. For (fully) strict computations
with $T_1(n)$ total work and $\tinf(n)$ work on the longest
path, it can be shown that
the computation can be completed in $O(T_1(n)/p+\tinf(n))$ expected
time steps by the work-stealing runtime system on a dedicated parallel
shared memory computing system with $p$ processors running the $p$
worker threads~\cite{AroraBlumofePlaxton01}. This is within a constant
factor of optimal. In this sense, \cilk comes with a provably
efficient runtime system. The \cilk runtime work-stealing algorithm
implements a randomized, greedy scheduling
strategy\index{greedy scheduling}.
A similar work-stealing algorithm is most likely also the
basis for the \openmp runtime system for executing \openmp tasks.
This, however, is deliberately not specified by the \openmp standard.

As seen with the \openmp\index{OpenMP} examples,
task parallel programs often follow
from recursive, divide-and-conquer algorithms where the recursive calls
are independent of each other. This was the case with the search
algorithm and the Quicksort\index{Quicksort} example.
Also, sorting by merging\index{merging} can be
expressed in this way. Runtime bounds for recursive algorithms, both
with regard to the total number of work, and the work of a single path
of recursive calls down to the base case, can often be expressed as
recurrence relations. In many cases, the solutions follow directly from
the Master Theorem~\ref{thm:master}\index{Master Theorem};
if not, the recurrence must be solved (by induction) by hand.

For standard implementations of Quicksort\index{Quicksort} and
Merge sort such analyses reveal that $T_1(n)=O(n\log n)$ and
$\tinf(n) = O(n)$. The
parallelism is modest in $O(\log n)$, meaning that
linear speed-up\index{speed-up!linear} can be
achieved only for a modest range of processor-cores\index{processor-core}
and threads. In the two cases, the
bottlenecks were the sequential partitioning step and
the sequential merge operation. To achieve more parallelism, parallel
algorithms for the bottleneck operations must be found.

In \Sec~\ref{sec:merging}, several parallel approaches were given
for merging\index{merging} in parallel in $O(n/p+\log n)$ time steps.
A drawback of
these algorithms for implementation as task parallel algorithms (with
no explicit notion of threads) is that the number of processors $p$ is
used and must be known. The final algorithm in this part of the
lecture notes is a different, recursive divide-and-conquer
merging algorithm that addresses these issues and
can readily be implemented with \cilk\index{Cilk} and \openmp\index{OpenMP}
tasks.

\begin{lstlisting}[style=SnippetStyle]
void par_merge(int A[], int n, int B[], int m, int C[])
{
  if (n<m) { // for the bounds, it must hold that n>=m
    int k;
    int *X;
    k = n; n = m; m = k;
    X = A; A = B; B = X;
  }
  if (m==0) {
    par_copy(C,A,n); // copy in parallel
    return;
  }

  int r = n/2; // it holds that n>=m
  int s = rank(A[r],B,m); // determine rank of A[m] in B
  C[r+s] = A[r];
  cilk_spawn par_merge(A,r,B,s,C);
  cilk_spawn par_merge(A+r+1,n-r-1,B+s,m-s,C+r+s+1);
  cilk_sync; // not necessary, implicit in Cilk
}
\end{lstlisting}

The algorithm ranks the middle element of the larger of the arrays in the
other array. It computes
$\rank(\texttt{A}[\floor{n/2}],\texttt{B})$ by binary search (see
\Sec~\ref{sec:ranking}), which gives two pairs of sufficiently smaller
subarrays that can be (recursively) merged together.
In case a pair has an array
without any elements, a (task) parallel copy operation is used to
copy the other array to the output array. For the parallel recursion
to terminate (and to avoid redundant computation on an element whose
position in the output is now known),
the element $\texttt{A}[\floor{n/2}]$, which is larger
than or equal to all previous elements in \texttt{A} and larger than
or equal to $\texttt{B}[s]$ and all previous elements in \texttt{B},
is written immediately to its correct position in the output array.

The recurrences, assuming for the two arrays
that $n=m$ with total input size $2n$, are as follows.
\begin{itemize}
\item
  Work $T_1(n)$:
  \begin{eqnarray*}
  T_1(2n) & = & T_1(n/2+\alpha n) + T_1(n/2+(1-\alpha)n)+O(\log n)
  \end{eqnarray*}
  for some $\alpha, 0\leq \alpha\leq 1$ that can vary throughout the
  evaluation of the recurrence and corresponds to where the rank in the smaller
  array is found. The $n/2$ term is the split index of the larger array.
\item
  Time $\tinf(n)$:
  \begin{eqnarray*}
    \tinf(2n) & \leq & \tinf(3/2n)+O(\log n)
  \end{eqnarray*}
  since the larger of the input arrays is always halved and in the worst case
  merged (recursively) with the other array of (at most) $n$ elements.
\end{itemize}
The second recurrence can be solved by the
Master Theorem\index{Master Theorem}
(Case 2 with $a=1, b=\frac{2n}{3/2n}=4/3, d=0, e=1$) to give
$\tinf(n) = O(\log^2 n)$,
whereas the first requires a direct induction proof to give
$T_1(n)=O(n)$. To see this, conjecture the solution to be
\begin{eqnarray*}
  T_1(n) & \leq & Cn-c\log_2 n
\end{eqnarray*}
for constants $C$ and $c$ where the
time to rank an element in a sequence of length $n$ is at most $c\log n$.
Using this as induction hypothesis, the recurrence relation now gives
\begin{eqnarray*}
  T_1(2n) & \leq & T_1(n/2+\alpha n) + T_1(n/2+(1-\alpha)n) + c\log_2 n \\
  & = & C(n/2+\alpha n)-c\log_2(n/2+\alpha n) + \\
  & & C(n/2+(1-\alpha)n)-c\log_2(n/2+(1-\alpha) n) + c\log_2 n \\
  & = & C2n + c\log_2(n/2+\alpha n) + c\log_2(n/2+(1-\alpha) n) + c\log_2 n 
  \quad .
\end{eqnarray*}
Assuming the worst case in both logarithmic terms; that is, $\alpha=1$ and
$\alpha=0$, respectively, gives
\begin{eqnarray*}
  & & C2n + c\log_2(n/2+\alpha n) + c\log_2(n/2+(1-\alpha) n) + c\log_2 n \\
  & = & C2n-2c\log_2(3/2 n) + c\log_2 n \\
  & = & C2n - 2c\log_2(3/2) - 2c\log_2 n + c\log_2 n \\
  & = & C2n - 2c\log_2 3 + 2c - c\log_2 n \\
  & = & C2n - 2c\log_2 3 + 2c - c(\log_2 2n - 1) \\
  & = & C2n - 2c\log_2 3 + c - c\log_2 2n \\
  & = & C2n - c(2\log_2 3 - 1) - c\log_2 2n \\
  & \leq & C2n - c\log_2 2n \\
\end{eqnarray*}
using $\log_2 2n = \log_2 2+\log_2 n=1+\log_2 n$ and
$2\log_2 3 -1>0$ which then
establishes the induction hypothesis.

We summarize in the following proposition.
\begin{proposition}
  The merging problem can be solved work-optimally with $T_1(n) = O(n)$ and
  $\tinf(n)=O(\log^2 n)$.
\end{proposition}

By computing instead the co-ranks\index{co-rank}
in \texttt{A} and \texttt{B} for index for $\floor{(n+m)/2}$
as described in \Sec~\ref{sec:corankmerge}
we could have found the exact indices of the parts of
\texttt{A} and \texttt{B} to merge to get exactly the two halves of
the resulting \texttt{C} array. This would have saved us from the
(nevertheless illustrative) induction proof and would give a possibly faster
algorithm. It is a good exercise to implement and compare the two
possibilities.

The \cilk merge algorithm can now be plugged into another recursive,
task parallel algorithm for sorting by merging\index{merging}.
The algorithm works
top-down by first merge sorting the two halves of the input array, and
then merging together the two sorted halves.

\begin{lstlisting}[style=SnippetStyle]
void par_mergesort(int A[], int n)
{
  if (n==1) return;
  
  cilk_spawn par_mergesort(A,n/2);
  cilk_spawn par_mergesort(A+n/2,n-n/2);
  cilk_sync; // necessary
  // allocate temporary array B
  par_merge(A,n/2,A+n/2,n-n/2,B);
  par_copy(A,B,n);
  // free B again (possibly inefficient)
}
\end{lstlisting}

The algorithm is not an \impi{in-place} algorithm since the
additional $n$-element work array \texttt{B} is needed for the
merging\index{merging}.
which also necessitates an explicit copy back operation.
For the complexity we have
\begin{eqnarray*}
  T_1(n) & = & 2T_1(n/2) + O(n)
\end{eqnarray*}
for the work and
\begin{eqnarray*}
  \tinf(n) & = & \tinf(n/2)+O(\log^2 n)
\end{eqnarray*}
for the parallel time complexity.
Both recurrences can be solved by Case 2 of the
Master Theorem~\ref{thm:master}\index{Master Theorem} to give
\begin{eqnarray*}
  T_1(n) & = & O(n\log n) \\
  \tinf(n) & = & O(\log^3 n) \quad . 
\end{eqnarray*}

We summarize in the following proposition.
\begin{proposition}
  An $n$-element array can be sorted work-optimally by Merge sort in
  $O(\frac{n\log n}{p}+\log^3 n)$ parallel time.
\end{proposition}

\section{Exercises}

\begin{enumerate}
\item
  Does the PRAM model presuppose a cache-memory?
\item
  \label{exe:sixmmm}
  Implement the six sequential, loop based matrix--matrix multiplication
  algorithms corresponding to the six possible permutations of
  the order of the loops. Write each variant as a function with
  function prototype
\begin{lstlisting}[style=SnippetStyle]
void mmm(const base_t* A, const base_t* B, base_t* C,
         const int m, const int l, const int n)
\end{lstlisting}
  for $m\times l$ matrix \texttt{A},
  for $l\times n$ matrix \texttt{B} and
  for $m\times n$ matrix \texttt{C} represented as one-dimensional arrays
  with a user defined base type
  \texttt{typedef int base\_t;} (where \texttt{int} can be replaced by
  any other arithmetic C datatype).
  Time the six variants for matrices of total size
  around $m\approx n \approx l\approx 1200$ elements with and without compiler
  optimization (\texttt{-O3} and other).
  Investigate the effects of different base datatypes, \eg,
  \texttt{char}, \texttt{int}, \texttt{float} and \texttt{double}.
  Allocate the matrices
  in linear storage with \texttt{malloc(m*n*sizeof(base\_t))}
  and typecast into a pointer to rows of the given number of elements.
\item
  Give and analyze a recursive algorithm for adding two $m\times n$ matrices.
\item
  \label{exe:recmmm}
  Implement the recursive, divide-and conquer matrix--matrix multiplication
  algorithm for multiplying two $m\times l$ and $l\times n$ matrices into
  an $m\times n$-matrix. Represent submatrices as blocks inside larger
  $m\times n$-matrices defined by row start index and number of rows $i0,m0$
  and column start index and number of columns $j0,n0$. Use, for
  instance, a C structure as shown below together with an
  iterative submatrix--submatrix multiplication procedure to handle
  the base cases.
  
  \begin{lstlisting}[style=SnippetStyle]
typedef struct {
  int m, n;   // rows and columns of matrix
  int i0, m0; // block start and size
  int j0, n0;
  base_t *M;  // the matrix elements, row after row
} matblk;

void mmmite(matblk A, matblk B, matblk C)
{
  // the matrices
  base_t (*a)[A.n] = (base_t(*)[A.n])A.M; // the matrix
  base_t (*b)[B.n] = (base_t(*)[B.n])B.M;
  base_t (*c)[C.n] = (base_t(*)[C.n])C.M;
  int i, j, k;

  for (i=0; i<A.m0; i++) {
    for (j=0; j<B.n0; j++) {
      c[C.i0+i][C.j0+j] = 0;
    }
  }
  for (i=0; i<A.m0; i++) {
    for (k=0; k<A.n0; k++) {
      for (j=0; j<B.n0; j++) {
        c[C.i0+i][C.j0+j] +=
          a[A.i0+i][A.j0+k]*b[B.i0+k][B.j0+j];
      }
    }
  }
}
  \end{lstlisting}

  The recursive algorithm can then be written as outlined below (with only
  two of the $8$ submatrix multiplications shown) where the matrices
  are cut roughly in half along the two dimensions. Complete the code
  and make sure it works (correctly) for any non-negative $m,l,n$.
  
  \begin{lstlisting}[style=SnippetStyle]
void mmmrec(matblk A, matblk B, matblk C)
{
  if (A.m0<=CUT||A.n0<=CUT||B.n0<=CUT) {
    mmmite(A,B,C);
  } else {
    matblk A00;
    matblk A01;
    matblk A10;
    matblk A11;

    matblk B00;
    matblk B01;
    matblk B10;
    matblk B11;

    // 4 intermediate results in two matrices
    matblk C0;
    matblk C1;

    C0.m = A.m0;
    C0.n = B.n0;
    C0.M = (base_t*)malloc(C0.m*C0.n*sizeof(base_t));
    base_t (*c0)[C0.n] = (base_t(*)[C0.n])C0.M;
    
    C1.m = C0.m;
    C1.n = C0.n;
    C1.M = (base_t*)malloc(C1.m*C1.n*sizeof(base_t));
    base_t (*c1)[C1.n] = (base_t(*)[C1.n])C1.M;
    
    A00.m = A.m;
    A00.n = A.n;
    A00.M = A.M;
    A00.i0 = A.i0;
    A00.m0 = A.m0/2;
    A00.j0 = A.j0;
    A00.n0 = A.n0/2;

    A01.m = A.m;
    A01.n = A.n;
    A01.M = A.M;
    A01.i0 = A.i0;
    A01.m0 = A.m0/2;
    A01.j0 = A.j0+A.n0/2;
    A01.n0 = A.n0-A.n0/2;

    B00.m = B.m;
    B00.n = B.n;
    B00.M = B.M;
    B00.i0 = B.i0;
    B00.m0 = B.m0/2;
    B00.j0 = B.j0;
    B00.n0 = B.n0/2;

    B10.m = B.m;
    B10.n = B.n;
    B10.M = B.M;
    B10.i0 = B.i0+B.m0/2;
    B10.m0 = B.m0-B.m0/2;
    B10.j0 = B.j0;
    B10.n0 = B.n0/2;

    C0.i0 = 0;
    C0.m0 = A00.m0;
    C0.j0 = 0;
    C0.n0 = B00.n0;

    C1.i0 = C0.i0;
    C1.m0 = C0.m0;
    C1.j0 = C0.j0;
    C1.n0 = C0.n0;
    
    mmmrec(A00,B00,C0);
    mmmrec(A01,B10,C1);

    // Analogous cases for the remaining C submatrices
    ...

    int i, j;
    base_t (*c)[C.n] = (base_t(*)[C.n])C.M;
    for (i=0; i<C.m0; i++) {
      for (j=0; j<C.n0; j++) {
	c[C.i0+i][C.j0+j] = c0[i][j]+c1[i][j];
      }
    }

    free(C0.M);
    free(C1.M);
  }
}
  \end{lstlisting}

  Experiment with good cut-off criteria for the base case; the one shown
  is just one, simple possibility. The explicit submatrix representations
  come with a certain redundancy: Improve the argument structure to get
  a leaner, more economic implementation; for instance, the matrix orders
  $m,l,n$ can be factored out and kept constant throughout the recursions.
  Do your code improvements make any difference in running time?
\item
  Implement the blocked, $6$-fold nested loop matrix--matrix multiplication
  algorithm. Compare to the best and worst of the $3$-fold nested loop
  implementations. Estimate a good block (tile) size based on the size of
  the last-level cache in your processor. Conduct experiments using
  different base datatypes.
\item
  Implement in the style of the previous exercises a \emph{fused-multiply-add}
  matrix--matrix operation working on $m\times l$ and $l\times n$ input
  matrices $A$ and $B$ and an input-output $m\times n$ result matrix $C$.
  The function shall compute $C\leftarrow C+AB$.

  \begin{lstlisting}[style=SnippetStyle]
void fmma(const base_t* C,
          const base_t* A, const base_t* B,
	  const int m, const int l, const int n) {
  base_t (*a)[l] = (base_t(*)[l])A; // the matrices
  base_t (*b)[n] = (base_t(*)[n])B;
  base_t (*c)[n] = (base_t(*)[n])C;

  int i, j, k;
  // ... best three loops for multiply-add
}
  \end{lstlisting}
  The code will be useful as a building block in both shared- and distributed
  matrix--multiplication implementations.
\item
  Complete and implement the false sharing example (update to elements in
  a C structure or a C array) from \Sec~\ref{sec:multicorecaches}
  with either \openmp or \pthreads.
  Compile with different optimization options (also entirely
  without optimization)
  and time the duration for a medium large number of updates,
  say \texttt{r=10000}. What is the observed difference in running time between
  the three cases (false sharing with struct, false sharing with array, no
  false sharing via local variables)? Why is the difference between the
  optimized and non-optimized versions so large? Explain by studying the
  generated assembler code.
\item
  Consider the false sharing example from
  \Sec~\ref{sec:multicorecaches} with an array of $16$ \texttt{int}
  elements. Eliminate false sharing by padding each element of the
  array to occupy a full cache line (on your processor), either by
  accessing the array in strides of $16$ (your cache line size) or by
  declaring the array to be of structures occupying a full cache
  line. Benchmark against the original version and comment on the
  results (performance differences).
\item
  The following two code snippets compute the in- and out-degree of a
  graph with $n$ vertices represented by its $0/1$ adjacency matrix
  \texttt{A[n][n]}.

\begin{lstlisting}[style=SnippetStyle]
int A[n][n]; // sample declaration; beware for large n
int indeg[n];

for (i=0; i<n; i++) indeg[i] = 0;
for (j=0; j<n; j++) {
  for (i=0; i<n; i++) indeg[j] += A[i][j];
}
\end{lstlisting}

\begin{lstlisting}[style=SnippetStyle]
int A[n][n]; // sample declaration; beware for large n
int outdeg[n];
  
for (i=0; i<n; i++) outdeg[i] = 0;
for (i=0; i<n; i++) {
  for (j=0; j<n; j++) outdeg[i] += A[i][j];
}
\end{lstlisting}

The code snippets are executed on a single processor with a directly mapped
cache and the adjacency matrix and all $n$ element arrays are larger
than what can fit in the cache, \ie, the cache capacity is less
than $n$ integers. The block size (cache line size) of the cache is
$64$ bytes and a cache block can therefore hold $16$ consecutive
integers starting from a block-aligned address in memory. For
simplicity, we assume that there is no hardware prefetching
mechanism.
What is the read cache miss rate of the in-degree computation? 
What is the read cache miss rate of the out-degree computation?
Can the read cache miss rate of either of the codes be improved?
Write the corresponding program(s).

Run the code snippets on your own computer and check for differences
in running time. Consider how to ensure that the matrix and the arrays
are not in cache prior to each time measurement.
\item
  Consider the following standard, sequential merging algorithm to be
  run on a single core with a certain cache system.
\begin{lstlisting}[style=SnippetStyle]
i = 0; j = 0; k = 0; 
while (i<n&&j<m) {
   C[k++] = (A[i]<=B[j]) ? A[i++] : B[j++]; 
}
while (i<n) C[k++] = A[i++]; 
while (j<m) C[k++] = B[j++];
\end{lstlisting}

The inputs are two large arrays of integers, $A$ and $B$, with $n$
elements each (both \texttt{n} and \texttt{m} set to $n$). The output
array $C$ contains $2n$ elements. All arrays are much
larger than can fit in the small cache. The block size of the cache is
$16$ integers \`a $4$ Bytes each, thus, $64$ Bytes in total. The cache is
directly mapped. The cache is write non-allocate, so that writes to
$C$ will not cause cache misses (this may not correspond to any
existing cache system).  It also means you can ignore array $C$ when
investigating the cache behavior.  The input arrays have been
allocated in such a way that each element $A[i]$ goes to the same cache line
as element $B[i]$ (same index $i$).

In order to analyze the cache behavior of the merging algorithm under
these conditions, consider different possible inputs.  Assume
that once an array element (\eg, \verb|A[i]|) has been accessed
(loaded from cache, with or without a cache miss), it will be in a
register.  Thus, subsequent accesses to a specific element (\eg,
\verb|A[i]|) will not be counted as a cache access. 
\begin{enumerate}
\item Construct a \emph{best case} input for $A$ and $B$ leading to
  the smallest number of cache misses (hint: how should the values
  inside $A$ and $B$ look like to avoid cache misses?). Give the cache
  miss rate for the $2n$ iterations of the merging algorithm.
\item
  Construct a \emph{worst case} input for $A$ and $B$ with the largest
  number of cache misses. Give the cache miss rate for the $2n$
  iterations of the merging algorithm.
\end{enumerate}
\item
  In \Sec~\ref{sec:createpthreads}, creating and starting
  $p$ \pthreads threads was done in
  a sequential loop and, thus, in $\Omega(p)$ time steps; this may be
  too slow for a large numbers of threads on a multi-core processor with
  a larger number of processor-cores, especially if the $p$ threads are
  started and stopped many times.  Show how to create the $p$ threads
  in $\Omega(\log p)$ operations by using a recursive, tree-like algorithm
  where each started thread creates two (or some other, constant number of)
  new threads. Do this in such a way that the threads are also correctly
  terminated (with \texttt{pthread\_exit()} and \texttt{pthread\_join()}.
\item
  Remove the race condition from the \pthreads \texttt{hello()} program
  by protecting the \texttt{printf()} call with a \pthreads mutex that
  can be accessed through the arguments to the \texttt{hello()} function.
  Enhance the program with a shared counter to let the threads also print
  the order in which they are activated (the order in which they acquire the
  mutex).
\item
  Implement a linear pipeline with \pthreads. Each stage in the pipeline
  is represented by a thread that gets a work item from the predecessor
  thread in the pipeline, processes the item and signals the successor
  thread that a work item is ready. The first and last threads are special. The
  first thread generates the work items or reads them from the input.
  The last thread consumes the final output for each work item and stores
  it as output. Termination can be detected by sending a null-work item through
  the pipeline. For inspiration, see, for instance, the book by Rauber and
  R\"unger~\cite{RauberRunger13}.
\item
  Complete and implement the \texttt{primes\_race()} and
  \texttt{primes\_atomic()} examples with \pthreads and a simple,
  pseudo-polynomial \texttt{isprime(i)} predicate-function (check for
  divisibility from $2,3,\ldots,\floor{\sqrt{i}}$).
  Run on your system with different (larger) numbers of threads and
  compare the outcome in number of primes found. You may verify against the
  \openmp prime-sieve implementation.
\item
  Implement a simple mutex (lock) with lock and unlock operations
  using the C TAS instruction \texttt{atomic\_flag\_test\_and\_set()}.
  Define a worst-case throughput benchmark in
  which threads for some allocated time slot aggressively lock and unlock
  but actually do nothing useful in their critical section. Compare the
  implemented lock to the \pthreads mutex with this benchmark for varying
  numbers of threads.
\item
  A potential improvement of the TAS lock from the previous exercise is
  to use the possibly expensive TAS operation speculatively by first
  checking with a non-atomic read operation whether the lock may be free:
\begin{lstlisting}[style=SnippetStyle]
volatile int locked;
while (!locked) {
  while (locked); // spin on lock non-atomically
  locked = atomic_flag_test_and_set(*lock);
}
\end{lstlisting}
Compare the performance of this lock against the TAS implementation with
increasing numbers of threads. Are there notable performance differences?
This technique is known as Test-And-Test-And-Set (TATAS). Why is the
\texttt{volatile} qualifier needed? What is it supposed to ensure or prevent?
\item
Compare the predefined \pthreads readers-writer mutex against the one written
using condition variables by creating a suitable benchmark: What is
the mutex throughput in number of operations over a certain time
slice?  Do the implementations behave similarly (number and
distribution of threads being granted mutual exclusion)? Make it possible to
configure the number of readers and writers and the distribution of the
operations to be performed.
\item
  In an \openmp program, some unit of \texttt{work()} per thread is to
  be repeated a number of times until some \texttt{done} condition is
  fulfilled. The \texttt{work()} function is assumed to be
  thread-safe. There are (at least) two natural ways of expressing
  this.

  The first alternative is:
  \begin{lstlisting}[style=SnippetStyle]
while (!done) {
#pragma omp parallel
  {
    int t = omp_get_thread_num();
    work(t); // do some work
  }
  done = ...; 
}
  \end{lstlisting}
  The second alternative is:
  \begin{lstlisting}[style=SnippetStyle]
#pragma omp parallel
 {
   int t = omp_get_thread_num();
   while (!done) {
     work(t); // do some work

     done = ...;
   }
 }
  \end{lstlisting}
  
  Explain the differences and discuss advantages and pitfalls of one
  alternative over the other (performance, thread activation, coordination
  between threads, barrier synchronization, race conditions, \etc). What
  are requirements for the \texttt{work()} function for either of the
  two styles?
\item
  Consider the following two \openmp program snippets:
  \begin{lstlisting}[style=SnippetStyle]
int i, n;
#pragma omp parallel private(i)
{
  for (i=0; i<n; i++) {
    // loop work O(1) per iteration
  }
}
  \end{lstlisting}
and
\begin{lstlisting}[style=SnippetStyle]
int i, n;
#pragma omp parallel for
for (i=0; i<n; i++) {
  // loop work O(1) per iteration
}
  \end{lstlisting}
  Explain the differences between the two cases, in particular which
  work is performed and how it is shared. Why is the
  \texttt{private(i)} declaration in the first case needed? 
  Assume that the work per iteration is constant $O(1)$. When executed
  with $p$ threads, what is the total work as a function of $n$ and $p$?
\item
  Consider the following \openmp program fragment.
\begin{lstlisting}[style=SnippetStyle]
int b[2000];
int i;  
for (i=0; i<2000; i++) b[i] = -i;
#pragma omp parallel private(b)
{
  int i;
  for (i=0; i<2000; i++)
    assert(b[i]==-i); // first assertion
  for (i=0; i<2000; i++) b[i] = omp_get_thread_num();
  for (i=0; i<2000; i++)
    assert(b[i]==omp_get_thread_num()); // second
}
\end{lstlisting}
Assume the program is executed with $p,p>1$ threads.
Explain which threads execute which iterations of the inner loops
in the \texttt{parallel} region. Explain why there are no race conditions
on the updates to \texttt{b[i]}.
Explain why the first assertion is violated. Propose, using only \openmp
clauses, a way to make the assertion hold, regardless of $p$. Explain
why the second assertion holds. Assume now, after your repair, that
the initialization in the master thread is changed to
\begin{lstlisting}[style=SnippetStyle]
  int c[2000];
  int *b = c;
\end{lstlisting}
Will the repair still work? Why is the second assertion violated?
\item
  \label{exe:openmpcorank}
  Let two ordered arrays of $n$ and $m$ elements, respectively, that
  have to be merged stably into a single, ordered array of $n+m$
  elements be given.  Use the co-ranking preprocessing idea of
  \Sec~\ref{sec:corankmerge} to let each of $p$ threads compute the
  co-rank indices in the two input arrays needed to make it possible
  for each thread to sequentially merge segments of the input arrays
  into a unique segment of the output array of roughly $\frac{n+m}{p}$
  elements. Implement this algorithm with \openmp using a single,
  parallel region. How does the parallel implementation compare
  against your best known, sequential merge implementation (which
  should be used as a subroutine in the parallel implementation)? How
  many co-rank computations are needed per thread? How many explicit
  (and implicit) barrier synchronization operations do your
  implementation need? With how few can you do?
\item
  \label{exe:loopschedules}
  In order to understand the \openmp schedules and the way loop
  iterations are assigned to threads, it is instructive to run the
  example code from the end of \Sec~\ref{sec:ompschedule} that records
  the threads assigned to iterations and counts the number of
  iterations performed per thread. Run the example with $1,2,4,5,9,30$
  threads by setting the \ompnumthreads environment variable. Run with
  schedules like \texttt{"static"}, \texttt{"static,1"},
  \texttt{"static,4"}, \texttt{"static,7"}, \texttt{"static,9"},
  \texttt{"dynamic"}, \texttt{"dynamic,2"}, \texttt{"dynamic,3"},
  \texttt{"dynamic,10"}, \texttt{"gui\-ded"} and \texttt{"guided,10"} by
  setting the \ompschedule environment variable, and check the outcome
  to verify your understanding. Try with a larger loop range;
  experiment with different loop starts and loop increments.
\item
  In order to understand the scheduling of sections, complete and run the
  following code with varying numbers of threads.
\begin{lstlisting}[style=SnippetStyle]
  int s, a, b, c, d; // shared counters
  int i, r = 10; // some number of repetitions
  int t = omp_get_max_threads();
  int thrd[t];
  for (i=0; i<t; i++) thrd[i] = 0;
  s = 0; a = 0; b = 0; c = 0; d = 0;
#pragma omp parallel private(i)
  {
    for (i=0; i<r; i++) {
#pragma omp sections reduction(+:s) nowait 
      {
#pragma omp section
	{
	  a++; s++;
	  thrd[omp_get_thread_num()]++;
	}
#pragma omp section
	{
	  b++; s++;
	  thrd[omp_get_thread_num()]++;
	}
#pragma omp section
	{
	  c++; s++;
	  thrd[omp_get_thread_num()]++;
	}
#pragma omp section
	{
	  d++; s++;
	  thrd[omp_get_thread_num()]++;
	}
      }
    }
  }
  for (i=0; i<t; i++) {
    printf("Thread %d: %d\n",i,thrd[i]);
  }
  assert(a+b+c+d==s);
  assert(s==4*r);  
\end{lstlisting}
\item
  Implement a parallel copy function
  \texttt{par\_copy(int a[], int b[], int n)} as a simple parallel
  loop.  
\begin{lstlisting}[style=SnippetStyle]
#pragma omp parallel for schedule(runtime)
for (i=0; i<n; i++) {
  a[i] = b[i];
}
\end{lstlisting}

Benchmark the performance with medium large, dynamically allocated
arrays of, say, $n=1\,000\,000$, $n=10\,000\,000$ and $n=100\,000\,000$
elements for different number of threads. Experiment with 
"static" and "static,1" schedules (and other schedules)
by setting the \ompschedule environment variable accordingly.
Compare outcomes and explain the differences. Run with $1, 4 8, 10, 100$
threads by setting the \ompnumthreads environment variable accordingly.
Try replacing the \texttt{parallel for} directive with
\texttt{taskloop}. What is a suitable \texttt{grainsize}?
Does the performance depend on how the arrays are initialized (first touched)?
More concretely, initialize both arrays before the copy loop each with
a simple, sequential loop setting each entry to, say, $-1$. Experiment with
sequential initialization and with parallel initialization with both the
same as well as with a different schedule from the one used in the copy
loop. Explain possible performance differences.
\item
  Repeat the previous exercise with gather and scatter loops (see
  \Sec~\ref{sec:arraycompaction}) instead of the simple copy loop.
  Use different index arrays, first
  permutations (identity permutation, pairwise swaps, reverse order,
  random permutation, \etc), second possibly surjective index arrays
  (all indices to $0$, \etc).

\begin{lstlisting}[style=SnippetStyle]
#pragma omp parallel for schedule(runtime)
for (i=0; i<n; i++) {
  a[i] = b[ix[i]];
}
#pragma omp parallel for schedule(runtime)
for (i=0; i<n; i++) {
  a[ix[i]] = b[i];
}
\end{lstlisting}
\item
  \label{exe:mv1}
Consider the following C function for doing matrix--vector multiplication.
\begin{lstlisting}[style=SnippetStyle]
int seq_mv(int *A, int m, int n, int *b, int *c)
{
  int (*a)[n] = (int(*)[n])A; // the matrix
  int i, j;

  for (i=0; i<m; i++) c[i] = 0;
  for (i=0; i<m; i++)  {
    for (j=0; j<n; j++) {
      c[i] += a[i][j]*b[j];
    }
  }
}
\end{lstlisting}
Parallelize this function with straightforward \texttt{omp parallel for} loop
parallelization. Which loops can be parallelized and which cannot? Can
both loops be parallelized and collapsed? Run the code for a number of
repetitions, say 30, determine the time per repetition
(use \ompwtime to time all repetitions, compute average),
for inputs with $m=600$ and $n=1003$ for different numbers of threads.
Try different loop schedules. Compute the speed-ups relative to the sequential
(best known?) code.
\item
  In oder to understand the transformation and scheduling of
  collapsed \openmp loops, write a simple, three-loop nest
  \begin{lstlisting}[style=SnippetStyle]
#pragma omp parallel for collapse(3) schedule(runtime)
for (i=0; i<m; i++) {
  for (j=0; j<n; j++) {
    for (k=0; k<l; k++) {
      // keep book: which thread did what?
    }
  }
}
  \end{lstlisting}
  Insert code to record which thread did which iteration $(i, j, k)$ and
  how many iterations each thread performed. Experiment with different
  schedules similarly to Exercise~\ref{exe:loopschedules}.
\item
  \label{exe:parmmm}
  Use \openmp to parallelize the six different, three loop algorithms
  for matrix--matrix multiplication from Exercise~\ref{exe:sixmmm} with
  \texttt{m}, \texttt{n} and \texttt{l} as the input size parameters.
  Which loops can be immediately parallelized? Which loops can be
  collapsed?  Benchmark the six implementations for input sizes around
  $1000$ and $2000$ elements per matrix dimension for varying numbers
  of threads. Perform a number of repetitions and compute either
  average or median values for the running times. Compute the speed-up
  relative to the best performing sequential variant (loop order). Is
  the cache-behavior different in the sequential and parallel cases?
\item
  Consider again the straightforward, two loop implementation of
  matrix--vector multiplication. In contrast to the parallelization in
  Exercise~\ref{exe:mv1}, now parallelize the innermost loop doing the
  actual product summations by using a parallel \texttt{reduction()}
  clause. Verify correctness against the sequential code. Benchmark
  and compare the running time for different numbers of threads and
  total matrix sizes to both the sequential implementation and the
  simple loops parallelizations from Exercise~\ref{exe:mv1}.
\item
  Run and benchmark the matrix--vector multiplication code
  of Exercise~\ref{exe:mv1} again with small $m$ (on the order
  of the number of threads) and (very) large $n$ and different numbers
  of threads. Try with schedules
  \texttt{"static,1"},
  \texttt{"static,16"} and the default \texttt{"static"}. Explain the
  observed differences in performance.
\item
  Implement a function \texttt{par\_max(a,n)} for finding the maximum
  of \texttt{n} numbers (of some C base type, \texttt{int},
  \texttt{double}, \ldots) in an array \texttt{a}. Use a simple loop and
  parallelize it with a \texttt{reduction(max:...)} clause. Is this better than
  the simple, sequential loop? For specific input sizes, what is the number
  of threads leading to the smallest running time?
\item
  Implement the task parallel function \texttt{occurs(x,a,n)} of
  \Sec~\ref{sec:omptask} for
  counting the number of occurrences of \texttt{x} in array \texttt{a}
  as a parallel loop. Which \openmp constructs can be used to do this
  correctly? Which of the two implementations can give speed-up with more
  than one thread, which is faster? What can you do to improve the
  performance of the task parallel implementation?
\item
  Show how a simple \texttt{for}-loop \texttt{for (i=0; i<n; i++) ...}
  in canonical form can be parallelized with \openmp tasks by giving a
  recursive algorithm for splitting the iteration range into single
  iterations each to be performed as a task.  The algorithm is
  analogous to the task parallel \texttt{occurs(x,a,n)} function of
  \Sec~\ref{sec:omptask}.  As in Exercise~\ref{exe:loopschedules},
  instrument the loop body with performance counters that record for
  each iteration executed as a task which thread executed that
  iteration/task and for each thread how many tasks (iterations) that
  thread executed. What is the overhead in terms of number of tasks that
  split loop ranges and do not directly perform iterations?
  Run with very large iteration ranges (be careful
  with output) and with different numbers of threads. Is the work well
  divided between the threads?
\item
  Consider the following \openmp program fragment.
  \begin{lstlisting}[style=SnippetStyle]
int cc;
#pragma omp parallel default(none) shared(cc)
{
  int r;
#pragma omp single
  cc = 0;
  for (r=0; r<omp_get_num_threads(); r++) {
#pragma omp single
    cc++;
    assert(cc-1==r);
#pragma omp barrier
  }
}
  \end{lstlisting}
  Explain why the assertion will hold. Explain why the explicit
  \texttt{omp barrier} is necessary. What happens if the \texttt{omp single}
  inside the loop is given a \texttt{nowait} clause? What about the
  \texttt{omp single} before the loop?
  What happens if \texttt{omp single} is replaced with \texttt{omp master}?
  What happens if \texttt{omp single} is replaced with \texttt{omp critical}?
  Formulate an assertion that will always hold for this last case.
\item
  The following program fragment uses \texttt{single} work sharing,
  critical sections, atomic updates and reductions to update global
  variables.  Apparently, the intention is that \texttt{a} is
  incremented once per iteration up to \texttt{iter}, and that the
  other variables are incremented by each thread in each iteration.
  \begin{lstlisting}[style=SnippetStyle]
int a, b, c, d;
a = 0; b = 0; c = 0; d = 0;
#pragma omp parallel private(a) reduction(+:d)
{
  int i;
  for (i=0; i<iter; i++) {
#pragma omp single
    a++;
#pragma omp critical
    b++;
#pragma omp atomic update
    c++;
    d++;
  }
}
  \end{lstlisting}
  What is are the values of \texttt{a}, \texttt{b}, \texttt{c} and \texttt{d}
  after \texttt{iter=20} iterations and $p=8$ threads? Does this correspond
  to the intended outcome? Is the program correct (no race conditions)? What
  must possibly be done to make the program correct with the intended outcome?
\item
  Define for a not necessarily ordered $n$-element array $A$
  $\rank(i,A,n)$ to be the number of elements $A[j]$ with either
  $A[j]<A[i]$ or $A[j]=A[i]\vee j<i$. Suggest an \openmp loop to parallelize
  the computation of $\rank(i,A,n)$. The function (not necessarily the
  parallelized version) is used in the following piece of code that moves
  the elements of $A$ to a new array $B$.
  \begin{lstlisting}[style=SnippetStyle]
for (i=0; i<n; i++) {
  int j = rank(i,A,n);
  B[j] = A[i];
}
\end{lstlisting}
  What does this code accomplish? What is the sequential complexity
  as a function of $n$? Find another parallelization instead
  of relying on the parallelized $\rank(i,A,n)$ function. Compare the
  resulting two parallel implementations, also in terms of performance.
  Can loop collapsing be used?
\item
  Implement the shared-counter solution for prime finding discussed
  in \Sec~\ref{sec:atomics} in \openmp in two versions.
  Let \texttt{int next;} be a shared counter initialized to $2$, the first
  prime. Activate the threads in a parallel region where they each iterate
  until the counter has reached the upper limit for the range to be checked.
  In the first solution, use \texttt{\#pragma omp atomic capture} to atomically
  read and update \texttt{next}. In the second solution, read and update
  under mutual exclusion using \texttt{\#pragma omp critical} to declare the
  critical section. Time the prime checking for up to $1\,000\,000$,
  $10\,000\,000$ and, if possible $100\,000\,000$ integers (repeat a number
  of times) for different numbers of threads, and compare the two solutions.
\item
  Implement the recursive inclusive prefix sums algorithm described in
  \Sec~\ref{sec:recursiveprefix} as a C program with
  \openmp. Benchmark against a best known sequential implementation
  with arrays of $n=100\,000$, $n=1\,000\,000$, and $n=10\,000\,000$
  elements (of C \texttt{int} and/or \texttt{double} type), respectively.
  What speed-up can be achieved? In case speed-up is below expectation,
  explain possible reasons.
\item
  Implement the iterative inclusive prefix sums algorithm
  described in \Sec~\ref{sec:iterativeprefix} as a C program with
  \openmp. Benchmark against a best known sequential implementation
  with arrays of $n=100\,000$, $n=1\,000\,000$, and $n=10\,000\,000$
  elements (of C \texttt{int} and/or \texttt{double} type), respectively.
  What speed-up can be achieved? In case speed-up is below expectation,
  explain possible reasons.
\item
  Implement the Hillis--Steele inclusive prefix sums described in
  \Sec~\ref{sec:hillissteele} as a C program with \openmp. Benchmark
  against a best known sequential implementation with arrays of
  $n=100\,000$, $n=1\,000\,000$, and $n=10\,000\,000$ elements (of C
  \texttt{int} and/or \texttt{double} type), respectively.  What
  speed-up can be achieved? In case speed-up is below expectation,
  explain possible reasons.
\item
  Improve either of the prefix sums algorithms implemented in the
  previous exercises by using blocking (see \Sec~\ref{sec:blocking}).
  This means that the $p$
  threads first sequentially, but in parallel, preprocess disjoint
  parts of the input array to arrive at a prefix sums problem of size
  $O(p)$ which is then solved by a parallel prefix sums
  function. Post-processing will in addition be necessary to arrive at
  the solution to the original prefix sums problem.  Benchmark against
  a best known sequential implementation with arrays of $n=100\,000$,
  $n=1\,000\,000$, and $n=10\,000\,000$ elements (of C \texttt{int}
  and/or \texttt{double} type), respectively.  Compared to your
  direct implementation from either of the previous exercises, how
  much improvement could be achieved?
\item
  Discuss algorithmic and architectural obstacles to obtaining full
  linear (perfect) speed-up for the \openmp prefix sums implementations.
\item
  Implement recursive matrix--matrix multiplication based on your
  sequential implementation from Exercise~\ref{exe:recmmm} using
  \openmp tasks to handle the recursive calls. Compare the results to
  the best parallelization of standard, three-loop matrix--matrix
  multiplication of Exercise~\ref{exe:parmmm}. Experiment with
  different matrix sizes (say, order $n=850, n=1000, n=1200$). Experiment
  with different recursion cut-off strategies and values. How good performance
  can you achieve in comparison? You can try to implement the cut-off with
  \openmp clauses.
\item
  The following part of a larger program shows two ways of computing the sum of
  $n$ numbers (here the numbers are all just ones).
  \begin{lstlisting}[style=SnippetStyle]
int jr = 0; // reduction-sum
int ja = 0; // atomic-sum
#pragma omp parallel for reduction(+:jr)
for (i=0; i<n; i++) {
  jr += 1;
#pragma omp atomic update
  ja += 1;
}
assert(jr==ja);
\end{lstlisting}
  Explain the differences between the two possibilities. Run the code
  and look for differences in performance.
\item
  The following part of a larger program shows two ways of computing
  indices and storing them in an array.
  \begin{lstlisting}[style=SnippetStyle]
int us[n];
int ua[n];
int jr = 0;
int ja = 0;
#pragma omp parallel for reduction(inscan,+:jr)
for (i=0; i<n; i++) {
  us[i] = jr;
#pragma omp scan exclusive(jr)
  jr += 1;
}
#pragma omp parallel for 
for (i=0; i<n; i++) {
#pragma omp atomic capture
  ua[i] = ja++;
}
assert(jr==ja);
\end{lstlisting}
  Explain the differences between the two possibilities. Run the code
  and look for differences in performance and outcome. You will be surprised.
\item
  Consider the following loop which is parallelized as a collection of tasks.
  The loop computes for each thread how many of the iterations were done by that
  thread (as done in \Sec~\ref{sec:ompschedule}).
  \begin{lstlisting}[style=SnippetStyle]
int t = omp_get_max_threads();
int iter[t];
for (i=0; i<t; i++) iter[i] = 0;
// with task loop
#pragma omp parallel
{
#pragma omp taskloop 
  for (i=0; i<n; i++) {
    iter[omp_get_thread_num()]++; // careful with large n
  }
}
nn = 0;
for (i=0; i<t; i++) nn += iter[i];
assert(nn==n); // all iterations done
  \end{lstlisting}
  Run with a large \texttt{n}. Why does the assertion not hold? Repair
  the program by inserting an additional, single work sharing construct.
  Compare the running time, now with large $n$, of the loop against
  the running time of the same loop
  parallelized with a standard, scheduled \texttt{omp parallel for} pragma.
  Try computing the total number of operations performed into a new,
  shared variable \texttt{int nnn = 0;} by using a \texttt{reduction()} clause.
  Does this have an effect on the time taken to execute the loop?
\item
  Use the Master Theorem~\ref{thm:master} as a blueprint for a program
  to explore the way \openmp may schedule tasks to threads. The following
  function follows the pattern of the recurrences covered by the
  Master Theorem for integer coefficients $a,b,d,e$ and generates tasks
  for each evaluation of the recurrence.
  \begin{lstlisting}[style=SnippetStyle]
void masterrecursion(const int a, const int b,
                     const int d, const int e,
		     int n, int ops[], int tsk[])
{
  tsk[omp_get_thread_num()]++;
  if (n<=1) {
    ops[omp_get_thread_num()]++;
  } else {
    int i, k;
    int nn = 1;
    for (i=0; i<d; i++)  nn *= n;
    k = 0;
    for (i=1; i<=n; i*=2) k++;
    int nk = 1;
    for (i=0; i<e; i++) nk *= k;
    ops[omp_get_thread_num()] += nn*nk;
    
    for (i=0; i<a; i++) {
#pragma omp task
      masterrecursion(a,b,d,e,n/b,ops,tsk);
    }
  }
}
\end{lstlisting}
  Execute the function in a parallel region with values for $a,b,d,e$ as
  found in the different recurrences discussed in the text, and different
  values for $n$.
  \begin{lstlisting}[style=SnippetStyle]
#pragma omp parallel
  {
#pragma omp single
    masterrecursion(a,b,d,e,n,ops,tsk);
  }
  \end{lstlisting}
  Extend the \texttt{masterrecursion} function to count atomically the
  total number of tasks and the total number of operations, and verify
  against the per thread computed values.
\item
  Is the usage of the \texttt{omp barrier} construct below legal?
  State and add an assertion on \texttt{j} that will hold after the loop.
  \begin{lstlisting}[style=SnippetStyle]
int i, j;
j = 0;
#pragma omp parallel private(i) firstprivate(j)
{
  for (i=0; i<500; i++) {
    if (i%2==0) {
#pragma omp barrier
      j++;
    } else {
      j--;
#pragma omp barrier
    }
  }
}
  \end{lstlisting}
\item
  Let \texttt{a[n]} be an array of \texttt{n} elements of some basic C
  type, and let \texttt{x} be an element of that type.  Write an
  \openmp program with inscan reductions to find the first and the
  last occurrence (indices) of \texttt{x} in \texttt{a[n]} and to
  calculate the number of occurrences.
\item
  Complete the 
  implementation of the task parallel Quicksort using \openmp tasks from
  \Sec~\ref{sec:omptask}. Benchmark your implementation with inputs of
  $n=1\,000\,000$, $n=10\,000\,000$ and $n=100\,000\,000$ elements of
  basetype \texttt{int} and \texttt{double} and different input permutations.
  What is the largest speed-up achievable compared to a sequential Quicksort
  implementation on your system and with how many threads? Experiment with
  different cut-off value to end the recursion earlier. Time permitting,
  experiment with different pivot-selection strategies.
\item
  Analyze the following variant of Quicksort where the partitioning
  step is done by a work-optimal parallel algorithm and the recursive
  calls are done sequentially. What cut-off is required in order to make
  the algorithm work- and cost-optimal?
  \begin{lstlisting}[style=SnippetStyle]
void Quicksort(int a[], int n)
{
  int i, j;
  int aa;

  if (n<CUTOFF) {
    // sort sequentially
    return;
  }

  // pivot selection
  int pivot = a[0]; // choose an element (non-randomly...)

  // partition
  // mark elements <pivot
  // array compaction, last index j
  // mark elements >=pivot
  // array compaction with pivot in a[j]

  Quicksort(a,j);
  Quicksort(a+j+1,n-j-1);
}
  \end{lstlisting}
  Assume that an exclusive prefix sums computation can be done in parallel
  by a function \texttt{exclusive\_prefix\_sums(a,n)} (see previous exercises)
  and use this to write out the Quicksort code. Implement and benchmark this
  algorithm against the task parallel formulation.
\item
  Implement a parallel partition function which partitions the elements
  into three blocks: those smaller than the given pivot, those equal to
  the pivot, and those larger than the pivot.
\item
  Show how to implement Quicksort partitioning in parallel using \openmp
  scan-reduction.  Benchmark your implementation against a best known
  sequential implementation and a parallel implementation using a
  ``hand-written'', adapted prefix sums computation.
\item
  Show how to implement Quicksort partitioning in parallel using \openmp
  atomics to maintain indices in arrays for the classes of elements (smaller
  than, larger than the pivot). Compare the performance of this implementation
  against an implementation with \openmp scan-reductions (previous exercise).
\item
  Use the parallel merge function implemented in
  Exercise~\ref{exe:openmpcorank} to devise a parallel Mergesort
  algorithm based on co-ranking. What is the asymptotic running time
  of your implementation? How does the cut-off at which we change to
  sequential merging have
  to be chosen in order to arrive at a cost-optimal algorithm? What is the
  parallel running time of your algorithm including (or excluding)
  the time for explicit and implicit barrier synchronizations? Implement
  your algorithm in \openmp and benchmark against your best, sequential
  Mergesort implementation.
\item
  Implement the prime sieve algorithm in the three versions discussed in
  \Sec~\ref{sec:primesieve} in \openmp: sequential, parallelization with
  ordered and parallelization with inscan-reduction. You may consider also
  using an own, hand-written prefix sums algorithm for the array compaction.
  Benchmark and compute the (relative) speed-up for finding primes up to
  $n=1\,000\,000$,  $n=10\,000\,000$ and  $n=100\,000\,000$ with different
  number of threads. Verify correctness by comparing to the work pool
  shared counter implementation.
\item
  Implement the task parallel merge from \Sec~\ref{sec:cilk} with \openmp
  (or with \cilk). Benchmark against a sequential merge implementation (see
  \Sec~\ref{sec:merging}). How large speed-up can be achieved with how many
  threads? What is a good cut-off for the recursion/task spawning?
\item
  Improve the task parallel merge implementation from the previous
  exercise by using the co-ranks of $n/2$ to split exactly the input
  arrays \texttt{A} and \texttt{B}. Benchmark against your previous
  implementation. Are there any improvements? How can an input arrays
  be constructed that lead to a worst possible behavior for the
  simple, rank-based algorithm?
\item
  Implement a Merge sort algorithm as in \Sec~\ref{sec:cilk} using the
  task parallel merge algorithm in either \openmp or \cilk. Benchmark
  against a standard, sequential, recursive, top-down Mergesort.  How
  large speed-up can be achieved with how many threads? What is a good
  cut-off for the recursion/task spawning? Can you eliminate or
  minimize the use of the additional array \texttt{B} with the extra,
  parallel copy operations?  Hint: It is possible to do with at most
  one copy operation.
\item
  Here is an implementation sketch of a sequential Breadth-First
  Search (BFS)\index{BFS} algorithm that assigns distance labels to
  all vertices of a graph starting from some given start vertex. The
  graph is represented by a collection of adjacency lists written in
  matrix notation. We assume that a FIFO \texttt{queue} data structure
  with \texttt{init()}, \texttt{enq()} and \texttt{deq()} operations
  and a \texttt{nonempty()} predicate has been implemented.
    \begin{lstlisting}[style=SnippetStyle]
int n, m; // size of graph
int i;
  
int s; // start vertex
int u, v;

int deg[n];    // vertex degrees
// space inefficient adjacency list representation
int adj[n][n]; 

int dist[n]; // labels to be assigned

queue Q, N; // queues to be implemented, simple array FIFO
init(Q);
init(N);

int l; // level

l = 0; dist[s] = l;
enq(s,Q); // put start vertex in Q

do {
  // process level l;
  do {
    deq(u,Q); // Q is not empty
    for (i=0; i<deg[u]; i++) {
      v = adj[u][i];
      if (dist[v]==-1) {
	dist[v] = l+1;
	enq(v,N);
      }
    }
  } while (nonempty(Q));
  Q = N; l++; // next level
} while (nonempty(Q));
  \end{lstlisting}
    Complete the sequential implementation. Write a simple graph
    generator in order to test and benchmark your implementation.
    Now, parallelize the sequential code with \openmp. Use critical
    sections or atomic operations in order to dequeue and/or enqueue
    elements from your FIFO queue. Consider allowing
    optimistic/speculative updates of the distance labels. Where is
    the parallelism in your parallel algorithm/implementation? What is
    the time complexity of your algorithm? Which properties of the
    input graph do you use in your complexity statement? Is your
    algorithm work-optimal? Benchmark with simple graphs using your
    input generator and compute the achieved speed-ups compared to
    your sequential (best possible) implementation.
\item
  \label{exe:fw}
  The Floyd-Warshall algorithm~\cite{Floyd62} for the
  \impi{all-pairs shortest path problem} (APSP)
  is based on the following observation. Let a
  weighted, directed graph $G=(V,E)$ be represented by an $n\times n$
  distance matrix $W[n,n]$ where $W[i,j]$ is the weight of the edge between
  vertex $i$ and vertex $j$, $i,j\in V$.
  Let $W^k[i,j]$ be the distance (weight of a shortest path) between
  vertices $i$
  and $j$ using only paths with vertices $0,1,\ldots,k-1$. It then holds
  that $W^0[i,j]=W[i,j]$,
  $W^k[i,j]=\min(W^{k-1}[i,j],W^{k-1}[i,k]+W^{k-1}[k,j])$ for $0<k<n$
  and that $W^{n-1}[i,j]$ is the length of a shortest path in $G$ between
  $i$ and $j$. The following function computes $W^{n-1}[i,j]$.

  \begin{lstlisting}[style=SnippetStyle]
int *fw_apsp2(int *w, int *wnext, int n) {
  int (*W)[n]     = (int(*)[n])w;
  int (*Wnext)[n] = (int(*)[n])(wnext);
  int (*WW)[n];
  
  int i, j, k;
  for (k=0; k<n; k++) {
    for (i=0; i<n; i++) {
      for (j=0; j<n; j++) {
	Wnext[i][j] = (W[i][j]>W[i][k]+W[k][j]) ?
                        W[i][k]+W[k][j] : W[i][j];
      }
    }
    WW = W; W = Wnext; Wnext = WW;
  }
  
  return (int*)W;
}
  \end{lstlisting}
  What is the complexity of the Floyd-Warshall algorithm on graphs $G$
  with $n$ vertices, $n=|V|$? Which of the loops can possibly be
  parallelized? Give a parallelization with \openmp.

  An alternative, less obvious implementation using only one matrix is
  given below.
  \begin{lstlisting}[style=SnippetStyle]
void fw_apsp(int *w, int n) {
  int (*W)[n] = (int(*)[n])w;
  
  int i, j, k;
  for (k=0; k<n; k++) {
    for (i=0; i<n; i++) {
      for (j=0; j<n; j++) {
	if (W[i][j]>W[i][k]+W[k][j]) {
          W[i][j] = W[i][k]+W[k][j];
        }
      }
    }
  }
}
  \end{lstlisting}
  Argue that the two implementations compute the same correct distance matrix.
  The second implementation has subtle dependencies and is less obviously
  parallelized. What can be done? Which of the implementations perform
  best? What is the speed-up that can be achieved on matrices of order $n=1000$
  with different numbers of threads on your system?
\end{enumerate}


\chapter{Distributed Memory Parallel Systems and \mpi}
\label{chp:distributedmemory}

The last third of the lectures on \parco deals with characteristics of
distributed memory systems where processors or shared memory
processor-nodes are interconnected via a communication network which
has to be used to exchange information between processes or
threads in any non-trivial parallel algorithm. The concept of
message-passing as a means to structure and implement algorithms is
introduced, concretely via \mpi which is dealt with extensively,
regarding both fundamental message-passing ideas but also the way
these have been realized in a concrete specification.

\section{Eighth block (1 lecture)}

This lecture block is an introduction to performance relevant aspects of
real, parallel, distributed memory systems.

A na\"ive, parallel distributed memory system model consists of a set of
$p$ processors each with local memory for program and data (MIMD
architecture\index{MIMD}). Processors execute independently and
asynchronously and exchange information through explicit
communication through an \impi{interconnection network}. Communication
is (significantly) more expensive than accessing data in local memory
and may be subject to additional constraints. The network may provide
means for synchronizing the processors.

In a corresponding distributed memory
programming model\index{programming model}, processes (or
threads) communicate explicitly by executing communication operations,
either pairwise, or in more complex collective patterns. Distributed
memory programming models also offer means for synchronizing
processes.

The concrete, distributed memory programming interface will be \mpi,
the \emph{Message-Passing Interface}, which is treated in depth in
the following parts of the lecture notes.

\subsection{Network Properties: Structure and Topology}
\label{sec:structureandtopology}

The distinguishing, new feature of distributed memory systems is the
interconnection network (sometimes just called the
\impi{interconnect}) needed for communication between processors,
which can be individual cores, multi-core CPUs, or larger entities
consisting of many multi-core CPUs, nowadays often enhanced with
GPUs\index{GPU} and other accelerators\index{accelerator}, see
\Sec~\ref{sec:hierarchical}.  These entities are physically
connected (electric or optical cables, or other, often just called
\impi{links}), and not all of these entities may be immediately,
directly connected with each other; typically, they are \emph{not}!
Also, some elements in the network may not be processors used for
computation, but simply \impi{network switches} serving communication
with other network elements. It is clear that both the physical and the
topological properties of the network (speed of the connections, processing
capabilities, the composition and structure of the network) play a
decisive role for the performance of algorithms and programs running
on distributed memory systems. It is also clear that without a
powerful interconnect, there can be no \parco: We are interested in
non-trivial problems requiring non-trivial communication and
interaction between processors.

An interconnect where the processors are also the communication
elements and in which there are no switches is called a
\impi{direct network}.  An interconnect, in which there are also
special switch elements (special communication processors with
connections to other elements) is, on the other hand, called an
\impi{indirect network}.

First, we are interested in investigating how structural properties of
the network influence the communication performance and the
capability to solve problems that we are interested in.

The structure or \impi{topology} of a communication network, both
direct and indirect, can be modeled as a(n un)directed, (un)weighted
graph $G=(V,E)$, where the vertices (nodes) $V$ denote processors or
network communication elements, and the edges $E$ model the immediate
connections or links between communication elements. Two elements
(processors or switches) $u,v\in V$ are immediately connected
(adjacent neighbors) if there is a (directed) edge (arc) $(u,v)\in
E$. For most communication networks, if network element $u$ can send
data directly to network element $v$ via a link $(u,v)$, then also $v$
can send data directly to $u$; that is, communication networks are
most often undirected (or bidirected) graphs, and the link $(u,v)$ can be
used in both directions. It can nevertheless sometimes be relevant
to consider directed graphs; and indeed there have been (few) examples
of real, parallel distributed memory systems built on directed
interconnection networks.

When two processors $u$ and $v$ are not
adjacent in the network, a path between $u$ and $v$ must be found
along which $u$ and $v$ can then communicate. Let a path between nodes
$u$ and $v$ have length $l$. Communicating some data from $u$ to $v$
along this path will take at least $l$ successive communication operations.

Recall that the \impi{diameter} of a graph $G=(V,E)$ is the maximum
over all shortest paths between pairs of nodes $u,v\in V$.

\begin{eqnarray*}
  \diam(G) & = & \max\{\dist(u,v) | u,v\in V\}
\end{eqnarray*}

Here, $\dist(u,v)$ denotes the distance in number of links that have
to be traversed to get from $u$ to $v$ in $G$. It is defined as
the length of a shortest path
from $u$ to $v$ measured in number of edges to traverse. The diameter is a
lower bound on the number of communication steps for communication
operations and algorithms that involve message transmission between
nodes $u$ and $v$ which have the longest distance in the communication
network. Note that we always take the diameter to be finite:
Disconnected networks cannot be used for \parco.

The out-$\degree(G)$ of a graph $G=(V,E)$ is the largest number of
outgoing edges from a node in $G$; that is, the largest node degree of
a node in $G$.
\begin{eqnarray*}
  \degree(G) & = & \max\{\degree(u) | u\in V\}
\end{eqnarray*}
where the node degree of $u\in V$ is given by $\degree(u) =
|\{v\in V | (u,v)\in E\}|$. The in-degree is defined analogously.

The \impi{bisection width} of a graph $G=(V,E)$ is the smallest number of
edges that must be removed in order for the graph to fall apart into
two roughly equal-sized parts (in number of vertices); that is,
to partition the vertices of $G$ into two disjoint subsets with no
edges between pairs of vertices in the two subsets. If two vertex
subsets $V',V''$ are roughly equal when $||V'|-|V''||\leq 1$, the formal
definition is as follows.

\begin{eqnarray*}
  \bisec(G) & = & 
  \min_{\begin{array}{c}
        V',V''\subset V\\
        V'\cup V''=V\\
        V'\cap V''=\emptyset\\
        \left||V'|-|V''|\right|\leq 1
        \end{array}
      }|\{(u,v)\in E, u\in V', v\in V''\}|
\end{eqnarray*}

While both $\diam(G)$ and $\degree(G)$ can be easily computed in
polynomial time for any given network topology graph $G$, $\bisec(G)$
can (most likely) not.  The problem of finding $\bisec(G)$ is
essentially the \emph{Graph Partitioning} problem, one of the
classical, standard NP-complete problems~\cite[ND14]{GareyJohnson79}.

The best possible communication network in terms of diameter and
bisection width is the \impi{fully connected network} $G=(V,E)$, where
$(u,v)\in E$ for all $u,v\in V$ (assume either $(u,u)\in E$ or
$(u,u)\not\in E$ as convenient). For a fully connected network $G$,
$\diam(G)=1$ and $\bisec(G)=|V|^2/4$ (for $|V|$ even). The significant
drawbacks of the fully connected network are the large number
of links, namely $|V|(|V|-1)$ and the high degree (number of links per node),
namely $\degree(G)=|V|-1$.

The worst possible communication networks that can support \parco are
the \impi{linear processor array} and the \impi{processor ring}, which
are graphs $A,R=(V,E)$ consisting of either a single path from vertex
$u\in V$ to vertex $v\in V$ both having degree $1$ with all other
vertices in-between having degree $2$ (processor array $A=(G,V)$), or
a single cycle spanning all vertices $v\in V$ each of which have
degree $2$ (processor ring $R$). For the linear array, $|E|=|V|-1$,
$\diam(A)=|V|-1$ and $\bisec(A)=1$, and for the ring $|E|=|V|$ (for
$|V|>2$), $\diam(R)=\floor{|V|/2}$ and $\bisec(R)=2$. A significant
advantage of linear arrays and rings is the small(est possible) number
of links (to keep the graph connected) and the low degree. A
\impi{tree network} $T=(V,E)$ likewise has $|E|=|V|-1$, $\bisec(T)=1$,
but typically $\diam(T)=O(\log |V|)$.

Number of communication edges (links) and node degrees entail concrete,
physical costs (money and space for cables and network connections)
when building \parco
systems with given network properties, as do other factors like, for
instance, the necessary physical lengths of cables. It is, therefore,
interesting, relevant, and highly challenging to find good compromises
between costs and structural network properties desirable for
supporting non-trivial \parco. Many different (with and
with no commercial potential) solutions have been given, see, for
instance, the aforementioned \url{http://www.top500.org}.

Numerous networks between the two extremes have been proposed and
studied, see, for instance~\cite{Leighton92}, and are not the topic of
these lectures. Only three classes of such communication networks shall be
mentioned, namely \emph{trees}, $d$-dimensional \emph{tori/meshes}, and
\emph{hypercubes}.

In a \impi{tree network}, the topology graph $T=(V,E)$ is a tree
(minimally connected graph over the nodes in $V$), most often with
logarithmic diameter as in balanced binary or $k$-ary trees, 
binomial trees, \etc (the linear array is a special case).
Being minimally connected, tree networks have
$\bisec(T)=1$, since removing any one link will make the network fall
apart. 

In a $d$-dimensional \impi{mesh network} with dimension sizes (or
orders) $r_0,\ldots,r_{d-1}$, the processors are identified with the
set of $d$-element integer vectors $V=\{(x_0,\ldots,x_{d-1}) | x_i\in
\{0,1,\ldots,r_i-1\}\}$. The number of processors in such a
$d$-dimensional mesh is therefore $|V|=\prod_{i=0}^{d-1}r_i$.  There
is a bidirected link $(u,v)$ between two processors
$u=(x_0,\ldots,x_{d-1})$ and $v=(y_0,\ldots,y_{d-1})$ if $|x_i-y_i|=1$
for some coordinate $i, 0\leq i<d$ and $x_j=y_j$ for all other
coordinates.  A \impi{torus network} or \impi{torus} is a mesh network
with additional ``wrap-around'' edges between processors at the
``borders'' of the mesh. That is, between two processors
$u=(x_0,\ldots,x_{d-1})$ and $v=(y_0,\ldots,y_{d-1})$ if $x_i=0$ and
$y_i=d_i-1$ for some $i$th coordinate and $x_j=y_j$ for all other
coordinates $j\neq i$. The diameter of a mesh $M=(V,E)$ is
$\diam(M)=\sum_{i=0}^{d-1} (r_i-1)$, and the degree is
$\degree(M)=2d$. The diameter of a torus $S=(V,E)$ is
$\diam(S)=\sum_{i=0}^{d-1} \floor{r_i/2}$, and the degree is likewise
$\degree(S)=2d$.

A uniform (symmetric, homogeneous) mesh or torus network has the same
order for all dimensions, $r=\sqrt[d]{p}$. The bisection width of a
symmetric mesh is $\bisec(M)=p^{\frac{d-1}{d}}=p/\sqrt[d]{p}=p/r$ and
of a symmetric torus $\bisec(S)=2p^{\frac{d-1}{d}}=2p/r$ (for $r$ even).

A \impi{hypercube network} $H=(V,E)$ is a special case of a uniform
torus (or mesh) network in which all coordinates are either $x_i=0$ or
$x_i=1$; note that in this case, mesh and torus coincide, the hypercube
torus has no more edges than the hypercube mesh.
Thus, the number of processors is
$p=2^d$ for some $d$, that is, a power of $2$, or the other way
around, the dimension of a $p$-processor hypercube is $d=\log_2
p$. Each processor has $d$ neighboring processors which for processor
$u=(x_0,\ldots,x_{d-1})$ are found by changing one of the $i$
coordinates from $x_i$ to $1-x_i$. This is the same as flipping
the $i$th bit in $u$ when viewed as a binary number.
Both the degree and the diameter of a
hypercube are $\degree(H)=\diam(H)=d=\log_2 p$. The bisection width is
$\bisec(H)=p/2$.

Modern high-performance systems are often built as torus networks of
$d=3,5,6$ dimensions or as indirect networks with multiple switches
of small, fully connected networks, often called \impi{multi-stage
  networks} of which there are many examples (\eg, InfiniBand).
Hypercube networks were once popular, but are currently not built
(what could some reasons be?).

\subsection{Communication Algorithms in Networks}
\label{sec:networkcomm}

Communication from a processor $u$ to another processor $v$ in a given
network $G=(V,E)$ requires at least $\dist(u,v)$ communication steps,
in which processor $u$ sends data to a neighboring processor that is
closer to $v$ (along an edge in $E$). This neighbor, in turn, sends data to
a neighboring processor that is closer to $v$ (along an edge in $E$),
\etc, until the data reaches $v$.  This is independent of the amount of
data to be transferred and the concrete costs incurred by sending and
receiving some amount of data (see later). It is relevant to study the
number of such communication steps that may be required for other,
more complex communication operations, apart from just the
transmission of information from one processor to another. We,
therefore, first assume that data to be communicated are all of some
small unit and that each communication step takes the same unit of
time.

In a communication step, a processor $u\in V$ can communicate with a
neighbor in the communication network $G=(V,E)$.  What exactly a
processor can do in a communication step depends on the
\emph{capabilities} of the communication system. We say that a
communication system is \impi{one-ported} (or \impi{single-ported}) if
a processor can engage in at most one communication operation in a
step. A communication system where a processor can be involved in up
to $k$ communication operations in the same step (that is,
concurrently) is called \impi{$k$-ported} or just \impi{multi-ported}.

If communication in a step between neighboring processors $u\in V$ and
$v\in V$ with $(u,v)\in E$ is only in one direction from $u$ to $v$ or
from $v$ to $u$, it is \impi{unidirectional}. The
communication system is said to be unidirectional if it can support
only unidirectional communication in a step. Communication in both
directions, from $u$ to $v$ and from $v$ to $u$, is bidirectional,
telephone-like (in an old sense of ``telephone'' where only
two parties could speak at the same time)\index{bidirectional telephone}. A
communication system that can support such communication is said to be
\impi{bidirectional}. Communication where a processor $u$ receives from a
processor $w$ and sends to a processor $v$ is said to be
general, send-receive, bidirectional communication.
A system that can support such communication in a
step is said to be \emph{bidirectional} in the general, \emph{send-receive}
sense\index{bidirectional send-receive}.

Most modern communication systems and networks can, approximately,
support general, send-receive, bidirectional communication. It can be
measured to what extent this is the case by system benchmarking, in
itself an interesting activity.  Systems with indirect, multi-stage
communication networks are often one-ported, whereas torus-based
systems are most often $2d$-ported and can approximately support
communication with all torus neighbors in a step.

Processors in a communication network can work independently and
concurrently. For the analysis of communication algorithms, we count the
total number of steps in which processors are communicating that are
required for solving the given problem, \ie, for the last processor to
finish. In each step, some or all of the processors in the network may
be involved. Sometimes, actually often, steps where many pairs of processors
are communicating are called \emph{rounds}.

Interesting communication problems often correspond to parallelization
patterns that are useful in complex
algorithms and applications, for instance, broadcasting data from one
processor to other processors, exchanging information between all
processors, \etc (see \Sec~\ref{sec:exchangepatterns}).
In any such communication pattern that involves
transmission of data from a processor $u$ to a processor $v$ where
$\dist(u,v)=\diam(G)$, an obvious lower bound on the number of steps (rounds)
required to complete the pattern operation is $\diam(G)$. One such
pattern is the broadcast operation which we formalize as the following
communication problem.

\begin{definition}[Broadcast problem]
  \index{broadcast problem}
  \label{def:broadcast}
  Let $G=(V,E)$ be a communication network, and $r\in V$ a given
  \emph{root processor} which has some indivisible unit of data that
  needs to be transmitted to all other processors $u\in V$. The
  \impi{broadcast problem} is to devise for a given network $G=(V,E)$
  and any root $r\in V$ an algorithm with the smallest possible number of
  communication rounds that transmits the data unit from $r$ to the other
  processors of $G$.
\end{definition}

Both the (structure and capabilities of the) network $G$ and the
chosen root processor $r$ are known to all processors and can be used
in the algorithm.  A solution to the broadcast problem for a given
class of communication networks is an algorithm that solves the
problem for the given root $r$ and describes the communication steps for
each processor, together with a proof that the algorithm completes in the
claimed number of communication rounds.  In particular, $G$ is not
part of the input but fixed (given) and can be used in the algorithm
design, whereas the root processor $r$ is usually taken to be an input
parameter which is, however, known to all processors.

In tree, torus, and hypercube networks, the diameter lower bound
argument gives a non-constant bound on the number of rounds
needed to solve the broadcast problem. It depends on the number of
processors in the network. But even
in a fully connected network with constant diameter one, the number of
communication rounds is non-constant, as captured by the following,
important statement.

\begin{theorem}
  \label{thm:broadcastlowerbound}
  In a fully connected, $p$-processor network $G=(V,E), p=|V|$
  with $k$-ported, unidirectional communication capabilities for $k\geq 1$,
  the number of communication rounds necessary and
  sufficient for solving the broadcast problem is $\ceiling{\log_{k+1}
    p}$.
\end{theorem}

The proof for the lower bound part of the claim is the following
information-theoretic argument. The best that an algorithm that solves
the broadcast problem can do is the following. In the first
communication step, only the root processor has the data which it can
disseminate to at most $k$ new processors that so far
did not have the data. In the next round, the best that each of the
$k+1$ processors that now have the data can do is to disseminate the
data to $k$ new processors that so far did not have the data. Therefore,
after the second round, the number of processors that have the data are 
at most $(k+1)+(k+1)k=(k+1)^2$. In summary,
from one communication round to the next, the best that an algorithm
can achieve is that a factor of $k+1$ more processors will have the
data. The smallest number of communication rounds $i$ that are
required for all processors to eventually receive the data is found by
solving $(k+1)^i\geq p$ which by taking the logarithm
on both sides gives $i\log(k+1)\geq \log p$. Thus, $i\geq
\ceiling{\log_{k+1}p}$ since the solution (number of rounds) must be
integral.

The argument almost immediately leads to an algorithm that matches
this lower bound. Partition the
communication network into $k+1$ pieces of roughly the same number of
processors. The root processor $r$ belongs to one of these pieces; for
the other pieces, a virtual root processor is chosen. The processors
must be able to do this with no communication, based on the
information they have on the identity of $r$ and the fact that $G$ is
fully connected. The root $r$ sends the data it has to the $k$
virtual roots. The broadcast problem has now been reduced to $k+1$
proportionally smaller broadcast problems, still on fully connected
networks, which can be solved recursively, in parallel. The
number of recursive steps needed for all pieces to have been reduced
to a single processor is $\ceiling{\log_{k+1}p}$ after which the data have
been broadcast.

Good and even optimal solutions for the broadcast problem, in the
sense of matching a known lower bound, are known for many types and classes of
networks, like trees, tori, and hypercubes (and many, many others),
but not always trivial. Efficient algorithms for broadcast and other
collective communication problems are not the subject of these lectures.

The broadcast problem for an arbitrary graph $G$, now as part of
the input, is NP-complete~\cite[ND49]{GareyJohnson79}.

The bisection width of a communication network gives a lower bound on
the number of communication rounds required for another important
communication problem.

\begin{definition}[All-to-all problem]
  \label{def:alltoall}
  Let $G=(V,E)$ be a communication network and assume that each
  processor $i\in V$ has, for each other processor in $G$, specific data that
  have to be sent to that processor.  The \impi{all-to-all problem} is
  to devise for a given network $G=(V,E)$ an
  algorithm with the smallest possible number of communication rounds that
  transmits all data from all processors to all other processors.
\end{definition}

The \impi{all-to-all problem}, also called personalized or individual
exchange, is a highly communication intensive problem. All processors
have distinct data for each of the other processors, so for each
processor $|V|-1$ data have to be sent and received. The
total communication volume is thus $|V|(|V|-1)$ data. What is the
smallest number of communication rounds required to handle this
volume?  Partition the set of processors into two roughly equal sized
sets of $|V|/2$ processors (for simplicity, we assume that $|V|$ is
even). The volume of data to be exchanged between the two sets is
$|V|^2/4$, independent of how the processors were partitioned, since the
all-to-all problem is symmetric. Now let the partition of the
processors be the partition that corresponds to the bisection width
$\bisec(G)$ of the network $G$. Since there are only $\bisec(G)$ links
connecting the two parts, each of which can carry data in a
communicating step, the required number of steps for any algorithm
solving the all-to-all problem is $\frac{|V|^2}{4\bisec(G)}$.

\begin{theorem}
  \label{thm:bisectionlowerbound}
  Let $G$ be a direct communication network with bisection width $\bisec(G)$.
  The number of communication rounds necessary to solve the all-to-all problem
  is at least $\frac{|V|^2}{4\bisec(G)}$.
\end{theorem}

For the fully connected network with the highest possible bisection
width, the all-to-all problem could possibly be solved in a single
communication round. This would, on the other hand, require that each
processor can communicate with all other processors in a single step,
which is not realistic. The bisection width lower bound alone is most
often too optimistic and not the sole limiting factor on achievable
all-to-all communication performance. On the other hand, a poor network
with constant bisection width (independent of the number of
processors) like a ring or a tree would need a quadratic number of
communication steps (in the number of processors) for all-to-all
communication, and there is nothing that can be done about that.

\subsection{Concrete Communication Costs}
\label{sec:communicationcosts}

Communication mostly involves not only small, indivisible units of
information, but (complex) data of some size $m$ (bytes, integers,
other relevant, but explicitly stated unit). What is the concrete,
real cost (in time) of transmitting such data between processors in
the network?

As a first shot, often a simple linear-affine time cost model is adopted.
The \impi{linear-affine transmission cost model}
states that transmitting $m$ units, $m\geq 0$,
from $u$ to $v$ in $G=(V,E)$ along a communication edge $(u,v)\in E$ takes
\begin{displaymath}
  \alpha+\beta m
\end{displaymath}
time units, where $\alpha$ is a fixed, \impi{start-up latency}
independent of $m$ (for the given network, so perhaps dependent on
$p$) and $\beta$ a \emph{time per unit} of data transmitted.

The linear-affine time cost model is a crude, first, and perhaps even
misleading approximation of the cost of communication between
processors in a network or distributed memory \parco
system. Nevertheless, for lack of better, such a model is (tacitly)
assumed in the analysis of the distributed memory algorithms in these
lecture notes.  The model correctly emphasizes that communication
takes time, both in terms of cost per transmitted unit and latency
and reminds us that both of these terms can be considerable.
In particular, since
\begin{displaymath}
  \alpha+\beta m \leq \sum_{i=0}^{k-1}(\alpha+\beta m_i) = k\alpha+\beta m 
\end{displaymath}
the model stresses that splitting a message of $m$ units into $k$
smaller messages of sizes $m_i,0\leq i<k$ with $m=\sum_{i=0}^{k-1}m_i$
can be detrimental to communication performance. Conversely, combining
smaller messages into larger ones can, whenever this is possible, be
of advantage by saving latencies (see next section).  The linear-affine
transmission cost model is a \emph{homogeneous} model and treats all
pairs of processors the same by ignoring their placement in the
network (distance in network, placement in shared memory compute
node). It also abstracts away routing and overall traffic (contention,
congestion) in the network, which will be treated next.

\subsection{Routing and Switching}
\label{sec:switching}

In a not fully connected network $G=(V,E)$ where not every processor
can communicate directly with any other processor, a general purpose
\impi{routing system} (\impi{routing algorithm}, \impi{routing protocol})
makes it possible for any processor $u\in V$ to send data
to any other processor $v\in V$ via some path of intermediate
processors in $V$. In a sense, the routing system turns a not fully
connected network of processors into a virtually fully connected
network, where any processor can communicate directly with any other
other processor, however, not necessarily at the same cost of
communication (see \Sec~\ref{sec:communicationcosts}). A routing
algorithm could be \emph{centralized}\index{routing!centralized}, but,
typically, routing is thought of as a \disco problem. A routing
protocol or scheme consists of a set of local, per processor/switch
algorithms, each making decisions on what to do with a received
message based on its own state and possibly the state of some of the
immediately adjacent processors or switches (local information). Sometimes,
parallel algorithms are designed without a routing system by
explicitly describing how processors communicate with each other
and along which paths. Such an approach can make it possible to give more
precise, better bounds on the expected running time but is not general
purpose and comes with a high design cost (specialized algorithm).  A
routing system may be realized in hardware, in software, or in a
combination of hard- and software. This is why the term routing
\emph{system} is used. Designing and analyzing routing algorithms for
different types of graphs is a typical \disco topic (recall
Definition~\ref{def:disco}\index{Distributed Computing}), but routing
systems and algorithms are not a topic of these lectures. A few terms
are useful, though.

The most important requirement to a routing system (algorithm,
protocol), is \impi{deadlock freedom}: A message sent from a processor
$u$ to a processor $v$ must eventually arrive correctly (and uncorrupted)
at processor $v$, \emph{regardless} of any other traffic in the
communication network. A deadlock\index{deadlock}
could arise when two processors or
network elements at the same time require a certain resource and mutually
block each other. For instance,
they could want to send data to the same processor or switch element,
possibly over the same network edge resulting in a conflict that
cannot be resolved. It may also be seen
as the task of the routing system to ensure \impi{reliable communication}
meaning that no data or parts of messages are lost,
no data are corrupted, and perhaps even that
data are delivered in some specific order (as must, for instance, be
guaranteed by \mpi, see \Sec~\ref{sec:pointtopoint}). This is
important, since network hardware does not always guarantee such
properties (think of reasons why this may be the case).

A routing system should be (as) fast (as possible). In the linear-affine
time cost model, routing data of $m$ units from processor $u$ to
processor $v$ along a path of length $l$ would take $l(\alpha+\beta
m)$ time units. For sufficiently large numbers of data units,
this can be improved by \impi{pipelining} as follows:
The $m$ units are separated into
smaller \impi{packets} of some maximum size of $b$ units (assuming
$m>b$) that are sent one after the other. The time for the last packet
to arrive at the destination processor $v$ would then be

\begin{eqnarray*}
l(\alpha+\beta b) + (\ceiling{m/b}-1)(\alpha+\beta b) & = & 
(l+\ceiling{m/b}-1)\alpha + \beta (l-1)b+\beta m \\
& = & (l-1)\alpha + \ceiling{m/b}\alpha+\beta (l-1) b + \beta m
\end{eqnarray*}

The first $l(\alpha+\beta b)$ term on the
left-hand side is the time for the first packet
to arrive at $v$. The $(\alpha + \beta b)$ factor in the second term
is the time for each following packet,
of which there are $\ceiling{m/b}-1$ remaining in total. Simplifying the
expression, we see that sending all
$\ceiling{m/b}$ packets has a cost of $\beta m$ since the last packet
may be smaller than $b$ units. If the packet size $b$ can be chosen
freely, a best possible packet size that minimizes the total transmission
time can be found by (calculus or) balancing the terms
$\ceiling{m/b}\alpha$ and $\beta (l-1) b$, which both depend on
$b$, against each other. That is, we solve $\ceiling{m/b}\alpha=\beta (l-1) b$
for $b$ which yields a best packet size $b$ of
\begin{eqnarray*}
  b & = & \sqrt{\frac{m}{l-1}}\sqrt{\frac{\alpha}{\beta}}
\end{eqnarray*}
and a shortest transmission time of
\begin{displaymath}
(l-1)\alpha + 2\sqrt{(l-1)m}\sqrt{\alpha\beta} +\beta m
\end{displaymath}
provided that $l>1$. The important result is that
with pipelining\index{pipelining}
the $\beta m$ term is not linearly dependent on the path
length $l$. Furthermore, the constant factor in the $\beta m$ term is
one, meaning that the network is utilized to the fullest extent: We pay only
the cost per unit $\beta$ per data element of $m$ once.
Routing with pipelining is sometimes called
\impi{packet switching}, whereas routing without pipelining is
called \impi{store-and-forward}. Both store-and-forward and packet
switching routing require some intermediate buffer space in the
routing system, either for all $m$ data units or for a block of up to
$b$ units. These and other terms are used somewhat differently in
different fields, depending on the level at which the network is
examined, the use (internet computing is different from \parco!),
tradition, and many other factors~\cite{TanenbaumWeatherall11}.

In a communication network there may be several, partially different
paths from a processor $u$ to a processor $v$. When data are 
sent from processor $u$ to processor $v$, the routing system chooses
an appropriate path. This choice, of course, depends on $u$ and $v$ and
the network topology $G=(V,E)$, but may also depend on the current
traffic in the system; that is, concurrent communication between other
processors.

With
\emph{deterministic (oblivious) routing}\index{deterministic routing}\index{oblivious},
the route (path to be taken) is determined
solely by the endpoints $u$ and $v$ and the structure of the network
$G$, whereas network traffic plays no role.
With \impi{adaptive routing}, the routing system takes other, concurrent
communication into account.
Thus, the route from $u$ to $v$ can be different from time to time.
A routing algorithm is said to use \impi{minimal routing} when
routing from $u$ to $v$ is always along a shortest path (of
length $\dist(u,v)$).  When
several paths are possible and pipelining\index{pipelining} (packet-switching)
is employed, it may be that different blocks (packets) are taking
different routes. In such cases, packets could potentially arrive at
the destination processor $v$ in a different order than the order in
which they were sent from the source processor $u$. It is then the
task of the routing system to assemble the packets in the
right order at the destination.

In the presence of traffic in the communication network due to many
pairs of processors communicating at the same time, the optimistic,
model based estimate of the transmission time from $u$ to $v$ will most likely
not hold, and data communication times will (for some pairs of
processors) be much higher. This can be due to network \impi{contention}, for
instance, on edges $(u,v)$ that occur in many paths and are needed
by several pairs of processors or to resource \impi{congestion} because of
a too high load on intermediate buffers or processors.
The best that can be hoped for in such cases is a serialization slowdown
proportional to the contention or congestion. The routing system
can apply different strategies to alleviate and control contention and
congestion, typically some form of \impi{flow control}.

\subsection{Hierarchical, Distributed Memory Systems}
\label{sec:hierarchical}

In modern \parco systems, the communication system has a
more complex, hybrid structure, consisting of communication networks
at different levels. Thus, a single, unweighted graph that alone
describes the topology of the whole system may not be adequate or
helpful.

A two-level hierarchical system, for instance, could consist of a
number of shared memory compute nodes, interconnected by a, typically,
indirect network. Thus, processor-cores\index{processor-core}
within the same shared memory
compute node may communicate with different communication characteristics than
processor-cores residing on different compute nodes. In particular, if
several processor-cores on the same compute node at the same time need
to communicate with processor-cores on other compute nodes, they will
have to share the network that connects the compute nodes. There will
be congestion on the node (connection to the network)
and possibly contention in the network as well. The simple, latency $\alpha$
and cost per unit $\beta$, linear-affine transmission
cost model\index{linear-affine transmission cost model} breaks down.

\subsection{Programming Models for Distributed Memory Systems}

Programming models\index{programming model}
for distributed memory systems usually abstract
away from concrete network properties as discussed in the previous
sections. They assume that the active entities of the model (processes,
threads, \ldots) can freely communicate as in a fully connected
network. Often, they come without a concrete cost model that says
at what cost (in time) active entities can communication with each other.
Processes are usually not synchronized, operate on local data
that are invisible to other processes (shared nothing) and follow
instructions of
local programs (SPMD\index{SPMD} or MIMD\index{MIMD}). They cooperate
(exchange data, synchronize) with the other processes by explicit or
implicit communication.  Distributed memory
programming models\index{programming model}
usually assume that message transmission between processes and threads
is deadlock free\index{deadlock free}, reliable\index{reliable communication}
and correct and typically ordered according to certain
constraints and rules. For the implementation of such programming
models, the runtime system and routing algorithms need to
ensure reliable message delivery between any processes in the
model. Distributed memory programming models sometimes provide means
for reflecting and exploiting properties of the underlying,
hierarchical communication system.  The programming model underlying
\mpi\index{MPI} is a good example.

A concrete implementation of a distributed memory
programming model\index{programming model},
like \mpi, has concrete costs in time (and in other factors) for the
different types of communication offered by the model.  If one
benchmarks communication performance (time) under different loads and
between processes residing in different parts of the system, network
and system properties will become manifest. Concrete communications
do usually not obey a simple, homogeneous cost model.  Such
system and implementation dependent differences may or, rather, may
not be reflected in the cost models for the programming model.  \mpi
does not come with a cost model at all. Therefore, strictly speaking,
all cost analysis of \mpi programs will be based on external
assumptions, benchmarking results, and known system properties. Simplified
assumptions are often made (homogeneous communication with linear-affine
transmission costs), though, and can be indicative of actual performance, but
can also be off target or even grossly misleading.

Distributed system programming models\index{programming model} can be
classified as either
\emph{data distribution centric}\index{distribution centric} or
\impi{communication centric}.  In a data distribution centric model,
the data structures defined by the model (arrays, multi-dimensional
arrays, vectors, matrices, tensors, complex objects, \ldots) are
distributed according to given rules across the processes (see
\Sec~\ref{sec:datadistributions}).  When a process accesses or updates
a part of a distributed data structure that resides with another
process, communication and possibly ``remote'' computation is
implied. When a process on the other hand accesses local data
``owned'' by itself, it can itself perform the specified
computation. This is often called the \impi{owner computes} rule.  A
communication centric model on the other hand usually does not define
distributed data structures. Such models instead focus on properties
of explicit communication and synchronization operations.

Examples of data distribution centric\index{distribution centric}
models are so-called
\emph{Partitioned Global Address Space} models (\impi{PGAS}). In such
models, data structures, typically simple $1,2,3,\ldots$-dimensional
arrays, can be distributed across threads (processes), and access to
non thread-local parts of arrays implicitly leads to communication.
Otherwise, computations are done following the owner computes rule. An
example implementation of a PGAS model is \impi{Unified Parallel C}
(\impi{UPC})~\cite{ElGhazawiCarlsonSterlingYelick05}. PGAS models and
languages will not be treated further in these lectures.
\mpi is, on the other hand, a communication centric model.

\section{Ninth block (3--4 lectures)}
\label{blk:mpi}

Our concrete example of a distributed memory programming interface
implementing a communication centric, distributed memory programming
model\index{programming model} is \impi{MPI}, the
\emph{Message-Passing Interface}~\cite{MPI-3.1,MPI-4.1,MPI96,MPI98:I,MPI98:II}.
\mpi\index{MPI} is an
older interface dating back to around 1992. It is (nevertheless, still)
widely used, especially in
HPC,\index{HPC} and relevant to study and learn because of the concepts it
introduces. \mpi is an interface for C
and Fortran (still an important programming language in HPC). \mpi is
maintained and developed further by the so-called \mpi Forum, an open
forum of academic institutions, laboratories, compute centers and
industry. Incidentally, many of the \mpi Forum members
are or were also part of the \openmp\index{OpenMP}
ARB\index{Architecture Review Board}. The standard is freely
available and can be found via~\url{www.mpi-forum.org}. These pages
also give information on the standardization process (currently
towards \mpi 4.2).

The reference for programming (and learning) \mpi is the latest
version of the standard~\cite{MPI-4.1}.  Some helpful reading are the
series of books on ``Using MPI''~\cite{MPI94,MPI99,MPI14}. Many
elementary textbooks on parallel programming, \eg, the books by Rauber
and R\"unger and
Schmidt~\etal~\cite{RauberRunger13,SchmidtGonzalezHundtSchlarb18} deal
extensively with aspects of \mpi.

This block of lectures gives an introduction to \mpi for \parco,
covering all its fundamental concepts and features. Some aspects of
\mpi will not be dealt with, most notably support for I/O, dynamic
management of processes (spawning and joining communication domains, see
later), and tool building (highly important, though).

\subsection{The Message-Passing Programming Model}

The message-passing programming model\index{programming model}
goes way back, at least to papers
by Dijkstra and Hoare in the 60ties and 70ties. The idea is that
parallel computations can be structured as sequential processes with no shared
information that communicate explicitly by sending and receiving
\emph{messages} between each other~\cite{Hoare78,Hoare85}. Restricting
interaction between the sequential processes to explicitly specified
(synchronous) communication operations was seen as a means
to develop provably correct, parallel and concurrent programs.  This 
message-passing model was called \emph{Communicating Sequential Processes}
(\impi{CSP}). CSP programs, in particular, cannot have data
races. The programming model\index{programming model}
that is implicitly behind \mpi is much
wider in scope than CSP. It incorporates both synchronous and
asynchronous point-to-point communication (CSP\index{CSP} focussed on
synchronous, handshaking communication), one-sided communication and
collective communication. It also provides features for data layout
description and interaction with the communication system and the external
environment (file system I/O).

Some main characteristics of the \mpi\index{MPI}
message-passing programming model\index{programming model} are:
\begin{enumerate}
\item
  Finite sets of processes in immutable
  \emph{communication domains}\index{communication domain}.
  Processes in the same domain can communicate with each
  other. The same process can belong to several communication domains.
\item
  In communication domains, processes are identified by their rank.
  Ranks are consecutive $0,\ldots p-1$,with $p$ being the number of
  processes in the communication domain (size). A process can
  have different ranks in different communication domains.
\item
  New communication domain are created from existing communication domains
  and a default domain consisting of all externally started processes.
\item
  Processes operate exclusively on local data.
  All communication between processes is explicit.
\item
  Communication is reliable\index{reliable communication} and ordered.
\item
  Communication is network oblivious\index{oblivious}
  and possible between all processes.
\item
  Three basic communication models:
  \begin{enumerate}
  \item
    Point-to-point communication between pairs of processes with different
    modes, non-local and local completion semantics.
  \item
    One-sided communication between one process and another with different
    synchronization mechanisms, local and non-local completion mechanisms.
  \item
    Collective communication, many different operations,
    non-local (and local) completion semantics.
  \end{enumerate}
\item
  Structure of communicated data is orthogonal to communication model and mode.
\item
  Communication domains may reflect the physical network topology and
  communication system.
\end{enumerate}

\mpi has no performance model, and there are no prescriptions in the
\mpi standard on how the many, many different \mpi constructs are to
be implemented nor on which algorithms are to be used, in particular,
not for the collective communication operations. Thus, detailed
(asymptotic) performance analysis of \mpi programs must make external
assumptions and informed guesses on how specific features are
implemented and how they perform.

However, \mpi is designed with the intention of making
high-performance implementations possible on wide ranges of \parco systems,
meaning that the functionality and semantics are
close to what an underlying communication system will offer, that
preprocessing and communication of meta-information is not necessary
(or strictly confined), and that memory required by library internals
is bounded and/or can be controlled. These design objectives explain
the concrete ``look-and-feel'' of the many \mpi functions\index{MPI}.

\subsection{The \mpi Standard}

The \mpi\index{MPI} standard is largely a well-reasoned, semantic
specification of the large set of \mpi operations.  The \mpi standard is
an open standard maintained by the so-called \mpi Forum, which in
principle anybody can join; see \url{mpi-forum.org} for the rules and
current discussions on the standard. The current version of the
standard is MPI-4.1~\cite{MPI-4.1}. The standardization efforts over
the past $20$ years have mostly resulted in extensions,
additions, and clarifications that maintain backward compatibility to
the original standard published in 1993. This may change.

\subsection{\mpi in C}

\mpi is defined and implemented as a library, and \mpi functionality
can be used by linking code against a concrete \mpi library\index{MPI}.
There are
several such libraries available, notably the open source libraries
\mpich, \mvapich, and \openmpi as well as vendor libraries, often for
specific High-Performance Computing systems.  C code using \mpi must
include the \mpi function prototype header with the
\texttt{\#include <mpi.h>}
preprocessor directive. All \mpi relevant functions and
predefined objects are prefixed with \texttt{MPI\_}, which identifies
the \mpi ``name space''.  It is considered illegal and is in any case very bad
practice to use the
\texttt{MPI\_} prefix for own functions or objects in the code. \mpi
programs are usually compiled with a special compiler (wrapper,
mostly) that takes care of proper linking against the \mpi library.
The typical example is \mpicc, which will also accept standard optimization
options and arguments.

We explain the \mpi functions by listing the C prototypes with the C
types for all arguments and explain the outcome for given inputs, loosely
a before-after explanation.

\mpi functions return an error code, and it is good practice to check
the error code (which is very often not done). The error code \mpisuccess
indicates success of the call.

\subsection{Compiling and Running \mpi Programs}

An \mpi\index{MPI} program is, unlike an \openmp\index{OpenMP} program,
simply a C
(or Fortran) program with library calls to \mpi functions. \mpi
programs can therefore be compiled with a standard C
compiler. Usually, \emph{an \mpi program} means a single program that
will be run by all started processes: Mostly, \mpi code follows the
SPMD\index{SPMD} paradigm.  It is possible, though, to let different
\mpi processes run different programs.  To facilitate linking against
the \mpi library, normally an \texttt{mpicc} compiler command that is
just a wrapper around the C compiler command is provided. It takes the
standard C compiler flags and options.

Running an \mpi program with a desired number of processes is somewhat
complex.  Resources, cores and compute nodes, for the processes must
be allocated and the processes started at the allocated compute
resources. For small, stand-alone systems (say, laptop or workstation,
small server) this is often done from the command line with a
command like \texttt{mpirun}. More commonly, and on larger systems, a
batch scheduling system like \texttt{slurm} is used.

When processes have been started, they become \mpi processes after
having initialized the \mpi library. In the \mpi context, processes
are most often bound (``pinned'') to specific
processor-cores\index{processor-core} or at
least to compute nodes.
This binding\index{pinning} is outside the control of \mpi.

It is usually possible to start more \mpi processes than there are
physical processor-cores in the system (which can be useful when
developing programs on a small system).
But as with \openmp\index{OpenMP} and \pthreads, \index{pthreads}
such \impi{oversubscription} must be used with care and is often
detrimental to performance.

\subsection{Initializing the \mpi Library}

After the processes are started on the system, the internal data
structures of the \mpi\index{MPI} library must be initialized. This
done by the \mpiinit call which takes the standard C argument count
and argument array as arguments. The C argument count and full
argument array are normally copied to all started \mpi processes, so
there is no need to transfer this information explicitly with \mpi
communication operations.  After use, all activity of the \mpi library
is completed and resources freed with an \mpifinalize call, which
should not be forgotten: The program may otherwise terminate
improperly. Prior to \mpiinit and after \mpifinalize, no \mpi calls
can be performed, except for the two check calls \mpiinitialized and
\mpifinalized that tell the caller (perhaps an application specific
library written with \mpi with its own initialization function)
whether \mpi has been initialized or completed. When the \mpi library
has been finalized, it cannot be initialized again within the same
program.

\begin{lstlisting}[style=SnippetStyle]
int MPI_Init(int *argc, char ***argv);
int MPI_Finalize(void);
int MPI_Finalized(int *flag);
int MPI_Initialized(int *flag);

int MPI_Abort(MPI_Comm comm, int errorcode);
\end{lstlisting}

The \mpiabort call can be used to force termination of the running \mpi
program in an emergency situation. 

An \mpi library can provide limited information about itself and
its environment by the following operations.

\begin{lstlisting}[style=SnippetStyle]
int MPI_Get_version(int *version, int *subversion);
int MPI_Get_library_version(char *version, int *resultlen);

int MPI_Get_processor_name(char *name, int *resultlen);
\end{lstlisting}

These calls illustrate the tediousness of \mpi being a library (and
the shortcomings of C for manipulating strings): For the strings
\texttt{version} and \texttt{name}, the user must reserve space of at
least \mpimaxversion and \mpimaxprocessor characters,
respectively. The strings are copied into these arrays, in C properly
terminated by a null character. The number of actual characters,
excluding the trailing null character, will be stored in
\texttt{resultlen}. Thus, in C, output arguments (result values) are
always of pointer type.

A process can read the wall-clock time\index{wall clock time}
from some time point in the past (in
seconds). The timers are local and (usually) not synchronized across
processes and processor-cores\index{processor-core}.
The call can be used to time process-local operations and is
heavily used for this.

\begin{lstlisting}[style=SnippetStyle]
double MPI_Wtime(void);
double MPI_Wtick(void);
\end{lstlisting}

Whether the timers are synchronized (global) can be queried by reading
an attribute. The attribute mechanism of \mpi is not covered in these
lectures, although it is important for library building with
\mpi~\cite{Traff23:attributes}. The existence of the attribute
mechanism in the \mpi standard illustrates how \mpi is intended to and
can support portable, application specific library building; but also
the tediousness of \mpi being a library and not an integrated part of
a programming language: Information must flow in and out of the \mpi
library to and from the application (specific library) explicitly
through the \mpi functions. Information that the compiler possesses is
not known to the \mpi library, but must be explicitly transferred.

\subsection{Failures and Error Checking in \mpi}
\label{sec:errorhandling}

The \mpi functions are, at first sight (do not be scared!), often
quite involved and sometimes take confusingly long lists of arguments
that all must be used correctly. If an argument is not as specified
according to the precondition of the operation, there is no guarantee
that the function will have the specified effect and produce the
desired outcome or any useful outcome at all! \mpi libraries perform
only rudimentary argument checks (whether preconditions are fulfilled),
but the extent of this is not specified in the standard and \mpi
libraries differ in the amount and kinds of such checks
done. Sometimes, tools or options can be used to perform more
extensive checking, which can, of course, be helpful in the development
phase of an application. But the programmer can most surely not rely
on the \mpi library to catch mistakes and errors. The \mpi standard
specifically states~\cite[page 340]{MPI-3.1} that ``An \mpi
implementation cannot or may choose not to handle some errors that
occur during \mpi calls. [\ldots] The set of errors that are handled by
\mpi is implementation dependent. [\ldots] Specifically, text [in the
  standard] that states that errors \emph{will} be handled should be
read as \emph{may} be handled'' (emphasis original). The most recent
MPI 4.1 standard takes the same stance~\cite[Page 449]{MPI-4.1}.

As mentioned, almost all \mpi functions return error codes. It can
make sense to check those and try to take action on certain error
return codes. But there is no guarantee that this will be possible:
The application may have crashed before returning from the operation
and no error code will ever be seen by the user. This is still the
most typical \mpi behavior on (programming) errors and (hardware) faults.
Therefore, \mpi programs typically do only a limited amount of error
code checking and only of certain functions.
In particular, communication failures due to
processor/node crashes or failures in the communication system are
typically not handled and will in most cases cause the whole
application to abort. The own, self-inflicted, and most common reason
for an application to crash is memory corruption through wrong use of
\mpi functions leading to memory being overwritten and/or wrongly
addressed. Here, memory diagnostic tools that check bounds and
accesses can be most helpful.

Part of the reason for \mpi not doing extensive error checking and
handling is that \mpi is designed to allow for high-performance
implementations. It, therefore, does not impose (expensive, extensive)
checks for errors and wrong usage of the \mpi functions.

\mpi aims to make it possible to control the response of the library
in case of failures.  This is accomplished through
\impi{error handlers} which are special functions that can be attached to
communicator objects (see next section) and are invoked by the \mpi
library when an error condition occurs in an \mpi call on that
communicator object. Error handlers are beyond the scope of these
lectures.  The quotes from the \mpi standard cited above still apply.

\begin{lstlisting}[style=SnippetStyle]
int MPI_Errhandler_create(MPI_Handler_function *function,
                          MPI_Errhandler *errhandler);
int MPI_Errhandler_set(MPI_Comm comm,
                       MPI_Errhandler errhandler);
int MPI_Errhandler_get(MPI_Comm comm,
                       MPI_Errhandler *errhandler);
int MPI_Errhandler_free(MPI_Errhandler *errhandler);
\end{lstlisting}

\subsection{\mpi Concept: Communicators}
\label{sec:communicators}

After processes have been started and the \mpi library initialized with
\mpiinit, the started processes are put into a \impi{communication domain}
called \mpicommworld. In addition,
each process is also put into a domain by itself called
\mpicommself. A communication domain represents an ordered set of
processes that can communicate with each other, each process with any
other process in the domain, and only in that domain. A domain has a
\emph{size}, which we often denote by $p$, which is the number of
processes it contains.  Each process has a unique, non-negative
\emph{rank} $r$ in the domain, $0\leq r<p$.  In \mpi, communication
domains are called \emph{communicators}. A \impi{communicator} is a
distributed object that can be operated on by all processes
belonging to the communicator. A communicator is referenced by a
\emph{handle} of type \mpicomm.  In particular, processes can look up
the \emph{size}, \ie, the number of processes, in any communicator
\texttt{comm} to which they belong, and determine their
own \emph{rank} in the communicator by the following functions.

\begin{lstlisting}[style=SnippetStyle]
int MPI_Comm_rank(MPI_Comm comm, int *rank);
int MPI_Comm_size(MPI_Comm comm, int *size);
\end{lstlisting}

Thus, the code snippet
\begin{lstlisting}[style=SnippetStyle]
int rank, size;
  
MPI_Comm_rank(MPI_COMM_WORLD,&rank);
MPI_Comm_size(MPI_COMM_WORLD,&size);
assert(0<=rank&&rank<size); // the condition on ranks
\end{lstlisting}
will, when executed by any of the started \mpi processes, identify the
process relative to all other started processes by its serial number
(rank) in the \mpicommworld communicator. As good as any \mpi program will
have such a code sequence somewhere after the \mpiinit call, and the
processes decide what to do based on \texttt{rank} and
\texttt{size}. Here, as almost always, the error return codes of the two
function calls are ignored.

This trivial piece of code illustrates a number of important \mpi
concepts and principles.
\begin{itemize}
\item
  Processes belong to communication domains, which are called
  communicators\index{communicator} in \mpi.
  In particular, they belong to the \mpicommworld
  communicator consisting of all externally started processes created
  by the \mpiinit call.
\item
  Processes have a \emph{rank} (serial number) in a communicator.
  Ranks are consecutive from $0$ to $p-1$ where $p$ is the size of the
  communicator ($0\leq\texttt{rank}<\texttt{size}$).
\item
  The rank of a given process in \mpicommworld is determined by
  external factors like how the processes were started and where in
  the system the process is placed (processor-core, shared memory
  compute node, particular processor in the network) relative to the other
  processes. However, the rank of a process in a communicator will never
  change.
\item
  Communicators are identified by handles of type \mpicomm, which are
  \impi{opaque} objects on which certain operations are defined (that is,
  their internal composition is not specified and cannot be know to the
  application programmer).
\item
  There can be several communicators in an application, and the same
  process can belong to many communicators, possibly with a different
  rank in each.
\item
  The communicator is the most fundamental object/concept in \mpi:
  All communication is relative to a communicator, all collective
  operations (see later) are relative to a
  communicator\index{collective operation}. In
  particular, processes in different communicators cannot
  communicate, and simultaneous communication on different
  communicators can never interfere.
\item
  \mpi objects are static objects. They cannot be changed (only 
  freed when no longer of use).
  New objects, for instance communicators, can be created
  from already existing ones by appropriate functionality.
\end{itemize}

For any communicator there is a special process rank \mpiprocnull
outside the range from $0$ to $p-1$ that can actually be
referenced and used for non-communication: Communication with
\mpiprocnull has no effect (see later).

The principle that communication is always with respect to a
communicator and that communication between
processes in one communicator can never
interfere with communication between processes in another 
is fundamental. It is what allows construction of
\impi{safe, parallel libraries}.
If each library used in an application uses its own
communicator(s), communication going on in different libraries can
never interfere.

For library construction, the essential operation on communicators
is the creation of a duplicate communicator. The duplicate represents
a communication domain with the same set of processes in the same
order, but is nevertheless a different domain. Thus, communication on a
communicator and its duplicate can never interfere. The \mpicommdup
operation is shown below. It is the first example of a so-called
\impi{collective operation}, meaning that it has to be called by all
processes in the communicator \texttt{comm}.

\begin{lstlisting}[style=SnippetStyle]
int MPI_Comm_dup(MPI_Comm comm, MPI_Comm *newcomm);
int MPI_Comm_dup_with_info(MPI_Comm comm, MPI_Info info,
                           MPI_Comm *newcomm);

int MPI_Comm_split(MPI_Comm comm, int color, int key,
                   MPI_Comm *newcomm);
int MPI_Comm_create(MPI_Comm comm, MPI_Group group,
                    MPI_Comm *newcomm);
int MPI_Comm_create_group(MPI_Comm comm, MPI_Group group,
                          int tag, MPI_Comm *newcomm);
int MPI_Comm_split_type(MPI_Comm comm,
                        int split_type, int key,
                        MPI_Info info, MPI_Comm *newcomm);
\end{lstlisting}

The \mpicommsplit and \mpicommcreate functions allow to create new
communicators from existing ones, possibly with fewer processes and
possibly with a different order. Both calls are collective: All
processes belonging to the communicator in the \texttt{comm} argument
have to make the call. When the calls return, each calling process
will be part of the new communicator \texttt{newcomm} and still also
of the old \texttt{comm} communicator which was used to coordinate all
the processes making the call. The \mpicommsplit operation takes an
integer \texttt{color} argument and all processes with the same color
(argument) will end up in the same \texttt{newcomm} communicator. The
\texttt{key} argument can be used to control the numbering of the
processes in the new communicator in the following way.  Processes
with the same color are sorted non-increasingly by the \texttt{key}
argument and this determines their ranks in \texttt{newcomm}. The
process with the smallest key will become rank $0$, the process with
the next smallest key rank $1$, and so on.  Processes with equal
\texttt{key} arguments are kept in their rank order\index{rank order}
in the \texttt{comm} communicator. The special \mpiundefined argument
as color, indicates that a process calling with this color is not
going to belong to any communicator. Thus, \mpicommsplit is a very
flexible operation for partitioning an existing communication domain into
new communication domains.  The discussion again illustrates some
fundamental principles.

\begin{itemize}
\item
  \mpi functions have input and output arguments. Output arguments in
  C have pointer type (we already  saw this with \mpirank and \mpisize).
\item
  There are functions in \mpi that are \emph{collective}, meaning that
  they have to be called eventually by all the processes belonging to the input
  communicator\index{communicator}.
  In \mpi, collective functions are \emph{always} called
  symmetrically: All processes (in the communicator) make
  the same call with possibly different argument values. The input
  arguments given by a process determine the role of that process in
  the call.
\item
  On return from an \mpicommsplit call, each calling process will
  have, in addition to the still existing, unchanged input
  communicator \texttt{comm}, a new communicator \texttt{newcomm} to
  which it belongs together with all the other processes that called
  with the same \texttt{color} argument. The rank is given by
  the position in the list of processes with the same color sorted
  by the \texttt{key} argument.
\item
  After completion of a communicator creating operation, each calling
  process will (in case of \mpicommsplit) belong to two communicators,
  \texttt{comm} and \texttt{newcomm}, possibly of different sizes and
  possibly with a different rank in each.
\item
  New processes are neither created nor started by these
  functionalities. The communicator creating functions operate on a
  given set of processes represented by their ranks in an input communicator.
  Only ranks and sizes may be different in the created communicators.
  The calling processes remain alive as they are, and retain their rank
  in the input communicator used in the call.
\end{itemize}

The \mpicommcreate call likewise allows to create arbitrary new
communicators from existing ones. This is based on process groups, a new
\mpi concept that will be explained briefly later. As with \mpicommsplit,
the \texttt{newcomm} returned to some processes can be an invalid \mpicommnull
communicator, a communicator on which no (communication) operations can
be performed, and that can mostly
not be used as input argument to \mpi functions. The last two
operations, \mpicommcreategroup and \mpicommsplittype, albeit both useful
and semantically interesting, are not treated in these lectures.

In the following, we will see concrete examples of the use of
\mpicommsplit and \mpicommcreate, for instance, in the implementations
of Quicksort-like algorithms\index{Quicksort},
stencil computations\index{stencil computation}
and matrix--matrix multiplication\index{matrix--matrix multiplication}.

After use, a communicator is freed by the \mpicommfree call. Since
communicators are distributed objects, all processes in the
communicator have to eventually call \mpicommfree on the
communicator. That is, this is also a collective
operation\index{collective operation}.  A program where some processes
are correctly calling \mpicommfree on a created communicator and
others not may not be able to complete properly.

\begin{lstlisting}[style=SnippetStyle]
int MPI_Comm_free(MPI_Comm *comm);
\end{lstlisting}

A communicator typically is a ``costly object'' in terms of
required memory space (depending on the quality of the \mpi library
implementation). Also for that reason it is \emph{always} good
practice to free \mpi objects that are no longer going to be used.

There is sometimes helpful functionality in \mpi for comparing
two communicators.

\begin{lstlisting}[style=SnippetStyle]
int MPI_Comm_compare(MPI_Comm comm1, MPI_Comm comm2,
                     int *result);
\end{lstlisting}

The possible outcomes are \mpiident, meaning that the two input
communicators are indeed referring to the same object, \mpicongruent,
meaning that the two input communicators represent the same processes
in the same rank order, \mpisimilar, meaning that the two input
communicators represent the same processes but not necessarily in the
same order, and \mpiunequal for anything else.  A communicator and its
duplicate would, thus, be \mpicongruent, but not \mpiident.  This
functionality is typically for use in application specific libraries
and more seldomly directly used in applications. 

To illustrate the concepts and functionality introduced so far, a
first (and our only) full-fledged \mpi program follows below; in the
following examples we will skip header-files, \texttt{main()}-function
definitions, mostly also \texttt{rank}- and \texttt{size}-lookup,
\etc. The program creates a duplicate of the \mpicommworld
communicator from which it splits off a communicator with processes
ranked in reverse order.  It next partitions the \texttt{comm}
communicator into communicators containing the processes with even
rank (in \texttt{comm}) and the processes with odd rank. All processes
at this point belong to three new communicators (plus \mpicommworld
and \mpicommself), partly with different ranks. Finally, the program
creates a subcommunicator in which the process with the highest rank
in the calling communicator has been excluded by giving this process
(which has rank equal to \texttt{size-1}) the special color
\mpiundefined. This type of subcommunicator can be useful for
master-worker applications\index{master-worker} (see
\Sec~\ref{sec:masterworker}), in which the worker processes need to
communicate between themselves, for instance, by collective operations
(see \Sec~\ref{sec:collective}), without involving the excluded
``master'' process\index{collective operation}.  Note that the
program, including the assertions, is constructed in such a way that
it can run for any number of started \mpi processes.

\begin{lstlisting}[style=SnippetStyle]
#include <stdio.h>
#include <stdlib.h>
#include <string.h>

#include <assert.h>

#include <mpi.h>

int main(int argc, char *argv[])
{
  int rank, size;
  MPI_Comm comm, mmoc, evodcomm, workcomm;
  int result;
  
  MPI_Init(&argc,&argv);
  
  comm = MPI_COMM_WORLD;
  MPI_Comm_compare(comm,MPI_COMM_WORLD,&result);
  assert(result==MPI_IDENT);
  
  MPI_Comm_dup(MPI_COMM_WORLD,&comm);
  MPI_Comm_compare(MPI_COMM_WORLD,comm,&result);
  assert(result==MPI_CONGRUENT);

  MPI_Comm_rank(comm,&rank);
  MPI_Comm_size(comm,&size);

  MPI_Comm_split(comm,0,size-rank,&mmoc);
  MPI_Comm_compare(comm,mmoc,&result);
  assert(size==1||result==MPI_SIMILAR);

  MPI_Comm_split(comm,rank%2,0,&evodcomm);
  MPI_Comm_compare(comm,evodcomm,&result);
  assert(size==1||result==MPI_UNEQUAL);

  MPI_Comm_free(&mmoc);
  MPI_Comm_free(&evodcomm);

  MPI_Comm_split(comm,(rank==size-1 ? MPI_UNDEFINED : 1),0,
		 &workcomm);
  if (workcomm!=MPI_COMM_NULL) {
    MPI_Comm_compare(comm,workcomm,&result);
    assert(result==MPI_UNEQUAL);

    MPI_Comm_free(&workcomm);
  }
  
  MPI_Comm_free(&comm);

  MPI_Finalize();
  
  return 0;
}
\end{lstlisting}

\subsection{Organizing Processes\marksec}
\label{sec:organizingprocesses}

We touch briefly on convenient functionality to give more structure to
the organization of \mpi processes than just the rank in a
communicator.

A running example through this part of the lectures is the stencil
computation\index{stencil computation} (see \Sec~\ref{sec:stencil}).
In a large $d$-dimensional matrix, all entries have
to be updated according to the same stencil rule\index{stencil rule}
for each entry, for
instance, an average over neighboring elements ``up, down, left, right,
front, rear'' ($3$-dimensional example)~\cite{Sourcebook03}. This
update is iterated a large number of times, until some convergence
criterion is met. In a distributed memory, message-passing setting,
the matrix is conveniently cut into rectangular block-submatrices, one
submatrix for each process, with all submatrices being of roughly the
same size. We will return to this example shortly (in
\Sec~\ref{sec:semanticterms} and~\ref{sec:onesidedstencil}).

For the communication that is needed for a parallel implementation of
the stencil update, it can be convenient to be able to think of the
processes as points in a $d$-dimensional grid with integer
coordinates. Dedicated \mpi communicator creation functionality makes it
possible to organize the processes into such a $d$-dimensional grid
with sizes $d_i$ such that $d_0\times d_1\times \cdots d_{d-1}=p$ by
giving each process a $d$-dimensional coordinate vector describing its
position in the grid. A communicator with an imposed grid structure is
called a \impi{Cartesian communicator}. Cartesian communicators
are created and used with the functionality listed below.

\begin{lstlisting}[style=SnippetStyle]
int MPI_Cart_create(MPI_Comm comm, int ndims,
                    const int dims[], const int periods[],
                    int reorder, MPI_Comm *cartcomm);

int MPI_Cartdim_get(MPI_Comm cartcomm, int *ndims);
int MPI_Cart_get(MPI_Comm cartcomm, int maxdims,
                 int dims[], int periods[], int coords[]);
int MPI_Cart_sub(MPI_Comm cartcomm, const int remain_dims[],
                 MPI_Comm *subcomm);

int MPI_Topo_test(MPI_Comm comm, int *status);

int MPI_Dims_create(int nnodes, int ndims, int dims[]);
\end{lstlisting}

A Cartesian communicator\index{Cartesian communicator}\index{communicator},
the \texttt{cartcomm} returned by the
\mpicartcreate call, is like any other communicator and can be used
wherever a ``normal'' communicator could, but carries additional
information about the size of the grid, namely the number of
dimensions $d$ and the size along each dimension. The number of
dimensions $d$ is given as input \texttt{ndims}. The sizes of the
dimensions are stored in the input array \texttt{dims[]} with $d$
entries and $\texttt{dims[}i\texttt{]}=d_i$.
It must hold that $\prod_{i=0}^{d-1}\texttt{dims[}i\texttt{]}\leq p$
where $p$ is the size of the old communicator \texttt{comm}. If
$\prod_{i=0}^{d-1}\texttt{dims[}i\texttt{]}<p$ some processes in
\texttt{comm} will not be part of the new \texttt{cartcomm}
communicator. These processes will be returned the value
\mpicommnull. Thus, in the \texttt{newcomm} communicator, the product of
the dimension sizes equals the size of the communicator.
The Cartesian grid is the set of integer vector coordinates
\begin{displaymath}
  \{(c_0,c_1,\ldots,c_{d-1}) | 0\leq c_i <\texttt{dims[}i\texttt{]}, 0\leq i<d\}
\end{displaymath}
and each process in \texttt{cartcomm} is uniquely associated with
one such vector. The association of processes with coordinate vectors is by
\emph{row-major} assignment (``last coordinate changes the fastest'').
More precisely, a process with coordinates $(c_0,c_1,\ldots,c_{d-1})$
has rank $r$ with
\begin{eqnarray*}
r & = & \sum_{i=0}^{d-1}c_{i}\prod_{j=i+1}^{d-1}d_{j}
\end{eqnarray*}
where $d_j=\texttt{dims[}j\texttt{]}$ and the empty product
$\prod_{j=i+1}^{d-1}d_{j}$ for $i=d-1$ being $1$. The rank $r$ can, of
course, be computed in $O(d)$ steps (better than the $O(d^2)$ steps
implied by the formula).  When stored in a C array \texttt{coords[]},
the coordinates are stored as $\texttt{coords[}i\texttt{]}=c_{i}$
for $0\leq i<d$, and $\texttt{coords[}d-1\texttt{]}$ is the fastest
changing coordinate for $r=0,1,\ldots,p-1$. 

The \texttt{periods} array is a Boolean (0/1) array indicating whether
the grid is periodic in the $i$th dimension, $0\leq i<d$. Periodic in
the $i$th dimension means that a coordinate vector
$(c_0,\ldots,d_i,\ldots,c_{d-1})$ is treated as
$(c_0,\ldots,0,\ldots,c_{d-1})$, and $(c_0,\ldots,-1,\ldots,c_{d-1})$
as $(c_0,\ldots,d_i-1,\ldots,c_{d-1})$. The grid ``wraps around'' in
the $i$th dimension. A full \emph{torus}\index{torus} is a grid that is
periodic in all $d$ dimensions.

The placement of the \mpi processes in a grid via the \mpicartcreate
operation by intention carries with it an implied,
preferred communication pattern,
namely that each process is likely (in the application) to communicate
with its immediate neighbors in the grid along the $d$ dimensions. It
is implied that a process with coordinate vector
$(c_0,\ldots,c_i,\ldots,c_{d-1})$ will most likely communicate (only,
in a preferred way) with the $2d$ processes $(c_0,\ldots,c_i\pm
1,\ldots,c_{d-1})$ for each $i, 0\leq i<d$. If the grid is not
periodic in dimension $i$, then some neighbors might not exist,
which is represented by \mpiprocnull, the non-existent process
mentioned in \Sec~\ref{sec:communicators}.

The \mpicartcreate takes a new type of argument, the \texttt{reorder}
flag. Setting this flag allows the \mpi library to attempt to rerank
(reorder) the processes in the output communicator, so as to better
reflect the process communication pattern that is implied by the
process grid organization. We say that an \mpi process has been
\emph{reranked} in a new communicator created from an existing one if
the ranks of the process in the two communicators are different.  The
reorder idea is that processes that are expected to communicate by
being neighbors in the Cartesian grid are reranked to processes on
processor-cores\index{processor-core}
in the physical system that are also close to each
other, for instance, by having a direct communication
link. Concretely, processes with ranks $i$ and $j$ in
\texttt{cartcomm} that are Cartesian neighbors may have different
ranks $i'\neq i$ and $j'\neq j$ in \texttt{comm} for processes $i'$
and $j'$ that are ``physically close''.  Whether, how and to what
extent an \mpi library does such a reranking and what the benefits
will be in concrete applications is entirely implementation and
situation dependent.

The \mpicartget operations are, again, mostly for library building
purposes and can be used to query a communicator created with
\mpicartcreate for information (but, perhaps surprisingly, not for
information on whether any reordering was done). Whether the
communicator is indeed Cartesian can be checked with the \mpitopotest
operation which will in that case return the value \mpicart.

For setting up Cartesian communicators over an existing
communicator of size $p$ (that is, with $p$ \mpi processes), the \mpidimscreate
function can be helpful for factoring $p$ into $d$ factors that are
(as) close to each other (as possible).
The factors are returned in non-increasing order
in the \texttt{dims} input/output array that must be initialized to
non-negative values. Positive entries indicate factors that are
already set and fixed, so only
\begin{displaymath}
  \frac{p}{\prod_{\texttt{dims[}i\texttt{]}>0}\texttt{dims[}i\texttt{]}}
\end{displaymath}
will be factored over the zero-entries in \texttt{dims}. The denominator
must divide $p$ exactly.

\begin{lstlisting}[style=SnippetStyle]
int MPI_Cart_rank(MPI_Comm cartcomm, const int coords[],
                  int *rank);
int MPI_Cart_coords(MPI_Comm cartcomm, int rank, int maxdims,
                    int coords[]);
int MPI_Cart_shift(MPI_Comm cartcomm, int direction, int disp,
                   int *rank_source, int *rank_dest);
\end{lstlisting}

The functions \mpicartrank and \mpicartcoords are used to translate
between ranks and coordinate vectors. Cartesian communicators, in
combination with \mpicommsplit, will be used later to facilitate the
implementation of the SUMMA matrix--matrix
multiplication\index{matrix--matrix multiplication} algorithm
(see \Sec~\ref{sec:la}). The shift operation \mpicartshift can be used
to compute the ranks of processes along the $i$th \texttt{direction}
(dimension) by giving an integer, not necessarily non-negative
displacement that takes coordinate $c_i$ into
$(c_i+\texttt{disp})\bmod p$. We will see an example in
\Sec~\ref{sec:groups}.

Here is now a part of an \mpi program for setting
up (and freeing) Cartesian communicators\index{Cartesian communicator}
for all dimensions $d,0\leq
d<p$ where $p$ is the number of processes in a given communicator
\texttt{comm}, and verifying the row-major placement of the \mpi
processes in each of the created, non-periodic Cartesian grids:

\begin{lstlisting}[style=SnippetStyle]
MPI_Comm_size(comm,&p);

int reorder = 0; // no reorder attempt; but try
for (d=1; d<=p; d++) {
  int dims[d], periods[d];
  int coords[d];

  MPI_Comm cartcomm;
  int size;
  int r, rr, dd, i;

  for (i=0; i<d; i++) dims[i] = 0;
  MPI_Dims_create(p,d,dims);

  for (i=0; i<d; i++) periods[i] = 0;
  MPI_Cart_create(MPI_COMM_WORLD,d,dims,periods,0,&cartcomm);
  assert(cartcomm!=MPI_COMM_NULL);

  MPI_Comm_size(cartcomm,&size);
  for (r=0; r<size; r++) {
    MPI_Cart_coords(cartcomm,r,d,coords);
    for (i=0; i<d; i++) {
      assert(0<=coords[i]&&coords[i]<dims[i]);
    }
    assert(coords[d-1]==r%dims[d-1]);
    rr = 0; dd = 1;
    for (i=d-1; i>=0; i--) {
      rr += coords[i]*dd;
      dd *= dims[i];
    }
    assert(rr==r);
  }

  MPI_Comm_free(&cartcomm);
}
\end{lstlisting}

The idea of specifying a likely pattern of most intense communication,
based on which the \mpi library can attempt to rerank processes, is
generalized with the so-called \emph{distributed graph
communicators}\index{distributed graph communicator}\index{communicator}.
Such communicators are created by specifying a communication graph of
possibly weighted communication edges between processes. Edge weights
could model, for instance, communication volume between two processes,
or frequency of communication, or other properties of the
application, but the meaning of the weights is not specified in the
\mpi standard.  The specified communication pattern is used for two
purposes by the \mpi library.  First, by setting the \texttt{reorder}
flag to \textbf{true} (\texttt{=1}), the \mpi library can attempt to rerank
the processes (see the above discussion) such that process ranks that
are adjacent in the communication graph by heavy communication edges
are placed on processes that are ``close'' to each other in the
calling communicator \texttt{comm}. Second, the communication graph
defines the so-called \impi{neighborhoods} for a special kind of
collective operations\index{collective operation},
the so-called \impi{neighborhood collectives}
that are explained briefly in \Sec~\ref{sec:neighborhoodcollectives}.
The functionality is a bit tedious at first sight and listed here for
completeness, but not treated further in these lectures. To use it, it
is necessary to consult the \mpi
standard~\cite[Chapter 8, Chapter 6]{MPI-4.1}.

\begin{lstlisting}[style=SnippetStyle]
int MPI_Dist_graph_create(MPI_Comm comm,
                          int n,
                          const int sources[],
                          const int degrees[],
                          const int destinations[],
                          const int weights[],
                          MPI_Info info, int reorder,
                          MPI_Comm *graphcomm);
int MPI_Dist_graph_create_adjacent(MPI_Comm comm,
                                   int indegree,
                                   const int sources[],
                                   const int sourceweights[],
                                   int outdegree,
                                   const int destinations[],
                                   const int destweights[],
                                   MPI_Info info, int reorder,
                                   MPI_Comm *graphcomm);

int MPI_Dist_graph_neighbors_count(MPI_Comm graphcomm,
                                  int *indegree,
                                  int *outdegree,
                                  int *weighted);
int MPI_Dist_graph_neighbors(MPI_Comm graphcomm,
                             int maxindegree,
                             int sources[],
                             int sourceweights[],
                             int maxoutdegree,
                             int destinations[],
                             int destweights[]);
\end{lstlisting}

A distributed graph communicator can, like the case was for Cartesian
communicators, be queried\index{Cartesian communicator}.
The \mpitopotest operation will return the value \mpidistgraph.

Process ranking (reordering) in \mpi (sometimes called
\impi{process mapping}) via \mpicommsplit, \mpicartcreate,
and \mpidistgraphcreate
is always realized in the following way. The \mpi processes are bound
to processor-cores and compute nodes in the system. Processes are
(almost always) statically bound to some part of the system and do not
move. Each \mpi process belongs to one or more communicators.  What can be
different from one communicator to another is only the rank that a
process may have. So, in \mpi, processes are not moved or remapped,
but the ranks can change from communicator to communicator.
Assume that two processes in the input communicator
\texttt{comm} have rank $i$ and rank $j$ and are adjacent
(neighbors) in a distributed graph or Cartesian grid. In the
resulting, reordered communicator, the ranks $i$ and $j$ may now be
the ranks of processes (in \texttt{comm}) that happen to be close to each other
in the system, for instance, by residing on the same compute node. Thus,
process reordering and process mapping are both misnomers. The \mpi
mechanisms are merely reordering ranks (reranking).

Since processes themselves do not move, this means that possibly data
from the process with rank $i$ in the input communicator
\texttt{comm} may have to be transferred to the process that
has rank $i$ in the a newly created (Cartesian or graph)
communicator\index{Cartesian communicator}\index{distributed graph communicator}.
Should such data transfer
be necessary, the application programmer must implement it
explicitly. Therefore, programs often do the process mapping early in
the application before the processes generate or read much data.

To support moving data between communicators where the same
process may have different ranks in the communicators, \mpi provides
mechanisms for translating between ranks in one communicator and
another. Some will be described in \Sec~\ref{sec:groups}. The communicator
comparing function \mpicommcompare may also be of some use here to detect
whether process ranks are different in two communicators.

\subsection{\mpi Concepts: Objects and Handles}

The most important \mpi object is the \impi{communicator}. A communicator is
the concrete representation of an ordered domain of \mpi processes that
can communicate with each other. It is a \impi{distributed object},
meaning that it can be accessed and used by all the
processes that have a reference to the object. \mpi objects are
referenced via predefined \mpi handle types, of which there are quite
a few, but not all that many. \mpi objects can, like the
communicators, be distributed and accessible by a whole set of
processes, or be \emph{local objects}\index{local object} that are
only accessible by the single process having the handle to the object.

Handles are mostly \impi{opaque} (with one important exception that will be
treated next). Their implementation is unspecified in the \mpi
standard.  An object referenced by a handle can be accessed and used
only through the functions defined on the corresponding type of
handle. The most important \mpi objects and corresponding handles are
the following:

\begin{itemize}
\item
  \mpicomm for communicators, distributed (\Sec~\ref{sec:communicators}).
\item
  \mpiwin for communication windows, represent a communication domain
  and associated pieces of memory, distributed (\Sec~\ref{sec:onesided}).
\item
  \mpidatatype for so-called datatypes that describe process local
  layout and structure of data to be communicated, local
  (\Sec~\ref{sec:datatypes}).
\item
  \mpigroup for ordered sets of processes as an object that can be
  manipulated by process local operations, local
  (\Sec~\ref{sec:groups}).
\item
  \mpistatus for information returned from a (point-to-point)
  communication operation, local. This is the exception to the opaqueness
  property of handles (see shortly).
\item
  \mpirequest for information about a pending, possibly not yet
  completed communication operation (mostly point-to-point, but also
  collective and one-sided), local.
\item
  \mpiop for binary operators for the reduction collectives, local.
\item
  \mpierrhandler for action to be taken on discovery of an error or
  failure, see remark on error handling in \mpi
  (\Sec~\ref{sec:errorhandling}), local, and not treated further in these
  lectures.
\item
  \mpiinfo for specifying additional information when creating
  (certain kinds of) objects like distributed graph communicators and
  communication windows. Local, and not treated in these lectures.
\end{itemize}

\subsection{\mpi Concept: Process Groups\marksec}
\label{sec:groups}

Process groups are local objects with handle type \mpigroup that
represent ordered sets of processes. No communication operations are
defined on process groups; the groups are for processes to locally
compute other ordered sets of processes. Groups are used as input to a
number of other, often collective \mpi functions that involve many
processes as arguments. The \mpicommcreate operation for partitioning
a communicator as specified by process local groups of processes was
one example (\Sec~\ref{sec:communicators}).

Initialization of the \mpi library does not construct any
process groups in the way that \mpicommworld is
constructed. Instead, a local group object can be extracted from a
distributed communicator object.  The \mpicommgroup call is a
local operation that a process can perform on a communicator. It
returns the ordered set of processes of the communicator as a local
group object. A process can query its rank in a group. If it does not
belong to the group, the special value \mpiundefined is returned.

\begin{lstlisting}[style=SnippetStyle]
int MPI_Comm_group(MPI_Comm comm, MPI_Group *group);

int MPI_Group_rank(MPI_Group group, int *rank);
int MPI_Group_size(MPI_Group group, int *size);
\end{lstlisting}

Operations on groups are somewhat set-like, but the order plays a role.

\begin{lstlisting}[style=SnippetStyle]
int MPI_Group_translate_ranks(MPI_Group group1,
                              int n, const int ranks1[],
                              MPI_Group group2, int ranks2[]);

int MPI_Group_union(MPI_Group group1, MPI_Group group2,
                    MPI_Group *newgroup);
int MPI_Group_intersection(MPI_Group group1, MPI_Group group2,
                           MPI_Group *newgroup);
int MPI_Group_difference(MPI_Group group1, MPI_Group group2,
                         MPI_Group *newgroup);

int MPI_Group_incl(MPI_Group group, int n, const int ranks[],
                   MPI_Group *newgroup);
int MPI_Group_excl(MPI_Group group, int n, const int ranks[],
                   MPI_Group *newgroup);

int MPI_Group_range_incl(MPI_Group group,
                         int n, int ranges[][3],
                         MPI_Group *newgroup);
int MPI_Group_range_excl(MPI_Group group,
                         int n, int ranges[][3],
                         MPI_Group *newgroup);

int MPI_Group_compare(MPI_Group group1, MPI_Group group2,
                      int *result);

int MPI_Group_free(MPI_Group *group);
\end{lstlisting}

We give three important examples of uses of \mpi process groups. The
first example shows how to create a communicator that does not contain
a certain, specified process. This is helpful and sometimes needed for
applications following the \impi{master-worker} pattern (see
\Sec~\ref{sec:masterworker}) where one \texttt{master} process,
determined by its rank, has a special role and should be excluded from
communication between the non-masters (worker processes). Such a
communicator was also created in the last example of
\Sec~\ref{sec:communicators}.

\begin{lstlisting}[style=SnippetStyle]
MPI_Group group, workers;
MPI_Comm work;
  
master = ...; // some arbitrary master (rank) in comm
MPI_Comm_rank(comm,&rank);
MPI_Comm_size(comm,&size);

MPI_Comm_group(comm,&group); // get the group
MPI_Group_excl(group,1,&master,&workers); // exclude master
MPI_Comm_create(comm,workers,&work);
if (rank==master) assert(work==MPI_COMM_NULL); // excluded!
else { // relative order of worker processes preserved
  int r;
  MPI_Comm_rank(work,&r);
  if (rank<master) assert(r==rank); else assert(r==rank-1);
}
MPI_Group_free(&group);
\end{lstlisting}

The group of processes from the given communicator \texttt{comm} is
extracted, each process computes a group excluding the given master
process (given as a process rank between $0$ and the size of
\texttt{comm}), and this group is used as input argument
to the \mpicommcreate function. Each process computes the same
group. The \texttt{master} process, that is not part of the group, is
returned the \mpicommnull value, whereas the workers are returned a
handle to a new \texttt{work} communicator. This communicator can now
be used for any kind of communication supported by \mpi between the worker
processes.  In the
example in \Sec~\ref{sec:communicators}, we saw the same effect
achieved less tediously with the \mpicommsplit
collective operation\index{collective operation}.

The second example computes, for each process in a $d$-dimensional
Cartesian grid (communicator), a group consisting of the $2d+1$
neighboring processes along the $d$ dimensions, including the process
itself.  It is assumed that the arrays \texttt{dims} and
\texttt{periods} have been correctly and sensibly initialized (see the
example in \Sec~\ref{sec:organizingprocesses}) prior to the
\mpicartcreate call by all processes. All other variables are likewise
assumed to have been declared and sensibly initialized.

\begin{lstlisting}[style=SnippetStyle]
MPI_Cart_create(comm,d,dims,periods,0,&cartcomm);
assert(cartcomm!=MPI_COMM_NULL);

MPI_Comm_group(cartcomm,&group);

MPI_Comm_rank(cartcomm,&r);
k = 0;
neighbors[k++] = r;
for (i=0; i<d; i++) {
  MPI_Cart_shift(cartcomm,i,1,&r1,&r2);
  if (r1!=MPI_PROC_NULL) neighbors[k++] = r1;
  if (r2!=MPI_PROC_NULL&&r1!=r2) neighbors[k++] = r2;
}
assert(k<=2*d+1);
MPI_Group_incl(group,k,neighbors,&neighborgroup);
// neighborgroup now ready for use
\end{lstlisting}

The \texttt{neighborgroup} computed for each process contains
the local, implied grid neighborhood. This will be used later for
synchronizing one-sided communication operations (\Sec~\ref{sec:onesided}).

The third and last example shows how to translate ranks between two
communicators. Assume that a new communicator
\texttt{comm\_new} has been created out of an old one
\texttt{comm\_old} (with \mpicommsplit, \mpicartcreate,
\mpidistgraphcreate or other operation), possibly with reranking
and possibly fewer processes. What we would need to know is
this: For the process with rank $i$ in the old communicator,
what is the rank $j$ in the old communicator
of the process that has rank $i$ in the new communicator? This
information is needed in case data have to be transferred from process
$i$ in the old communicator to the process that now has rank $i$ in
the new communicator.

\begin{lstlisting}[style=SnippetStyle]
int i; // misused as array of one element in translation
MPI_Comm_rank(comm_old,&i);
MPI_Comm_group(comm_old,&group_old);
MPI_Comm_group(comm_new,&group_new);

MPI_Group_translate_ranks(group_new,1,&i,group_old,&j);

MPI_Group_free(&group_old);
MPI_Group_free(&group_new);
\end{lstlisting}

Since communication in \mpi must always be done between processes
belonging to the same communicator with that communicator as a handle
to the communication operations either the new or the old communicator
must be used for the data transfer.  After the rank translation call,
process $i$ in the old communicator can send its data to process $j$
(also in the old communicator), because process $j$ is the process
that has rank $i$ in the new communicator \texttt{comm\_new}.

\subsection{Point-to-Point Communication}
\label{sec:pointtopoint}

Processes that belong to the same communication domain by having a
handle to the same communicator can communicate with each other within
(or: relative to) that communicator.
We first describe the classical \mpi message-passing model of
point-to-point communication between pairs of processes.

It is important that \mpi communication between processes in a
communicator has no connectivity restrictions. Any process can
communicate with any other process as if the processes would be
running on processors in a fully connected network (See
\Sec~\ref{sec:structureandtopology}). It is the task of the \mpi
library and runtime (routing) system\index{routing system} to
facilitate such communication.  Recall that \mpi does not provide a
cost model for communication between processes and does not provide
any performance guarantees on commutation or other operations. In
particular, it cannot without further ado be assumed that
communication costs are homogeneous, that is, similar for any pair of
communicating processes regardless of their ranks in their
communicator.  Neither can it be assumed that communication costs are
independent of the overall traffic in the system, that is, concurrent
communication between other \mpi processes possibly in other
communicators (and even entirely unrelated processes from different
programs running concurrently on the system). Communication costs must
therefore be measured by appropriate benchmarks from which reasonable
assumptions may follow.

It is also important that communication in \mpi is always
reliable. This means that a transmitted message can \emph{always} be
assumed to arrive uncorrupted and in full. In case the \parco system
and communication network on which the \mpi program is running are not
reliable, it is again the task of the \mpi library and runtime system
to ensure reliable communication\index{reliable communication}.

Finally, point-to-point communication is \emph{ordered}. This means
that a sequence of messages sent from one process to another will
(eventually) become available at the receiving process in that
order.

In point-to-point communication, two processes are explicitly
involved. A sending process belonging to a communication domain
(communicator) specifies an amount of data to be sent to an explicitly given
\emph{determinate} receiving process which must be prepared to receive
at least the sent amount of data. The next two functions are the basic
\mpi point-to-point communication operations.

\begin{lstlisting}[style=SnippetStyle]
int MPI_Send(const void *buf, int count, MPI_Datatype datatype,
             int dest, int tag, MPI_Comm comm); 
int MPI_Recv(void *buf, int count, MPI_Datatype datatype,
             int source, int tag, MPI_Comm comm,
             MPI_Status *status);
\end{lstlisting}

Data to be sent and received are specified by the first three
arguments: A buffer address pointing to the part of memory where data
are (to be) located, an element count, and an argument describing the
structure of each element (see \Sec~\ref{sec:datatypes}). Data
buffers are always managed by the user and have to be allocated before
use. In C, buffers are represented by a pointer (address) to the start
of the buffer, and there are in general no restrictions on where and
how buffers to be used with \mpi are allocated. Larger buffers should
always be allocated dynamically in heap storage with a memory
allocator like \texttt{malloc()}. It is likewise the users
responsibility to \texttt{free()} allocated storage after
use. Problems with non-allocated, wrongly allocated and prematurely or
never freed storage are the most frequent and tiresome for the C
programmer.

By posting the \mpisend call, a sending process initiates and
completes sending data (a \texttt{count} number of elements of given
structure specified by the \texttt{datatype}) to the receiving
process. The sending process returns from the call when the data are
safely on their way (in the communication system) and the send buffer
can again be used for other data. By posting the \mpirecv call, a
receiving process declares itself ready to receive up to the described
amount of data (\texttt{count} number of elements of given structure
specified by the \texttt{datatype}) from a sending process. The call
completes when the data sent have been received correctly and without
loss (see discussion above). Thus, for point-to-point communication to
take place, both sending and receiving process are explicitly
involved. The receiving process must have allocated enough buffer
space (the \texttt{buf} argument) to hold the data that are being
sent. For communication to take place, the sending and the receiving
process must give the same integer \impi{message tag} to the
message. The sending process must give the rank of the receiving
process and this receiving process must be prepared to receive from that
process. However, wildcards for process rank and tag are possible, as
will be discussed later.  Thus, sending of messages is
\emph{determinate}, but receiving is not.

The send-receive functionality illustrates another important \mpi
principle. All(most all) space for \mpi data, notably data buffers but
also argument lists \etc, is in \emph{user space} and managed by the
application programmer. It is important to always have allocated
enough buffer space for data that are being sent and received and to
later free this space to avoid running out of memory. This is
sometimes forgotten, with dire consequences. Memory corruption due to
insufficient buffer space is one of the most frequent errors in \mpi
programs, frustrating and often hard to find, since memory corruption
(and program crash!) may become manifest only later in the program
execution and not immediately at the function call that caused the
memory corruption.

To illustrate point-to-point communication between processes in a
communicator, here is an \mpi implementation of the broadcast
operation described in Definition~\ref{def:broadcast} and discussed
intensively later (\Sec~\ref{sec:collective}). The process with rank
\texttt{root} is the process having the data, and \texttt{count} is
the number of elements.  The elements that are being communicated are
simple C integers of type \texttt{int}, which in \mpi are described by
the \mpi datatype \mpiint. The program is written to work for any number
of processes larger than one.

\begin{lstlisting}[style=SnippetStyle]
#define TAG 1000
  
MPI_Comm_rank(comm,&rank);
MPI_Comm_size(comm,&size);
assert(size>1);

int *buffer = ...; // allocate and free
if (rank==root) {
  MPI_Send(buffer,count,MPI_INT,(rank+1)%size,TAG,comm);
} else if (rank==(root-1+size)%size) {
  MPI_Status status;
   
  MPI_Recv(buffer,count,MPI_INT,(rank-1+size)%size,TAG,
           comm,&status);
} else {
  MPI_Status status;

  MPI_Recv(buffer,count,MPI_INT,(rank-1+size)%size,TAG,
           comm,&status);
  MPI_Send(buffer,count,MPI_INT,(rank+1)%size,TAG,comm);
}
\end{lstlisting}

The processes in this algorithm are organized as a processor
ring\index{processor ring}, and the number of dependent communication
steps (rounds) for the algorithm is $p-1$, where $p$ is the number of
\mpi processes. See \Sec~\ref{sec:stepcomplexity} for more on possible
analysis of message-passing algorithms. The $i$th process in the ring
(counting from the root process) needs to wait for process $i-1$ to
have received the data, \etc. Theorem~\ref{thm:broadcastlowerbound}
tells that this algorithm is poor. Here is another implementation of
the broadcast operation that is likewise poor, but not equally so and
not for the same reasons (why?).

\begin{lstlisting}[style=SnippetStyle]
#define TAG 1000
  
MPI_Comm_rank(comm,&rank);
MPI_Comm_size(comm,&size);
  
int *buffer = ...; // allocate and free
if (rank==root) {
  int i;

  for (i=0; i<size; i++) {
    if (i==root) continue;
    MPI_Send(buffer,count,MPI_INT,i,TAG,comm);
  }
} else {
  MPI_Recv(buffer,count,MPI_INT,root,TAG,comm,
           MPI_STATUS_IGNORE);
}
\end{lstlisting}

This example shows the use of a special value as \texttt{status}
argument, namely \mpistatusignore. This value can be used when the
return \texttt{status} of the receive operation is not needed.
Otherwise, the \texttt{status} object, referenced through
the corresponding handle, contains
information on the completion of the receive operation, \ie, whether
an error occurred, from which process the data were received, and how
much data was received.

Since receive calls like \mpirecv can specify (in their \texttt{count}
argument) more elements than actually sent in a send call,
functionality is needed for the receiving process to find out how much
data was actually received (and, by implication, sent).
This information is available in the process local status object. The
following two functions \mpigetcount and \mpigetelements that operate
on \texttt{status} objects are defined for this purpose. The
\texttt{datatype} is an input argument, which imposes an
interpretation of the received data and is needed in order to
correctly compute the number (\texttt{count}) of units of such
\texttt{datatype} that were
received (\mpigetcount). For complex datatype units that comprise many
simple datatype elements, the \mpigetelements
call instead computes the number of 
\impi{basic datatype} elements
that were received (see \Sec~\ref{sec:datatypes}). For
basic (simple, non-complex) datatypes like \mpiint as used in the code
examples above, the \mpigetcount and \mpigetelements calls are
equivalent and will return the same \texttt{count}.

\begin{lstlisting}[style=SnippetStyle]
int MPI_Get_count(const MPI_Status *status,
                  MPI_Datatype datatype, int *count);
int MPI_Get_elements(const MPI_Status *status,
                     MPI_Datatype datatype, int *count);
\end{lstlisting}

The \texttt{status} object/handle is peculiar in \mpi. Handles were
said to be opaque, but handles of type \mpistatus are only half
so. Status objects have three predefined fields, namely \mpisource,
\mpitag, and \mpierror. These are important for non-determinate
communication as will be explained in \Sec~\ref{sec:nondeterminate}.

Algorithms often desire or even require that a process can both send
and receive a message in a single communication round as, for
instance, permitted in the one-ported, fully bidirectional
send-receive communication model (see \Sec~\ref{sec:networkcomm}).
The example below, where the processes are organized in a ring like in
the first broadcast implementation above, will obviously lead to a
\impi{communication deadlock}\index{deadlock}.  Each process is
waiting to receive data from its predecessor process in the ring, but
these data cannot be sent, since this process is also waiting to
receive data from its predecessor process, \etc.

\begin{lstlisting}[style=SnippetStyle]
#define TAG 1000

MPI_Status status;
int *a = ...;
int *b = ...;

// receive from predecessor
MPI_Recv(a,count,MPI_INT,(rank-1+size)%size,TAG,
         comm,&status);
// send to successor
MPI_Send(b,count,MPI_INT,(rank+1)%size,TAG,comm);
\end{lstlisting}

\mpi provides an \mpisendrecv operation to handle such situations. It
combines the functionality and parameters of a blocking send and a
blocking receive operation. With \mpisendrecv as declared below,
a process can at the
same time in the same, single communication operation send and receive
data to and from two other
processes in the communicator --- that could actually be the same
process or even the process itself --- without the risk of
deadlocking. When an \mpisendrecv call
returns, data have left the send buffer (as with \mpisend), which can
then be reused for other data and received into the receive buffer
(as with \mpirecv). The status of the receive part is recorded in the
\texttt{status} object.

\begin{lstlisting}[style=SnippetStyle]
int MPI_Sendrecv(const void *sendbuf,
                 int sendcount, MPI_Datatype sendtype,
                 int dest, int sendtag,
                 void *recvbuf,
                 int recvcount, MPI_Datatype recvtype,
                 int source, int recvtag,
                 MPI_Comm comm, MPI_Status *status);
int MPI_Sendrecv_replace(void *buf,
                         int count, MPI_Datatype datatype,
                         int dest, int sendtag,
                         int source, int recvtag,
                         MPI_Comm comm, MPI_Status *status);
\end{lstlisting}

Send and receive buffers must not overlap in any way, since this would
lead to an indeterminate situation (a race
condition\index{race condition}): Did the send part take place, in
part or in total, before or after the receive part? Which data were actually
sent? It is the
programmers responsibility to make sure that the non-overlapping buffer
property is indeed
guaranteed. Neither compiler nor \mpi library will or can check this.
Such unintentionally overlapping buffers are another common source of
often very hard to find errors in \mpi programs.  In case data should
be sent from some buffer and (later) be received into the same
buffer, the \mpisendrecvreplace operation can be used. It most
likely will allocate some intermediate space for the receive part and
later copy the received data back, therefore entailing potentially significant
extra costs. As always, the \mpi standard neither prescribes nor forbids
any particular implementation. If one needs to know, only benchmarking
and \mpi library code inspection, if possible (as it would be for an
open source library), can help.

With \mpisendrecv the deadlock situation from above is resolved:

\begin{lstlisting}[style=SnippetStyle]
#define TAG 1000

MPI_Status status;
int *a = ...;
int *b = ...;

MPI_Sendrecv(b,count,MPI_INT,(rank+1)%size,TAG,
             a,count,MPI_INT,(rank-1+size)%size,TAG,
             comm,&status);
\end{lstlisting}

\subsection{Determinate \vs Non-Determinate Communication}
\label{sec:nondeterminate}

A sending process always specifies a determinate, specific receiver by
its rank in the communicator. A sending process also gives each
message sent a specific tag. In \mpi, a \impi{message tag} is just a
non-negative integer that is attached as a label to a message (up to a
specified upper bound given by \mpitagub). The
message tag can be used by the receiver to distinguish one kind of
tagged message from other kinds of tagged messages and to select
which message is to be received by an \mpirecv call in case more than
one message has been sent from one or more other processes.

As seen above, a receiving process can explicitly specify the rank of
the sending process from which it wants to receive a message
with a specific tag. In contrast to sending processes, receiving
processes can also receive from a \emph{non-determinate} process. This is
done by specifying a wildcard \mpianysource for the rank argument.
This will enable the receiving process to receive the message from any of
the processes in the communicator. Likewise, the tag argument can be
given a wildcard \mpianytag.

Whereas programs using only determinate ranks in the communication
operations are communication deterministic, programs using the
\mpianysource wildcard can be non-deterministic. Non-deterministic
programs can, not surprisingly, cause problems not encountered with
deterministic programs. The following examples illustrate 
these points.

Point-to-point communication is ordered. If messages are sent
by a sequence of \mpisend (or \mpisendrecv) operations with the same tag from
the same process, the messages will become ready to be received by the
destination process in exactly that order. This is referred to as
\emph{ordered communication} in \mpi. The program below illustrates
the advantages of the ordering constraint. Data from two buffers with
different numbers of elements and different element types, the first
with $500$ integers (\mpiint) and the second with 100 doubles (\mpidouble),
are sent from process $0$ to the last process with rank $p-1$ ($p$ being
the number of processes in communicator \texttt{comm}). It is good
SPMD\index{SPMD} style to always write \mpi programs so that they
work for any number of processes, which is the case here for any number of
processes larger than one, as asserted. An open \texttt{else} instead
of the \texttt{else if (rank==size-1)} conditional would lead to a
communication deadlock\index{deadlock} when the number of processes
is larger than two.

\begin{lstlisting}[style=SnippetStyle]
MPI_Comm_rank(comm,&rank);
MPI_Comm_size(comm,&size);
assert(size>1);
  
if (rank==0) {
  int buf1[500];
  double buf2[100];
    
  MPI_Send(buf1,500,MPI_INT,size-1,TAG,comm);
  MPI_Send(buf2,100,MPI_DOUBLE,size-1,TAG,comm);      
} else if (rank==size-1) {
  MPI_Status status;
    
  int buf1[1000];
  double buf2[200];
  int cc;
    
  MPI_Recv(buf1,1000,MPI_INT,0,TAG,comm,&status);
  MPI_Get_elements(&status,MPI_INT,&cc);
  assert(cc<1000);
  assert(cc==500);
    
  MPI_Recv(buf2,200,MPI_DOUBLE,0,TAG,comm,&status);
  MPI_Get_elements(&status,MPI_DOUBLE,&cc);
  assert(cc<200);
  assert(cc==100);	     
}
\end{lstlisting}

Since less data than expected by the receiving process are sent with
each message, the exact number of data elements received in each of
the messages has to be computed by the \mpigetelements operation. The
assertions assert that the (stack) allocated
buffers\index{stack allocation} are not overflowing.
For the \mpirecv operation, the
\texttt{count} argument is an upper bound on the number of elements
that can be received. This upper bound should, of course, be no larger
than the actual number of elements in the buffer used for
reception. Again, the compiler can and will not check this: It is
entirely the programmer's responsibility to ensure that buffers are
not overwritten. Overwriting will most likely cause segmentation
faults (crash!) at some point in the program execution.  It is also
worth noticing that the message tag has nothing to do with the type of
the messages being communicated: the same tag is used for both the
\mpiint and the \mpidouble messages. The message tag is a label
associated with a message that can be used to ensure that a receive
operation indeed receives the message intended for that
operation. Stack allocation, especially of variable sized arrays (in
C99 terms called variable length arrays) instead of heap allocation
with \texttt{malloc()}, is an often convenient and sometimes
defensible practice in C programs, but it should be used with caution.
The stack space is not as large as the heap and can easily be
exhausted, especially with recursive algorithms. The compiler or C
runtime will not notice: a program crash inevitably ensues.

In the next example, the data to be sent to process $p-1$ come from
two different processes. In order to avoid waiting times, the
receiving process uses \mpianysource to be able to receive the message
from whichever source process becomes ready first. Here, both buffers
contain C integers, and both sending processes use the same message
tag. Since the two sent messages have different numbers of elements,
the receiving process must ensure that both receive buffers are large
enough to hold the number of elements in the largest message. This is
the price of non-determinacy. The \mpisource field of the \mpistatus object
is used to distinguish the messages based on the source of origin.

\begin{lstlisting}[style=SnippetStyle]
MPI_Comm_rank(comm,&rank);
MPI_Comm_size(comm,&size);
assert(size>2); // at least three processes

if (rank==0) {
  int buf1[500];
  
  MPI_Send(buf1,500,MPI_INT,size-1,TAG,comm);      
} else if (rank==1) {
  int buf2[100];
    
  MPI_Send(buf2,100,MPI_INT,size-1,TAG,comm);
} else if (rank==size-1) {
  MPI_Status status;
    
  int buf1[1000];
  int buf2[1000];
  int cc;

  // receive from "first" sender
  MPI_Recv(buf1,1000,MPI_INT,MPI_ANY_SOURCE,TAG,
           comm,&status);
  MPI_Get_elements(&status,MPI_INT,&cc);
  assert(cc<1000);
  if (status.MPI_SOURCE==0) {
    assert(cc==500);
  } else {
    assert(status.MPI_SOURCE==1);
    assert(cc==100);
  }

  // and then from second sender
  MPI_Recv(buf2,1000,MPI_INT,MPI_ANY_SOURCE,TAG,
           comm,&status);
  MPI_Get_elements(&status,MPI_INT,&cc);
  assert(cc<1000);
  if (status.MPI_SOURCE==0) {
    assert(cc==500);
  } else {
    assert(status.MPI_SOURCE==1);
    assert(cc==100);
  }
}
\end{lstlisting}

In this program, two processes send messages, and one process receives
both messages: all messages sent are indeed received. \mpi programs
should maintain this property, and it is good \mpi programming
discipline to always keep this in mind. Cleanup of sent but unreceived
messages will, as far as they have not caused a
deadlock\index{deadlock}, be done by the \mpifinalize call.

Non-determinacy can easily lead to incorrect, possibly crashing
programs.  In the next, erroneous program, the sending processes send
different types and numbers of elements (\mpiint and \mpidouble).
For the receiving process, however, it has been forgotten that the two
messages may arrive in any order depending on the relative timing of
the two sending processes (and possibly other factors).

\begin{lstlisting}[style=SnippetStyle]
MPI_Comm_rank(comm,&rank);
MPI_Comm_size(comm,&size);
assert(size>2); // at least three processes

if (rank==0) {
  int buf1[500];
  
  MPI_Send(buf1,500,MPI_INT,size-1,TAG,comm);      
} else if (rank==1) {
  double buf2[100];
  
  MPI_Send(buf2,100,MPI_DOUBLE,size-1,TAG,comm);
} else if (rank==size-1) {
  MPI_Status status;
  
  int    buf1[1000];
  double buf2[200]; // buffer space could be insufficient
  int cc;

  // receive "first" message: data type could be wrong
  MPI_Recv(buf1,1000,MPI_INT,MPI_ANY_SOURCE,TAG,
           comm,&status);
  MPI_Get_elements(&status,MPI_INT,&cc);
  assert(cc<1000);
  if (status.MPI_SOURCE==0) {
    assert(cc==500);
  } else {
    assert(status.MPI_SOURCE==1);
    assert(cc==100);
  }
  
  // receive "second" message: data type could be wrong, again
  MPI_Recv(buf2,200,MPI_DOUBLE,MPI_ANY_SOURCE,TAG,
           comm,&status);
  MPI_Get_elements(&status,MPI_DOUBLE,&cc);
  assert(cc<1000);
  if (status.MPI_SOURCE==0) {
    assert(cc==500);
  } else {
    assert(status.MPI_SOURCE==1);
    assert(cc==100);
  }
}
\end{lstlisting}

The program may crash, possibly with an \mpi error message that a received
message has been truncated or that one of the assertions was violated.
Note that data may also be incorrectly received: the datatype argument
will not always correspond to the type of the sent data (see later).

The correct order of the received messages can be enforced by using
different tags for the two messages. Of course, this sacrifices the
potential performance advantage of non-determinacy. The program
below is correct. The first \mpirecv operation by process $p-1$ can
only receive a message with tag \texttt{TAG0}. Such a message will
eventually be sent by process $0$. The next to be received message
must have tag \texttt{TAG1} and, also, such a message will eventually be
sent by process $1$.
  
\begin{lstlisting}[style=SnippetStyle]
#define TAG0 500
#define TAG1 501

if (rank==0) {
  int buf1[500];
  
  MPI_Send(buf1,500,MPI_INT,size-1,TAG0,comm);      
} else if (rank==1) {
  double buf2[100];
  
  MPI_Send(buf2,100,MPI_DOUBLE,size-1,TAG1,comm);
} else if (rank==size-1) {
  MPI_Status status;
  
  int    buf1[1000];
  double buf2[200];
  int cc;
  
  MPI_Recv(buf1,1000,MPI_INT,MPI_ANY_SOURCE,TAG0,
           comm,&status);
  MPI_Get_elements(&status,MPI_INT,&cc);
  assert(cc==500);
  
  MPI_Recv(buf2,200,MPI_DOUBLE,MPI_ANY_SOURCE,TAG1,
           comm,&status);
  MPI_Get_elements(&status,MPI_DOUBLE,&cc);
  assert(cc==100);
}
\end{lstlisting}

Message tags are a specialty of point-to-point communication. They can
be highly useful when it is necessary to be able to distinguish
between different kinds of messages and can provide an extra piece of
information ``on top of'' the messages themselves that would otherwise
have to be incorporated manually as part of each message.  One-sided
communication (\Sec~\ref{sec:onesided}) and collective communication
(\Sec~\ref{sec:collective}) do not provide message or communication
tags.

In order to find out whether a message from a determinate or
non-determinate source with a given or wildcard tag is ready to be
received, \mpi provides calls to probe for such possible messages.
The following operation sets a status object for an ``incoming''
message that is ready to be received.

\begin{lstlisting}[style=SnippetStyle]
int MPI_Probe(int source, int tag, MPI_Comm comm,
              MPI_Status *status);
\end{lstlisting}

A probe call returns when a message with the specified
characteristics (source and tag) is ready to be received but the
message is not received by the probe and the data are not accessible.
To actually receive the message and make the data available in a
receive buffer, an \mpirecv or other point-to-point message
receive operation must be executed.

Advanced note: This separation into the probing for a message and the
actual reception of the message can cause problems
(race conditions\index{race condition})
when \mpi is used in a multi-threaded program, for instance, with
\openmp\index{OpenMP} or \pthreads\index{pthreads}
where many threads perform \mpi operations and a
different thread from the one doing the probe can have received the
message. The \mpi specification provides means to deal with this
situation, but this is beyond these lectures.

\subsection{Point-to-Point Communication Complexity and Performance}
\label{sec:stepcomplexity}

When two \mpi processes become ready to communicate at the same time,
one process sending $m$ units of data and the other being ready to receive
(at least) $m$ units, the time for transmitting the $m$ data units may
naively be modeled as $\alpha+\beta m$ for
a given, constant start-up \emph{latency} $\alpha$ and a cost per
unit $\beta$ (see \Sec~\ref{sec:communicationcosts}).
In a more refined model, $\alpha$ and $\beta$ might
depend on the placement of the two communicating processes in the
system, the \emph{mapping} of the communicator to the processors, the
total number of processes in the communicator, and on the overall
traffic in the communication system during the communication
operation.

Alternatively, we can account for this data transmission as one
\impi{communication step}, irrespectively of the amount of data being
transmitted. Several independent pairs of processes can, if the
underlying communication network is strong enough by having enough
bisection width\index{bisection width}, communicate independently, and
if all processes communicate (roughly) the same amount of data $m$, we
count such a set of concurrent communication operations between many
pairs of processes as one communication step. In the case where many
processors are communicating more or less at the same time and
performing their respective communication steps concurrently we
commonly use the term \impi{communication round} for all these
concurrent steps.

The \impi{communication round complexity} for a message-passing
algorithm can be accounted for as the number of such
\emph{communication rounds} required from the start of all processes
until the last process has completed its last communication step.  In
a bit more general terms, this amounts to finding the longest
(weighted) path of communication operations from a first to a last
process in a communication DAG (Directed Acyclic Graph)\index{DAG}
that describes the communication operations of the algorithm.  In the
communication DAG, there is an edge from process $i$ to process $j$
when process $i$ sends a message that is received by process $j$, that
is, when both processes are \emph{ready} to communicate.  The
communication round complexity is the length of a longest path of
dependent send and receive communication operations in the DAG
corresponding to the execution of the algorithm. The DAG way of
accounting for the complexity of a message passing (\mpi) algorithm
can also be used for the case where the amount of data communicated
between pairs of processes is not the same in each round in which case
the DAG edges are assigned weights reflecting the cost of the
corresponding communication step according to some model of pairwise
communication costs.

The linear array broadcast implementation from
\Sec~\ref{sec:pointtopoint} was claimed to take $p-1$ communication
steps. This can be seen by an inductive argument. If there are only two
processes, one communication step is obviously required and
suffices. With $p>2$ processes, the root process first sends the data
to the next process in one communication step. This process now
behaves like a root for $p-1$ processes. By induction, the broadcast
can now be completed in $p-2$ steps, for a total of
$(p-2)+1=p-1$ steps.

Below is a better broadcast algorithm that completes in
$\ceiling{\log_2 p}$ steps, matching the lower bound on the number of
communication rounds for broadcast of
Theorem~\ref{thm:broadcastlowerbound}. The algorithm assumes that broadcast
is from the process with rank $0$ in the communicator and to ensure this it
uses a virtual ranking \texttt{virt} of the processes.
This trick of reranking by subtracting a \texttt{root} rank is often useful
and will perform better than creating a new communicator with the desired
property.

\begin{lstlisting}[style=SnippetStyle]
MPI_Comm_rank(comm,&rank);
MPI_Comm_size(comm,&size);

virt = (rank-root+size)%size; // rerank by hand

// buffer of some type from somewhere
if (virt!=0) {
  int d = 1;
  while (virt>=d) {
    dist = d; d <<= 1; // multiply by two
  }
  // rerank back from virt to real rank
  MPI_Recv(buffer,count,datatype,(virt-dist+root+size)%size,
           TAG,comm,MPI_STATUS_IGNORE);
  dist = d;
} else dist = 1;
while (virt+dist<size) {
  MPI_Send(buffer,count,datatype,(virt+dist+root)%size,
           TAG,comm);
  dist <<= 1; // multiply by two
}
\end{lstlisting}

\subsection{\mpi Concepts: Semantic Terms}
\label{sec:semanticterms}

The simple send and receive operations, as well as all the other
operations discussed in the previous sections, are
\impi{blocking}. This is an \mpi\index{MPI} specific semantic term,
which means that the call returns when the operation is complete,
locally, from the calling process' point of view. With \mpisend in
particular, this (just) means that data are out of the send-buffer,
which can now be reused for other purposes, for instance, as buffer
for the next \mpisend operation. Also, other resources have been given
free and can be reused. When, \eg, \mpicommsplit returns, a new
communicator has been created and is ready for use from the calling
process' point of view.  The \impi{blocking} property does not imply
anything about what other processes have done. It simply and only
means that an operation has been completed locally by the calling
process according to the semantics of the operation. For instance,
return from an \mpisend call does not mean that the data have been
received by the receiving process which might not even have posted its
\mpirecv call. Data might not even be anywhere near the receiving
process. They could, for instance, have been buffered somewhere by the
\mpi library, a technique that is often used for small messages and
can in some situations have some performance advantages. On the other
hand, for a blocking operation to complete, some action by other
processes \emph{may} be necessary. For instance, very large data in
\mpisend calls are typically not buffered by \mpi libraries. In such
cases, point-to-point communication can complete only when sending and
receiving processes are both active. Obviously, a blocking \mpirecv
call cannot complete before the data have been sent, so action by the
sending process can in this case indeed be inferred.

As counterpart, \mpi defines operations that are \impi{nonblocking}.
Such operations will return \emph{immediately} (whatever
that means, how fast is immediate?)
and \emph{always} independently of actions being taken by other
processes. They are therefore also called \impi{immediate operations}
and prefixed with an \texttt{I} (Capital ``I'') after \texttt{MPI\_} in \mpi.

An \mpi operation is said to have \impi{local completion} if it can
always complete independently of action by other processes. Trivial
examples seen so far are the blocking operations \mpirank and
\mpisize. The \mpisend operation is blocking but does not have local
completion since action by the receiving process \emph{may} be required.  The
same holds for the \mpirecv operation, which always requires action by
a sending process.  The collective \mpicommcreate operation is blocking
and does not have local completion: Some action may be
and mostly is required by the other processes in order to create the new
communicator and make it possible to complete the call.
A process calling a blocking collective operation\index{collective operation}
will in most cases not be able to complete and
return from the call before at least some of the other processes in
the communicator has at least performed the corresponding call.

The counterpart to local completion is \impi{non-local completion},
which means that, in order for an operation to complete, action by
other processes \emph{may} be needed. Here, \emph{action} means that
other processes perform \mpi calls that enable the operation
to complete. Per definition, nonblocking\index{nonblocking} \mpi
calls have local completion.

As discussed for the blocking \mpisend operation, an implementation of
\mpi that does intermediate buffering may make it possible for an
\mpisend call to complete and return even without the receiving
process having posted a matching \mpirecv call. But it may
also not. Relying on such implementation specific behavior is bad and
dangerous practice, since it makes programs non-portable. The program
may run on one machine with one \mpi library, but it may stop working
on the next machine with a different \mpi library. Or it may work
for small data and problem sizes and then suddenly and mysteriously stop
working when problem and data get larger.
The practice of (perhaps unbeknownst to the programmer)
relying on implementation
dependent behavior of \mpi operations is called \impi{unsafe programming}.

Concretely, for blocking communication with \mpisend and \mpirecv, one
should always write the application under the assumption that
completion is indeed non-local and such that there will always,
eventually, be a matching \mpirecv call for any \mpisend call executed
by some process.

Here is a typical example of unsafe communication with processes
communicating in a rank ordered ring pattern. All processes initiate a
(blocking) \mpisend call to the next process in the ring, after which
they receive data from the previous process in the ring. The \mpisend
may --- or may \emph{not} --- be able to complete, depending on the
message count and on implementation details of the \mpi library. If it
cannot complete, a deadlock ensues. The nasty thing about this kind of
code is that it may well work under the right circumstances and then
suddenly stop working (deadlock) when conditions change.  That is why
this style of programming is called \emph{unsafe}. Unsafe programs are
in particular not portable.

\begin{lstlisting}[style=SnippetStyle]
#define TAG 1000

MPI_Status status;
int *a = ...;
int *b = ...;

assert(size%d==0); // even number of processes required
MPI_Send(b,count,MPI_INT,(rank+1)%size,TAG,comm);
MPI_Recv(a,count,MPI_INT,(rank-1+size)%size,TAG,
         comm,&status);
\end{lstlisting}

In some of the examples above, message tags were used to enforce a certain
order on received messages. This usage of tags can easily result in an
unsafe program
as the example below shows\index{unsafe programming}. The
first \mpisend call may not be able to complete.

\begin{lstlisting}[style=SnippetStyle]
#define TAG1 100
#define TAG2 101

if (rank==0) {
  int buf1[500];
  double buf2[100];
  
  MPI_Send(buf2,100,MPI_DOUBLE,size-1,TAG2,comm);
  MPI_Send(buf1,500,MPI_INT,size-1,TAG1,comm);
  
} else if (rank==size-1) {
  // order, buf2 smaller than buf1, but no overflow
  MPI_Status status;
  
  int buf1[1000];
  double buf2[200];
  int cc;
  
  MPI_Recv(buf1,1000,MPI_INT,0,TAG1,comm,&status);
  MPI_Get_elements(&status,MPI_INT,&cc);
  assert(cc<=1000);
  assert(cc==500);
  
  MPI_Recv(buf2,200,MPI_DOUBLE,0,TAG2,comm,&status);
  MPI_Get_elements(&status,MPI_DOUBLE,&cc);
  assert(cc<=200);
    assert(cc==100);
  }
\end{lstlisting}

Care is needed to ensure that a program is not
unsafe\index{unsafe programming}.  Sometimes this can be difficult. For a more
interesting example, we return to the
stencil pattern\index{stencil computation} as discussed in
\Sec~\ref{sec:organizingprocesses} and \Sec~\ref{sec:stencil}. The
communication part of a two-dimensional stencil code below
shows\index{stencil computation}.  Here, the processes have been
organized as a two-dimensional Cartesian communicator, see
\Sec~\ref{sec:organizingprocesses}. For each process, the ranks of the
(up to four) neighboring processes, \texttt{left}, \texttt{right},
\texttt{up} and \texttt{down}, are computed with the \mpicartshift
functionality (some of these ranks may be \mpiprocnull).  Each process
has out and in buffers for its four neighboring processes from where
it needs to both send and receive data. The stencil update and the
four send and receive operations are repeated until some convergence
criterion is fulfilled and the \texttt{done} flag is set to
\textbf{true} (\texttt{=1}).  This will most likely depend on the not
shown computations within each iteration (there could also be a fixed,
predetermined number of iterations). A first attempt could look as
follows.

\begin{lstlisting}[style=SnippetStyle]
#define STENTAG 11
  
int left, right;
int up,   down;
  
MPI_Cart_shift(cartcomm,1,1,&left,&right);
MPI_Cart_shift(cartcomm,0,1,&up,  &down);

double *out_left, *out_right, *out_up, *out_down;
double *in_left,  *in_right,  *in_up,  *in_down;

... // allocate and set buffers

int done = 0;
while (!done) { // iterate until convergence
  ... // the stencil update (computation)
  
  MPI_Send(out_left, n,MPI_DOUBLE,left, STENTAG,cartcomm);
  MPI_Send(out_right,n,MPI_DOUBLE,right,STENTAG,cartcomm);
  MPI_Send(out_up,   n,MPI_DOUBLE,up,   STENTAG,cartcomm);
  MPI_Send(out_down, n,MPI_DOUBLE,down, STENTAG,cartcomm);
   
  MPI_Recv(in_left,  n,MPI_DOUBLE,left, STENTAG,cartcomm,
           MPI_STATUS_IGNORE);
  MPI_Recv(in_right, n,MPI_DOUBLE,right,STENTAG,cartcomm,
           MPI_STATUS_IGNORE);
  MPI_Recv(in_up,    n,MPI_DOUBLE,up,   STENTAG,cartcomm,
           MPI_STATUS_IGNORE);
  MPI_Recv(in_down,  n,MPI_DOUBLE,down, STENTAG,cartcomm,
           MPI_STATUS_IGNORE);

  done = 1; // some termination criterion
}
\end{lstlisting}

Depending on the completion semantics, the four send operations may
not be able to complete before the corresponding receive operations
have been initiated, which in that case will not be possible: the
program deadlocks\index{deadlock}.
It is a good exercise to reflect on this example
and on how the code can be made safe and portable. We will return to
this example several times.

\subsection{\mpi Concept: Specifying Data with (Derived) Datatypes}
\label{sec:datatypes}

\begin{table}[t!]
  \caption{Some C datatypes and their corresponding \mpidatatype.}
  \label{tab:datatypes}
  \begin{center}
    \begin{tabular}{ll}
      \toprule
      C language type & Corresponding \mpi datatype \\
      \midrule
      \texttt{char} & \mpichar \\ 
      \texttt{short} & \mpishort \\ 
      \texttt{int} & \mpiint \\ 
      \texttt{long} & \mpilong \\ 
      \texttt{float} & \mpifloat \\
      \texttt{double} & \mpidouble \\
      \bottomrule
    \end{tabular}
  \end{center}
\end{table}

Data to be communicated in \mpi are always specified the
same way. A block of elements is described by a triple consisting of a
starting address (or offset) in memory (buffer), a number of elements
(count), and a structure/layout of the elements (datatype). As a
mnemonic for the \mpi communication operations it is helpful to keep
in mind that data are always triples of
\texttt{buffer,count,datatype}; this greatly reduces the number of
arguments one has to think of and makes it easy to guess/reconstruct
the signature of many \mpi operations.

The third argument in the triple, the \mpidatatype, describes the
structure or layout of the data elements to be communicated (sent or
received) locally, for calling the process. For basic, simple, non-complex
objects like the \texttt{int}s and \texttt{double}s in a C program,
there are corresponding, predefined \mpi handles like \mpiint and
\mpidouble that describe to the \mpi library that the bits and bytes
in a data buffer represent these kinds of objects.

For the simple, and most common case of elements from a consecutive
buffer, for instance, an array of elements of some simple (programming
language) type being communicated, the datatype argument just tells the
\mpi library that the bytes are to be interpreted as the corresponding
programming language type is represented in memory. There is, therefore,
an \mpi datatype for each simple, elementary programming language
datatype. Some of the corresponding \mpi datatypes for C types are shown in
\Tab~\ref{tab:datatypes}. Fortran has Fortran-like names for the
corresponding \mpi types.

Correct \mpi programs require that data elements (corresponding to
some programming language type)
that are sent as a sequence of \mpi datatypes are
received as a sequence of the same \mpi datatypes. Observing this
requirement ensures that the bits and bytes that are sent and received
are interpreted and handled in the intended way, both by the sending
and by the receiving process. It is important to understand that the
programming language type of objects (buffers) are not and cannot be
\emph{a priori} known to the \mpi library.
Therefore, the library has to be instructed in each communication
operation by each involved process.
The \mpi datatype information is \emph{not} in
any way part of the transmitted data. It is entirely the programmer's
responsibility to ensure that all communicated data are given the
right \mpi datatype by both sending and receiving processes. Neither
compiler nor \mpi library can and will (for performance reasons) check
this. For this same reason, \mpi does not perform type conversion (as
known from, \eg, C). If a data buffer is sent as a sequence of \mpiint
objects and received as a sequence of \mpifloat objects, no useful
outcome can be expected. Most certainly, the \texttt{int}s will not be
converted to \texttt{float}s in a semantically meaningful way!

The next three small examples illustrate this. In the first example,
some \texttt{long}s are sent correctly as \mpilong, but wrongly
received as \mpidouble and stored in a (large enough, presumably)
buffer of \texttt{double}s.

\begin{lstlisting}[style=SnippetStyle]
if (rank==0) {
  long a[n];
  MPI_Send(a,n,MPI_LONG,size-1,TAG,comm);
} else if (rank==size-1) {
  double a[n];
  MPI_Recv(a,n,MPI_DOUBLE,0,TAG,comm,MPI_STATUS_IGNORE);
}
\end{lstlisting}

In the second example, \texttt{double}s sent correctly are received
as a sequence of \mpibyte elements. This may or may not give correct results
but is in any case a dangerously incorrect \mpi programming style.

\begin{lstlisting}[style=SnippetStyle]
double a[n];
if (rank==0) {
  MPI_Send(a,n,MPI_DOUBLE,size-1,TAG,comm);
} else if (rank==size-1) {
  MPI_Recv(a,n*sizeof(double),MPI_BYTE,0,TAG,
           comm,MPI_STATUS_IGNORE);
}
\end{lstlisting}

In the third and last example, the objects are sent and received as
streams of uninterpreted bytes. This is not technically wrong, but any
type information on how \texttt{double}s are to be handled (\eg,
Endianness) is lost.

\begin{lstlisting}[style=SnippetStyle]
double a[n];
if (rank==0) {
  MPI_Send(a,n*sizeof(double),MPI_BYTE,size-1,TAG,comm);
} else if (rank==size-1) {
  MPI_Recv(a,n*sizeof(double),MPI_BYTE,0,TAG,
           comm,MPI_STATUS_IGNORE);
}
\end{lstlisting}

The next purpose of \mpi datatypes is to be able to describe layouts
of complex, often non-consecutive data in process local memory.
This gives the \mpi library
the possibility to read and write data elements from specific
locations and not necessarily as a consecutive stream of elements in a
simple, linear buffer (array).  Natural examples are the columns of 
two-dimensional matrices, submatrices of larger, $d$-dimensional matrices,
complex C structures with different component types, \etc. The
\mpi concept of a datatype is, thus, somewhat different from the same-named,
semantic programming language concept. In \mpi, a datatype
describes the (local, spatial) structure of data objects to be
communicated and carries little, further semantics.

The idea of the \mpi \impi{user-defined datatype} or
\impi{derived datatype} mechanism is to be able to encapsulate such complex
data layouts into a single unit which can then be used as the unit of
communication in \emph{all} \mpi communication operations. A derived
datatype represents an ordered list of simple, basic datatypes (as we
have seen: \mpiint, \mpidouble, \mpichar, \etc) together with a
displacement or relative offset for each simple element. The offset
for an element gives the linear position of the element in memory
relative to a given base address, \eg, the buffer argument supplied in
the \mpi communication calls.

An explicit list of basic element datatypes with their displacements
is in \mpi terms called a \impi{type map}. The type map, which is an
opaque construct that is not directly accessible to the programmer, is
used locally by communicating \mpi processes to access the basic
elements in local memory in the order and displacement implied by the
list, regardless of whether processes are sending or receiving data. A
type map is, thus, a purely process local construct and the type maps
of one process are not known to any other processes.  Identical type
maps for different processes can, of course, be constructed by the
programmer, but \mpi itself cannot and does not exchange type maps or
any other type information. Datatypes and type maps are not
\emph{first-class citizens} in \mpi.

Communication in \mpi can be thought of as a stream of elements
described by a corresponding stream of simple, basic datatypes. This
stream of simple, basic datatypes can be thought of as the type map
stripped of the displacements. Such an ordered sequence of basic
datatypes is, in \mpi terms, called a \impi{type signature}. When two
processes are communicating with point-to-point send and receive
operations, the signature of the data that are sent must be a prefix
of the signature of the data that the receiving process is prepared to
receive (since the receive operation can specify a larger element
count than the send operation).  Again, the signatures are not part of
the data that are being communicated.  It is purely the programmers
responsibility to guarantee that the signature rule is obeyed. By this
choice, it is possible with the help of the programmer to do type safe
communication in \mpi without the burden (and performance
disadvantage) of having to communicate additional type meta information.

In \mpi, neither type maps nor type signatures are represented explicitly
by lists of basic datatypes and displacements.
Instead, \mpi provides a number of
constructors for compactly describing more and more irregular layouts
of data in memory. Layouts described by this mechanism are called
\emph{derived}\index{derived datatypes} or \emph{user-defined
datatypes}\index{user-defined datatype}. In
\Sec~\ref{sec:deriveddatatypes} the \mpi datatype constructors will be
briefly explained. A derived datatype can be used in any \mpi
communication operation and in all operations that take \mpidatatype
arguments.

A type map, as represented by a derived datatype, is a complex object
encompassing possibly many different, basic datatypes together with their
displacements. The \emph{size} of a derived datatype is the number of
bytes required (locally, for the process) to represent all the basic
datatypes in the derived datatype. The \impi{extent} of a derived
datatype is a quantity in bytes associated with a derived datatype
which is necessary to define what happens when a derived datatype is
used in communication operations with an element count larger than one.
The signature of the derived datatype is the unit of communication. A
count $c, c>1$ tells that more than one such unit is to be
communicated. The $i$th unit, $0\leq i<c$, is taken from relative
offset $i\cdot\texttt{extent}$ from the given communication buffer address,
where \texttt{extent} is the extent of the datatype.  The following
\mpi calls return the size and extent of both simple, basic,
predefined datatypes and user-defined datatypes\index{user-defined datatype}.

\begin{lstlisting}[style=SnippetStyle]
int MPI_Type_size(MPI_Datatype datatype, int *size);
int MPI_Type_extent(MPI_Datatype datatype, MPI_Aint *extent);

int MPI_Type_get_extent(MPI_Datatype datatype,
                       MPI_Aint *lb, MPI_Aint *extent);
int MPI_Type_get_true_extent(MPI_Datatype datatype,
                            MPI_Aint *true_lb,
                            MPI_Aint *true_extent);
\end{lstlisting}

Often, but not always, the extent of a derived datatype corresponds to
the ``footprint'' in memory of the type layout described by that
datatype. This is the linear difference between the simple element
(basic datatype) with the smallest displacement and the simple element
with the largest displacement plus the size of that element. The
datatype constructors all have associated rules for how the extent of
the resulting derived datatype is computed. There are, however,
special type constructors for creating datatypes with a different
(arbitrary) extent, a feature that is extremely powerful for advanced
usage of \mpi. Therefore, the extent is not simply the ``memory
footprint'' of the layout. However, the memory footprint is needed when
new memory for some complex layout needs to be allocated. For
this, the special call \mpitypetrueextent is defined. Unfortunately,
even this is not always sufficient for computing the right amount of
memory space. Memory allocation and derived datatypes in \mpi need
care.

The calls returning an extent have arguments of type pointer to
\mpiaint. This argument type is not an \mpi handle, but the type of an
object that can represent an \emph{address-sized integer}. In many
cases (compilers, systems), an \mpiaint is indeed different from a C
\texttt{int} (64 versus 32 Bits). The \mpiaint type is used for many
\mpi operations where it is important that an argument is a process
local address; but is not used very consistently in the \mpi
standard.

\subsection{\mpi Concept: Matching Communication Operations}

In order for point-to-point communication between two processes to be
successful, the \mpisend and \mpirecv operations must \emph{match}.
First of all, the two processes must make their calls on the same
communicator\index{communicator}:
In \mpi, communication on one communicator can \emph{never}
interfere with communication on another communicator. In particular,
communication with an \mpisend on one communicator and an otherwise
correct \mpirecv operation on another communicator will never take place and
will result in a deadlock\index{deadlock}.
The destination rank given by the sending process must match the rank
given by the receiving process.  Either the receiving process gives
the rank of the sending process explicitly or the \mpianysource
wildcard. Likewise, the message tags must be the same or the receiving
operation must use the \mpianytag wildcard. As mentioned, it is
perfectly legal for a process to communicate with itself. However,
with blocking operations only, it is not possible to do this in a safe
way; at least one of either the send or receive operations has to be
nonblocking. Alternatively, the \mpisendrecv operation can be used.
Also, in such cases care has to be taken that receive and send buffer (which are
on the same process) do not overlap in anyway. Otherwise, the
result would depend on the exact order in which data elements are
received and sent and put into the respective buffers: a kind of race
condition\index{race condition} that is not allowed in correct \mpi
programs.

Second, the amount of data sent in the \mpisend operation must be at
most the amount of data that the \mpirecv operation is prepared to
receive, as specified by its count and datatype arguments. The \mpi types
of the sent and received elements must correspond. Technically, this
means that the signature of the sent data must be a prefix of the
signature of the data specified in the receive call. As discussed,
\mpi cannot (without considerable overhead) and does not check for this.

When an \mpisend and an \mpirecv call match, communication can take
place and the \mpi implementation guarantees that data are eventually
correctly received. There is no need for low-level consistency or
correctness checks on behalf of the user code.

Communication with the special \mpiprocnull process always matches
but has no effect, neither in an \mpisend nor in an \mpirecv
operation.

\subsection{Nonblocking Point-to-Point Communication}
\label{sec:nonblockingpoint}

\mpi defines nonblocking\index{nonblocking} point-to-point
communication counterparts
for the simple \mpisend, \mpirecv, and \mpiprobe operations.

\begin{lstlisting}[style=SnippetStyle]
int MPI_Isend(const void *buf,
              int count, MPI_Datatype datatype,
              int dest, int tag,
              MPI_Comm comm, MPI_Request *request);
int MPI_Irecv(void *buf, int count, MPI_Datatype datatype,
              int source, int tag,
              MPI_Comm comm, MPI_Request *request);
int MPI_Iprobe(int source, int tag, MPI_Comm comm, int *flag,
               MPI_Status *status);
\end{lstlisting}

With the MPI 4.0 version of the \mpi standard, also
nonblocking\index{nonblocking} versions of the
\mpisendrecv and \mpisendrecvreplace
operations have been introduced.

\begin{lstlisting}[style=SnippetStyle]
int MPI_Isendrecv(const void *sendbuf,
                  int sendcount, MPI_Datatype sendtype,
                  int dest, int sendtag,
                  void *recvbuf,
                  int recvcount, MPI_Datatype recvtype,
                  int source, int recvtag,
                  MPI_Comm comm,
                  MPI_Request *request);
int MPI_Isendrecv_replace(void *buf,
                          int count, MPI_Datatype datatype,
                          int dest, int sendtag,
                          int source, int recvtag,
                          MPI_Comm comm,
                          MPI_Request *request);
\end{lstlisting}

All these operations return ``immediately'': What exactly this means and
how fast ``immediate'' is, is, by the nature of the \mpi standard
specification, not defined. The important point is that the
operations have entirely local completion semantics and return
independently of any \mpi actions taken by any other
processes. Nonblocking point-to-point operations can, therefore, be
used to avoid situations that might otherwise lead to a deadlock
(unsafe code) with blocking communication\index{unsafe programming}.

These nonblocking send and receive operations take the same input
parameters as their blocking counterparts, but have a new output
argument, the \mpirequest object (handle). The \mpirequest object can
be used to query the completion status of the corresponding operation
and to enforce completion. A nonblocking \mpiisend call with ensuing,
enforced completion has the same effect (semantics) as a blocking
\mpisend call. That is, enforced completion means only that the send
operation has been completed from the process' point of view and does
not imply anything about the receiving process, not even that is
has reached the matching receive call. The nonblocking counterpart of
the probe operation, the \mpiiprobe, does, curiously, not return an \mpirequest
object. Instead, the completion of the probe for a matching, incoming
message is indicated in the \texttt{flag} return argument (pointer).

There is a whole repertoire of operations for checking and enforcing
completion of immediate, pending \mpiisend and \mpiirecv communication
operations. These calls can either test whether an operation, referred
to by an \mpirequest object, is complete, which is signalled in a
\texttt{flag} return argument, or enforce (by waiting) the completion
of an operation. There are calls that operate on a list of request
objects, rather than a single object. They can test for or enforce
completion of either some single (arbitrary) operation, some, or all
operations in the set of requests (given as an input array). For
complete operations, their status is returned in corresponding
\mpistatus objects, just as was the case for the blocking \mpisend and
\mpirecv calls.

\begin{lstlisting}[style=SnippetStyle]
int MPI_Wait(MPI_Request *request, MPI_Status *status);
int MPI_Test(MPI_Request *request,
             int *flag, MPI_Status *status);

int MPI_Waitany(int count, MPI_Request requests[], int *indx,
                MPI_Status *status);
int MPI_Testany(int count, MPI_Request requests[], int *indx,
                int *flag, MPI_Status *status);
int MPI_Waitall(int count, MPI_Request requests[],
                MPI_Status statuses[]);
int MPI_Testall(int count, MPI_Request requests[],
                int *flag, MPI_Status statuses[]);
int MPI_Waitsome(int incount, MPI_Request requests[],
                 int *outcount,
                 int indices[], MPI_Status statuses[]);
int MPI_Testsome(int incount, MPI_Request requests[],
                 int *outcount,
                 int indices[], MPI_Status statuses[]);

int MPI_Request_free(MPI_Request *request);
\end{lstlisting}

The nonblocking communication operations separate the
initialization and the completion of an operation and can be
most convenient for writing safe programs that cannot deadlock in any
possible situation. For instance, the \mpisendrecv operation is
equivalent to either

\begin{lstlisting}[style=SnippetStyle]
MPI_Request request;
MPI_Status  status;
MPI_Isend(sendbuf,sendcount,sendtype,dest,sendtag,
          comm,&request);
MPI_Recv(recvbuf,recvcount,recvtype,source,recvtag,
         comm,&status);
MPI_Wait(&request,MPI_STATUS_IGNORE);
\end{lstlisting}
or
\begin{lstlisting}[style=SnippetStyle]
MPI_Request request;
MPI_Status  status;
MPI_Irecv(recvbuf,recvcount,recvtype,source,recvtag,
          comm,&request);
MPI_Send(sendbuf,sendcount,sendtype,dest,sendtag,comm);
MPI_Wait(&request,&status);
\end{lstlisting}
or even
\begin{lstlisting}[style=SnippetStyle]
MPI_Request request[2];
MPI_Status  status[2];
MPI_Irecv(recvbuf,recvcount,recvtype,source,recvtag,
          comm,&request[0]);
MPI_Isend(sendbuf,sendcount,sendtype,dest,sendtag,
          comm,&request[1]);
MPI_Waitall(2,request,status);
\end{lstlisting}
where, for the last code snippet, the status of the receive operation
is in \texttt{status[0]}.

The unsafe, two-dimensional stencil code\index{stencil computation}
built from blocking \mpisend
and \mpirecv operations (\Sec~\ref{sec:semanticterms}) can now be
made safe and deadlock-free by simply using nonblocking send and
receive operations\index{unsafe programming}. This may have the additional
performance advantage that communication can take place in the order in
which processes become ready and not in the fixed order given
by the sequence of blocking \mpisend and \mpirecv operations.

\begin{lstlisting}[style=SnippetStyle]
MPI_Request request[8];
  
int done = 0;
while (!done) { // iterate until convergence
  ... // the stencil update (computation)

  int k = 0;
  MPI_Isend(out_left, n,MPI_DOUBLE,left, STENTAG,cartcomm,
            &request[k++]);
  MPI_Isend(out_right,n,MPI_DOUBLE,right,STENTAG,cartcomm,
            &request[k++]);
  MPI_Isend(out_up,   n,MPI_DOUBLE,up,   STENTAG,cartcomm,
            &request[k++]);
  MPI_Isend(out_down, n,MPI_DOUBLE,down, STENTAG,cartcomm,
            &request[k++]);
    
  MPI_Irecv(in_left,  n,MPI_DOUBLE,left, STENTAG,cartcomm,
            &request[k++]);
  MPI_Irecv(in_right, n,MPI_DOUBLE,right,STENTAG,cartcomm,
            &request[k++]);
  MPI_Irecv(in_up,    n,MPI_DOUBLE,up,   STENTAG,cartcomm,
            &request[k++]);
  MPI_Irecv(in_down,  n,MPI_DOUBLE,down, STENTAG,cartcomm,
            &request[k++]);

  MPI_Waitall(k,request,MPI_STATUSES_IGNORE);

  done = 1; // some termination criterion
}
\end{lstlisting}

The special value \mpistatusesignore argument indicates that all statuses
should be ignored and no output status array is given.

The stencil computation\index{stencil computation}
can also be made safe and deadlock-free with
the combined \mpisendrecv operation. The trick is to communicate along
the dimensions of the process grid, one after the other, in a ring-like
fashion, receiving from the left process and sending to the right process
\etc. Since all processes perform their communication operations in a
symmetric fashion, no deadlocks can occur. For a possible performance
advantage, \mpiisendrecv could have been used as in the above code.

\begin{lstlisting}[style=SnippetStyle]
int done = 0;
while (!done) { // iterate until convergence
  ... // the stencil update (computation)
  
  MPI_Sendrecv(out_left, n,MPI_DOUBLE,left, STENTAG,
               in_right, n,MPI_DOUBLE,right,STENTAG,cartcomm,
               MPI_STATUS_IGNORE);
  MPI_Sendrecv(out_right,n,MPI_DOUBLE,right,STENTAG,
               in_left,  n,MPI_DOUBLE,left, STENTAG,cartcomm,
               MPI_STATUS_IGNORE);
  MPI_Sendrecv(out_up,   n,MPI_DOUBLE,up,   STENTAG,
               in_down,  n,MPI_DOUBLE,down, STENTAG,cartcomm,
               MPI_STATUS_IGNORE);
  MPI_Sendrecv(out_down, n,MPI_DOUBLE,down, STENTAG,
               in_up,    n,MPI_DOUBLE,up,   STENTAG,cartcomm,
               MPI_STATUS_IGNORE);

  done = 1; // some termination criterion
}
\end{lstlisting}

\subsection{Exotic Send Operations\marksec}

\mpi provides a few more send operations with additional semantic
content. These operations come in both blocking and
nonblocking\index{nonblocking}
variants. There is a \impi{synchronous send} operation, where local
completion implies that the receiving process has indeed started
reception of the message by a matching receive operation. There is a
\impi{buffered send} operation, where data are explicitly stored in a
local buffer in order to provide local completion semantics. The local
buffer is allocated in user space which, for this use, needs to be
explicitly attached to the \mpi library. Finally, there is a
\impi{ready send} operation, which can be used provided that a
matching receive operation has already been posted before the buffered
send. Send-receive communication can possibly be implemented more
efficiently under this precondition. The ready send operation was
included in \mpi to enable such implementations. Using it correctly
requires additional explicit or implicit synchronization and is rather
left to experts.

These more exotic send functions are listed below, in order of
exoticness, but are not covered further in these lectures.

\begin{lstlisting}[style=SnippetStyle]
int MPI_Ssend(const void *buf,
              int count, MPI_Datatype datatype,
              int dest, int tag, MPI_Comm comm);
int MPI_Bsend(const void *buf,
              int count, MPI_Datatype datatype,
              int dest, int tag, MPI_Comm comm);
int MPI_Rsend(const void *buf,
              int count, MPI_Datatype datatype,
              int dest, int tag, MPI_Comm comm);

int MPI_Buffer_attach(void *buffer, int size);
int MPI_Buffer_detach(void *buffer_addr, int *size);
\end{lstlisting}

The nonblocking counterparts are listed below.
\begin{lstlisting}[style=SnippetStyle]
int MPI_Issend(const void *buf,
               int count, MPI_Datatype datatype,
               int dest, int tag,
               MPI_Comm comm, MPI_Request *request);
int MPI_Ibsend(const void *buf,
               int count, MPI_Datatype datatype,
               int dest, int tag,
               MPI_Comm comm, MPI_Request *request);
int MPI_Irsend(const void *buf,
               int count, MPI_Datatype datatype,
               int dest, int tag,
               MPI_Comm comm, MPI_Request *request);
\end{lstlisting}

Any type of send operation can match with any type of receive
operation, whether blocking or nonblocking\index{nonblocking}.
There is only one kind of receive operation in \mpi and completion of
a receive operation signifies that data have been received correctly
from a matching, sending process.

For completeness, we mention that it is/should be technically possible
to cancel a message. However, the semantics and guarantees of this operation
are not clear and relying on this functionality is never recommended in \mpi
programs.

\begin{lstlisting}[style=SnippetStyle]
int MPI_Cancel(MPI_Request *request);
int MPI_Test_cancelled(const MPI_Status *status, int *flag);
\end{lstlisting}

\subsection{\mpi Concept: Persistence\marksec}
\label{sec:persistentpoint}

An additional, recently extended \mpi concept, which we do not cover in
these lectures, is 
\emph{persistent (point-to-point) communication}\index{persistent communication}.
The idea is to be able to separate the initialization of a communication operation
(argument parsing, reservation of memory and communication resources,
algorithmic preprocessing) from the operation itself to make it
possible to execute the operation many times with the same arguments.
Persistent operations aim to make it possible to \emph{amortize}
possibly expensive set-up costs over many uses of the same operation.

Concretely, \mpi reuses \mpirequest handles as objects
to store the precomputed information for a persistent communication
operation. The \mpi standard defines a persistent counterpart for all
the different types of send operations as well as for the receive
operation. New operations are introduced to (re)start any
single or a whole set of persistent communication operations.

\begin{lstlisting}[style=SnippetStyle]
int MPI_Send_init(const void *buf,
                  int count, MPI_Datatype datatype,
                  int dest, int tag,
                  MPI_Comm comm, MPI_Request *request);
int MPI_Ssend_init(const void *buf,
                   int count, MPI_Datatype datatype,
                   int dest, int tag,
                   MPI_Comm comm, MPI_Request *request);
int MPI_Bsend_init(const void *buf,
                   int count, MPI_Datatype datatype,
                   int dest, int tag,
                   MPI_Comm comm, MPI_Request *request);
int MPI_Rsend_init(const void *buf,
                   int count, MPI_Datatype datatype,
                   int dest, int tag,
                   MPI_Comm comm, MPI_Request *request);
int MPI_Recv_init(void *buf,
                  int count, MPI_Datatype datatype,
                  int source, int tag,
                  MPI_Comm comm, MPI_Request *request);

int MPI_Start(MPI_Request *request);
int MPI_Startall(int count, MPI_Request requests[]);
\end{lstlisting}

Both of the start calls are local and nonblocking, although the init
calls may take a non-trivial amount of time depending on the amount
of preprocessing that can be or is done (\mpi libraries may vary).
Thus, the persistent communication
operations behave like the corresponding nonblocking
operations. Completion can be checked or enforced with the same
operations on the \mpirequest object as explained in
\Sec~\ref{sec:nonblockingpoint}.

\subsection{More on User-defined, Derived Datatypes\marksec}
\label{sec:deriveddatatypes}

The datatype argument of the communication operations seen so far
describes the process local unit of communication and the count
argument the number of such units. The units we have seen in the small
examples hitherto corresponded to the basic C datatypes like
\texttt{int}s to the \mpidatatype \mpiint, \etc
(\Tab~\ref{tab:datatypes}). A process local communication unit can be
more complex, though, and describe a whole sequence of basic datatypes
together with their relative displacements in memory. Such a
description is called the \impi{type map} of a memory layout.

Type maps are represented by \mpidatatype objects (handles) in a more
concise form than by explicit, possibly very long lists of basic
datatypes and their displacements. \mpi provides a set of constructors for
constructing new, more complex datatypes out of already existing ones:
\mpi objects cannot be changed, only new objects can be created from
existing ones. Datatype objects are called \impi{derived datatypes}
and are means to describe the structure (layout in memory) of complex
data in the local memory of a process.

A set of fundamental constructors are listed below in order of
increasing generality. That is, the structures that can be described
by one constructor can also be described by the following ones.
These, on the other hand, can describe layouts that cannot be described
by a previous one.

\begin{lstlisting}[style=SnippetStyle]
int MPI_Type_contiguous(int count, MPI_Datatype oldtype,
                        MPI_Datatype *newtype);
int MPI_Type_vector(int count, int blocklength, int stride,
                    MPI_Datatype oldtype,
                    MPI_Datatype *newtype);
int MPI_Type_create_indexed_block(int count, int blocklength,
                                  const int displacements[],
                                  MPI_Datatype oldtype,
                                  MPI_Datatype *newtype);
int MPI_Type_indexed(int count, const int blocklengths[],
                     const int displacements[],
                     MPI_Datatype oldtype,
                     MPI_Datatype *newtype);
int MPI_Type_create_struct(int count, const int blocklengths[],
                           const MPI_Aint displacements[],
                           const MPI_Datatype types[],
                           MPI_Datatype *newtype);
\end{lstlisting}

Note that the naming of these type creating functions is somewhat
inconsistent.  This has historical reasons, and the \mpi archeologist can
mine out which.

Before a derived datatype can be used in communication operations, it
must be \emph{committed} to the \mpi library. The \mpitypecommit
operation is a designated point in the program execution where the
\mpi library can perform optimizations on the type map description.
Such optimizations that can be costly can hopefully be amortized
over many uses of the same, derived datatype. As with other \mpi
objects, derived datatypes should be freed after use: They may take
up (rarely, but sometimes considerable) resources.

\begin{lstlisting}[style=SnippetStyle]
int MPI_Type_commit(MPI_Datatype *datatype);

int MPI_Type_free(MPI_Datatype *datatype);
\end{lstlisting}

As with all user-created \mpi objects,
derived datatypes explicitly created in the application must be freed. The
predefined datatypes \mpiint, \mpidouble, \etc, cannot be freed.

The constructors listed above describe the following sorts of data layouts.
As can be seen from the interface listings, all constructors take (various
kinds of) repetition counts, lists of displacements, and previously
defined units of communication described as derived datatypes.

\begin{enumerate}
\item
  A \impi{contiguous type} describes a contiguous repetition of an
  already described unit, where one unit follows immediately after the
  previous one. More formally, with $c$ units of extent $e$, the $i$th
  unit has displacement $ie, 0\leq i<c$.
\item
  A \impi{vector type} describes a regularly strided (spaced)
  repetition of blocks of an already described unit.
  More formally, with $c$ blocks of extent $e$ and stride $s$,
  the $i$th unit has displacement $ise, 0\leq i<c$.
\item
  A \impi{block index type} describes a sequence of contiguous blocks
  of previously described units, each with a specific, relative
  displacement. All blocks have the same size in number of units.  More
  formally, with $c$ blocks of $b$ units with extent $e$ and displacement
  $d_i$ for the $i$th block, the elements have displacements
  $d_ie+je, 0\leq i<c,0\leq j<b$.
  \item
  An \impi{index type} describes a sequence of blocks of previously
  described units, each with a specific, relative displacement; blocks
  may have different sizes in number of units.  More formally, with
  $c$ blocks of $b_i$ units with extent $e$ and displacement $d_i$ for
  the $i$th block, the elements have displacements $d_ie+je, 0\leq
  i<c$ and $0\leq j<b_i$.
\item
  A \impi{structured type} describes a sequence of blocks of
  previously described units, each with a specific, relative
  displacement, blocks may have different sizes in number of units,
  and the units of the blocks may be different, previously described
  units. In contrast to the previous constructors, displacements $d_i$
  are in bytes. Since the units may be different, they have possibly
  different extents $e_i$.  More formally, with $c$ blocks of $b_i$
  units with extent $e_i$ and displacement $d_i$ for the $i$th block,
  the elements have displacements $d_i+je_i, 0\leq i<c$ and $0\leq
  j<b_i$.
\end{enumerate}

The elements in contiguous blocks of units are spaced from each other
by the \impi{extent} of the unit, see \Sec~\ref{sec:datatypes}.
Likewise, all relative displacements are multiples of the extent of the
unit. Only for structured types, the displacements are 
given in bytes: A single extent multiplier in the displacement makes no sense
since the different blocks can have different units.
The extent of a newly constructed derived datatype (unit) is
the linear distance from the beginning of the first block to the end
of the last block in the unit.

It is worth noticing that with the types of constructors described
above, it is indeed possible to construct type maps where some data
elements have the same displacement. Such type maps are not \emph{per se}
illegal or disallowed. A type map with this property is said to have
\emph{overlapping entries}. The rules for matching communication are
intended to enforce that the outcome of a communication operation is
determinate. Thus, in particular datatypes used as arguments for
receive buffers in receive operations must not have overlapping
entries. For datatypes used as send arguments, this is not a problem
and allowed; whether this is good programming practice is a
different matter. Such usages should be carefully deliberated.

A first example illustrates the probably most common, often convenient
and efficient use of the vector datatype. For this, we elaborate on
the stencil example introduced in \Sec~\ref{sec:semanticterms},
where the placement of data and communication buffers was left open
until now. In the distributed stencil computation, a large matrix
$M[m,n]$ is subdivided into $p$ smaller submatrices each with 
roughly the same number of elements. The
stencil computation\index{stencil computation} updates
each matrix element $M[i,j]$ by a function (for instance, an average)
over the neighboring elements. A common stencil is the
$5$-point stencil, where $M[i,j]$ is updated by a function of the
five elements $M[i,j],M[i,j+1],M[i,j-1],M[i+1,j],M[i-1,j]$.

Now, for each \mpi process, let \texttt{matrix} be the submatrix for the
process. We implement a weakly scaling\index{weakly scaling} version of
the stencil computation\index{stencil computation},
in which the size of the local matrix is kept
constant. We let \texttt{m} and \texttt{n} be the number of local
rows and local columns, respectively. Thereby, the total size of the
matrix $M$ over all the $p$ processes
is $p(\texttt{m}\times \texttt{n})$. It is convenient to
actually think of the matrix (and the submatrices) as having two
additional rows and two additional columns, thus being of size
$M[m+2][n+2]$, and such that row and column
elements $M[-1,j],M[m,j],M[i,-1]$ and $M[i,n]$,
for $-1\leq j<n+1$ and $-1\leq i<m+1$, can be addressed in the
stencil computation\index{stencil computation}.
These extra rows and columns are called the
(sub)matrix \impi{halo} (see also the similar code in
\Sec~\ref{sec:stencil}).

In C, each submatrix with its halo can be allocated dynamically by declaring a
pointer to rows of size $n+2$ and then allocating space for $m+2$ such
rows. This is shown in the code below, which also shows how the
address of matrix element \texttt{[0][0]} is shifted, such that the
halo rows and columns can be addressed by indices \texttt{-1} and
\texttt{m} and \texttt{n}, respectively. Be careful when later freeing
this dynamically allocated memory.

\begin{lstlisting}[style=SnippetStyle]
m = ...;
n = ...; // small weak scaling example
  
double (*matrix)[n+2];
matrix = (double(*)[n+2])malloc((m+2)*(n+2)*sizeof(double));
// shift matrix; be careful with free
matrix =
  (double(*)[n+2])((char*)matrix+(n+2+1)*sizeof(double));

// initialize matrix including halo
for (i=-1; i<m+1; i++) {
  for (j=-1; j<n+1; j++) {
    matrix[i][j] = ...;
  }
}
\end{lstlisting}

We have already seen how the \mpi processes can be organized 
into a two-dimensional mesh with the \mpicartcreate operation
(\Sec~\ref{sec:organizingprocesses}), such that each process has
neighboring processes in the left, right, up and down directions (some
of which are possibly \mpiprocnull).  The halos of the process local
submatrices represent rows and columns of the full matrix that are
present at two processes. In that sense, the process submatrices have
overlapping rows and columns and this overlap has to be kept consistent.
Thus, the local stencil
updates can be performed for all matrix entries $M[i,j],0\leq
i<m,0\leq j<n$, provided that the halo rows and columns have been
filled in advance with the corresponding elements from the submatrices
at the neighboring processes. The halo column $M[i,-1]$ must be filled
with elements from the rightmost column of the left neighbor process,
the halo row $M[-1,j]$ analogously with elements from the bottom row of
the top neighbor,
and so on. Since the matrices in C are in
row major order\index{row-major order}, the rows
for the up and down neighbors are consecutive, one-dimensional arrays
in memory and can readily be sent and received. The columns, however,
are not consecutive but consist of the first element of each
row. With the row length being $n+2$ elements, this layout of data in
memory can be described as an \mpi vector type with element blocks of
one element that are strided $n+2$ elements apart. A corresponding
datatype for communication of such layouts is created by the
\mpitypevector constructor and committed for use with \mpitypecommit.
The addresses of the communication buffers of the rows and columns to
be sent to neighboring processes are now the addresses of the matrix
element $M[0,0]$ (for left and up neighbor) and $M[0,n-1]$ (for the
right neighbor) and $M[m-1,0]$ (for the down neighbor). The addresses of
the rows and columns to be received and stored in the halo rows and columns are
$M[0,-1]$ (left), $M[0,n]$ (right), $M[-1,0]$ (up) and $M[m,0]$ (down). 

\begin{lstlisting}[style=SnippetStyle]
int left, right;
int up,   down;
  
MPI_Cart_shift(cartcomm,1,1,&left,&right);
MPI_Cart_shift(cartcomm,0,1,&up,  &down);

MPI_Datatype column;

MPI_Type_vector(m,1,n+2,MPI_DOUBLE,&column);
MPI_Type_commit(&column);
  
double *out_left, *out_right, *out_up, *out_down;
double *in_left, *in_right, *in_up, *in_down;

out_left  = &matrix[0][0];
out_right = &matrix[0][n-1];
out_up    = &matrix[0][0];
out_down  = &matrix[m-1][0];

in_left   = &matrix[0][-1];
in_right  = &matrix[0][n];
in_up     = &matrix[-1][0];
in_down   = &matrix[m][0];

MPI_Request request[8];
  
int done = 0;
while (!done) { // iterate until convergence
    ... // the stencil update (computation)

  int k = 0;
  MPI_Isend(out_left, 1,column,    left, STENTAG,cartcomm,
            &request[k++]);
  MPI_Isend(out_right,1,column,    right,STENTAG,cartcomm,
            &request[k++]);
  MPI_Isend(out_up,   n,MPI_DOUBLE,up,   STENTAG,cartcomm,
            &request[k++]);
  MPI_Isend(out_down, n,MPI_DOUBLE,down, STENTAG,cartcomm,
            &request[k++]);
    
  MPI_Irecv(in_left,  1,column,    left, STENTAG,cartcomm,
            &request[k++]);
  MPI_Irecv(in_right, 1,column,    right,STENTAG,cartcomm,
            &request[k++]);
  MPI_Irecv(in_up,    n,MPI_DOUBLE,up,   STENTAG,cartcomm,
            &request[k++]);
  MPI_Irecv(in_down,  n,MPI_DOUBLE,down, STENTAG,cartcomm,
            &request[k++]);

  MPI_Waitall(k,request,MPI_STATUSES_IGNORE);

  done = 1; // some termination criterion
}

MPI_Type_free(&column);
\end{lstlisting}

Alternatively to the vector type, a resized double datatype with an
extent of $n+2$ doubles could have been used (see the discussion
below). It is an instructive exercise to work this out in detail and
to compare it against the solution just described.

In the next example the \mpitypevector constructor is used to
describe an $n$ column submatrix of an $m\times (np)$ matrix with $m$
rows and $np$ columns, where $p$ is the number of \mpi processes.  In
the program, all processes have a matrix of this size and send their
first $n$ columns to the process with rank $0$. This process reconstructs
a full $m\times (np)$ matrix by putting the $p$ column submatrices together
one after the other. Matrices are
maintained per hand in row-major order\index{row-major order}.
The elements corresponding to
$n$ consecutive columns are, thus, blocks of $n$ elements starting at
each multiple $inp$ of $np$ for $i, 0\leq i<m$. The resulting, full
$m\times (np)$ matrix is stored at process $0$ in a separately
allocated, new matrix. Thus, it cannot happen that a process sends and
receives data from overlapping memory regions.

\begin{lstlisting}[style=SnippetStyle]
int m, n; int i, j;
    
m = ...;
n = ...;

double *matrix;
matrix = (double*)malloc(m*size*n*sizeof(double));

MPI_Datatype cols;
MPI_Type_vector(m,n,n*size,MPI_DOUBLE,&cols);
MPI_Type_commit(&cols);

MPI_Request request;
MPI_Isend(matrix,1,cols,0,MATTAG,comm,&request);
if (rank==0) {
  double *newmatrix;
  newmatrix = (double*)malloc(m*size*n*sizeof(double));

  for (i=0; i<size; i++) {
    MPI_Recv(newmatrix+i*n,1,cols,i,MATTAG,
             comm,MPI_STATUS_IGNORE);
  }
}  
MPI_Wait(&request,MPI_STATUS_IGNORE);

MPI_Type_free(&cols);
\end{lstlisting}

In this example where communication is of the individual $m\times n$
submatrices and the receiving process gives the displacement
for each received submatrix explicitly, the extent of the vector
datatype does not play a role. This is not always so. Sometimes,
the default extent of a derived datatype is not what is effectively
needed in order to access and store
the data correctly in the right locations. An
important type creating function for controlling the extent of a
datatype, outside the scope of these lectures, is the resizing function.
It allows to create a new datatype with arbitrary extent from an
existing, derived datatype.

\begin{lstlisting}[style=SnippetStyle]
int MPI_Type_create_resized(MPI_Datatype oldtype,
                            MPI_Aint lb, MPI_Aint extent,
                            MPI_Datatype *newtype);
\end{lstlisting}

Should displacements in multiples of the extent of the \mpidatatype
old unit not be sufficient(ly expressive), also constructors where all
strides and displacements are given in bytes are provided.

\begin{lstlisting}[style=SnippetStyle]
int MPI_Type_hvector(int count,
                     int blocklength, MPI_Aint stride,
                     MPI_Datatype oldtype,
                     MPI_Datatype *newtype);
int MPI_Type_hindexed(int count, const int blocklengths[],
                      const MPI_Aint displacements[],
                      MPI_Datatype oldtype,
                      MPI_Datatype *newtype);
int MPI_Type_create_hvector(int count,
                            int blocklength, MPI_Aint stride,
                            MPI_Datatype oldtype,
                            MPI_Datatype *newtype);
int MPI_Type_create_hindexed_block(int count, int blocklength,
                                   const MPI_Aint displacements[],
                                   MPI_Datatype oldtype,
                                   MPI_Datatype *newtype);
int MPI_Type_create_hindexed(int count,
                             const int blocklengths[],
                             const MPI_Aint displacements[],
                             MPI_Datatype oldtype,
                             MPI_Datatype *newtype);
\end{lstlisting}

Complex, composite layouts corresponding to distributed arrays and
subarrays can be described with the following two composite derived
datatype constructors that are also well beyond the scope of these lectures
(but sometimes seen in applications).

\begin{lstlisting}[style=SnippetStyle]
int MPI_Type_create_darray(int size, int rank, int ndims,
                           const int gsizes[],
                           const int distribs[],
                           const int dargs[],
                           const int psizes[],
                           int order, MPI_Datatype oldtype,
                           MPI_Datatype *newtype);

int MPI_Type_create_subarray(int ndims,
                             const int sizes[],
                             const int subsizes[],
                             const int starts[],
                             int order, MPI_Datatype oldtype,
                             MPI_Datatype *newtype);
\end{lstlisting}

We finally mention that \mpi provides a special datatype for opaque,
compact storage of data described by derived datatypes. The datatype
for such data is \mpipacked. Three functions make it possible to
pack and unpack data into this format. This functionality should
ideally never be needed.

\begin{lstlisting}[style=SnippetStyle]
int MPI_Pack(const void *inbuf,
             int incount, MPI_Datatype datatype,
             void *outbuf, int outsize, int *position,
             MPI_Comm comm);
int MPI_Unpack(const void *inbuf, int insize, int *position,
               void *outbuf,
               int outcount, MPI_Datatype datatype,
               MPI_Comm comm);
int MPI_Pack_size(int incount, MPI_Datatype datatype,
                  MPI_Comm comm, int *size);
\end{lstlisting}

\subsection{\mpi Concept: Progress}

When is (point-to-point) communication that is eventually to happen,
for instance, by a pair of correctly matching send and receive
operations, actually happening? What are the guarantees in \mpi that
an application will \emph{progress} and eventually complete?  The
na\"ive and expected (non-)answer is ``as fast and efficiently as possible
for the underlying communication network, but possibly depending on
the overall load of the system''.

The \mpi\index{MPI} standard does not prescribe how the communication
system (hardware and software) is to be implemented. It only loosely
states the \impi{progress rule} that correct communication that could
happen eventually should happen, at the very latest when \mpifinalize
or some other \mpi operation is invoked. This gives a lot of freedom
to \mpi library implementers, and implementers indeed exploit this
freedom. There are three basic implementation alternatives to ensure
progress in \mpi.  Progress can be enforced by either:
\begin{enumerate}
\item
  hardware, meaning communication network and network
  processor or network interface card,
\item
  separate progress thread in the \mpi runtime system or
\item
  \mpi library calls which interact with the runtime system and advance
  not yet completed communication operations.
\end{enumerate}

Since \mpi library implementations rely, to different extents, on all
three mechanisms, it is usually good advice and good practice to make
\mpi calls regularly in the application to ensure that the
communication in the application is progressing.

\subsection{One-Sided Communication}
\label{sec:onesided}

With the two-sided, point-to-point communication model seen so far,
the two communicating processes are both explicitly involved in the
communication to take place, one specifying where data to be sent are
located in process local memory and how they are structured, the other
specifying where the data to be received are to go in that process'
local memory and how they are structured.  Communication can take
place when both processes have posted their respective calls. The
processes complete according to the semantics described so far for the
communication operations and modes that are being used.

In contrast, with \mpi \impi{one-sided communication}, one process
alone explicitly initiates the communication. It, therefore, has to
specify what is happening at both sides. \mpi provides one-sided
communication operations for retrieving data (\mpiget) from another
process, for transferring data to another process (\mpiput), for
transferring data to and performing an operation (an \mpiop, see
later) at another process (\mpiaccumulate), as well as a number of
special, atomic operations\index{atomic operation} on data at other
processes. These communication initiating operations are all
nonblocking\index{nonblocking}.
The process that initiates the communication operation
is, in \mpi terms, referred to as the \impi{origin process} and the
process to or from which data are transferred or retrieved
as the \impi{target process}. In order to ensure that a data transfer
has taken place and is completed, whether at origin or at target
process, an explicit synchronization must be performed. This can
involve both origin and target processes. With one-sided
communication, synchronization is, thus, explicit and decoupled from the
communication operation. This was different for point-to-point
communication. There, synchronization and completion is coupled to the
communication operation, regardless of whether it is blocking or
nonblocking. In contrast to point-to-point communication, all
one-sided communication calls are \emph{nonblocking} in the \mpi
sense.

In the distributed memory programming model\index{programming model},
processes do not share
address spaces in any way. An address (pointer) at one process has no
meaning at another process. Thus, means are needed to make it possible
for an origin process to address data at a target process. In \mpi,
processes participating in one-sided communication expose parts of
their memory in a special, distributed data structure called a
\impi{communication window}. Communication windows are represented by
a new kind of handle of type \mpiwin.  Data at target processes are
referenced by non-negative, relative displacements and translated into actual
addresses in the exposed memory. \mpi provides the collective
\mpiwincreate operation for creating communication windows. In the call,
each process supplies the process local address of the memory to expose,
the size (in bytes) of the memory it will expose, and a process local
displacement unit to be used when translating displacements into
addresses. The \mpi operations for managing windows and memory are
shown below.

\begin{lstlisting}[style=SnippetStyle]
int MPI_Win_create(void *base, MPI_Aint size, int disp_unit,
                   MPI_Info info, MPI_Comm comm,
                   MPI_Win *win);
int MPI_Win_free(MPI_Win *win);
                   
int MPI_Win_get_group(MPI_Win win, MPI_Group *group);

int MPI_Win_allocate(MPI_Aint size, int disp_unit,
                     MPI_Info info,
                     MPI_Comm comm, void *baseptr,
                     MPI_Win *win);
int MPI_Win_allocate_shared(MPI_Aint size, int disp_unit,
                           MPI_Info info, MPI_Comm comm,
                           void *baseptr, MPI_Win *win);
int MPI_Win_shared_query(MPI_Win win,
                        int rank,
                        MPI_Aint *size, int *disp_unit,
                        void *baseptr);
int MPI_Win_create_dynamic(MPI_Info info, MPI_Comm comm,
                           MPI_Win *win);
int MPI_Win_attach(MPI_Win win, void *base, MPI_Aint size);
int MPI_Win_detach(MPI_Win win, const void *base);
\end{lstlisting}

Window creation is a collective operation\index{collective operation}
for the processes in the communicator used in the call.
This means that all processes in the
communicator must eventually call \mpiwincreate. Memory per process
that is to be exposed to other processes must have been allocated in
advance, either with a C standard memory allocator like
\texttt{malloc()} or with a special, dedicated memory-allocator that
is defined in the \mpi specification and implemented by the library.
Using stack allocated
data in a communication window is dangerous practice since this memory
can disappear before the window is freed: a very subtle source of
memory bugs. The rationale for having special allocators is that a HPC
system\index{HPC} may have special regions
of memory that are particularly well suited to one-sided
communication, \eg, can be read and written by other processors with
special instructions or can be efficiently shared between some
\mpi processes (\eg, processes on the same shared memory compute node). The
special allocator (with its special free operation) makes it possible
to enforce the use of such memory regions in a portable way. Window
objects should, as always, be freed when no longer used in the
application which is done by the collective \mpiwinfree call.
However, allocated and exposed memory must be freed
explicitly; freeing memory is not taken care of by \mpiwinfree for
windows created with \mpiwincreate.

The \mpiinfo object makes it possible to provide additional
information on the use of the communication window to the \mpi
library. The special \mpiinfonull value is always a
valid argument and is the only
type of \mpi ``info'' that we will consider in these lectures.

\begin{lstlisting}[style=SnippetStyle]
int MPI_Alloc_mem(MPI_Aint size, MPI_Info info, void *baseptr);
int MPI_Free_mem(void *base);
\end{lstlisting}

\begin{lstlisting}[style=SnippetStyle]
int MPI_Win_get_info(MPI_Win win, MPI_Info *info_used);
int MPI_Win_set_info(MPI_Win win, MPI_Info info);
\end{lstlisting}

The one-sided communication operations are listed below.

\begin{lstlisting}[style=SnippetStyle]
int MPI_Get(void *origin_addr,
            int origin_count, MPI_Datatype origin_datatype,
            int target_rank, MPI_Aint target_disp,
            int target_count, MPI_Datatype target_datatype,
            MPI_Win win);
int MPI_Put(const void *origin_addr,
            int origin_count, MPI_Datatype origin_datatype,
            int target_rank, MPI_Aint target_disp,
            int target_count, MPI_Datatype target_datatype,
            MPI_Win win);

int MPI_Accumulate(const void *origin_addr,
                   int origin_count,
                   MPI_Datatype origin_datatype,
                   int target_rank, MPI_Aint target_disp,
                   int target_count,
                   MPI_Datatype target_datatype,
                   MPI_Op op, MPI_Win win);
int MPI_Get_accumulate(const void *origin_addr,
                       int origin_count,
                       MPI_Datatype origin_datatype,
                       void *result_addr,
                       int result_count,
                       MPI_Datatype result_datatype,
                       int target_rank, MPI_Aint target_disp,
                       int target_count,
                       MPI_Datatype target_datatype,
                       MPI_Op op, MPI_Win win);

int MPI_Fetch_and_op(const void *origin_addr,
                     void *result_addr,
                     MPI_Datatype datatype,
                     int target_rank, MPI_Aint target_disp,
                     MPI_Op op, MPI_Win win);
int MPI_Compare_and_swap(const void *origin_addr,
                        const void *compare_addr,
                        void *result_addr,
                        MPI_Datatype datatype,
                        int target_rank, MPI_Aint target_disp,
                        MPI_Win win);
\end{lstlisting}

The \mpiget and \mpiput calls are the two basic one-sided
communication calls. Each specifies data for the operation at the
calling \impi{origin process} in the usual form of a base address, an
element count, and a datatype that describes the kind and structure of
the elements (\Sec~\ref{sec:datatypes} and
\Sec~\ref{sec:deriveddatatypes}). What is to happen at the
\impi{target process} is likewise specified with the operation in the form
of a relative displacement, an element count, and a datatype. Data at
both origin and target processes can be arbitrarily structured, and
any predefined or committed user-defined derived
datatype\index{user-defined datatype} can be used
for both \texttt{origin\_datatype} and \texttt{target\_datatype}. The
two datatypes can even be different. However, for a one-sided
communication call to be correct, the signature of the data to be
transmitted must be a prefix of the signature of the data to be
received. Thus, for \mpiget, the sequence of data elements
described by \texttt{target\_count} and
\texttt{target\_datatype} must be a prefix of the data elements described by
\texttt{origin\_count} and \texttt{origin\_datatype}. For \mpiput, it is
the other way around. This is the same as the rule for point-to-point
communication (\Sec~\ref{sec:datatypes}). As with point-to-point
communication, also \mpiprocnull can be used as rank for the target
process: no communication will take place.

The one-sided communication calls are like the nonblocking
point-to-point operations: They only indicate that communication
eventually is to take place. When this exactly happens depends on
the synchronization mechanisms that will be used and to a very large
extent on the \mpi library implementation. In order to be able to
write provably correct programs, \mpi poses strict conditions on which
data elements can be written where. These rules, in effect, state that no
data element may possibly be (over)written by more than one one-sided
communication operation before synchronization has taken place;
programs that violate this rule are simply erroneous. As with so many
other things in \mpi, it is solely the programmer's responsibility to
ensure that this cannot happen. Thus, two or more \mpiput operations
are not allowed to put any data to the same target address. Two or
more \mpiget operations are not allowed to retrieve data to the same
origin address. Concurrent \mpiget and \mpiput operations that
reference the same address are also not allowed; this situation is a
classical \impi{data race}. Different one-sided communication
operations cannot be kept separate from each other by means of message
tags\index{message tag} as was the case for point-to-point communication.

A one-sided communication operation that accesses data at a target
process with some displacement \texttt{disp}, will access the address
\begin{displaymath}
  \texttt{base}+\texttt{disp}\cdot\texttt{disp\_unit}
\end{displaymath}
where both \texttt{base} and \texttt{disp\_unit} are the values
provided by the target process in the \mpiwincreate call. In standard
uses of one-sided communication, all processes give the same
\texttt{disp\_unit}.

The \mpiaccumulate call is like an \mpiput operation, but will apply
the supplied \mpi binary \mpiop operator (see later)
on the supplied origin and the stored target elements.
The \mpiaccumulate operation is an exception to the stated
rules: several concurrent operations can update the same
elements. Such concurrent updates are performed like atomic
operations\index{atomic operation}
but are atomic only per element. The \mpigetaccumulate
retrieves the old element values from the exposed target memory
before doing the accumulation. Only the predefined \mpiop operators and
no user-defined operators can be used (think about why this is the case).

The atomic \impi{Fetch-And-Operate (FAO)} and
\impi{Compare-And-Swap (CAS)} operations provide
atomic operation functionality to \mpi and
can be used (only) on single elements of a predefined datatype. An
efficient \mpi library implementation may be able to execute these
calls by native, atomic operations, at least under some circumstances.

\subsection{One-Sided Completion and Synchronization}
\label{sec:onesidedcompletion}

A one-sided communication operation by itself is nonblocking and
neither determines when data are transferred between origin and target
processes nor when data will be available at either of the
processes. This must be enforced by explicit synchronization
operations.

In order to understand, work with, and reason about one-sided
communication, \mpi employs a so-called \impi{communication epoch}
model.  From each process' point of view, one-sided communication
takes place in disjoint epochs. Epochs are opened and closed by
synchronization operations. A process that wants to access the window
memory of some other process must open a next epoch for \emph{access}
to that process (\impi{access epoch}).  A process whose window memory
may be accessed by another process must open an epoch for
\emph{exposure} to that process (\impi{exposure epoch}).

The \mpi one-sided communication model provides two kinds of
synchronization operations for opening epochs: With
\impi{active synchronization}, both origin and target processes actively open
their respective access and exposure epochs. With
\impi{passive synchronization}, the origin process alone will open an epoch for
access (at the origin process) and exposure (at the target process).
Epochs must be explicitly closed. When an origin process closes its
access epoch, all one-sided communication operations will be completed
from the origin process' point-of-view. In particular, all data
elements retrieved by \mpiget or \mpigetaccumulate operations will be
available for use. When a target process closes its exposure epoch,
all one-sided communication operations on that target will be complete
at the target. In particular, data transferred with \mpiput will be
available for use. Operations for closing epochs are thus blocking.

\mpi provides two kinds of operations for active synchronization. The
\mpiwinfence is a collective operation\index{collective operation}
over all processes belonging to
the window. An \mpiwinfence will close a preceding epoch and
open an access epoch with access to all other processes
and an exposure epoch giving exposure to all other
processes for
each of the processes.
The \mpiwinfence operation has non-local completion semantics
and may thus have to wait for other processes to perform their corresponding
\mpiwinfence call.

Dedicated, more specific control over access and exposure is provided
by the \mpiwinstart and \mpiwinpost operations. The first 
provides access to a group of processes (represented as \mpigroup
objects, see \Sec~\ref{sec:groups}), the second one grants
exposure to a group of processes. Access and exposure epochs are
explicitly closed with \mpiwincomplete and \mpiwinwait, respectively.
The \mpiwintest operation is a nonblocking version of \mpiwinwait.
The \mpiwinstart operation has non-local completion semantics and
may thus have to wait for the processes that are to be accessed
to perform their \mpiwinpost call. The \mpiwinpost operation has local
completion semantics.  Therefore, in the frequent case where a process
both seeks access and grants access to other processes, the
\mpiwinpost call should be performed before the \mpiwinstart call. The
other order is \emph{unsafe}\index{unsafe} and the program may
deadlock.

\begin{lstlisting}[style=SnippetStyle]
int MPI_Win_fence(int assert, MPI_Win win);

int MPI_Win_post(MPI_Group group, int assert,
                 MPI_Win win); // for exposure
int MPI_Win_start(MPI_Group group, int assert,
                  MPI_Win win); // for access
int MPI_Win_complete(MPI_Win win);
int MPI_Win_wait(MPI_Win win);
int MPI_Win_test(MPI_Win win, int *flag);

int MPI_Win_lock(int lock_type, int rank, int assert,
                 MPI_Win win);
int MPI_Win_unlock(int rank, MPI_Win win);

int MPI_Win_lock_all(int assert, MPI_Win win);
int MPI_Win_unlock_all(MPI_Win win);
\end{lstlisting}

The \mpiwinlock and \mpiwinunlock operations passively open a target
exposure epoch and an origin access epoch. A target can be opened for
exclusive access by the locking origin process alone by providing the
\mpilockexclusive lock type. A target can be opened for shared,
concurrent access by more than one \mpi process by providing the
\mpilockshared lock type.

These operations have nothing to do with locks (mutexes\index{mutex})
in the sense seen so
far (see \Sec~\ref{sec:melocks}): They do not and cannot provide
mutual exclusion\index{mutual exclusion}\index{Lock}.
When a target process is
``locked'' exclusively, data can indeed be accessed by the one-sided
communication operations but since the \mpiput and \mpiget operations
are nonblocking\index{nonblocking}, nothing can be done with this
data. The exception is, of course, the \mpiaccumulate operations.  In
order to use the data from \mpiget, access and exposure epochs have to
be closed by the \mpiwinunlock call. When this happens, another
process may come between and ``lock'' the target and change the data.
Read (get), compute and update (put) under mutual exclusion is, thus, not
provided.

\subsection{Example: One-Sided Stencil Updates}
\label{sec:onesidedstencil}

As an example we now implement the stencil update that we saw before
with blocking and nonblocking point-to-point communication using
one-sided communication instead\index{stencil computation}. For this,
a window is created from the Cartesian communicator that was created
for defining the neighborhoods, see \Sec~\ref{sec:semanticterms}.
An advantage of this implementation over the point-to-point
implementations is flexibility: It could be that not
all four neighbors have to be updated in some iteration.
With the one-sided implementation, the corresponding \mpiget calls could
simply be dropped. We here give a full-fledged implementation also using a
vector datatype (see \Sec~\ref{sec:deriveddatatypes}).

First, we implement the stencil update with active, collective \mpiwinfence
synchronization for opening access and exposure epoch on all processes, for
all processes.

\begin{lstlisting}[style=SnippetStyle]
int left, right;
int up,   down;
  
MPI_Cart_shift(cartcomm,1,1,&left,&right);
MPI_Cart_shift(cartcomm,0,1,&up,  &down);

MPI_Datatype column;
  
MPI_Type_vector(m,1,n+2,MPI_DOUBLE,&column);
MPI_Type_commit(&column);
  
double *out_left, *out_right, *out_up, *out_down;
double *in_left, *in_right, *in_up, *in_down;

out_left  = &matrix[0][0];
out_right = &matrix[0][n-1];
out_up    = &matrix[0][0];
out_down  = &matrix[m-1][0];

in_left   = &matrix[0][-1];
in_right  = &matrix[0][n];
in_up     = &matrix[-1][0];
in_down   = &matrix[m][0];
  
MPI_Win win;

MPI_Win_create((double*)matrix-(n+2+1),
               (m+2)*(n+2)*sizeof(double),sizeof(double),
               MPI_INFO_NULL,cartcomm,&win);
  
int disp_left, disp_right, disp_up, disp_down;
  
disp_left  = (n+2)+n;
disp_right = (n+2)+1; 
disp_up    = m*(n+2)+1;
disp_down  = (n+2)+1;

int done = 0;
while (!done) { // iterate until convergence
    ... // the stencil update (computation)

  MPI_Win_fence(MPI_MODE_NOPRECEDE,win);

  MPI_Get(in_left,   1,column,left,
          disp_left, 1,column,win);
  MPI_Get(in_right,  1,column,right,
          disp_right,1,column,win);
  MPI_Get(in_up,     n,MPI_DOUBLE,up,
          disp_up,   n,MPI_DOUBLE,win);
  MPI_Get(in_down,   n,MPI_DOUBLE,down,
          disp_down, n,MPI_DOUBLE,win);

  MPI_Win_fence(MPI_MODE_NOSUCCEED,win);
  // data available
    
  done = 1; // some termination criterion
}
  
MPI_Win_free(&win);
MPI_Type_free(&column);
\end{lstlisting}

Because of the collective nature of the \mpiwinfence operations, the processes
are ``more synchronized'' than needed. Each process needs to access
window memory of its at most four neighboring processes and likewise
provide exposure to these processes. For such situations, the
dedicated synchronization mechanism could be more efficient, providing
a looser form of synchronization.

\begin{lstlisting}[style=SnippetStyle]
int neighbors[4];
  
MPI_Group group;
MPI_Group accessexposure;

MPI_Comm_group(cartcomm,&group);
int k = 0;
if (left!=MPI_PROC_NULL)  neighbors[k++] = left;
if (right!=MPI_PROC_NULL) neighbors[k++] = right;
if (up!=MPI_PROC_NULL)    neighbors[k++] = up;
if (down!=MPI_PROC_NULL)  neighbors[k++] = down;
MPI_Group_incl(group,k,neighbors,&accessexposure);

int done = 0;
while (!done) { // iterate until convergence
    ... // the stencil update (computation)

  MPI_Win_post(accessexposure,0,win);
  MPI_Win_start(accessexposure,0,win);

  MPI_Put(out_left,  1,column,left,
          disp_left, 1,column,win);
  MPI_Put(out_right, 1,column,right,
          disp_right,1,column,win);
  MPI_Put(out_up,    n,MPI_DOUBLE,up,
          disp_up,   n,MPI_DOUBLE,win);
  MPI_Put(out_down,  n,MPI_DOUBLE,down,
          disp_down, n,MPI_DOUBLE,win);
   
  MPI_Win_complete(win);
  MPI_Win_wait(win);

  done = 1; // some termination criterion
}
\end{lstlisting}

\subsection{Example: Distributed Memory Binary Search}
\label{sec:onesidedbinsearch}

The following binary search example illustrates a situation where one-sided
communication is a more suitable model than two-sided point-to-point
communication with \mpisend and \mpirecv. The situation here is that a
process needs data from some other process, which is, however, 
not aware of that need. One-sided communication makes it possible
for the process that knows to alone do the communication!

Let \texttt{a} be a distributed array with local blocks, all of the
same size \texttt{n}. Assume that the distributed array is ordered:
Within each process local block, the elements are
ordered, and the elements of the block of some process are smaller
than or equal to the elements of the local block of the next (higher
ranked) process. The total number of array elements over all processes
is \texttt{n*p} where \texttt{p} is the number of processes
in our communication window. We want to do binary search in such an array.
Each process should be allowed to initiate a search for some element
\texttt{x}.  The result shall be a global index \texttt{i}, such that
$\texttt{a[i]}\leq \texttt{x}<\texttt{a[i+1]}$. From the global index,
the process where the element \texttt{x} was found and the relative
index in the block of that process can easily be computed.

In the code, we assume that the window \texttt{win} has already been created
and that the local \texttt{a} arrays are of C type \texttt{float}.

\begin{lstlisting}[style=SnippetStyle]
int l, u, m;
int target, locali;

float ma;
  
l = -1; u = n*p; // total size of distributed array
  
do {
  m =(l+u)/2;

  target = m/n; // locate middle element
  locali = m%n;

  MPI_Win_lock(MPI_LOCK_SHARED,target,0,win);
  MPI_Get(&ma,1,MPI_FLOAT,
          target,locali,1,MPI_FLOAT,win); // get middle element
  MPI_Win_unlock(target,win);

  if (x<ma) u = m; else l = m;
} while (l+1<u);
\end{lstlisting}

Binary search takes $O(\log n)$ iterations in each of which the
searching process passively synchronizes (with \mpiwinlock) with a
target process, which is determined by dividing the index \texttt{m}
to be accessed with the block size. The displacement to be accessed is
the index modulo the block size. Since the target process only reads
elements, \mpilockshared exposure at the target is sufficient and can
allow other MPI processes to search concurrently.

Merging\index{merging} by co-ranking\index{co-rank}
can be implemented by similar considerations. It is a good exercise to do this.

\subsection{Additional One-Sided Communication Operations\marksec}

The one-sided communication model provides communication operations
that return an \mpirequest object that can be used for individually
testing or enforcing completion of that operation, similar to the nonblocking
point-to-point communication operations. They are listed here for
completeness.

\begin{lstlisting}[style=SnippetStyle]
int MPI_Rput(const void *origin_addr,
             int origin_count, MPI_Datatype origin_datatype,
             int target_rank, MPI_Aint target_disp,
             int target_count, MPI_Datatype target_datatype,
             MPI_Win win, MPI_Request *request);
int MPI_Rget(void *origin_addr,
             int origin_count, MPI_Datatype origin_datatype,
             int target_rank, MPI_Aint target_disp,
             int target_count, MPI_Datatype target_datatype,
             MPI_Win win, MPI_Request *request);
int MPI_Raccumulate(const void *origin_addr,
                    int origin_count,
                    MPI_Datatype origin_datatype,
                    int target_rank, MPI_Aint target_disp,
                    int target_count,
                    MPI_Datatype target_datatype, MPI_Op op,
                    MPI_Win win,MPI_Request *request);
int MPI_Rget_accumulate(const void *origin_addr,
                        int origin_count,
                        MPI_Datatype origin_datatype,
                        void *result_addr,
                        int result_count,
                        MPI_Datatype result_datatype,
                        int target_rank, MPI_Aint target_disp,
                        int target_count,
                        MPI_Datatype target_datatype,
                        MPI_Op op,
                        MPI_Win win, MPI_Request *request);
\end{lstlisting}

\subsection{\mpi Concept: Collective Semantics}

So far, we have seen many examples of \mpi\index{MPI} operations that are
collective in the sense that they have to be called by all processes
belonging to the input communicator\index{communicator}.
More concretely, if a collective operation\index{collective operation}
$C$ on a communicator \texttt{comm} is called by
some process in \texttt{comm}, then all other processes in \texttt{comm} must
also eventually call $C$ and no other collective operation $C'$ before $C$ on
\texttt{comm}. By this rule, for each communicator the application
programmer must ensure that all collective calls are done in the same
order by all processes in the communicator. As with other calls and
operations in \mpi, disregarding this rule and doing something else is
plain wrong and the outcome undefined. Concretely, this means that any
behavior is possible: deadlock, memory corruption, immediate program
crash, and even successful completion with apparently sensible
results. The latter is the most misleading and dangerous behavior!

Collective operations\index{collective operation}
like $C$ are always invoked \emph{symmetrically}.
That is, the same function $C$ is called by all processes, but the
processes can give different parameters, and the arguments can have a
different meaning on the different processes (see shortly). For all
collectives, arguments must be given
\emph{consistently}\index{consistent arguments} over the calling
processes. This means different things for different collectives.
For instance, for the \mpicommcreate collective
operation (see \Sec~\ref{sec:communicators}), there are rules for
the input group arguments, namely that all processes that belong to a
group given as input by some process must call with an equivalent
\texttt{group} argument.  Recall that groups are process local
objects; in the collective call, all processes in the \texttt{group}
must have created a group for the same set of processes in the same
order. Disregarding such rules on consistent arguments is erroneous. There is
no guarantee on how an \mpi library may react (deadlock, crash, weird
results, \ldots). 

Here are two further examples illustrating the consistency rules, anticipating
the collective operations\index{collective operation}
to be discussed in the next
section. The \mpibcast operation\index{broadcast operation}
broadcasts a buffer of some number of
elements from a \texttt{root} process to all other processes in the
communicator. It is a consistency requirement that all processes
specify the same \texttt{root} process and exactly the same number of
elements (adhering to the type signature rules). In the first example (below),
the non-root processes inadvertently give a larger element count than
the root process. The program may well run with some \mpi libraries,
but the outcome will sooner or later prove fatal: the last, fourth
element in the \texttt{dims} array has never been received by the
non-root processes. The \texttt{dims[3]} element could be anything.

\begin{lstlisting}[style=SnippetStyle]
MPI_Comm_rank(comm,&rank);
MPI_Comm_size(comm,&size);

if (rank==root)
  MPI_Bcast(&dims[0],3,MPI_INT,root,comm);
} else {
  MPI_Bcast(&dims[0],4,MPI_INT,root,comm);
}
\end{lstlisting}

In the second example, the non-roots give the fixed root value $0$ for
the fourth argument of the \mpibcast call. The consistency requirement
for \mpibcast is, however, that all processes must give the
same value for the \texttt{root} argument. The program will most
likely hang with most \mpi libraries when \texttt{root} is \emph{not}
process $0$ in the communicator (deadlock!).

\begin{lstlisting}[style=SnippetStyle]
MPI_Comm_rank(comm,&rank);
MPI_Comm_size(comm,&size);

if (rank==root)
  MPI_Bcast(&dims[0],4,MPI_INT,root,comm);
} else {
  MPI_Bcast(&dims[0],4,MPI_INT,0,comm);
}
\end{lstlisting}

Note that the broadcast collective operation\index{collective operation}
is invoked symmetrically
by all processes making their \mpibcast call. The different,
asymmetric outcomes for the different processes (root and non-roots)
are determined by the supplied input arguments.  In contrast,
point-to-point communication (and also one-sided communication) is
\emph{non-symmetric}: There are distinct send operations, like
\mpisend, \mpiisend, \ldots, and different, distinct receive
operations, like \mpirecv and \mpiirecv.  A process determines its
role (sender or receiver) in the communication operation by the
appropriate, different calls.  One-sided communication is the extreme
case: Only one side makes a communication call at all. Also some of
the synchronization operations are asymmetric. A sometimes seen
beginner's mistake is
to try to perform a broadcast by letting the root process call \mpibcast
while non-roots try to get the data by calling \mpirecv. Such programs
(almost) never work (if they do, by luck).

Collective operations\index{collective operation}
seen so far, and also those that will be
introduced in the next section, are all \impi{blocking} in the \mpi
sense. When a process returns from a collective call $C$, the
operation has been completed from that process' point of view. All
resources needed for the call have been given free by
the call and can be reused. In collective operations for exchanging
information between processes, this in particular means that data are
out of the process' send buffers, and have been delivered in its
receive buffers. Send buffers can again be used freely to store new data
for the following communication operations, and values in receive buffers
can be used for computation by the process.

Like for point-to-point communication, also some nonblocking
collective operations\index{collective operation} have been defined in
\mpi. The semantic rules are slightly different than those for
nonblocking point-to-point communication\index{nonblocking}.
Nonblocking collective operations are beyond these lectures (some will
be mentioned for completeness, though, see
\Sec~\ref{sec:nonblockingcollectives}).

Blocking collective operations\index{collective operation}
have \impi{non-local completion}. This
means (as for point-to-point communication) that for a process to
complete a collective call, it may require, and in most cases does
require(!), that the other processes in the communicator actively
engage in the operation. The rules for correct usage of collective
operations exactly ensure that for any collective call $C$ made by
some process eventually all processes in the communicator
will have made the collective call to $C$. At the latest at that point, $C$
can be completed for the processes.

On the other hand, collective operations\index{collective operation}
are and should indeed be
thought of as \impi{non-synchronizing} by the application
programmer. A process returning from its blocking collective call $C$
cannot make any inference about what any of the other processes have
done or not done. Some processes may not even have reached the point
in their code where they perform the $C$ call! There is one
conspicuous, obvious exception to this rule (think ahead).

A program using collective operations\index{collective operation}
that relies on synchronizing
behavior or makes any such assumptions is called
\impi{unsafe}. We stress again:
Unsafe programming is a pernicious practice. An unsafe program
may well run under some circumstances (\mpi library, system, number of
compute-nodes, problem size, \ldots) and then suddenly not run
anymore (or produce wrong results) when circumstances
change\index{unsafe programming}.  Unsafe programs are non-portable
programs!

\subsection{Collective Communication and Reduction Operations}
\label{sec:collective}

Collective communication in \mpi, the third important communication
model (for thoughtful amusement, see~\cite{Gorlatch04}),
more specifically refers to the small set of $17$ functions or
patterns for data exchange and reductions over all processes in a
communicator (see \Sec~\ref{sec:exchangepatterns} and \Sec~\ref{sec:barrier}).
These $17$ collective operations\index{collective operation}
are what is commonly meant by the term (\mpi)
\impi{collectives}.

The \mpi collectives are broadly of the following types
(see \Sec~\ref{sec:exchangepatterns}):

\begin{itemize}
\item
  A \impi{barrier operation} ensures that all processes have reached
  a certain point in their execution.
\item
  A \impi{broadcast operation} transfers the same data from one
  designated process to all other processes.
\item
  A \impi{gather operation} collects data from all processes on one
  designated process.
\item
  A \impi{scatter operation} transfers different, individual data from one
  designated process to each of the other processes.
\item
  An \impi{allgather operation}, also known as all-to-all broadcast,
  gathers the same data to all processes, or, equivalently,
  broadcasts data from each process to all other processes.
\item
  An \impi{all-to-all operation}, also known as \impi{personalized exchange}
  or \emph{transpose}, transfers different, individual data from each
  process to each of the other processes.
\item
  A \impi{reduction operation} applies a binary, associative operator
  in order to data contributed by all processes and makes the result available
  to one or all processes in total or in part.
\item
  A \impi{scan operation}\index{scan} performs a prefix sums\index{prefix sums}
  computation in rank order\index{rank order} on data contributed by the processes.
\end{itemize}

The designated process for the broadcast, gather and scatter
operations is called the \impi{root process} or just \emph{root}. The
operations exist in different variants according to the amount of data
that are supplied and collected by the processes. Variants
where each process either receives or sends the same amount
of data to other processes are called \emph{regular}\index{regular collective}.
Variants where
different pairs of processes may send and/or receive amounts of data that are
different from other processes' amounts are called
\emph{irregular}\index{irregular collective}.
For historical reasons, the irregular variants of
the \mpi collective operations\index{collective operation}
are sometimes (but not always) called
``vector'' (and ``vee'') variants.
Data are always specified as blocks of elements,
each block by a count and (derived) datatype argument. It is sometimes
helpful, especially for the reduction and scan\index{scan} operations, to think of
input and output as mathematical vectors of elements,
most often of the same, basic
datatype like \mpiint, \mpifloat, \etc.

\begin{table}[t!]
  \caption{Classification of the \mpi collectives along the
    dimensions of pairwise data regularity and rootedness (symmetry).}
  \label{tab:collectiveclassification}
\begin{center}
\begin{tabular}{ccc}  
  \toprule
  & Regular & Irregular (vector) \\
  & \mpibarrier & \\
  \midrule
  \multirow{4}{*}{Rooted (asymmetric)}
  & \mpibcast & \\
  & \mpigather & \mpigatherv \\
  & \mpiscatter & \mpiscatterv \\
  & \mpireduce & \\
  \midrule
  \multirow{7}{*}{Non-rooted (symmetric)}
  & \mpiallgather & \mpiallgatherv \\
  & \mpialltoall & \mpialltoallv \\
  & & \mpialltoallw \\
  & \mpiallreduce & \\
  & \mpireducescatterblock & \mpireducescatter \\
  & \mpiscan & \\
  & \mpiexscan & \\
  \bottomrule
\end{tabular}
\end{center}
\end{table}

The usage of the terms is not always consistent and different people
sometimes mean different things or use different words.  It is
maybe helpful as a mnemonic to classify the collectives based on
the regularity of data exchanged and whether some
process has a special role: Regular \vs irregular (``vector'') and
rooted (asymmetric) \vs non-rooted (symmetric). See
\Tab~\ref{tab:collectiveclassification} for such a classification
using the names given to the collective operations\index{collective operation}
by \mpi.

The performance and concrete implementation of the collectives are, as for
everything else in \mpi, \emph{not} specified by the \mpi standard.
In order to say something about what can be expected, in particular, to
make performance predictions, assumptions have
to be imposed from the outside.

\begin{table}[t]
  \caption{Time complexity of the \mpi
    collective operations\index{collective operation} in the
    linear-affine cost communication model under fully-connected network (and
    one-ported) communication assumptions. The total problem size is
    $m$ and the number of processes $p$.}
  \label{tab:collectivecomplexity-fully}
\begin{center}
  \begin{tabular}{cc}
    \toprule
    Collective & Time complexity $T_{p}(m)$ \\
    \midrule
    \mpibarrier & $O(\log p)$ \\
    \midrule
  \mpibcast & $O(m+\log p)$ \\
  \mpigather & $O(m+\log p)$ \\
  \mpiscatter & $O(m+\log p)$ \\
  \mpiallgather & $O(m+\log p)$ \\
  \mpialltoall & Between $O(m+p)$ and $O(m\log p)$ \\
  \midrule
  \mpireduce & $O(m+\log p)$ \\
  \mpiallreduce & $O(m+\log p)$ \\
  \mpireducescatterblock & $O(m+\log p)$ \\
  \mpiscan & $O(m+\log p)$ \\
  \mpiexscan & $O(m+\log p)$ \\
  \bottomrule
\end{tabular}
\end{center}
\end{table}

Time complexities of the regular collectives in a simple, homogeneous,
linear-affine cost transmission model (see
\Sec~\ref{sec:communicationcosts}) on fully connected networks with
one-ported communication capabilities with $p$ processors and total
data $m$ are as stated in \Tab~\ref{tab:collectivecomplexity-fully}.
On networks that are not fully connected, having diameter larger than
one (see \Sec~\ref{sec:structureandtopology}), the time complexities are
as stated in \Tab~\ref{tab:collectivecomplexity-diameter}.  Finding
the algorithms that achieve these bounds is not at all trivial. A good
starting point for the interested reader
is~\cite{ChanHeimlichPurkayasthavandeGeijn07} and~\cite{Bruck97} with
interesting trade-offs for all-to-all communication. For collective
algorithms, it is important that the dominating terms in the upper
bound, which often correspond to the number of communication rounds or
critical path length, have small constants. Analyzing (and
improving) these constant terms is important.

\begin{table}[t]
  \caption{Time complexity of the \mpi
    collective operations\index{collective operation} in the
    linear-affine cost communication model under non-fully connected network
    assumptions. The total problem size is $m$ and the number of
    processes $p$ and the network diameter $d$.}
  \label{tab:collectivecomplexity-diameter}
\begin{center}
  \begin{tabular}{cc}
    \toprule
    Collective & Time complexity $T_p(m)$ \\
    \midrule
    \mpibarrier & $O(d)$ \\
    \midrule
  \mpibcast & $O(m+d)$ \\
  \mpigather & $O(m+d)$ \\
  \mpiscatter & $O(m+d)$ \\
  \mpiallgather & $O(m+d)$ \\
  \mpialltoall & $O(m+pd)$ \\
  \midrule
  \mpireduce & $O(m+d)$ \\
  \mpiallreduce & $O(m+d)$ \\
  \mpireducescatterblock & $O(m+d)$ \\
  \mpiscan & $O(m+d)$ \\
  \mpiexscan & $O(m+d)$ \\
  \bottomrule
\end{tabular}
\end{center}
\end{table}

The interface specifications for the regular communication/data
exchange collectives are listed below. The \mpibarrier operation is
special: It does not communicate any data but has the sole effect of
logically synchronizing the processes. All processes in the
communicator must eventually call the barrier operation,
and no process is allowed to return from this blocking call before all other
processes have made their call to \mpibarrier. This is the only
collective with synchronizing behavior where a process
that returns from its call can infer and rely on (all) other processes
also having made the call. For all other blocking collectives, the
return from a call by a process means only that the operation has been
completed from that process' point of view. It is not possible to
infer anything about the other processes in general and some may not even
have made the corresponding call. Relying on synchronizing behavior of
collectives is \impi{unsafe programming} and
can lead to unpleasant surprises with errors that can be very
hard to debug.

We now give the signatures for and discuss the individual \mpi collectives.

\begin{lstlisting}[style=SnippetStyle]
int MPI_Barrier(MPI_Comm comm);

int MPI_Bcast(void *buffer, int count, MPI_Datatype datatype,
              int root, MPI_Comm comm);

int MPI_Gather(const void *sendbuf,
               int sendcount, MPI_Datatype sendtype,
               void *recvbuf,
               int recvcount, MPI_Datatype recvtype,
               int root, MPI_Comm comm);
int MPI_Scatter(const void *sendbuf,
                int sendcount, MPI_Datatype sendtype,
                void *recvbuf,
                int recvcount, MPI_Datatype recvtype,
                int root, MPI_Comm comm);
                 
int MPI_Allgather(const void *sendbuf,
                  int sendcount, MPI_Datatype sendtype,
                  void *recvbuf,
                  int recvcount, MPI_Datatype recvtype,
                  MPI_Comm comm);
int MPI_Alltoall(const void *sendbuf,
                 int sendcount, MPI_Datatype sendtype,
                 void *recvbuf,
                 int recvcount, MPI_Datatype recvtype,
                 MPI_Comm comm);
\end{lstlisting}

For the \mpibcast operation, the designated \emph{root process} (the
process with rank equal to \texttt{root}) transfers the data stored
at the address \texttt{buffer} to the other processes in the
communicator used in the call. Data consists of \texttt{count} elements
of type and structure described by the \texttt{datatype} argument. The
processes can give different datatype and count arguments, but all
processes must specify \emph{exactly} the same type
signature\index{type signature}: the same lists
of elements of a basic datatype. The collective rule is, thus, stricter than
the signature rules for point-to-point and one-sided
communication. Also, all processes must give the same value for the
\texttt{root} argument; if they do not, a deadlock\index{deadlock}
is likely to occur.
Whether this actually happens depends on the concrete \mpi library
implementation and on the context of the call.

For the other collectives, similar rules apply. Data leaving a process
are specified in the send buffer arguments and data to be received by
a process in the receive buffer arguments. Again, signatures between
processes where a data transfer is to take place must be
\emph{identical}. For the rooted collectives, all processes must give the
same root argument. In all the
collective operations\index{collective operation}, count and
datatype arguments together describe one \emph{block} of data. As we
will see, some collectives send and/or receive $p$ blocks of data on
communicators with $p$ processes. It is solely the programmers
responsibility to ensure that in such cases, enough memory space has been
allocated. Forgetting this is a common mistake, the consequence of
which is almost always memory corruption and program crash.

The \mpigather operation collects different data from all $p$ processes to the
designated root process. The data to be stored at the root process are
stored starting at the \texttt{recvbuf} address. The data from each
process will consist of \texttt{recvcount} elements, all of the type
and structure described by the \texttt{recvtype} (derived)
datatype. The data from the processes are stored in \impi{rank order},
with the data from process $i$ at the address
\begin{displaymath}
  \texttt{recvbuf}+i\cdot\texttt{recvcount}\cdot\texttt{extent}
\end{displaymath}
where \texttt{extent} is the extent in bytes of the
\texttt{recvtype} datatype, as can be found by the \mpitypeextent call
(explained \Sec~\ref{sec:datatypes}). Thus, $p$ blocks with the same
structure are received, offset from the address given by \texttt{recvbuf}.
The \texttt{recvbuf} must have
been allocated large enough for these $p$ blocks. The data that a process
contributes are stored starting at the \texttt{sendbuf}
address. Each process contributes \texttt{sendcount} elements of
type and structure given by \texttt{sendtype}. Each process' send
signature must be identical to the signature of the received data. For
all non-root processes, the receive buffer arguments are not
significant. All processes contribute data to the root, including the
root itself! The data from the root to the root are stored at the
address $\texttt{recvbuf}+\texttt{root}\cdot
\texttt{recvcount}\cdot\texttt{extent}$. This may incur a memory copy
operation at the root process. Such a perhaps costly (perhaps not)
memory copy can be avoided by letting the root process give the
special address argument \mpiinplace for the \texttt{sendbuf}
argument. This means that the root process does not care about data from
itself and nothing will be copied into the \texttt{recvbuf} from the root.
Many other collective operations\index{collective operation}
have the same ``problem'',
and the \mpiinplace argument can be applied in many cases.

The \mpiscatter operation is the counterpart (some say ``dual'')
of the \mpigather operation.
Data blocks stored at the root process in rank order\index{rank order}
in \texttt{sendbuf} are transmitted to the other processes 
from this buffer. The data for process $i$ are stored
at the address
\begin{displaymath}
  \texttt{sendbuf}+i\cdot\texttt{sendcount}\cdot\texttt{extent}
\end{displaymath}
where \texttt{extent} is the extent (in bytes) of the \texttt{sendtype}
(derived) datatype. Same rules and considerations as for \mpigather
apply, including the caveat on sufficient buffer space.
Also here, the \mpiinplace argument can be given as the
\texttt{recvbuf} argument at the root to prevent that data are copied
from the send buffer to the receive buffer at the root.

Here is an example illustrating the use of the \mpigather collective
together with derived datatypes. An $m\times (np)$ matrix is to be put
together from column submatrices of $n$ columns (out of $np$ columns in
total) at the root process. This is done by gathering the column
submatrices at the root. It is a good exercise to recap the extent
rules for \mpigather and figure out why it is necessary to modify the
extent of the receive datatype (by creating a new datatype with the
\mpityperesized operation, see \Sec~\ref{sec:deriveddatatypes}).

\begin{lstlisting}[style=SnippetStyle]
double (*matrix)[n];
matrix = (double(*)[n])malloc(m*n*size*sizeof(double));

MPI_Datatype vec, cols;
MPI_Type_vector(m,n,n*size,MPI_DOUBLE,&vec);
MPI_Type_create_resized(vec,0,n*sizeof(double),&cols);
MPI_Type_commit(&cols);
  
double (*fullmatrix)[size*n];
if (rank==root) {
  fullmatrix =
    (double(*)[size*n])malloc(m*n*size*sizeof(double));
}
  
MPI_Gather(matrix,m*n,MPI_DOUBLE,fullmatrix,1,cols,root,comm);

MPI_Type_free(&vec);
MPI_Type_free(&cols);

free(matrix);
if (rank==root) free(fullmatrix);
\end{lstlisting}

The \mpiallgather operation has the same effect as if each process would
be the root in an \mpigather operation and would send the same data in each
of these \mpigather operations;
that is, the same effect as $p$ (the number of \mpi
processes in the communicator) \mpigather operations
with root arguments $i=0,\ldots,p-1$ and the same other arguments.
Equivalently, \mpiallgather
has the same effect as if each process $i, 0\leq i<p$
would copy its data from its \texttt{sendbuf} to the address
$\texttt{recvbuf}+i\cdot\texttt{recvcount}\cdot\texttt{extent}$ and
perform a broadcast operation from this buffer of \texttt{recvcount}
elements described by the \texttt{recvtype}
datatype. In other
words, data from all processes are gathered in rank order\index{rank order}
by all processes.
The \mpiinplace argument can be used to indicate that the data from
a process are already in the correct position in that process'
\texttt{recvbuf}. Thus, the copy operation above could be saved.
The \mpi rules for \mpiinplace for \mpiallgather are strict,
though, and require that if some process give \mpiinplace as
\texttt{sendbuf} argument, then all processes must do so.

Finally, in the \mpialltoall operation, each process has individual
(``personalized'') data to transmit to each other process.  The data
for process $i, 0\leq i<p$ are stored starting from address
\begin{displaymath}
  \texttt{sendbuf}+i\cdot\texttt{sendcount}\cdot\texttt{sendextent}
\end{displaymath}
and the data from process $j$ are received and stored starting at
address
\begin{displaymath}
  \texttt{recvbuf}+j\cdot\texttt{recvcount}\cdot\texttt{recvextent}
  \quad .
\end{displaymath}

The data sent to each process consist of \texttt{sendcount} elements
of type and structure described by \texttt{sendtype}, and the data
received of \texttt{recvcount} elements are as described by
\texttt{recvtype}. As can be seen, the \mpialltoall operation has the
same effect as $p$ \mpiscatter operations with roots $i=0,\ldots,p-1$
or as $p$ \mpigather operations with roots $i=0,\ldots,p-1$. For
completeness, we mention that the \mpiinplace
argument can also be used with \mpialltoall,
but with a quite different meaning and flavor:
The \mpiinplace argument can be given for the \texttt{sendbuf}
argument in which cases data are sent from and received (replaced) in the
same \texttt{recvbuf} address (in rank order\index{rank order}). If used,
all processes must call with the \mpiinplace argument.

For the gather, scatter, allgather, and all-to-all operations, also
so-called irregular or ``vector'' variants are defined in \mpi.
The interface specifications for these irregular
communication/data exchange collectives are listed below.

\begin{lstlisting}[style=SnippetStyle]
int MPI_Gatherv(const void *sendbuf,
                int sendcount, MPI_Datatype sendtype,
                void *recvbuf,
                const int recvcounts[],
                const int recvdispls[],
                MPI_Datatype recvtype, int root,
                MPI_Comm comm);
int MPI_Scatterv(const void *sendbuf,
                 const int sendcounts[],
                 const int senddispls[],
                 MPI_Datatype sendtype,
                 void *recvbuf,
                 int recvcount, MPI_Datatype recvtype, int root,
                 MPI_Comm comm);

int MPI_Allgatherv(const void *sendbuf,
                   int sendcount, MPI_Datatype sendtype,
                   void *recvbuf,
                   const int recvcounts[],
                   const int recvdispls[],
                   MPI_Datatype recvtype,
                   MPI_Comm comm);
                   
int MPI_Alltoallv(const void *sendbuf,
                  const int sendcounts[],
                  const int senddispls[],
                  MPI_Datatype sendtype,
                  void *recvbuf,
                  const int recvcounts[],
                  const int recvdispls[],
                  MPI_Datatype recvtype,
                  MPI_Comm comm);
int MPI_Alltoallw(const void *sendbuf,
                  const int sendcounts[],
                  const int senddispls[],
                  const MPI_Datatype sendtypes[],
                  void *recvbuf,
                  const int recvcounts[],
                  const int recvdispls[],
                  const MPI_Datatype recvtypes[],
                  MPI_Comm comm);
\end{lstlisting}

Each of these operations perform the same kind of communication/data
exchange operations as their regular counterpart, but the amount of
data contributed can vary between processes. For instance, the
\mpigatherv operation transfers data from all processes to a given
root process. Data to be transferred are specified by the send buffer
argument triple (\texttt{sendbuf}, \texttt{sendcount}, and
\texttt{sendtype}) and the processes may, in contrast to the
\mpigather operation, specify different numbers of elements to be
transferred. The root process has a vector (hence the ``vector''
suffix \texttt{v} to these operations) of counts where
\texttt{recvcounts[$i$]} specifies the count of elements (of type
\texttt{recvtype}) from process $i$. The send signature of process $i$
specified by the \texttt{sendcount} and \texttt{sendtype}
arguments must be identical to the signature at the root process given
by \texttt{recvcounts[$i$]} and \texttt{recvtype}. At the root the
data are gathered starting at memory address \texttt{recvbuf}. More
precisely, the data from process $i$ are stored starting at address
\begin{displaymath}
  \texttt{recvbuf}+\texttt{recvdispls[}i{]}\cdot\texttt{extent}
\end{displaymath}
where \texttt{extent} is the extent (in bytes) of the
\texttt{recvtype} derived datatype.  Thus, the displacement vector
\texttt{recvdispls} is the relative offset or displacement of the data
from each process in units of the extent of the receive type.

The \mpiscatterv, \mpiallgatherv, and \mpialltoallv operations are
similar. Where several blocks of data are to be transferred to other processes,
there are \texttt{sendcounts} and send \texttt{senddispls} vectors in the
argument lists, and where several data blocks are to be transferred from other
processes there are \texttt{recvcounts} and receive \texttt{recvdispls}
vectors in the argument lists. There are a single send and
a single receive datatype argument, \texttt{sendtype} and \texttt{recvtype},
respectively, describing the type
and structure of all data sent or received. The \mpialltoallw
operation is different in this respect. This special collective has a
separate datatype argument for each data block
to and from each of the other processes.

Using irregular collectives can be tedious. Assume a root process has
to gather different amounts of data from the other processes, like the
column vector \mpigather application above, but now with possibly
different numbers of columns for each process. The root may, however,
not know in advance how much data it is going to receive from
each of the other processes. Since the \mpigatherv collective needs
the \texttt{recvcounts} and \texttt{recvdispls} vectors to be set up
correctly, the element counts must first be collected from all
processes. For this, the regular \mpigather operation can be
used. So, first the element counts are gathered at the root and stored
in the \texttt{recvcounts} vector, based on which appropriate
displacements are computed (in the example, data are stored
consecutively, but this must not necessarily always be so). 
Finally, the data can be correctly collected with the \mpigatherv
operation.

\begin{lstlisting}[style=SnippetStyle]
// gather counts from all processes
MPI_Gather(&sendcount,1,MPI_INT,recvcounts,1,MPI_INT,root,
           comm);
if (rank==root) {
  // compute displacements, on root only
  recvdispls[0] = 0;
  // data to be received consecutively (prefix sums)
  for (i=1; i<size; i++) { 
    recvdispls[i] = recvdispls[i-1]+recvcounts[i-1];
  }
}
// gather the possibly different amounts of data 
MPI_Gatherv(sendbuf,sendcount,sendtype,recvbuf,
            recvcounts,recvdispls,recvtype,root,comm);
\end{lstlisting}

The \mpiinplace argument can also be used
for the irregular communication/data exchange collectives.
Sometimes, this is convenient, and it can sometimes
even give a performance benefit.

The \emph{reduction collectives} additionally perform computation on
the data supplied by the processes making the collective call. Here, it
is convenient to think of the processes as supplying vectors of
some count number of elements of a basic datatype (like \mpiint, \mpifloat,
\mpilong, \mpidouble, \etc), although derived datatypes can be used in
some circumstances. These vectors are reduced element by element using a
binary operator supplied in the call and result in a result vector with
the same number of elements.  The interface specifications
for the reduction type collectives are listed below.

\begin{lstlisting}[style=SnippetStyle]
int MPI_Reduce(const void *sendbuf, void *recvbuf,
               int count, MPI_Datatype datatype,
               MPI_Op op, int root, MPI_Comm comm);
int MPI_Allreduce(const void *sendbuf, void *recvbuf,
                  int count, MPI_Datatype datatype,
                  MPI_Op op, MPI_Comm comm);
int MPI_Reduce_scatter_block(const void *sendbuf,
                             void *recvbuf,
                             int recvcount,
                             MPI_Datatype datatype,
                             MPI_Op op, MPI_Comm comm);
int MPI_Reduce_scatter(const void *sendbuf,
                       void *recvbuf, const int recvcounts[],
                       MPI_Datatype datatype,
                       MPI_Op op, MPI_Comm comm);

int MPI_Scan(const void *sendbuf,
             void *recvbuf, int count, MPI_Datatype datatype,
             MPI_Op op, MPI_Comm comm);
int MPI_Exscan(const void *sendbuf,
               void *recvbuf, int count, MPI_Datatype datatype,
               MPI_Op op, MPI_Comm comm);
\end{lstlisting}

Let $\oplus$ be an associative, binary operator operating elementwise
on vectors $x$ and $y$ with the same number of elements $c$. The
reduction collective operations\index{collective operation}
perform a reduction like
\begin{eqnarray*}
 z & = & x_0\oplus x_1\oplus \cdots \oplus x_{p-1}
\end{eqnarray*}
where $x_i$ is the vector supplied by \mpi process $i$ and $p$ the number
of processes. Brackets can be left away due to associativity; $x\oplus
(y\oplus z)= (x\oplus y)\oplus z$. Operators are not assumed to be
commutative but many commonly used operators are commutative ($+,\max,\ldots$).
If operator $\oplus$ is commutative, then
\begin{eqnarray*}
  z & = & x_0\oplus x_1\oplus \cdots \oplus x_{p-1} \\
  & = & x_{\pi(0)}\oplus x_{\pi(1)}\oplus \cdots \oplus x_{\pi(p-1)} \\
\end{eqnarray*}
for any permutation $\pi: \{0,\ldots,p-1\}\rightarrow\{0,\ldots p-1\}$.
This can possibly be exploited by the reduction algorithms
underlying an \mpi library implementation and sometimes is. However,
reductions are preferred to be
performed in \impi{rank order} and \mpi libraries normally try to respect
this as far as possible. The special \mpi query operation \mpiopcommutative
can be used to find out whether a given (user-defined) operator is commutative.

\begin{table}[t!]
  \caption{Binary operators for collective reduction operations.}
  \label{tab:binaryreductionoperators}
\begin{center}
  \begin{tabular}{cc}
    \toprule
    Operator & \mpi \\
    \midrule
    Sum & \mpisum \\
    Product & \mpiprod \\
    Minimum & \mpimin \\
    Maximum & \mpimax \\
    Logical (wordwise) and, or, exclusive or & \mpiland, \mpilor, \mpilxor\\
    Bitwise and, or, exclusive or & \mpiband, \mpibor, \mpibxor\\
    \midrule
    Minimum with location & \mpiminloc \\
    Maximum with location & \mpimaxloc \\
    \bottomrule
  \end{tabular}
\end{center}
\end{table}

\mpi provides a number of predefined operators working on vectors of
basic datatypes stored consecutively in send and receive buffers with
a count of elements. Operators are identified by the \mpiop handle. It
is also possible for the application programmer to define own
operators by attaching a function with a predefined signature to an
operator handle, but this is beyond the scope of these lectures.  The
standard \mpi operators are listed in
\Tab~\ref{tab:binaryreductionoperators}. All these operators are
(mathematically) commutative and associative.

In the reduction and scan\index{scan} collectives, all processes must give the
same \mpiop argument, otherwise the results are undefined (as can be
imagined). All processes must give input vectors with the same number
of elements (of the same basic datatype).

Elementwise binary reduction by some operator $\oplus$ on two input
vectors of $c$ elements means, for instance, that
\begin{eqnarray*}
  \left(\begin{array}{c}
    x_{c-1} \\
    \vdots \\
    x_1 \\
    x_0
  \end{array}\right)
  +
  \left(\begin{array}{c}
    y_{c-1} \\
    \vdots \\
    y_1 \\
    y_0
  \end{array}\right)
  & = &
  \left(\begin{array}{c}
    x_{c-1}+y_{c-1} \\
    \vdots \\
    x_1+y_1 \\
    x_0+y_0
  \end{array}\right)
\end{eqnarray*}
for when $\oplus$ is the $+$ operator \mpisum, and
\begin{eqnarray*}
  \min\left\{
  \left(\begin{array}{c}
    x_{c-1} \\
    \vdots \\
    x_1 \\
    x_0
  \end{array}\right),
  \left(\begin{array}{c}
    y_{c-1} \\
    \vdots \\
    y_1 \\
    y_0
  \end{array}\right)
  \right\}
  & = &
  \left(\begin{array}{c}
    \min\{x_{c-1},y_{c-1}\} \\
    \vdots \\
    \min\{x_1,y_1\} \\
    \min\{x_0,y_0\}
  \end{array}\right)
\end{eqnarray*}
when $\oplus$ is the minimum operator \mpimin.

The reduction collectives differ in the way the output vector is
stored. For the \mpireduce operation, which takes a root argument, the
computed $c$-element result vector $z$ is stored in the receive buffer at
the root. The \texttt{recvbuf} argument is significant only for
the root process. For the \mpiallreduce operation, all processes
receive the computed result $z$ in their respective receive
buffers. With the \mpireducescatterblock and \mpireducescatter
operations, the $c$-element result vector $z$ is split into subvectors
$z^0, z^1,\ldots z^{p-1}$ of $c_0,c_1,\ldots c_{p-1}$ elements,
respectively, with $c=\sum_{i=0}^{p-1}c_i$, and the vector $z_i$
stored in the receive buffer at process $i$. For
\mpireducescatterblock, all $c_i$ are equal and so subvectors have the
same number of elements given by a single \texttt{count} argument,
whereas for \mpireducescatter the $c_i$
counts are stored in the input vector \texttt{recvcounts} with
$\texttt{recvcounts[}i\texttt{]}=c_i$. All processes must give the
same \texttt{recvcounts} vector as input. The \mpireducescatter
operation is the irregular (``vector'' variant) and
\mpireducescatterblock the regular variant of this collective
operation\index{collective operation}
(see \Tab~\ref{tab:collectiveclassification}).
The \mpiinplace argument can be given as \texttt{sendbuf}
argument in some cases. For \mpireduce, the root process (only)
can specify that the input vector is to be taken from the \texttt{recvbuf}
address where the result of the reduction is
also stored by giving \mpiinplace as \texttt{sendbuf} argument. For
\mpiallreduce, \mpireducescatterblock, and \mpireducescatter, if
one process gives the \mpiinplace argument, then all
processes must give the \mpiinplace argument.

A simple, but very common application of collective reduction
operations is checking for agreement on some Boolean outcome. Say, all
processes need to agree on some convergence criterion which follows
from all processes having locally satisfied some criterion.  Agreement
can be checked by performing a reduction with a Boolean (logical)
``and'' operation and then making sure that all processes receive the
result. The case could occur in a
stencil computation\index{stencil computation}, which is
iterated until convergence by all processes is reached (see
\Sec~\ref{sec:stencil}). It could be implemented with an \mpiallreduce
operation with the logical ``and'' operation \mpiland; the \mpiinplace
argument is convenient here.

\begin{lstlisting}[style=SnippetStyle]
while (!done) {
  ... // the stencil update (computation)

  int k = 0;
  MPI_Isend(out_left,c,MPI_DOUBLE,left,TAG,cartcomm,
            &request[k++]);
  MPI_Isend(out_right,c,MPI_DOUBLE,right,TAG,cartcomm,
            &request[k++]);
  MPI_Isend(out_up,c,MPI_DOUBLE,up,TAG,cartcomm,
            &request[k++]);
  MPI_Isend(out_down,c,MPI_DOUBLE,down,TAG,cartcomm,
           &request[k++]);
  
  MPI_Irecv(in_left,c,MPI_DOUBLE,right,TAG,cartcomm,
            &request[k++]);
  MPI_Irecv(in_right,c,MPI_DOUBLE,left,TAG,cartcomm,
            &request[k++]);
  MPI_Irecv(in_up,c,MPI_DOUBLE,down,TAG,cartcomm,
            &request[k++]);
  MPI_Irecv(in_down,c,MPI_DOUBLE,up,TAG,cartcomm,
            &request[k++]);
  
  MPI_Waitall(k,request,MPI_STATUSES_IGNORE);
  
  done = 1; // some local convergence criterion
  MPI_Allreduce(MPI_IN_PLACE,&done,1,MPI_INT,MPI_LAND,
                cartcomm);
  // global agreement, same number of iterations
}
\end{lstlisting}

The two scan\index{scan} collective operations\index{collective operation}
\mpiscan and \mpiexscan implement
the \emph{inclusive prefix sums} and
\emph{exclusive prefix sums}\index{prefix sums}
operations (elementwise, on $c$-element vectors), respectively, see
\Sec~\ref{sec:prefixsums}. The $i$th elementwise inclusive or
exclusive prefix sum is stored at process $i$.  Processes can use the
\mpiinplace argument to indicate that input is to be taken from the
\texttt{recvbuf} address (where the result is also placed).

An important, later addition to \mpi, is the capability to locally
apply a binary operator on two input vectors. The operator can be any
of the predefined \mpiop operators (or even a user-defined
operator). This local operation is shown below; the second argument is
both the second input and the address where the result is stored. This
is sometimes convenient and sometimes not; there is (unfortunately)
no three-argument version $a=b+c$ of this local operation in
\mpi~\cite{Traff23:attributes}.

\begin{lstlisting}[style=SnippetStyle]
int MPI_Reduce_local(const void *inbuf, mvoid *inoutbuf,
                     int count, MPI_Datatype datatype,
                     MPI_Op op);
int MPI_Op_commutative(MPI_Op op, int *commute);
\end{lstlisting}

Below is an implementation of a $p-1$ communication round algorithm
for \mpiscan, which illustrates the use of \mpireducelocal. A copy
of the input in the receive buffer to the send buffer is needed and
implemented by an \mpisendrecv operation, where each process sends the
input data to itself~\cite{Traff23:attributes}.
Here, this operation is done on the special
\mpicommself communicator which is a predefined singleton
communicator that consists of
the process itself only. This copy would be unnecessary if the
\mpiinplace argument had been given to the \mpiscan operation. 

\begin{lstlisting}[style=SnippetStyle]
MPI_Sendrecv(sendbuf,c,MPI_FLOAT,0,SCANTAG,
             recvbuf,c,MPI_FLOAT,0,SCANTAG,MPI_COMM_SELF,
             MPI_STATUS_IGNORE);
if (rank>0) {
  MPI_Recv(tempbuf,c,MPI_FLOAT,rank-1,SCANTAG,comm,
           MPI_STATUS_IGNORE);
  MPI_Reduce_local(tempbuf,recvbuf,c,MPI_FLOAT,MPI_SUM);
}
if (rank<size-1) {
  MPI_Send(recvbuf,c,MPI_FLOAT,rank+1,SCANTAG,comm);
}
\end{lstlisting}

The algorithm is linear in the number of \mpi processes and not fast.
It is a good exercise to consider in which ways the algorithm is
inefficient (cost) and how it can be improved.

For \mpicommself, the following holds.
\begin{lstlisting}[style=SnippetStyle]
int rank, size;
  
MPI_Comm_rank(MPI_COMM_SELF,&rank);
MPI_Comm_size(MPI_COMM_SELF,&size);
assert(size==1);
assert(rank==0);
\end{lstlisting}

As mentioned, it is possible for the application programmer to define
and register own, binary functions as \mpiop operations. The
functionality for this is listed below.

\begin{lstlisting}[style=SnippetStyle]
int MPI_Op_create(MPI_User_function *user_fn, int commute,
                  MPI_Op *op);
int MPI_Op_free(MPI_Op *op);
\end{lstlisting}

\subsection{Complexity and Performance of Applications with Collective Operations}
\label{sec:collcomplexity}

Many message-passing algorithms can be expressed entirely or almost
entirely in terms of
collective operations~\cite{Gorlatch04}\index{collective operation}.  This
often holds for algorithms following a loosely Bulk Synchronous
Parallel\index{BSP} pattern described in
\Sec~\ref{sec:bsppattern}: A sequence of steps where processes perform
local computations followed by collective operations that summarize
and redistribute data over the processes. Level-wise Breadth-First
Search (BFS)\index{BFS} would follow this pattern. At each level,
processes locally explore new vertices reachable from vertices from
the current level and then either exchange vertices (by one of the
\mpialltoall collectives) or computes the set of vertices for the next
level (by \mpiallreduce or other reduction collective).

Such algorithms could be analyzed in terms of the local computations
and the collective operations performed\index{collective operation}.
Assume than an input graph
$G=(V,E)$ with $n$ vertices and $m$ edges is given, with depth $K$
from the given start vertex $s\in V$. An implementation maintaining a
set of new vertices for each level, represented as a bit-map, would
take $K$ iterations and should do at most $m$ vertex updates in total
over the $K$ iterations. By a good distribution of the edges $E$ over
the processes this can presumably be parallelized to take $O(m/p)$
time steps in total with the $p$ processes.  At the end of each
iteration, an \mpiallreduce operation is done, for a total time of
\begin{displaymath}
  O(m/p) + K\ T_{\mpiallreduce(p)}(n)
\end{displaymath}
where $T_{\mpiallreduce(p)}(n)$ denotes the time of
an \mpiallreduce operation on
$n$-element input sets on a communicator with $p$ processes.
Collective operations are expensive so the analysis could therefore
focus on the exact number of these operations. The actual runtime and
complexity of the collectives, for instance as stated in
Table~\ref{tab:collectivecomplexity-fully} and
Table~\ref{tab:collectivecomplexity-diameter}, would not need to be
known.  Since \mpi does not make any performance guarantees or
prescribes specific algorithms for the collective operations, the
actual complexity for some \mpi library cannot be known a priori. Note
that there is no claim made that the outlined BFS\index{BFS} implementation
is in any way best possible or even a (very) good one.

\subsection{Examples: Elementary Linear Algebra}
\label{sec:la}

Matrix--vector multiplication and
matrix--matrix multiplication\index{matrix--matrix multiplication} are two
elementary operations in linear algebra.
The collective operations\index{collective operation} we
have seen in the preceding sections are convenient for solving these
problems in parallel without relying on shared memory access to the
input and output matrices and vectors.

In such operations, the input matrices and vectors are distributed in
some way over the available processes. The output is likewise
distributed over the processes in some (possibly other) way. The
distribution of input and output should be considered part of the
\impi{problem specification} and an algorithm/implementation for
solving any such problem must respect the prescribed distribution. If
the distribution is different, either another algorithm must be
developed, or the distribution must be changed (by some
algorithm). Distributions are most often balanced, meaning that with
$p$ processes, each process will posses $1/p$ of the total input and
compute $1/p$ of the total output. Often, somewhat complex
(block cyclic\index{data distribution!block cyclic},
see~\ref{sec:datadistributions}) distributions
have to be used to achieve a good load balance between the processes.
It is obvious that no efficient, parallel algorithm can be allowed to gather
the full input or the full output (Amdahl's Law\index{Law!Amdahl's Law}).

We first give two implementations of algorithms for performing
matrix--vector multiplication for two different input and output
distributions. The full input is a real-valued (\texttt{double})
$m\times n$ matrix $M$ and a real-valued $n$ element vector $x$.
The output is a real-valued $m$ element vector $y$ with $y=Mx$. For
simplicity, we assume that $p$, the number of processes, divides
both $m$ and $n$. It is, of course, a good exercise to generalize the
implementations to arbitrary input sizes $m$ and $n$.

In the first example, the input matrix is distributed
row-wise\index{data distribution!row-wise},
meaning that each process has $m/p$ full, consecutive rows of the
matrix $M$. Process $0$ the first such $m/p$ rows, process $1$ the
next $m/p$ rows, and so on. The input vector $x$ is likewise
distributed in pieces of $n/p$ consecutive elements. The output vector
$y$ is to be distributed in the same manner with $m/p$ consecutive
elements per process.

Let $M_i$ be the $(m/p)\times n$ part of the matrix of process
$i$. The part of the output for process $i$ can be computed as
$y_i=M_ix$. In order to do this computation, the full $x$ vector must
be available at all processes which can be accomplished with an
\mpiallgather operation. The rest is easy.

\begin{lstlisting}[style=SnippetStyle]
MPI_Comm_rank(comm,&rank);
MPI_Comm_size(comm,&size);

assert(m%size==0); // regular only
assert(n%size==0);

double *fullvector;
fullvector = (double*)malloc(n*sizeof(double));
  
MPI_Allgather(vector,n/size,MPI_DOUBLE,
              fullvector,n/size,MPI_DOUBLE,comm);
for (i=0; i<m/size; i++) {
  result[i] = matrix[i][0]*fullvector[0];
  for (j=1; j<n; j++) {
    result[i] += matrix[i][j]*fullvector[j];
  }
}
free(fullvector);
\end{lstlisting}

The run time complexity of this first algorithm can easily be analyzed
as follows. As stated in \Tab~\ref{tab:collectivecomplexity-fully},
the allgather operation can be done in
\begin{displaymath}
  T_{\mpiallgather(p)}(n)=O(n+\log p)
\end{displaymath}
time. The process
local matrix--vector product computation takes $O((m/p)n)$ time, for a
total of $O((m/p)n+n+\log p)$ time steps. This is cost-optimal with
$p$ in $O(m)$ processors if we assume that $n>\log p$ since sequential
matrix--vector multiplication takes $O(mn)$ time steps.

In the second example, the input matrix is distributed
column-wise\index{data distribution!column-wise},
meaning that each process has $n/p$ consecutive columns with $m$ rows of the
matrix $M$. Process $0$ the first such $n/p$ columns, process $1$ the
next $n/p$ columns, and so on. The input vector $x$ is likewise
distributed in pieces of $n/p$ consecutive elements. The output vector
$y$ is to be distributed in the same manner with $m/p$ consecutive
elements per process.

Let $M'_i$ be the $m\times (n/p)$ part of the matrix of process
$i$. The full output vector $y$ can be computed as
$y=\sum_{i=0}^{p-1}M'_ix_i$ and then be distributed into the
parts $y_i$ of $m/p$ consecutive elements per process. The summation
and subsequent distribution of the parts can be accomplished by an
\mpireducescatterblock operation.

\begin{lstlisting}[style=SnippetStyle]
MPI_Comm_rank(comm,&rank);
MPI_Comm_size(comm,&size);

assert(m%size==0); // regular only
assert(n%size==0);

double *partial;
partial = (double*)malloc(m*sizeof(double));

for (i=0; i<m; i++) {
  partial[i] = matrix[i][0]*vector[0];
  for (j=1; j<n/size; j++) {
    partial[i] += matrix[i][j]*vector[j];
  }
}

MPI_Reduce_scatter_block(partial,result,m/size,MPI_DOUBLE,
                         MPI_SUM,comm);
free(partial);
\end{lstlisting}

The run time complexity of the second algorithm can easily be analyzed
as follows: The process local work for the initial matrix--vector
multiplication is in $O(m(n/p))$. According to
\Tab~\ref{tab:collectivecomplexity-fully}, the reduce-scatter
operation can be done in
\begin{displaymath}
  T_{\mpireducescatter(p)}(m)=O(m+\log p)
\end{displaymath}
time, for a total of $O(m(n/p)+m+\log p)$ time steps. This is
cost-optimal for $p$ in $O(n)$ processors if we assume that $m>\log
p$.

Summarizing, we have found the following.
\begin{theorem}
  \label{alg:matrixvector}
  Matrix--vector multiplication of an $m\times n$ matrix with an $n$
  element vector can be done work-optimally on a $p$ processor
  distributed memory system with message-passing communication in
  $O(mn/p+\min(m,n)+\log p)$ time steps.
\end{theorem}

Which of the two algorithms performs better in practice depends on the
actual quality of the implementation of the \mpiallgather and
\mpireducescatterblock operations, and on the magnitude of $m$ and
$n$. Keep in mind that the two algorithms assume different
distributions of the input matrix! A more scalable algorithm, one for
which more processors can be employed with linear speed-up\index{speed-up},
can be given by combining the two ideas (with a different distribution of the
input). It is a good exercise to extend the two algorithms to work
also for the case where $p$ divides neither $m$ nor $n$. The irregular
collectives \mpiallgatherv and \mpireducescatter will be of help and
actually do most of the (conceptual) work.

The more challenging operation to perform without having the
matrices stored in shared memory and being accessible to every thread
(process) is matrix--matrix multiplication\index{matrix--matrix multiplication}.
Given an $m\times l$ input
matrix $A$, an $l\times n$ input matrix $B$, compute the $m\times n$
output matrix $C$ as $C=AB$. For simplicity, we assume that the number
of processes $p$ is a square (which is not entirely without loss of
generality), that is $p=\sqrt{p}\sqrt{p}$ for an integer $\sqrt{p}$,
and that $\sqrt{p}$ divides all of $m,l,n$. The input distribution is
balanced such that each process has input submatrices of
$(m/\sqrt{p})\times (l/\sqrt{p})=ml/p$ and $(l/\sqrt{p})\times
(n/\sqrt{p})=ln/p$ elements, respectively. The algorithm produces an
output submatrix of $(m/\sqrt{p})\times (n/\sqrt{p})=mn/p$ elements for
each of the $p$ processes.

We organize the processes in a quadratic, 2-dimensional mesh and give
each processor a coordinate $(i,j)$, for instance, by creating a
Cartesian communicator with \mpicartcreate as shown in
\Sec~\ref{sec:organizingprocesses}. The submatrices for process
$i,j$ are denoted by $A_{ij}, B_{ij}$ and $C_{ij}$, respectively. Each
output submatrix $C_{ij}$ is computed straight ahead by
\begin{eqnarray*}
  C_{ij} & = & \sum_{k=0}^{\sqrt{p}-1}A_{ik}B_{kj}
\end{eqnarray*}

We observe that on each row of processes, the same $A_{ik}$
submatrices and on each column of processes, the same $B_{kj}$
submatrices are needed by all processes. This can be accomplished by
$\sqrt{p}$ broadcast operations on the rows and on the columns of
processes. To implement this conveniently with \mpi, communicators for
the processes in same the rows and the same columns are
needed. Fortunately, creating communicators for processes with the
same row coordinate and processes with the same column coordinate can
be done with the proper \mpicommsplit operations.
Naturally, this potentially expensive communicator creation should be
done once and for all (and reused over many matrix--matrix
multiplications\index{matrix--matrix multiplication}).
The initial communicator (with a square number of processes) is \texttt{comm}.

\begin{lstlisting}[style=SnippetStyle]
MPI_Comm_size(comm,&size);

int rc[2];     // row-column factorization
int period[2];
int coords[2]; // coordinates of process
int reorder;

rc[0] = 0; rc[1] = 0;
MPI_Dims_create(size,2,rc);
assert(rc[0]==rc[1]); // number of processes must be square
  
period[0] = 0;
period[1] = 0;
reorder = 0;
  
MPI_Cart_create(comm,2,rc,period,reorder,&cartcomm);
MPI_Cart_coords(cartcomm,rank,2,coords);

MPI_Comm_split(cartcomm,coords[0],0,&rowcomm);
MPI_Comm_split(cartcomm,coords[1],0,&colcomm);
  
int rowrank, colrank;
MPI_Comm_rank(rowcomm,&rowrank);
MPI_Comm_rank(colcomm,&colrank);
assert(rowrank==coords[1]);
assert(colrank==coords[0]);
\end{lstlisting}

The matrix--matrix multiplication\index{matrix--matrix multiplication}
can now easily
be implemented as shown in the code below. The row and column communicators
are \texttt{rowcomm} and \texttt{colcomm}. The multiplication and
summation of submatrices is done by an efficient, sequential
implementation which we have black-box encoded in the
fused-matrix--multiply-add procedure \texttt{fmma()}. We assume that the
matrices are represented by a pointer to an array of the elements in
row-major order\index{row-major order}.

\begin{lstlisting}[style=SnippetStyle]
int rowsize, colsize;
MPI_Comm_size(rowcomm,&rowsize);
MPI_Comm_size(colcomm,&colsize);
assert(rowsize==colsize); // size is square

double *Atmp, *Btmp;
// allocate space for temporary matrices

int i;
for (i=0; i<rowsize; i++) {
  double *AA, *BB;

  AA = (i==rowrank) ? A : Atmp;
  MPI_Bcast(AA,m/rowsize*l/rowsize,MPI_DOUBLE,i,rowcomm);

  BB = (i==colrank) ? B : Btmp;
  MPI_Bcast(BB,l/rowsize*n/rowsize,MPI_DOUBLE,i,colcomm);

  fmma(C,AA,BB,m/rowsize,l/rowsize,n/rowsize);
}
\end{lstlisting}

The parallel running time\index{parallel time} of the matrix--matrix
multiplication\index{matrix--matrix multiplication} implementation
can be analyzed as follows: As building block, a sequential
matrix--matrix multiplication algorithm is used. We assume it
takes $M(m,l,n)$ operations to multiply an $m\times l$ matrix with
an $l\times n$ matrix. The cost of adding two matrices is asymptotically
much smaller. The algorithm performs $2\sqrt{p}$ \mpibcast operations
of matrices
with $(m l)/p$ and $(l n)/p$ elements, respectively. According to
\Tab~\ref{tab:collectivecomplexity-fully}, this can be done in
\begin{eqnarray*}
  O(\sqrt{p} \frac{ml + ln}{p} +\log\sqrt{p}) & = &
  O(\frac{l(m + n)}{\sqrt{p}} +\log p)
\end{eqnarray*}
time steps. The number of process local matrix--matrix multiplications
is $\sqrt{p}$, each of which takes
$M(m/\sqrt{p},l/\sqrt{p},n/\sqrt{p})$ time steps. The sequential
matrix--matrix multiplication algorithm we have seen takes
$M(m,l,n)=O(mln)$ steps. Using this algorithm gives
\begin{eqnarray*}
\sqrt{p}\ O((m/\sqrt{p})(l/\sqrt{p})(n/\sqrt{p})) & = & O(\frac{mln}{p})
\end{eqnarray*}
with linear speed-up\index{speed-up!linear} for the multiplication work.

Summarizing, with the standard sequential matrix--matrix
multiplication\index{matrix--matrix multiplication}
algorithm as plug-in, we have the following:

\begin{theorem}
  \label{alg:matrixmatrix}
  Matrix--matrix multiplication can be done in $O(mln/p+l(m +
  n)/\sqrt{p} +\log p)$ time steps on a $p$ processor
  system with message-passing communication which is
  cost-optimal compared to a sequential
  $M(m,l,n)=O(mln)$ matrix--matrix multiplication algorithm.
\end{theorem}
Speed-up is linear\index{speed-up!linear}
as long as $p$ is in $O((\frac{mn}{m+n})^2)$, assuming
that both the first and second term dominate the last $\log p$ term.

This algorithm for matrix--matrix multiplication doing broadcast
operations on rows and columns of processes (and improvements thereof)
is called SUMMA (Scalable Universal Matrix Multiplication
Algorithm)~\cite{VandeGeijnWatts97}.

\subsection{Examples: Sorting Algorithms}
\label{sec:mpisorting}

The Quicksort\index{Quicksort} algorithm idea lends itself well also
to parallel implementation by point-to-point and collective communication.
There are two natural variants. As in the preceding lectures, we assume
that good pivots can be found by some means, which is of course
crucial for both the theoretical and practical performance. We, however,
ignore this aspect here and leave it to others, for instance~\cite{AxtmannSanders17,AxtmannWiebigkeSanders18,Roughgarden17,SandersLammHuebschleSchradeDachsbacher18}.

For a distributed memory implementation, we assume that the input data
(elements from some totally ordered set, like integers, floating point
numbers, objects, \etc) have been evenly distributed over the available
processes. For input of $n$ elements in total, each process will, thus,
have (approximately) $n/p$ elements. The elements are to be globally
sorted in such a way that, preferably, each process will have
approximately $n/p$ elements of the output.
The output must fulfill that, for each process, the elements
in the process' part of the output are sorted, and that the elements of
process $i$ are all larger than or equal to the elements of process
$i-1$ (for $i>0$) and smaller than or equal to the elements of process
$i+1$ (for $i<p-1$).

For the parallel Quicksort\index{Quicksort},
we assume that the number of processes $p$
is a power of two, $p=2^k$ for some $k, k\geq 0$. We formulate the
algorithm recursively, but recurse on the number of processes. Each
recursive call will split (exactly) its number of processes into two
halves until one process is left with some array of elements to sort
sequentially. An implementation for $p, p>1$ \mpi
processes in a communicator \texttt{comm}
would go as follows.

\begin{enumerate}
\item
  Select a global pivot for the $n$ elements and distribute this
  pivot to all $p$ processes.
\item
  Processes locally partition their set of elements into elements
  smaller than or equal to the global pivot and elements larger than
  or equal to the global pivot.
\item
  The processes pairwise exchange elements, such that half the
  processes will have elements smaller than or equal to the global
  pivot and the other half of processes will have element larger than
  or equal to the global pivot. Concretely, this will be done such
  that processes with rank $i, i<p/2$ will have the smaller elements,
  and processes $i, i\geq p/2$ will have the larger elements.
\item
  The communicator \texttt{comm} with the $p$ processes is split into
  two communicators with processes smaller than $p/2$ and processes
  larger than or equal to $p/2$, respectively.
\item
  Each process recursively calls Quicksort on the new communicator of
  $p/2$ processes to which it belongs.
\end{enumerate}

With only one process, $p=1$, a sequential Quicksort\index{Quicksort}
is used to sort the process' $n/p=n$ elements.
With such an implementation and a
best known implementation of sequential Quicksort, absolute and
relative speed-up\index{speed-up!absolute}\index{speed-up!relative}
of the implementation will coincide.

Step 1 will most likely involve one or more collective operations,
\eg, \mpibcast\index{collective operation}.
For the local computation in Step 2, a
best known sequential implementation for partitioning (in-place)
should be used. See, for
instance~\cite{Sedgewick77,Sedgewick78,SedgewickWayne11}. We note that
the global pivot for the processes may actually not be in the set of
input elements for any one process.  For Step 3, point-to-point
communication is used, for example like this (for elements of C type
\texttt{double}):

\begin{lstlisting}[style=SnippetStyle]
double *a; // input elements
double *b; // temporary array for communication

int n;      // size of local array a
int nn;     // local pivot index computed by partition function
int nl, ns; // number of larger and smaller elements

int half = size/2;
if (rank<half) {
  // will receive elements smaller than pivot
  nl = n-nn;
  MPI_Sendrecv(&nl,1,MPI_INT,rank+half,QTAG,
               &ns,1,MPI_INT,rank+half,QTAG,
               comm,MPI_STATUS_IGNORE);
  n = nn+ns;
  b = (double*)malloc(n*sizeof(double));
  assert(n==0||b!=NULL);
  
  MPI_Sendrecv(a+nn,nl,MPI_DOUBLE,rank+half,QTAG,
               b+nn,ns,MPI_DOUBLE,rank+half,QTAG,
               comm,MPI_STATUS_IGNORE);
  memcpy(b,a,nn*sizeof(double)); 
} else {
  // will receive elements larger than pivot
  ns = nn;
  MPI_Sendrecv(&ns,1,MPI_INT,rank-half,QTAG,
               &nl,1,MPI_INT,rank-half,QTAG,
               comm,MPI_STATUS_IGNORE);
  n = n-nn+nl;
  b = (double*)malloc(n*sizeof(double));
  assert(n==0||b!=NULL);
  
  MPI_Sendrecv(a,ns,MPI_DOUBLE,rank-half,QTAG,
               b,nl,MPI_DOUBLE,rank-half,QTAG,
               comm,MPI_STATUS_IGNORE);
  memcpy(b+nl,a+ns,(n-nl)*sizeof(double)); 
}
// split communicator and recurse
// free(b); when done
\end{lstlisting}

In Step 2, the partitioning function has locally partitioned the
\texttt{a} array and computed the pivot index \texttt{nn} that separates
larger and smaller elements. The processes exchange elements pairwise.
The processes with ranks smaller than $p/2$
are to receive the smaller elements, while the higher ranked processes are
to receive the larger elements. The first \mpisendrecv operation
exchanges the number of small and large elements needed for this,
based on which the temporary communication array \texttt{b} can be allocated.
The element exchange itself is now done by the second \mpisendrecv
operation. The elements for each process for the recursive call are in
the newly allocated \texttt{b} array. Some care has to be taken to
make sure such intermediate arrays are properly freed.

Step 4 is again a typical case for the \mpicommsplit operation. This may
introduce overhead that can affect overall performance, and it may be
worthwhile to consider whether explicit communicator splitting can be
avoided.

Assuming that pivots are selected perfectly and lead to even
partitions at all levels of the recursions, the running time can be
asymptotically estimated with the following recurrence relation. The
$O(\log p)$ term is for the collective operations for pivot selection and
the $O(n/p)$ term for the element exchange.
\begin{eqnarray*}
  T(n,p) & = & O(\log p) + O(n/p) + T(n/2,p/2) \\
  T(n,1) & = & O(n\log n)
\end{eqnarray*}
Since $(n/2)/(p/2)=n/p$, each level of the recursion will contribute the
$O(n/p)$ term, and since $\log_2 p$ recursive calls are needed ($p$ is a
power of two), the solution is
\begin{eqnarray*}
  T(n,p) & = & O(\log^2 p) +(\log_2 p) O(n/p) + O(n/p \log(n/p)) \\
  & = & O(\log^2 p) +O(\frac{n\log p}{p} + \frac{n\log n-n\log p}{p})) \\
  & = & O(\log^2 p) + O((n/p) \log n)
\end{eqnarray*}
with linear speed-up\index{speed-up!linear}
when $n$ is sufficiently large compared to $\log^2_2 p$.

For well-behaved inputs and pivot selection, this implementation can
work well in practice, but it does not guarantee that the output is
balanced as blocks of $n/p$ elements per process. It is a
good exercise to consider how bad the algorithm can behave, and how
worst-case inputs may look, also under different assumptions on the
pivot selection.

Another common parallel Quicksort\index{Quicksort} implementation variant,
which is
sometimes referred to as \impi{HyperQuicksort}~\cite{Wagar87}, is to
let the processes first sort their $n/p$ elements; this makes perfect
pivot selection per process trivial. It possibly also makes it easier
to find a good overall pivot. Local arrays are kept sorted through the
recursive calls, and in order to maintain sorted order, a merge step
is needed after the element exchange.  These variants, and others that
rely solely on collective communication operations for exchanging data
are discussed further and implemented in~\cite{Traff18:quicksort}.  A
drawback of Quicksort as implemented here is that the number of
processes must be a power of two. This is quite a restriction, and it
is worthwhile thinking about whether this can be alleviated.

A completely different idea for sorting (non-negative) integers is
\impi{counting sort} (or \impi{bucket sort}) which can also be given a
parallel, distributed memory implementation. Stable counting sort is a
building block in \impi{radix sort}. Given input of $n$ elements (with
integer keys), the idea is to count the number of occurrences of each
key by using the keys as indices into a counting array. After
counting, the counting array can be used to reserve space for buckets
in the right (increasing) order of the right sizes for each of the
occurring keys. Finally, the elements are put into their corresponding
buckets.  This can all be done in time proportional to the key range
and the number of elements $n$.  When the key range is no larger than
$O(n)$ this is linear in $O(n)$.

In a distributed memory setting, each process will have $n/p$ of the
elements available. Processes locally compute the sizes of the
buckets.  For each bucket, the processes must all know the total number
of elements for that bucket.
This can be computed by an allreduce operation over the
\texttt{bucketsize} vectors. Each process must also know, for each
bucket, how many elements on smaller ranked processes will go into
that bucket. This is a natural application of an
exclusive prefix sums\index{prefix sums} computation,
again over the computed bucket sizes. Here is a part of such a
counting sort (bucket sort) implementation.

\begin{lstlisting}[style=SnippetStyle]
int n = ...; // key range, number of buckets
int bucketsize[n];
int allsize[n]; // global size of buckets
int presize[n]; // size of buckets in smaller ranked processes

// local counting
for (i=0; i<n; i++) bucketsize[i] = 0;
for (i=0; i<n; i++) bucketsize[key[i]]++;

MPI_Allreduce(bucketsize,allsize,n,MPI_INT,MPI_SUM,comm);
MPI_Exscan   (bucketsize,presize,n,MPI_INT,MPI_SUM,comm);
\end{lstlisting}

The counts in the \texttt{presize} and \texttt{allsize} vectors can
now be used to compute which elements are to be sent to other
processes and how many elements each process has to receive from
other processes. The final element exchange can be done with \mpialltoall and
\mpialltoallv operations. To complete, local sorting or reordering is
needed. It is a good exercise to try to implement this idea in detail.

\subsection{Nonblocking and Persistent Collective Operations\marksec}
\label{sec:nonblockingcollectives}

The 17 standard collectives explained in the last section are all
blocking in the \mpi semantic sense. Recent additions to \mpi are a whole set
of corresponding, nonblocking
and also persistent collective operations\index{collective operation}.
Nonblocking
collectives are not part of the material of these lectures, but the
operations are listed here for completeness. The operations complete
``immediately'', irrespective of any action taken by the other
processes in the communicator
(which is what nonblocking means)\index{nonblocking}.
They return an \mpirequest object that can be used to query for and enforce
completion of any given operation, just as was the case with the
nonblocking point-to-point communication operations
(\Sec~\ref{sec:nonblockingpoint}).

A highly important difference to nonblocking point-to-point
communication is that blocking and nonblocking collectives cannot be
combined in the sense that some processes invoke the blocking variant and
other processes the nonblocking variant and expect a sensible outcome.
The reason for this is that blocking and nonblocking
implementations may use (completely) different algorithms. Therefore,
the steps taken by a process doing a broadcast with \mpiibcast may not
match the steps taken by another process doing the broadcast with
\mpibcast.

The nonblocking, regular exchange operations are the following.

\begin{lstlisting}[style=SnippetStyle]
int MPI_Ibarrier(MPI_Comm comm, MPI_Request *request);
  
int MPI_Ibcast(void *buffer, int count, MPI_Datatype datatype,
               int root, MPI_Comm comm, MPI_Request *request);
int MPI_Igather(const void *sendbuf,
                int sendcount, MPI_Datatype sendtype,
                void *recvbuf,
                int recvcount, MPI_Datatype recvtype,
                int root, MPI_Comm comm, MPI_Request *request);
int MPI_Iscatter(const void *sendbuf,
                 int sendcount, MPI_Datatype sendtype,
                 void *recvbuf,
                 int recvcount, MPI_Datatype recvtype,
                 int root, MPI_Comm comm, MPI_Request *request);
int MPI_Iallgather(const void *sendbuf,
                   int sendcount, MPI_Datatype sendtype,
                   void *recvbuf,
                   int recvcount, MPI_Datatype recvtype,
                   MPI_Comm comm, MPI_Request *request);
int MPI_Ialltoall(const void *sendbuf,
                  int sendcount, MPI_Datatype sendtype,
                  void *recvbuf,
                  int recvcount, MPI_Datatype recvtype,
                  MPI_Comm comm, MPI_Request *request);
\end{lstlisting}

The nonblocking, regular reduction collectives are the following.
\begin{lstlisting}[style=SnippetStyle]
int MPI_Ireduce(const void *sendbuf, void *recvbuf,
                int count, MPI_Datatype datatype,
                MPI_Op op, int root, MPI_Comm comm,
                MPI_Request *request);
int MPI_Iallreduce(const void *sendbuf, void *recvbuf,
                   int count, MPI_Datatype datatype,
                   MPI_Op op, MPI_Comm comm,
                   MPI_Request *request);
int MPI_Ireduce_scatter_block(const void *sendbuf,
                              void *recvbuf,
                              int recvcount,
                              MPI_Datatype datatype,
                              MPI_Op op, MPI_Comm comm,
                              MPI_Request *request);

int MPI_Iscan(const void *sendbuf, void *recvbuf,
              int count, MPI_Datatype datatype, MPI_Op op,
              MPI_Comm comm, MPI_Request *request);
int MPI_Iexscan(const void *sendbuf, void *recvbuf,
                int count, MPI_Datatype datatype, MPI_Op op,
                MPI_Comm comm, MPI_Request *request);
\end{lstlisting}

The irregular, nonblocking data exchange operations are the
following.

\begin{lstlisting}[style=SnippetStyle]
int MPI_Igatherv(const void *sendbuf,
                 int sendcount, MPI_Datatype sendtype,
                 void *recvbuf,
                 const int recvcounts[],
                 const int recvdispls[],
                 MPI_Datatype recvtype,
                 int root,  MPI_Comm comm,
                 MPI_Request *request);
int MPI_Iscatterv(const void *sendbuf,
                  const int sendcounts[],
                  const int senddispls[],
                  MPI_Datatype sendtype,
                  void *recvbuf,
                  int recvcount, MPI_Datatype recvtype,
                  int root, MPI_Comm comm,
                  MPI_Request *request);
int MPI_Iallgatherv(const void *sendbuf,
                    int sendcount, MPI_Datatype sendtype,
                    void *recvbuf,
                    const int recvcounts[],
                    const int senddispls[],
                    MPI_Datatype recvtype,
                    MPI_Comm comm, MPI_Request *request);
int MPI_Ialltoallv(const void *sendbuf,
                   const int sendcounts[],
                   const int senddispls[],
                   MPI_Datatype sendtype,
                   void *recvbuf,
                   const int recvcounts[],
                   const int recvdispls[],
                   MPI_Datatype recvtype,
                   MPI_Comm comm, MPI_Request *request);
int MPI_Ialltoallw(const void *sendbuf,
                   const int sendcounts[],
                   const int senddispls[],
                   const MPI_Datatype sendtypes[],
                   void *recvbuf,
                   const int recvcounts[],
                   const int recvdispls[],
                   const MPI_Datatype recvtypes[],
                   MPI_Comm comm, MPI_Request *request);
\end{lstlisting}

Finally, there is the single, irregular nonblocking reduce-scatter operation.
\begin{lstlisting}[style=SnippetStyle]
int MPI_Ireduce_scatter(const void *sendbuf, void *recvbuf,
                        const int recvcounts[],
                        MPI_Datatype datatype, MPI_Op op,
                        MPI_Comm comm, MPI_Request *request);
\end{lstlisting}

A nonblocking communicator duplicate operation is also
included in \mpi.

\begin{lstlisting}[style=SnippetStyle]
int MPI_Comm_idup(MPI_Comm comm,
                  MPI_Comm *newcomm, MPI_Request *request);
\end{lstlisting}

The repertoire of nonblocking collective operations in \mpi may grow
with time. A most recent addition was for instance a complete set of $17$
persistent collective operations
(see \Sec~\ref{sec:persistentpoint}). These operations have (almost)
the same interfaces as the corresponding nonblocking operations and
binds all input parameters in a request object that can be (re)used
as many times as desired. We do not list the operation interfaces here.

\subsection{Sparse Collective Communication: Neighborhood Collectives\marksec}
\label{sec:neighborhoodcollectives}

A recent addition to \mpi is a number of collective communication
operations that perform data exchanges not over all processes but only
among subsets of the processes. These so-called
\impi{neighborhood collectives}
are not treated in these lecture notes, but the
functionality is mentioned here for completeness.

The idea of sparse (in contrast to dense),
neighborhood collective communication is that each
process can perform a data exchange operation with a small set of
neighboring processes. What a neighboring process is, is defined by
defining the set of neighborhoods, collectively, for all
processes. In \Sec~\ref{sec:organizingprocesses}, two ways of
defining neighborhoods by creating new communicators
with associated neighborhoods were discussed,
in detail \mpicartcreate and briefly
touched upon \mpidistgraphcreate.

The collective operations\index{collective operation} on sparse
neighborhoods are of the allgather and all-to-all type and come in
both regular and irregular variants, as well as in blocking and
nonblocking (and persistent) variants. All neighborhood collectives
are strictly collective, that is they have to be called by all
processes in the communicators, and no synchronization behavior is
implied.

Note that the signatures of these operations are identical to those of
the standard collective operations. This can be helpful for
remembering how these functions look and what they
do~\cite{Traff21:orthogonality}.

The regular, blocking and nonblocking variants are listed below.
\begin{lstlisting}[style=SnippetStyle]
int MPI_Neighbor_allgather(const void *sendbuf,
                           int sendcount, MPI_Datatype sendtype,
                           void *recvbuf,
                           int recvcount, MPI_Datatype recvtype,
                           MPI_Comm comm);
int MPI_Neighbor_alltoall(const void *sendbuf,
                          int sendcount, MPI_Datatype sendtype,
                          void *recvbuf,
                          int recvcount, MPI_Datatype recvtype,
                          MPI_Comm comm);

int MPI_Ineighbor_allgather(const void *sendbuf,
                            int sendcount, MPI_Datatype sendtype,
                            void *recvbuf,
                            int recvcount, MPI_Datatype recvtype,
                            MPI_Comm comm, MPI_Request *request);
int MPI_Ineighbor_alltoall(const void *sendbuf,
                           int sendcount, MPI_Datatype sendtype,
                           void *recvbuf,
                           int recvcount, MPI_Datatype recvtype,
                           MPI_Comm comm, MPI_Request *request);
\end{lstlisting}

The irregular (``vector''), blocking and nonblocking variants are
listed below.

\begin{lstlisting}[style=SnippetStyle]
int MPI_Neighbor_allgatherv(const void *sendbuf,
                            int sendcount, MPI_Datatype sendtype,
                            void *recvbuf,
                            const int recvcounts[],
                            const int recvdispls[],
                            MPI_Datatype recvtype,
                            MPI_Comm comm);
int MPI_Neighbor_alltoallv(const void *sendbuf,
                           const int sendcounts[],
                           const int senddispls[],
                           MPI_Datatype sendtype,
                           void *recvbuf,
                           const int recvcounts[],
                           const int recvdispls[],
                           MPI_Datatype recvtype,
                           MPI_Comm comm);
int MPI_Neighbor_alltoallw(const void *sendbuf,
                           const int sendcounts[],
                           const MPI_Aint senddispls[],
                           const MPI_Datatype sendtypes[],
                           void *recvbuf,
                           const int recvcounts[],
                           const MPI_Aint recvdispls[],
                           const MPI_Datatype recvtypes[],
                           MPI_Comm comm);

int MPI_Ineighbor_allgatherv(const void *sendbuf,
                             int sendcount,
                             MPI_Datatype sendtype,
                             void *recvbuf,
                             const int recvcounts[],
                             const int recvdispls[],
                             MPI_Datatype recvtype,
                             MPI_Comm comm,
                             MPI_Request *request);
int MPI_Ineighbor_alltoallv(const void *sendbuf,
                            const int sendcounts[],
                            const int senddispls[],
                            MPI_Datatype sendtype,
                            void *recvbuf,
                            const int recvcounts[],
                            const int recvdispls[],
                            MPI_Datatype recvtype,
                            MPI_Comm comm,
                            MPI_Request *request);
int MPI_Ineighbor_alltoallw(const void *sendbuf,
                            const int sendcounts[],
                            const MPI_Aint senddispls[],
                            const MPI_Datatype sendtypes[],
                            void *recvbuf,
                            const int recvcounts[],
                            const MPI_Aint recvdispls[],
                            const MPI_Datatype recvtypes[],
                            MPI_Comm comm,
                            MPI_Request *request);
\end{lstlisting}

\subsection{MPI and Threads\marksec}

\mpi can be and often is used together with thread interfaces like
\openmp or \pthreads.\index{pthreads}
The idea is, for systems with shared memory
multi-core nodes that are interconnected by a communication network,
to let cores on the shared memory node compute as threads and let only
a single or a few \mpi processes on the shared memory node perform
communication with processes on other nodes using \mpi.  This is a
two-level, heterogeneous, hierarchical,
programming model\index{programming model}. A limited
number of processes per shared memory node communicate with other
processes using \mpi and threads inside the processes use a thread
model to compute in parallel. The threads are the active entities
inside the processes.  Therefore, such a two-level model raises the
question which threads can or are allowed to perform \mpi operations
(in order to avoid race conditions\index{race condition},
deadlocks\index{deadlock} or other deadly issues)?

\mpi answers the question by defining the level of thread support that
an \mpi library implementation can provide. There are four defined
levels of thread support. With \mpithreadsingle, only a single thread
is allowed to execute which essentially means that thread parallel
programming cannot be used! With \mpithreadfunneled threads can be
used, but only a designated, single \emph{main} or \emph{master}
thread can perform \mpi calls.  With \mpithreadserialized all threads
are allowed to perform \mpi calls, but only one at a time. It is the
programmer's responsibility to ensure that this is the case, for
instance, by using critical sections and other mechanisms provided by
the thread model that is used. With \mpithreadmultiple, all threads
can perform \mpi calls and may do so concurrently, in parallel. The
levels of thread support are ordered as
$\mpithreadsingle<\mpithreadfunneled<\mpithreadserialized<\mpithreadmultiple$,
meaning that a program that assumes a higher level of thread support,
\eg, \mpithreadserialized, may not run correctly if the \mpi library
supports only a lower level.

Threads levels are controlled and queried by a special initialization
function to be used instead of \mpiinit. With \mpiinitthread, the user
gives a required thread level, and the function returns a thread level
that can be supported. If the required thread level cannot be
supported, the provided level is the highest thread level of
the \mpi library implementation.  If the required thread level can be
supported, the provided level returned is larger than or equal to the
required level.

\begin{lstlisting}[style=SnippetStyle]
int MPI_Init_thread(int *argc, char ***argv,
                    int required, int *provided);
int MPI_Is_thread_main(int *flag);
int MPI_Query_thread(int *provided);
\end{lstlisting}

\subsection{MPI Outlook}

A number of important aspects and parts of the huge \mpi standard
were deliberately not treated in these bachelor lecture notes. These
include a whole model for input-output and communication with the
external file system (MPI-IO), dynamic process management (spawning
new \mpi processes from an application, connecting running \mpi
processes), so-called \impi{inter-communicators} (that are important
for process management), \mpi attributes (a very useful mechanism for
library building by which information can be attached to \mpi
objects~\cite{Traff23:attributes}),
the profiling and tools interfaces (important for library and performance
analysis tool building), partitioned point-to-point
communication and a few other things. The treatment stayed
within the so-called ``world model'', in which externally started processes are
grouped together within the \mpicommworld communicator. We did not at
all cover the alternative ``sessions model'', in which this is not the
case and processes initially have to create the
communicator they want to belong to.

The most recent, at the time of writing, version of the \mpi standard
is MPI 4.1 (November 2nd, 2023).
The \mpi forum is actively preparing a next version with
further additions and corrections to the standard.  Some of the
important recent additions were and are persistent collective
operations\index{collective operation}
(see \Sec~\ref{sec:persistentpoint}), the sessions
model, so-called partitioned (point-to-point) communication,
additional support for portably adapting applications to specifics of
system topologies (\mpicommsplittype is one function of this kind),
and further provisioning for fault tolerant \mpi programming.

\section{Exercises}

\begin{enumerate}
\item
  Give a polynomial-time algorithm to compute the diameter $\diam(G)$
  of a given, directed or undirected graph $G=(V,E)$. Hint: Reapply BFS.
\item
  Devise an algorithm for the broadcast problem for $d$-dimensional
  hypercubes with $p=2^d$ processors. What is the number of
  communication rounds taken by your algorithm? How does that relate
  to the diameter lower bound for the broadcast problem? Is your
  algorithm optimal?
\item
  Argue why the communication round complexity for a semantic barrier
  operation in fully connected, one-ported $p$-processor
  communication systems is in $\Theta(\log p)$.
\item
  Consider a high-performance computing system consisting of a (large)
  number of shared memory multi-core processor nodes interconnected
  with a complex communication network. Assume that some processor $i$
  is sending a (large) number of message packets $b=b_0,b_1,b_2,\ldots$
  one after the other to some other, different processor $j$ in the
  system.  What might be reasons that packets are not necessarily
  delivered in the sent order to processor $j$?  What would an \mpi
  library need to do in order to guarantee that messages that are sent
  in sequence are indeed received (seen by the receiving process) in
  the same order? What if individual packets (that could be parts of
  larger messages) are lost or corrupted? What would an \mpi library
  implementation have to do?
\item
  On your favorite system, run the communicator creation example from
  \Sec~\ref{sec:communicators} instrumented with print-statements
  to show the process ranks in old and new communicators. Develop
  assertions to express the relations between old and new ranks in all
  the communicators.  Extend the example with a partition of the
  \texttt{comm} communicator duplicate of \mpicommworld into two
  communicators consisting of the processes with rank smaller than
  some given rank \texttt{split} and the processes with rank larger
  than or equal to the \texttt{split} process. Create the same
  communicators by using the process group functionality of
  \Sec~\ref{sec:groups}.  Verify by assertions and use of
  \mpicommcompare and \mpigroupcompare that the created communicators
  are indeed equivalent.
\item
  The following program has the intention of collecting information
  at a given \texttt{root} process from all processes, somewhat like
  the \mpigather operation can do. The code has numerous safety issues and
  obviously does not work. Pinpoint the problems and repair the code;
  there are several possibilities for a ``correct'' solution, since the
  outcome has not been explicitly specified.
  \begin{lstlisting}[style=SnippetStyle]
MPI_Comm_rank(comm,&rank);
MPI_Comm_size(comm,&size);

if (rank==root) {
  int all[size];
  MPI_Status status;
    
  MPI_Send(&rank,1,MPI_INT,root,1000,comm);
  for (i=0; i<size; i++) {
    MPI_Recv(&all[i],1,MPI_INT,MPI_ANY_SOURCE,1000,
             comm,&status);
  }
  for (i=0; i<size; i++) assert(all[i]==i);
} else {
  MPI_Send(&rank,1,MPI_INT,root,1000,comm);
}
  \end{lstlisting}
\item
  A root process identified by a \texttt{root} rank that is not
  necessarily $0$ is given for a communicator. For some application, a
  new communicator with the same processes is needed where the root
  process has rank $0$, the process with the next rank has rank $1$, the
  process with the next to next rank has rank $2$ and so on. Use
  \mpicommsplit to create a new communicator where processes have been
  reranked towards root $0$.
  How costly is explicit communicator creation compared to manual
  reranking using a virtual rank \verb|virt = (rank-root+size)%size|
  in the application?
\item
  Implement the unsafe ring and the unsafe stencil communication
  patterns from \Sec~\ref{sec:semanticterms} using blocking
  \mpisend and \mpirecv operations. Use a simple, $5$-point average
  element stencil update rule. Devise an experiment to determine
  at which buffer sizes deadlocks occur (on the system and \mpi
  library available to you). Are these sizes different in the two
  cases?
\item
  Implement (incorrect!) programs as in \Sec~\ref{sec:datatypes}
  where a process sends data as a sequence of \mpilong to another
  process that receives the data as a sequence of \mpidouble, and vice
  versa, and examine the outcome. Are there interesting differences between
  the two cases? Is the outcome of such communication meaning- or
  useful?
\item
  \label{exe:stencil}
  Implement the two-dimensional stencil computation safely and
  correctly with \mpisendrecv as described at the end of
  \Sec~\ref{sec:nonblockingpoint}; use any non-trivial $5$-point
  stencil update rule (for instance, a simple average).  For the
  communication of submatrix columns, copy the elements of leftmost
  and rightmost columns into intermediate, consecutive
  buffers. Likewise, receive the column elements in intermediate,
  consecutive buffers and copy these into their desired positions in
  the matrix columns after the communication.  Verify correctness by
  comparison to a sequential implementation.  Iterate the stencil
  computation a number of times (for instance, an input
  parameter). Repeat and time the whole computation using \mpiwtime
  over a number of repetitions and compute the average and best
  completion time (best time for the slowest \mpi process over the
  repetitions). Experiment with different matrix sizes and different
  numbers of \mpi processes in different configurations.  Present
  strong scaling results where the total matrix size is kept
  independent of the number of processes and weak scaling results
  where the submatrix size per process is kept constant.
\item
  \label{exe:stencil2}
  Repeat Exercise~\ref{exe:stencil} using instead 
  nonblocking \mpiisend and \mpiirecv communication as explained in
  \Sec~\ref{sec:nonblockingpoint}.
\item
  Repeat now Exercise~\ref{exe:stencil2} using instead of nonblocking
  the persistent \mpisendinit and \mpirecvinit operations that were
  briefly explained in \Sec~\ref{sec:persistentpoint}. How does the
  performance of the nonblocking and persistent implementations differ?
  Use a sufficiently large number of stencil iterations.
\item
  Given an $m\times n$ matrix in process-local memory for some
  process.  Implement a process local matrix transposition into an
  $n\times m$ matrix using the \mpitypevector datatype to describe
  columns of either input or output matrix. The exercise illustrates
  the problem of doing process-local, \mpi type correct data
  reorganization and the power of \mpi datatypes for effecting
  this. For the process-local copy, use communication on \mpicommself
  and use either \mpisendrecv communication or a collective operation
  like \mpiallgather (see the end of \Sec~\ref{sec:collective}).
\item
  Repeat Exercise~\ref{exe:stencil} eliminating the intermediate,
  consecutive buffers by using instead an \mpitypevector datatype to
  describe the strided layout of a submatrix column. Benchmark and compare
  the results to your results from Exercise~\ref{exe:stencil}, and discuss
  (notable) differences. Instead of using an \mpitypevector, try 
  \mpityperesized to create a special \texttt{double} datatype with the
  extent of a full row. What might the advantages of this solution be
  compared to the (less flexible) \mpitypevector solution?
\item
  The stencil implementations suggested in
  \Sec~\ref{sec:semanticterms} \etc use a decomposition of the
  $n\times n$ matrix into smaller $n_r\times n_c$ submatrices where
  $r$ and $c$ are the numbers of row and columns of the \mpi
  process grid. Why is this distribution beneficial compared to, say,
  a row-wise or a column-wise distribution as used for the
  matrix--vector multiplication algorithms? For your answers, consider
  the ratio of communication volume to computation done per \mpi
  process.
\item
  Implement a two-dimensional, $9$-point stencil computation where the
  update rule (say, average; sometimes used in image processing
  applications) for a matrix-element $M[i,j]$ depends on $9$
  neighboring elements (including the element itself), $M[i,j-1],
  M[i+1,j-1], M[i+1,j], M[i+1,j+1], M[i,j+1], M[i-1,j+1], M[i-1,j],
  M[i-1,j-1]$ and $M[i,j]$ itself. Some of these elements may be
  undefined, instead their values are given by border (boundary)
  conditions. Partition the full matrix into roughly square matrices
  over a \mpi process grid. Implement the communication (horizontal,
  vertical, diagonal) with \mpisendrecv. How can you avoid deadlocks?
  What is the communication volume as a function of the input matrix
  size $nm$?  What is the amount of local computation (element
  updates) per stencil iteration? What is the ratio of computation
  steps to communication volume? What is the parallel running time per
  stencil iteration? To how many processors will the implementation
  scale?  A well-known trick can reduce the number of communication
  operations per process from $8$ to $4$. What is the idea? Does such
  an optimization make a difference in performance?
\item
  Repeat Exercise~\ref{exe:stencil} using instead one-sided \mpiget or
  \mpiput communication as explained in
  \Sec~\ref{sec:onesidedstencil}. Try with both \mpiwinfence and
  with \mpiwinpost-\mpiwinstart-\mpiwincomplete-\mpiwinwait synchronization.
  Compare the performance of the one-sided implementations against each
  other, and compare to either of the solutions with point-to-point
  communication. You may or may not use \mpitypevector to ease
  communication with left and right neighbors.
\item
  Complete the implementation of the binary search operation with one-sided
  communication outlined in \Sec~\ref{sec:onesidedbinsearch}. Each
  process contributes an ordered array (of, say, \texttt{float}s) of
  \texttt{n} elements for the window. Each process can perform binary search
  in the array by calling a search function on the window. Return values
  should be the rank of the process where the element belongs and the relative
  index (displacement) in the window of that process.
  Benchmark your implementation with one process and with all
  processes performing search operations. Consider worst and best cases.
\item
  Give a full, distributed memory implementation of merging by
  co-ranking as described in \Sec~\ref{sec:corankmerge}. Use one-sided
  communication with \mpiwinlock and \mpiwinunlock for implementing a
  \texttt{corank()} function (see the previous exercise). You can use
  either one or two windows for storing the ordered, distributed
  input arrays $A$ and $B$ of $n$ and $m$ elements per process,
  respectively. The output should be, for each process, an ordered array
  $C$ of size $n+m$ elements such that all elements at some process
  $i, 0<i$ are equal to or larger than all elements at process $i-1$.
  Benchmark (weak scaling) your implementation for larger and larger
  $n$ and $m$ and different numbers of processes $p$ (not only powers-of-two).
\item
  \label{exe:corankmpi}
  As in the previous exercise, give a full, distributed memory
  implementation of merging by co-ranking (\Sec~\ref{sec:corankmerge})
  where now the input arrays $A$ and $B$ of $n$ and $m$ elements in
  total are distributed as follows. Assume for simplicity that both
  $n$ and $m$ are divisible by $p$, the number of \mpi
  processes. Divide the total input into blocks of size roughly
  $(n+m)/p$. The $A$ array is divided into $\frac{np}{n+m}$ blocks
  which are assigned to the first \mpi processes $0,1,\ldots
  \frac{np}{n+m}-1$.  The $B$ array is divided into $\frac{mp}{n+m}$
  blocks and assigned to the remaining processes
  $\frac{np}{n+m},\ldots,p-1$. Assume here that $n,m,p$ are chosen
  such that all fractions are nice numbers. This way, each process has
  a part of the input of (roughly) the same size as all other
  processes. The total input can be kept in an \mpiwin window.  The
  size of the output per \mpi process in the $C$ array should be of
  the same size as the input for each process, ad each process local
  $C$ array should be ordered and the $C$ elements at any process $i$
  be smaller than or equal to the $C$ element of the next process
  $i+1$. Implement a \texttt{corank()} function using passive
  \mpiwinlock and \mpiwinunlock synchronization that can work with
  this input distribution. Benchmark your implementation as in the
  previous exercise. This \texttt{corank()} function could be a
  building block of an \mpi mergesort implementation.
\item
  In the following, typical \mpi benchmarking loop for benchmarking a single
  problem instance for some algorithm use \mpireduce to
  find the parallel time(s) for each problem size for each repetition
  (recall: defined as the
  time spent by the slowest process) and store the result at root
  process $0$.  Use as few \mpireduce calls as possible. Extend your
  implementation to also compute the load imbalance (defined as the
  difference between the slowest and the fastest process), again with
  as few \mpireduce calls as possible.
  \begin{lstlisting}[style=SnippetStyle]
double start, stop, spent;
for (r=0; r<REPETITIONS; r++) { // do some repetitions
  MPI_Barrier(comm);
  MPI_Barrier(comm);

  start = MPI_Wtime();

  ... // operation on comm to be benchmarked 

  stop = MPI_Wtime();
  spent = stop-start;
  ... // store spent time for this repetition somewhere
}
... // post-process: minimum, average?
... // collect results at process 0
  \end{lstlisting}
  Post-processing can either be done on the processes or, centrally,
  at process $0$ which then does the statistics. Consider both
  options. What might the pros and cons be?
\item
  Write a series of small programs that illustrate the semantics of
  the collective operations. Each program should allocate proper send
  and receive buffers of, say, \mpiint type, at all processes, either
  of a small constant number of elements or proportional to $p$, the
  number of processes in the communicator. Initialize all buffers with
  values that make it easy to verify that a) values are exchanged (and
  reduced) properly with the right results in the receive buffers and
  b) no send buffers have been modified. Instrument the program first
  with print statements, and verify by inspection with
  $p=1,2,4,5,7,\ldots$ \mpi processes.  Then formulate assertions that
  make it possible to verify exhaustively at larger scale that the
  collective operations do as claimed.

  Start with the simple, regular collectives \mpibcast, \mpigather, \mpiscatter,
  \mpiallgather, \mpialltoall. Proceed to the regular reduction collectives
  \mpireduce, \mpiallreduce, \mpireducescatterblock, \mpiscan, and \mpiexscan.
  Time and interest permitting, extend your analysis to the irregular
  counterparts of these collective operations.

  Here is an example:
  \begin{lstlisting}[style=SnippetStyle]
int rank, size;
MPI_Comm_rank(comm,&rank);
MPI_Comm_size(comm,&size);

int n = 2;
int buffer[n+1];

int root = size-1;
if (rank==root) {
  buffer[0] = size;
  buffer[1] = 0;
  buffer[2] = -rank-1;
} else {
  buffer[0] = -rank-1;
  buffer[1] = -rank-1;
  buffer[2] = -rank-1;
}
MPI_Bcast(buffer,n,MPI_INT,root,comm);

assert(buffer[0]==size);
assert(buffer[1]==0);
assert(buffer[2]==-rank-1);

if (rank==0) {
  printf("Rank %d: buffer=[%d,%d,%d]\n",
         rank,buffer[0],buffer[1],buffer[2]);
}
  \end{lstlisting}
\item
  Implement an own vector-scan operation with the same interface and
  semantics as \mpiscan using the (blocked) Hillis--Steele algorithm of
  \Sec~\ref{sec:hillissteele} (do not call your interface \mpiscan!).
  Make sure that the implementation is \texttt{safe} by using the
  proper point-to-point communication operations. What is the number
  of communication rounds? What is the run time complexity of the
  implementation as a function of the number of processes $p$ and the
  number of vector elements $n$ per process? Note that no barrier
  synchronization (like \mpibarrier) is needed.
\item
  The following two collective \mpi calls are supposed to implement a
  barrier operation in the same way as \mpibarrier does.
  \begin{lstlisting}[style=SnippetStyle]
MPI_Gather(NULL,0,MPI_INT,NULL,0,MPI_INT,0,comm);
MPI_Scatter(NULL,0,MPI_INT,NULL,0,MPI_INT,0,comm);
  \end{lstlisting}
  Explain why this is not necessarily semantically equivalent to an
  \mpibarrier, and why using this implementation as a barrier
  substitute will be unsafe.  Can you come up with code demonstrating
  that the implementation can go wrong?  Give a simple fix, still in
  terms of \mpigather and \mpiscatter.  Is your fix efficient in the
  volume of data communicated?
\item
  In \Sec~\ref{sec:collective}, it was described how to locally copy
  data on an \mpi process from a send buffer with a send datatype to
  a receive buffer with a receive datatype using \mpisendrecv on the special
  \mpicommself communicator. Show how to implement the same local copy
  operation with a collective operation on \mpicommself. Hint: consider
  \mpiallgather or \mpialltoall.
\item
  Implement \mpiallgather by a series of \mpibcast operations.
  Implement \mpiallgather by a series of \mpigather operations.
  Repeat the exercise for \mpiallgatherv. Time permitting, time the
  new implementations in comparison to \mpiallgather for different \mpi
  process configurations and input block sizes (counts).
\item
  Implement \mpialltoall by a series of \mpigather operations.
  Implement \mpialltoall by a series of \mpiscatter operations.
  Repeat the exercise for \mpialltoallv and \mpialltoallw.  Time
  permitting, time the new implementations in comparison to
  \mpialltoall and variants for different \mpi process configurations
  and input block sizes (counts).
\item
  Implement \mpireducescatterblock by a series of \mpireduce
  operations.  Implement \mpireducescatterblock by an \mpireduce
  operation followed by an \mpiscatter operation.  Repeat the exercise
  for \mpireducescatter.  Time permitting, time the new
  implementations in comparison to \mpireducescatterblock for
  different \mpi process configurations and input block sizes
  (counts).
\item
  MPI does not have a collective operation for computing element-wise
  suffix-sums (the ``opposite'' of prefix-sums) over input vectors provided
  by the processes in a communicator. Show how to easily compute
  the inclusive suffix-sums, analogously to what \mpiscan does, by using
  \mpicommsplit and \mpiscan as building blocks.
\item
  Show how to implement \mpiscan in terms of \mpiexscan and
  \mpireducelocal.  Show how to implement \mpiexscan in terms of
  \mpiscan; here additional communication may be necessary.
\item
  Devise an \mpi program using collective operations for computing the
  scalar (dot) product of two distributed $n$-element vectors
  \texttt{a} and \texttt{b}, \ie, the sum
  $\sum_{i=0}^{n-1}\texttt{a[}i\texttt{]}\texttt{b[}i\texttt{]}$. The
  vectors are represented as disjoint blocks of consecutive elements
  of roughly $n/p$ elements, and each process has two such blocks of
  \texttt{a} and \texttt{b} elements, respectively. Give two variants
  of the program, one that stores the result (dot product) at a
  designated \texttt{root} process and one that stores the result at
  all processes. The programs should work correctly regardless of
  whether $p$ divides $n$, $p$ being the number of available \mpi
  processes, preferably also for the case where $n<p$.
\item
  Finite sets can be represented by bitmaps of $n$ bits where $n$ is
  the maximum cardinality of such a set: An element is in the set if
  and only if the corresponding bit is set. Union and intersection of
  such sets can then easily be computed by ``bitwise or'' and
  ``bitwise and'' operations.  Now, let some maximum cardinality $n$
  be given, and let sets be represented by $m$-element arrays of
  \mpilong integers with $n=64m$ (assuming that
  $\texttt{sizeof(long)}=64$). Give collective calls for computing,
  for all $p$ processes in a communicator \texttt{comm}, first the
  union and second the intersection of $p$ such sets, with the
  resulting set stored at all $p$ processes. Assume now instead that
  the resulting set from a union or intersection operation is to be
  stored in a distributed fashion, with roughly $n/p$ bits per
  process. Give also for this case collective calls for computing the
  union and intersection of $p$ such sets with the resulting set stored
  in a distributed fashion. Each of the $p$ input sets, one for each
  process, is a full set of $n$ bits. Assume first that $p$ divides $m$.
  Give also a solution where $m$ is not necessarily divisible by $p$.
\item
  Many of the \mpi collectives can relatively easily and conveniently
  be expressed and implemented in terms of other \mpi collectives
  without any further ado like copying data and doing local
  computations (see previous exercises).  Why do you think \mpi offers
  as many collectives as it does? Which ones would you think of as
  redundant? Are there collectives that are not easily or at all
  reducible to other collectives?
\item
  Which, if any, of the following three \mpi programs are correct? They
  are all assumed to broadcast a value from the last process in the
  communicator and do a barrier. Explain your answers and explain possible
  outcomes.

  \begin{enumerate}
  \item

\begin{lstlisting}[style=SnippetStyle]
int rank, size;
MPI_Comm_rank(comm,&rank);
MPI_Comm_size(comm,&size);

int i, j;
if (rank%2==0) {
  MPI_Bcast(&i,1,MPI_INT,size-1,comm);
  MPI_Barrier(comm);
} else {
  MPI_Bcast(&j,1,MPI_INT,size-1,comm);
  MPI_Barrier(comm);
}
\end{lstlisting}

\item

\begin{lstlisting}[style=SnippetStyle]
int rank, size;
MPI_Comm_rank(comm,&rank);
MPI_Comm_size(comm,&size);

int i, j;
if (rank%2==0) {
  MPI_Bcast(&i,1,MPI_INT,size-1,comm);
  MPI_Barrier(comm);
} else {
  MPI_Barrier(comm);
  MPI_Bcast(&j,1,MPI_INT,size-1,comm);
}
\end{lstlisting}

\item

\begin{lstlisting}[style=SnippetStyle]
int rank, size;
MPI_Comm_rank(comm,&rank);
MPI_Comm_size(comm,&size);

int i, j;
if (rank%2==0) {
  MPI_Bcast(&i,1,MPI_INT,size-1,comm);
} else {
  MPI_Barrier(comm);
}
\end{lstlisting}
\end{enumerate}
\item
  Define a collective operation for computing all prefix sums of a
  distributed array. More precisely, each process contributes an array
  $a$ of $n$ elements ($n$ may be different for different processes). These
  arrays together make up a large (virtual) array formed by concatenating the
  $p$ arrays in rank order ($p$ is the number of processes). Your operation
  should compute the (inclusive or exclusive) prefix sums on this array.
  Which collective operation(s) might be convenient as building blocks?
  Analyze and state the performance of your implementation as a function of
  $n$ and $p$ and the complexity of the (collective) \mpi operations you
  use.
\item
  Implement matrix--vector multiplication for row-wise distributed matrices
  using \mpisendrecv on a ring of processes to gather the full input vector
  at all processes.
\item
  Implement matrix--vector multiplication for row-wise distributed matrices
  with the same number of full rows per process with \mpiallgather as
  described in \Sec~\ref{sec:la}. Perform strong and weak scalability
  experiments for matrices with $m$ rows and $n$ columns where $p|m$ for
  different, not too small values of $m,n,p$.
\item
  Implement matrix--vector multiplication for row-wise distributed matrices
  with possibly different numbers of full rows per process
  with \mpiallgatherv. See the description in \Sec~\ref{sec:la}.
  Perform strong and weak scalability
  experiments for matrices with $m$ rows and $n$ columns for
  different, not too small values of $m,n,p$.
\item
  Implement matrix--vector multiplication for column-wise distributed matrices
  with the same number of full columns per process with \mpireducescatterblock
  as described in \Sec~\ref{sec:la}.
  Perform strong and weak scalability
  experiments for matrices with $m$ rows and $n$ columns where $p|n$ for
  different, not too small values of $m,n,p$.
\item
  Implement matrix--vector multiplication for column-wise distributed matrices
  with possibly different numbers of full columns per process with
  \mpireducescatter. See the description in \Sec~\ref{sec:la}.
  Perform strong and weak scalability
  experiments for matrices with $m$ rows and $n$ columns for
  different, not too small values of $m,n,p$.
\item
  Consider the three matrix-distributions, row-wise, column-wise, and
  block-wise, discussed and used in \Sec~\ref{sec:la}. Assume that a
  full matrix needs to be collated at some single, given root process
  by putting the submatrices from the $p$ processes together. Try to
  accomplish this with a single \mpigather call.
  Datatypes (\mpitypevector and \mpityperesized) may be
  useful (and possibly needed) in order to avoid process-local
  reorganizations of submatrices at either root or non-root
  processes. It may likewise be that \mpigatherv is needed.
\item
  Complete the implementation of the SUMMA matrix--matrix multiplication
  described in \Sec~\ref{sec:la}.
  Perform strong and weak scalability
  experiments for matrices with $m$ rows and $n$ columns for
  different, not too small values of $m,n,p$. Your implementation will most
  likely require that $p$ is a square, and that both $p|m$ and $p|n$. You may
  assume that $n=m$. The less such restrictions, the better.
\item
  Devise and implement a distributed memory Breadth-First Search
  operation in a bulk synchronous way as outlined in
  \Sec~\ref{sec:bsppattern}.  The input graph, given as a collection
  of edges stored as adjacency lists (arrays) is distributed in some
  form over the $p$ \mpi processes. The processes explore the graph
  iteratively in a level-by-level way. In each iteration, each process
  starts with a set of new vertices that has been reached in the
  previous iteration. The processes explore their vertices and
  compute, locally, a collection of vertices that can now be reached
  for the next iteration and have not been seen so far. At the end of
  the iteration, these local collections of new vertices are put
  together and distributed over the processes for the next
  iterations. The last iteration is reached when no new vertices are
  discovered. Thus, the total number of iterations equals the largest
  distance from the given start vertex to another vertex in the graph.
  The output should be for each vertex the distance of that vertex from
  the given start vertex and, if possible, a reference to the parent of that
  vertex in a BFS-tree rooted at the start vertex.

  Consider a suitable distribution of the input graph. Either the
  vertices are distributed roughly evenly over the processes such that
  a process that has a vertex also has all the edges connected to that
  vertex. Alternatively, the edges might be distributed roughly
  evenly across the processes. Use bitmaps to represent seen vertices,
  and allow each process to maintain a full seen/not seen bitmap for all
  vertices of the input graph. Use also bitmaps to represent the collections
  of new vertices considered by the processes in each iteration. Use 
  \mpiallreduce to compute the union of all local bitmap-represented sets.

  Analyze the complexity of your algorithm in terms of the size of the
  input graph ($n$ vertices and $m$ edges), the depth of a BFS tree rooted
  at the given start vertex,
  and the number of \mpi processes. For \mpiallreduce,
  you can use the estimates from Table~\ref{tab:collectivecomplexity-fully}.
  Benchmark your algorithm with different, randomly generated input graphs
  for different numbers of \mpi processes. What speed-up can you achieve
  compared to your own, best, sequential BFS implementation?
\item
  Complete the distributed memory implementation of the Quicksort
  algorithm discussed in \Sec~\ref{sec:mpisorting}. Assuming that bad pivots
  are chosen throughout, how skewed can the resulting output distribution
  (in terms of numbers of output elements per process) be?
  What would be the worst-case
  running time assuming that the final, sequential sorting is done optimally
  in $O(n\log n)$ time steps.
\item
  Design and implement a parallel, distributed memory bottom-up Merge
  sort algorithm. The input per process is an unsorted array of $n$
  elements and the output per process should be a sorted array of the
  same size such that the output elements at any one process are
  smaller than or equal to the output elements of the next (higher
  ranked) process. The output elements over all processes must of
  course be a permutation of the input elements over all processes.
  One approach is to recursively split the input communicator down to
  communicators with just a single process (as done for Quicksort in
  the previous exercise). Each process then (merge) sorts its input
  elements (using a best possible Merge sort implementation) after
  which the processes merge their elements with those of other
  processes going up the hierarchy of communicators. Use the merging
  by co-ranking algorithm and implementation of
  Exercise~\ref{exe:corankmpi} for the merge steps.  You may at first
  assume that the number of processes $p$ is a power of two. What is
  the parallel running time of your algorithm as a function of $p$ and
  $n$?  Conduct strong and weak scaling experiments with your
  implementation and compute the speed-up and parallel efficiency
  relative to your best, sequential Merge sort (already used for the
  initial sorting per process; thus, absolute and relative speed-up
  will coincide). You may be able to improve your implementation by
  not explicitly creating the communicators as you recursively
  decrease $p$ and instead stay with only the given input communicator
  with the $p$ processes. This may in addition make it possible to
  stay with only one \mpiwin (input and output) window. You may be
  able to generalize this implementation to work well with any number
  of processes. In that case, what is the parallel running time as a
  function of $n$ and $p$? Be as exact as possible. How do the
  concrete, parallel running time of your improved implementation
  compare against your first try?
\item
  Complete an implementation of a distributed counting sort as
  outlined in \Sec~\ref{sec:mpisorting}. Each process has input of $n$
  integer elements in a range $[0,r]$ stored in an array. The output
  should be a sorted array segment of $n$ elements of the total input
  array of $pn$ elements. The elements of some rank $i, i>0$ must all
  be equal to or larger than the elements of rank $i-1$. Your
  algorithm has to count the number of elements of each key $k\in
  [0,r]$ over the processes and use this to redistribute the elements. A
  final, process-local sort or merge may/will be necessary. Which
  collective operations are you using? What is the estimated running
  time of your implementation as a function of $n$ nd $p$? Benchmark
  and compare against a sequential counting sort for $n=1\,000\,000$,
  $n=10\,000\,000$ and $n=100\,000\,000$ elements per
  process. Depending on $p$, it may not be able to sort sequentially
  (in a reasonable amount of time).
\item
  The following code is a sequential implementation of the
  Floyd-Warshall algorithm discussed in Exercise~\ref{exe:fw} for
  Chapter~\ref{chp:sharedmemory} for solving the all-pairs shortest
  path problem on a weighted graph given by an initial weight matrix
  $W[n,n]$.
  \begin{lstlisting}[style=SnippetStyle]
void fw_apsp(int *w, int n) {
  int (*W)[n] = (int(*)[n])w;
  
  int i, j, k;
  for (k=0; k<n; k++) {
    for (i=0; i<n; i++) {
      for (j=0; j<n; j++) {
	if (W[i][j]>W[i][k]+W[k][j]) {
          W[i][j] = W[i][k]+W[k][j];
        }
      }
    }
  }
}
  \end{lstlisting}
  The exercise is to give a distributed memory implementation with \mpi
  following the idea of the SUMMA matrix--matrix multiplication
  algorithm presented in \Sec~\ref{sec:la}. More concretely, assume
  that a square number of \mpi processes organized into a
  Cartesian $\sqrt{p}\times\sqrt{p}$ communicator is available. Create
  communicators of the processes belonging to the same row and
  processes belonging to the same column of processes in the
  Cartesian communicator.  Assume further that $p$ divides $n$,
  $p|n$, and that the weight matrix is distributed cyclically over the
  processes as $(n/\sqrt{p})\times (n/\sqrt{p})$ submatrices.  The
  algorithm perform $\sqrt{p}$ iterations. In iteration $k$, the $k$th
  process in each row and in each column broadcasts its weight matrix
  to the processes in the row and the column, respectively. The
  processes can then locally perform $n/\sqrt{p}$ iterations of the
  Floyd-Warshall update operation (innermost two loops).

  Write out this algorithm in detail and state the parallel running
  time under reasonable assumptions on the broadcast time complexity.
  Implement your algorithm with \mpi, and perform benchmark experiments
  for a number of larger $n$ values and different numbers of \mpi processes.
  How much speed-up can you achieve compared to your best, sequential
  implementation of the Floyd-Warshall algorithm?
\end{enumerate}

\appendix


\chapter{Proofs and Supplementary Material}
\section{A Frequently Occurring Sum}

One of the most frequently occurring (finite) sums in \parco
is the \impi{geometric series} $1+q+q^2+q^3+\cdots +q^n =
\sum_{i=0}^{n}q^i$. The geometric series is the sum of the elements of
the geometric progression $1,q,q^2, q^3,\ldots,q^n$, where each element
of the sequence except the first follow from the previous by
multiplying with the common ratio $q$.  For $q=1$, obviously
$\sum_{i=0}^{n}q^i = (n+1)$ (since also $0^0=1$).  For any other $q,q\neq 1$,
it is well-known (and easy to see, even without using
induction) that
\begin{eqnarray}
  \sum_{i=0}^{n}q^i & = & \frac{q^{n+1}-1}{q-1} \\ \nonumber
  & = & \frac{1-q^{n+1}}{1-q} \quad .
\end{eqnarray}
When $|q|<1$, the geometric series is convergent, and we can write
\begin{eqnarray}
  \sum_{i=0}^{\infty}q^i & = & \frac{1}{1-q}  \quad .
\end{eqnarray}

For instance, with $q=2$, $\sum_{i=0}^{n}q^i = 2^{n+1}-1$, and with
$q=\frac{1}{2}$, $\sum_{i=0}^{n}q^i = 2-\frac{1}{2^n}$ (and
$\sum_{i=1}^{n}q^i = 1-\frac{1}{2^n}$).  For other elementary sums and
series occurring in standard analysis of algorithms,
see~\cite{CormenLeisersonRivestStein22,GrahamKnuthPataschnik94} and
other textbooks.

\section{Logarithms Reminder}

The logarithm $\log_b x$ with base $b, b>0, b\neq 1$ of some $x,x>0$
is the inverse of exponentiation with base $b$, that is $x = \log_b
b^x$ and $x=b^{\log_b x}$. When clear from context (or not relevant,
see the following), the base is left out and the logarithm function is
written $\log x$.  By convention, $\log^c x$ is notation for $(\log
x)^c$ and is not the iterated application of the logarithm function,
which is denoted $\log^{(n)} x$ and for integer $n\geq 0$ defined by
$\log^{(n)} x = \log\log^{(n-1)} x$ for $n>0$ and $\log^{(0)}
x=x$. For constant $c,c\geq 1$, functions in $O(\log^c x)$ are called
``poly-logarithmic'' by being polynomials in $(\log x)$.

It follows that $\log_b 1 = 0, \log_b b = 1$. Let $x=b^a$ and $y=b^c$.
Then from the laws of exponentiation, $\log_b xy = \log_b(b^a b^c) =
\log_b b^{a+c}=a+c=\log_b x+\log_b y$. Similarly, it follows that
$\log_b\frac{x}{y} = \log_b x - \log_b y$.
Also $\log_b x^d = \log_b (b^a)^d = \log_bb^{ad} = ad = da = d\log_b x$.
It now follows that for any other other base $e, e>0, e\neq 1$,
$\log_e x = \log_e b^{\log_b x} =  \log_b x \log_e b$, so any two logarithms with
different constant bases differ only by a constant factor.

Common logarithm bases in \parco are $b=2$, $b=2.718281828459\ldots$, $b=10$,
$b=(k+1)$ for some positive integer $k$. For all of these, $\log_b
x$ is in $O(\log n)$. A sometimes useful observation for graphs with
$n$ vertices and $m$ arcs is that here $O(\log m)=O(\log n)$ since
$m\leq n^2$.

\section{The Master Theorem}
\label{sec:masterproof}

The ``Master Theorem''\index{Master Theorem}, Theorem~\ref{thm:master},
gives closed form solutions for a range of divide-and-conquer
recurrences of the following form, for constants $a\geq 1, b>1, d\geq
0, e\geq 0$ (the $c$ is omitted to avoid any confusion with constants
hidden behind the $O$) that very often occur in the analysis of (parallel)
algorithms:

\begin{eqnarray*}
  T(n) & = & aT(n/b)+O(n^d \log^e n) \\
  T(1) & = & O(1)
\end{eqnarray*}

The theorem claims a closed-form solution in either of three forms:
\begin{enumerate}
  \item
  $T(n)=O(n^d\log^e n)$ if $a/b^d < 1$ (equivalently $b^d/a>1$),
\item
  $T(n) = O(n^d\log^{e+1} n)$ if $a/b^d = 1$ (equivalently $b^d/a=1$), and
\item
  $T(n) = O(n^{\log_b a})$ if $a/b^d > 1$ (equivalently $b^d/a<1$).
\end{enumerate}

Let $C$ be a constant at least as large as the leading constant in either
of $O(1)$ or $O(n^d \log^e n)$. Then, the recurrence takes the form

\begin{eqnarray*}
  T(n) & \leq & aT(n/b)+C(n^d \log^e n) \quad .
\end{eqnarray*}

First, assume $n=b^k$. With this, $\log^e n = (\log b^k)^e=k^e$ and
the recurrence takes the form
\begin{eqnarray*}
  T(b^k) & \leq & aT(b^k/b)+C(b^{kd} k^e) \quad .
\end{eqnarray*}

Expanding the recurrence for the first few values of $k$, $k=1,2,3$ yields:
\begin{eqnarray*}
  T(b) & \leq & Ca+C(b^d 1^e) \\
  T(b^2) & \leq & Ca^2+Ca(b^d 1^e) + C(b^{2d} 2^e)\\
  T(b^3) & \leq & Ca^3+Ca^2(b^d 1^e) + Ca(b^{2d} 2^e) + C(b^{3d} 3^e) \quad .
\end{eqnarray*}

From this, we conjecture that
\begin{eqnarray*}
  T(b^k) & \leq & C a^k(1+\sum_{i=1}^k \left(\frac{b^d}{a}\right)^i i^e) \quad .
\end{eqnarray*}

The claim is easily verified by induction. The base case $T(1) \leq C$
holds, since the sum is void (no summands, per definition $0$), by
the choice of the constant $C$. Assuming the claim for $k-1$ yields:

\begin{eqnarray*}
  T(b^k) & \leq & aT(b^k/b)+C(b^{kd} k^e) \\
  & = & aT(b^{k-1})+C(b^{kd} k^e) \\
  & = & a(C a^{k-1}(1+\sum_{i=1}^{k-1} \left(\frac{b^d}{a}\right)^i i^e)) +
  C(b^{kd} k^e) \\
  & = & C a^{k}(1+\sum_{i=1}^{k-1} \left(\frac{b^d}{a}\right)^i i^e)) +
  Ca^k(\left(\frac{b^{d}}{a}\right)^k k^e) \\
  & = & C a^k(1+\sum_{i=1}^k \left(\frac{b^d}{a}\right)^i i^e)
\end{eqnarray*}
since $C(b^{kd} k^e)=Ca^k(\left(\frac{b^{d}}{a}\right)^k k^e)$ by multiplying
and dividing again by $a^k$ leading to the last, $k$th term in sum.

We now distinguish three cases for bounding the sum
$\sum_{i=1}^k \left(\frac{b^d}{a}\right)^i i^e$
from above.

\begin{enumerate}
\item $b^d/a>1$:
  \begin{eqnarray*}
    \sum_{i=1}^k \left(\frac{b^d}{a}\right)^i i^e & \leq &
    k^e \sum_{i=1}^k \left(\frac{b^d}{a}\right)^i \\
    & = & O(k^e \left(\frac{b^d}{a}\right)^{k+1})
  \end{eqnarray*}
  since the sum is a geometric series and $i^e\leq k^e$ for $i\leq k$ which can
  be factored out. Therefore,
  \begin{eqnarray*}
    T(b^k) & = & O(a^k \left(\frac{b^d}{a}\right)^{k+1} k^e) \\
    & = & O(b^{kd} \left(\frac{b^d}{a}\right) k^e) \\
    & = & O(n^d \log^e n)  \quad .
    \end{eqnarray*}
\item $b^d/a=1$:
  \begin{eqnarray*}
    \sum_{i=1}^k \left(\frac{b^d}{a}\right)^i i^e & = & \sum_{i=1}^k i^e\\
    & \leq & k^{e+1}
  \end{eqnarray*}
  Therefore, by using the bound on $T(b^k)$
  \begin{eqnarray*}
    T(b^k) & = & O(a^k k^{e+1}) \\
    & = & O(b^{kd}k^{e+1}) \\
    & = & O(n^d \log^{e+1} n) \quad .
  \end{eqnarray*}
\item $b^d/a<1$: In this case, we use the fact that an exponential
  function $f^i$ for $f>1$ grows faster than the (any) polynomial
  $i^e$. We choose a constant $f, f>1$ with
  $\left(\frac{b^d}{a}\right)f<1$. Then, for some constant $k'$, it
  holds that $i^e<f^i$ for $i\geq k'$.
  \begin{eqnarray*}
    \sum_{i=1}^k \left(\frac{b^d}{a}\right)^i i^e & \leq &
    \sum_{i=1}^{k'-1} \left(\frac{b^d}{a}\right)^i i^e +
    \sum_{i=k'}^k \left(\frac{b^d}{a}\right)^i i^e \\
    & \leq & \sum_{i=1}^{k'-1} \left(\frac{b^d}{a}\right)^i i^e +
    \sum_{i=k'}^{\infty} \left(\frac{b^d}{a}\right)^i f^i \\
    & = & \sum_{i=1}^{k'-1} \left(\frac{b^d}{a}\right)^i i^e +
    \sum_{i=k'}^{\infty} (\left(\frac{b^d}{a}\right) f)^i \quad .
  \end{eqnarray*}
  The first sum is finite. The second sum, which is a geometric
  series with a quotient smaller than one, is convergent (to a constant).
  Therefore
  \begin{eqnarray*}
    T(b^k) & = & O(a^k) \\
    & = & O(a^{\log_b n}) \\
    & = & O(n^{\log_b a}) \quad .
  \end{eqnarray*}
\end{enumerate}

When $n$ is not a power of $b$, it holds that for some $k$,
$b^{k-1}<n<b^k=n'$. Since $T(n)$ is monotone, we have for the three cases
\begin{enumerate}
  \item
\begin{eqnarray*}
  T(n) \leq T(n') & = & O(n'^d \log^e n')\\
  & = & O((n'/n)^d n^d \log^e ((n'/n)n)) \\
  & = & O((n'/n)^d n^d (\log^e (n'/n) + \log^e n) \\
  & = & O(n^d \log^e n)
\end{eqnarray*}
since $n'/n<b$ can be upper bounded by the constant $b$.
\item
\begin{eqnarray*}
  T(n) \leq T(n') & = & O(n'^d \log^{e+1} n')\\
  & = & O(n^d \log^{e+1} n)
\end{eqnarray*}
with the same calculation and argument as in Case 1.
\item
\begin{eqnarray*}
  T(n) \leq T(n') & = & O(n'^{\log_b a}) \\
  & = & O((n'/n)^{\log_b a} n^{\log_b a}) \\
  & = & O(b^{\log_b a} n^{\log_b a}) \\
  & = & O(n^{\log_b a})
\end{eqnarray*}
since $n'/n<b$ and also $b^{\log_b a}$ is constant.
\end{enumerate}

The theorem therefore holds for any $n, n\geq 1$. The bounding
arguments do not give any useful estimates of the constants incurred
by the recurrence; but it can be shown that the bounds are
asymptotically tight for recurrences of the form
\begin{eqnarray*}
  T(n) & = & aT(n/b)+\Theta(n^d \log^e n) \\
  T(1) & = & O(1)
\end{eqnarray*}

The Master Theorem can be improved to give closed-form solutions also
for negative values of $e, e<0$.

\bibliographystyle{plain}
\bibliography{traff,parallel}

\printindex

\end{document}